%% file: main.tex
\documentclass[runningheads,a4paper]{llncs}

\usepackage[T1]{fontenc}
\usepackage{graphicx}
\usepackage{amsmath, amssymb, mathtools}
\usepackage{supertabular}

\usepackage{ottlayout}

\usepackage{mathbbol}

\usepackage{bookmark}

\usepackage{xurl}

\usepackage{todonotes}

\usepackage{bussproofs}

\usepackage{fontawesome}

\usepackage{here}

\usepackage{comment}

\usepackage{xspace}

\usepackage{tabularx}

\usepackage{stmaryrd}

\excludecomment{longver}
\includecomment{shortver}

\hypersetup{unicode,bookmarksnumbered=true,hidelinks,final}

\begin{document}

\include{math}
% include Ott-generated tex file
\include{lamfb}

\title{Contextual Modal Type Theory \\ with Polymorphic Contexts}

% Change it to anonymoustrue if you want to make it anonymous
\newif\ifanonymous\anonymousfalse

\ifanonymous
\author{Anonymous}
\else
\author{Yuito Murase\inst{1} \and
    Yuichi Nishiwaki\inst{2} \and
    Atsushi Igarashi\inst{1}}
\authorrunning{Y. Murase et al.}
\institute{Kyoto University, Kyoto, Japan\\
    \email{\{murase@fos.kuis.kyoto-u.ac.jp, igarashi@kuis.kyoto-u.ac.jp\}} \and
    Independent Researcher
    \email{yuichi.nishiwaki@icloud.com}}
\fi

\maketitle

\begin{abstract}
  
  Modal types---types that are derived from proof systems of modal logic---have been studied as theoretical foundations of metaprogramming, where program code is manipulated as first-class values.
 In modal type systems, modality corresponds to
 a type constructor for code types and controls free variables and their types in code values.
  Nanevski et al.\ have proposed \emph{contextual modal type theory}, which has modal types with fine-grained information on free variables: modal types are explicitly indexed by \emph{contexts}---the types of all free variables in code values.

This paper presents \lamfb, a novel extension of contextual modal type theory with \emph{parametric polymorphism over contexts}.  Such an extension has been studied in the literature but, unlike earlier proposals, \lamfb is more general in that multiple parts of a single context can be abstracted.  We formalize \lamfb with its type system and operational semantics given by $\beta$-reduction and prove its basic properties
including subject reduction, strong normalization, and confluence.
Moreover, to demonstrate the expressive power of polymorphic contexts, we show a type-preserving embedding from a two-level fragment of Davies' \lamcirc, which is based on linear-time temporal logic, to \lamfb.

    \keywords{Contextual modal types, Fitch-style modal lambda-calculi, Metaprogramming, Polymorphic contexts}
\end{abstract}

% \AI{AI should read the proofs carefully.  Change line breaks in the source file.  Write a full version.  Other thoughts: Extend the embedding to \(n\)-levels and/or refined environment classifiers, show preservation of semantics (by logical relations, if a reduction-based argument is hard).}

\input{1-introduction}
\input{2-motivation}
\input{3-fcmtt}
\input{4-polyctx}
\input{5-basicprops}
\input{6-embedlttt}
\input{7-relatedwork}
\input{8-conclusion}

\bibliographystyle{splncs04}
\bibliography{mybib}

\clearpage
\appendix
\input{fulldef}

\input{artifact}

\end{document}

%% file: math.tex
\newcommand{\lamcirc}{$\lambda_{\bigcirc}$\xspace}
\newcommand{\lamcirctwo}{$\lambda_{\bigcirc \le 1}$\xspace}
\newcommand{\lambra}{$\lambda_{[]}$\xspace}
\newcommand{\lamfb}{$\lambda_{\forall[]}$\xspace}
\newcommand{\diff}{\stackrel{\mathrm{def.}}{\iff}}
\newcommand{\lampolyopen}{$\lambda^{poly}_{open}$\xspace}
\newcommand{\nubox}{$\nu^{\Box}$\xspace}
\newcommand{\lamctx}{$\lambda^{ctx}$\xspace}
\newcommand{\lamctxi}{$\lambda^{ctx}_{I}$\xspace}
\newcommand{\lambrasub}{$\lambda^{[]}_{<}$\xspace}
\newcommand{\lamalpha}{$\lambda^{\alpha}$\xspace}
\renewcommand{\qed}{\hfill\blacksquare}

%% file: lamfb.tex
% generated by Ott 0.32 from: lamfb.ott
\newcommand{\lfbdrule}[4][]{{\displaystyle\frac{\begin{array}{l}#2\end{array}}{#3}\quad\lfbdrulename{#4}}}
\newcommand{\lfbusedrule}[1]{\[#1\]}
\newcommand{\lfbpremise}[1]{ #1 \\}
\newenvironment{lfbdefnblock}[3][]{ \framebox{\mbox{#2}} \quad #3 \\[0pt]}{}
\newenvironment{lfbfundefnblock}[3][]{ \framebox{\mbox{#2}} \quad #3 \\[0pt]\begin{displaymath}\begin{array}{l}}{\end{array}\end{displaymath}}
\newcommand{\lfbfunclause}[2]{ #1 \equiv #2 \\}
\newcommand{\lfbnt}[1]{\mathit{#1}}
\newcommand{\lfbmv}[1]{\mathit{#1}}
\newcommand{\lfbkw}[1]{\mathbf{#1}}
\newcommand{\lfbsym}[1]{#1}
\newcommand{\lfbcom}[1]{\text{#1}}
\newcommand{\lfbdrulename}[1]{\textsc{#1}}
\newcommand{\lfbcomplu}[5]{\overline{#1}^{\,#2\in #3 #4 #5}}
\newcommand{\lfbcompu}[3]{\overline{#1}^{\,#2<#3}}
\newcommand{\lfbcomp}[2]{\overline{#1}^{\,#2}}
\newcommand{\lfbgrammartabular}[1]{\begin{supertabular}{llcllllll}#1\end{supertabular}}
\newcommand{\lfbmetavartabular}[1]{\begin{supertabular}{ll}#1\end{supertabular}}
\newcommand{\lfbrulehead}[3]{$#1$ & & $#2$ & & & \multicolumn{2}{l}{#3}}
\newcommand{\lfbprodline}[6]{& & $#1$ & $#2$ & $#3 #4$ & $#5$ & $#6$}
\newcommand{\lfbfirstprodline}[6]{\lfbprodline{#1}{#2}{#3}{#4}{#5}{#6}}
\newcommand{\lfblongprodline}[2]{& & $#1$ & \multicolumn{4}{l}{$#2$}}
\newcommand{\lfbfirstlongprodline}[2]{\lfblongprodline{#1}{#2}}
\newcommand{\lfbbindspecprodline}[6]{\lfbprodline{#1}{#2}{#3}{#4}{#5}{#6}}
\newcommand{\lfbprodnewline}{\\}
\newcommand{\lfbinterrule}{\\[5.0mm]}
\newcommand{\lfbafterlastrule}{\\}
\newcommand{\lfbmetavars}{
\lfbmetavartabular{
 $ \lfbmv{x} ,\, \lfbmv{y} $ & \lfbcom{variables} \\
 $ \gamma ,\, \delta $ & \lfbcom{context variables} \\
 $ \mathbb{x} ,\, \mathbb{y} $ & \lfbcom{variable-set variables} \\
 $ \lfbmv{i} $ & \lfbcom{lttt level invex variables} \\
}}

\newcommand{\lfbk}{
\lfbrulehead{\lfbnt{k}}{::=}{\lfbcom{transitions}}\lfbprodnewline
\lfbfirstprodline{|}{\lfbsym{0}}{}{}{}{\lfbcom{no transition}}\lfbprodnewline
\lfbprodline{|}{\lfbnt{k}  \lfbsym{+}  \lfbsym{1}}{}{}{}{\lfbcom{move one world}}\lfbprodnewline
\lfbprodline{|}{\lfbnt{k_{{\mathrm{1}}}}  \lfbsym{-}  \lfbnt{k_{{\mathrm{2}}}}} {\textsf{M}}{}{}{\lfbcom{pred}}\lfbprodnewline
\lfbprodline{|}{\lfbnt{k_{{\mathrm{1}}}}  \lfbsym{+}  \lfbnt{k_{{\mathrm{2}}}}} {\textsf{M}}{}{}{\lfbcom{sum}}\lfbprodnewline
\lfbprodline{|}{\lfbsym{1}} {\textsf{M}}{}{}{\lfbcom{one}}\lfbprodnewline
\lfbprodline{|}{\lfbsym{2}} {\textsf{M}}{}{}{\lfbcom{two}}\lfbprodnewline
\lfbprodline{|}{ \mathsf{count} ( \lfbnt{k} , \sigma ) } {\textsf{M}}{}{}{\lfbcom{lock replacement}}\lfbprodnewline
\lfbprodline{|}{\lfbsym{(}  \lfbnt{k}  \lfbsym{)}} {\textsf{S}}{}{}{}}

\newcommand{\lfbbase}{
\lfbrulehead{\iota}{::=}{\lfbcom{base types}}\lfbprodnewline
\lfbfirstprodline{|}{\textrm{int}}{}{}{}{\lfbcom{int type}}\lfbprodnewline
\lfbprodline{|}{\textrm{str}}{}{}{}{\lfbcom{str type}}}

\newcommand{\lfbT}{
\lfbrulehead{\lfbnt{T}  ,\ \lfbnt{S}}{::=}{\lfbcom{types}}\lfbprodnewline
\lfbfirstprodline{|}{\iota}{}{}{}{\lfbcom{base type}}\lfbprodnewline
\lfbprodline{|}{\lfbnt{S}  \rightarrow  \lfbnt{T}}{}{}{}{\lfbcom{function types}}\lfbprodnewline
\lfbprodline{|}{\lfbsym{[}  \lfbnt{C}  \vdash  \lfbnt{T}  \lfbsym{]}}{}{}{}{\lfbcom{contextual modal types}}\lfbprodnewline
\lfbprodline{|}{ \forall \gamma . \lfbnt{T} }{}{}{}{\lfbcom{polymorphic context types}}\lfbprodnewline
\lfbprodline{|}{\lfbnt{T}  \lfbsym{[}  \Sigma  \lfbsym{]}} {\textsf{M}}{}{}{}\lfbprodnewline
\lfbprodline{|}{\lfbsym{(}  \lfbnt{T}  \lfbsym{)}} {\textsf{S}}{}{}{}\lfbprodnewline
\lfbprodline{|}{ \lfbnt{T} } {\textsf{S}}{}{}{}\lfbprodnewline
\lfbprodline{|}{ \llbracket  T^{1}  \rrbracket } {\textsf{M}}{}{}{\lfbcom{embedding of lamcirctwo object-level type}}\lfbprodnewline
\lfbprodline{|}{ \llbracket  T^{0}  \rrbracket_{ \lfbnt{C} } } {\textsf{M}}{}{}{\lfbcom{embedding of lamcirctwo meta-level type}}}

\newcommand{\lfbC}{
\lfbrulehead{\lfbnt{C}  ,\ \lfbnt{D}}{::=}{\lfbcom{contexts}}\lfbprodnewline
\lfbfirstprodline{|}{ \bullet }{}{}{}{\lfbcom{empty context}}\lfbprodnewline
\lfbprodline{|}{\lfbnt{C}  \lfbsym{,}  \lfbnt{T}}{}{}{}{\lfbcom{assumption}}\lfbprodnewline
\lfbprodline{|}{\lfbnt{C}  \lfbsym{,}  \gamma}{}{}{}{\lfbcom{polymorphic context}}\lfbprodnewline
\lfbprodline{|}{ \lfbnt{T} } {\textsf{M}}{}{}{}\lfbprodnewline
\lfbprodline{|}{ \gamma } {\textsf{M}}{}{}{}\lfbprodnewline
\lfbprodline{|}{\lfbnt{C}  \lfbsym{[}  \Sigma  \lfbsym{]}} {\textsf{M}}{}{}{}\lfbprodnewline
\lfbprodline{|}{ \lfbnt{C} \lfbsym{,} \lfbnt{D} } {\textsf{M}}{}{}{\lfbcom{append contexts}}\lfbprodnewline
\lfbprodline{|}{ \mathsf{rg} ( \Gamma ) } {\textsf{M}}{}{}{\lfbcom{region of an environment}}\lfbprodnewline
\lfbprodline{|}{\lfbsym{(}  \lfbnt{C}  \lfbsym{)}} {\textsf{S}}{}{}{}\lfbprodnewline
\lfbprodline{|}{\mathsf{diff} \, \lfbsym{(}  \lfbmv{x}  \lfbsym{,}  \tilde{\Gamma}  \lfbsym{)}} {\textsf{M}}{}{}{}}

\newcommand{\lfbM}{
\lfbrulehead{\lfbnt{M}  ,\ \lfbnt{N}}{::=}{\lfbcom{terms}}\lfbprodnewline
\lfbfirstprodline{|}{\lfbmv{x}}{}{}{}{\lfbcom{variables}}\lfbprodnewline
\lfbprodline{|}{ \lambda \lfbmv{x} ^{ \lfbnt{T} }. \lfbnt{M} }{}{\textsf{bind}\; \lfbmv{x}\; \textsf{in}\; \lfbnt{M}}{}{\lfbcom{abstractions}}\lfbprodnewline
\lfbprodline{|}{\lfbnt{M} \, \lfbnt{N}}{}{}{}{\lfbcom{applications}}\lfbprodnewline
\lfbprodline{|}{ \lfbkw{quo} \langle \Gamma \rangle \lfbnt{M} }{}{}{}{\lfbcom{quotations}}\lfbprodnewline
\lfbprodline{|}{ \lfbkw{unq} _{ \lfbnt{k} } \lfbnt{M} [  \theta  ] }{}{}{}{\lfbcom{unquotations}}\lfbprodnewline
\lfbprodline{|}{ \Lambda \gamma . \lfbnt{M} }{}{}{}{\lfbcom{context abstcations}}\lfbprodnewline
\lfbprodline{|}{ \lfbnt{M} @ \lfbnt{C} }{}{}{}{\lfbcom{context applications}}\lfbprodnewline
\lfbprodline{|}{\lfbnt{M}  \lfbsym{[}  \sigma  \lfbsym{]}} {\textsf{M}}{}{}{}\lfbprodnewline
\lfbprodline{|}{ \lfbnt{M} \lfbsym{[} \Sigma \lfbsym{;} \bar{\sigma} \lfbsym{]} } {\textsf{M}}{}{}{}\lfbprodnewline
\lfbprodline{|}{ \lfbnt{M} \lfbsym{[} \Sigma \lfbsym{;} \bar{\sigma} \lfbsym{]} _{ G } } {\textsf{M}}{}{}{}\lfbprodnewline
\lfbprodline{|}{\lfbsym{(}  \lfbnt{M}  \lfbsym{)}} {\textsf{S}}{}{}{}\lfbprodnewline
\lfbprodline{|}{ \lfbnt{M} } {\textsf{S}}{}{}{}\lfbprodnewline
\lfbprodline{|}{ \llbracket  M^{1}  \rrbracket_{ \tilde{\Gamma} } } {\textsf{M}}{}{}{\lfbcom{embedding of lamcirctwo object-level term}}\lfbprodnewline
\lfbprodline{|}{ \llbracket  M^{0}  \rrbracket_{ \tilde{\Gamma} } } {\textsf{M}}{}{}{\lfbcom{embedding of lamcirctwo meta-level term}}}

\newcommand{\lfbMM}{
\lfbrulehead{\theta  ,\ \eta}{::=}{\lfbcom{explicit substitutions}}\lfbprodnewline
\lfbfirstprodline{|}{ \bullet }{}{}{}{\lfbcom{empty explicit substitution}}\lfbprodnewline
\lfbprodline{|}{\theta  \lfbsym{,}  \lfbnt{M}}{}{}{}{\lfbcom{assumption}}\lfbprodnewline
\lfbprodline{|}{\theta  \lfbsym{,}  \mathbb{x}}{}{}{}{\lfbcom{variable-set variables}}\lfbprodnewline
\lfbprodline{|}{ \lfbnt{M} } {\textsf{M}}{}{}{}\lfbprodnewline
\lfbprodline{|}{ \overrightarrow{x} } {\textsf{M}}{}{}{}\lfbprodnewline
\lfbprodline{|}{ \theta \lfbsym{,} \overrightarrow{x} } {\textsf{M}}{}{}{}\lfbprodnewline
\lfbprodline{|}{ \theta_{{\mathrm{1}}} \lfbsym{,} \theta_{{\mathrm{2}}} } {\textsf{M}}{}{}{}\lfbprodnewline
\lfbprodline{|}{\theta  \lfbsym{[}  \sigma  \lfbsym{]}} {\textsf{M}}{}{}{}\lfbprodnewline
\lfbprodline{|}{ \theta \lfbsym{[} \Sigma \lfbsym{;} \bar{\sigma} \lfbsym{]} } {\textsf{M}}{}{}{}\lfbprodnewline
\lfbprodline{|}{ \theta \lfbsym{[} \Sigma \lfbsym{;} \bar{\sigma} \lfbsym{]} _{ G } } {\textsf{M}}{}{}{}\lfbprodnewline
\lfbprodline{|}{\lfbsym{(}  \theta  \lfbsym{)}} {\textsf{S}}{}{}{}\lfbprodnewline
\lfbprodline{|}{ \mathsf{dom} \lfbsym{(} \Gamma \lfbsym{)} } {\textsf{M}}{}{}{}\lfbprodnewline
\lfbprodline{|}{ \lfbnt{M} } {\textsf{M}}{}{}{\lfbcom{single explicit substitution}}\lfbprodnewline
\lfbprodline{|}{ \overrightarrow{x} } {\textsf{M}}{}{}{\lfbcom{variable series as explicit substitution}}\lfbprodnewline
\lfbprodline{|}{ \mathbb{x} } {\textsf{M}}{}{}{\lfbcom{series variable as explicit substitution}}}

\newcommand{\lfbxs}{
\lfbrulehead{\overrightarrow{x}  ,\ \overrightarrow{y}}{::=}{\lfbcom{variable series}}\lfbprodnewline
\lfbfirstprodline{|}{ \bullet }{}{}{}{\lfbcom{empty variable series}}\lfbprodnewline
\lfbprodline{|}{\overrightarrow{x}  \lfbsym{,}  \lfbmv{y}}{}{}{}{}\lfbprodnewline
\lfbprodline{|}{\overrightarrow{x}  \lfbsym{,}  \mathbb{y}}{}{}{}{}\lfbprodnewline
\lfbprodline{|}{ \lfbmv{x} } {\textsf{M}}{}{}{}\lfbprodnewline
\lfbprodline{|}{ \mathbb{x} } {\textsf{M}}{}{}{}\lfbprodnewline
\lfbprodline{|}{ \mathsf{gensyms} _{ G } \lfbsym{(} \lfbnt{C} \lfbsym{;} V \lfbsym{)} }{}{}{}{}\lfbprodnewline
\lfbprodline{|}{ (  \overrightarrow{x}  ) } {\textsf{S}}{}{}{}}

\newcommand{\lfbGElm}{
\lfbrulehead{\lfbnt{GElm}}{::=}{}\lfbprodnewline
\lfbfirstprodline{|}{\lfbmv{x}  \colon  \lfbnt{T}}{}{}{}{}\lfbprodnewline
\lfbprodline{|}{\mathbb{x}  \colon  \gamma}{}{}{}{}\lfbprodnewline
\lfbprodline{|}{\overrightarrow{x}  \colon  \lfbnt{C}}{}{}{}{}\lfbprodnewline
\lfbprodline{|}{\text{\faLock}}{}{}{}{}\lfbprodnewline
\lfbprodline{|}{ \mathbb{x} \colon \lfbnt{C} }{}{}{}{}}

\newcommand{\lfbG}{
\lfbrulehead{\Gamma  ,\ \Delta}{::=}{\lfbcom{environment}}\lfbprodnewline
\lfbfirstprodline{|}{ \bullet }{}{}{}{\lfbcom{empty environment}}\lfbprodnewline
\lfbprodline{|}{\Gamma  \lfbsym{,}  \lfbnt{GElm}}{}{}{}{\lfbcom{assumption}}\lfbprodnewline
\lfbprodline{|}{ \lfbnt{GElm} }{}{}{}{}\lfbprodnewline
\lfbprodline{|}{ \Gamma \lfbsym{,} \Delta }{}{}{}{\lfbcom{append}}\lfbprodnewline
\lfbprodline{|}{\mathsf{head} \, \lfbsym{(}  \Gamma  \lfbsym{)}} {\textsf{M}}{}{}{\lfbcom{head environment}}\lfbprodnewline
\lfbprodline{|}{\Gamma  \uparrow  \lfbnt{k}} {\textsf{M}}{}{}{\lfbcom{pop worlds k times}}\lfbprodnewline
\lfbprodline{|}{ \Gamma   \lfbsym{[}   \Sigma   \lfbsym{;}   \bar{\sigma}   \lfbsym{]} } {\textsf{M}}{}{}{\lfbcom{context subsitution}}\lfbprodnewline
\lfbprodline{|}{\lfbsym{(}  \Gamma  \lfbsym{)}} {\textsf{S}}{}{}{}\lfbprodnewline
\lfbprodline{|}{ \Gamma } {\textsf{S}}{}{}{}\lfbprodnewline
\lfbprodline{|}{ \lvert  \tilde{\Gamma}  \rvert_0 } {\textsf{M}}{}{}{\lfbcom{level-0 named context from intermediate named context}}\lfbprodnewline
\lfbprodline{|}{ \lvert  \tilde{\Gamma}  \rvert_1 } {\textsf{M}}{}{}{\lfbcom{level-1 named context from intermediate named context}}\lfbprodnewline
\lfbprodline{|}{ \lfbnt{GElm} }{}{}{}{\lfbcom{single element}}}

\newcommand{\lfbSubElm}{
\lfbrulehead{\lfbnt{SubElm}}{::=}{}\lfbprodnewline
\lfbfirstprodline{|}{\lfbmv{x}  \coloneqq  \lfbnt{M}}{}{}{}{}\lfbprodnewline
\lfbprodline{|}{\mathbb{x}  \coloneqq  \mathbb{y}}{}{}{}{}\lfbprodnewline
\lfbprodline{|}{\Gamma  \coloneqq  \theta}{}{}{}{}\lfbprodnewline
\lfbprodline{|}{\overrightarrow{x}  \coloneqq  \theta}{}{}{}{}\lfbprodnewline
\lfbprodline{|}{ \text{\faLock} _{ \lfbnt{k} } }{}{}{}{}}

\newcommand{\lfbSub}{
\lfbrulehead{\sigma}{::=}{\lfbcom{substitution}}\lfbprodnewline
\lfbfirstprodline{|}{ \bullet }{}{}{}{\lfbcom{empty substitution}}\lfbprodnewline
\lfbprodline{|}{\sigma  \lfbsym{,}  \lfbnt{SubElm}}{}{}{}{\lfbcom{append substitution}}\lfbprodnewline
\lfbprodline{|}{ \lfbnt{SubElm} }{}{}{}{}\lfbprodnewline
\lfbprodline{|}{ \sigma_{{\mathrm{1}}} \lfbsym{,} \sigma_{{\mathrm{2}}} } {\textsf{M}}{}{}{\lfbcom{concat substitution}}\lfbprodnewline
\lfbprodline{|}{ id _{ \Gamma } } {\textsf{M}}{}{}{\lfbcom{identity substiuttion on G}}\lfbprodnewline
\lfbprodline{|}{\mathsf{head} \, \lfbsym{(}  \sigma  \lfbsym{)}} {\textsf{M}}{}{}{\lfbcom{head of global substitution}}\lfbprodnewline
\lfbprodline{|}{\sigma  \uparrow  \lfbnt{k}} {\textsf{M}}{}{}{\lfbcom{pop transitions by k}}\lfbprodnewline
\lfbprodline{|}{\lfbsym{(}  \sigma  \lfbsym{)}} {\textsf{S}}{}{}{\lfbcom{paren}}\lfbprodnewline
\lfbprodline{|}{ \lfbnt{SubElm} } {\textsf{M}}{}{}{\lfbcom{single element}}}

\newcommand{\lfbCSub}{
\lfbrulehead{\Sigma}{::=}{\lfbcom{context subsitution}}\lfbprodnewline
\lfbfirstprodline{|}{ \bullet }{}{}{}{\lfbcom{empty context subsitution}}\lfbprodnewline
\lfbprodline{|}{\Sigma  \lfbsym{,}  \gamma  \coloneqq  \lfbnt{C}}{}{}{}{\lfbcom{append context subsitutiton}}\lfbprodnewline
\lfbprodline{|}{ \gamma \coloneqq \lfbnt{C} }{}{}{}{}\lfbprodnewline
\lfbprodline{|}{\lfbsym{(}  \Sigma  \lfbsym{)}} {\textsf{S}}{}{}{}\lfbprodnewline
\lfbprodline{|}{ \gamma \coloneqq \lfbnt{C} }{}{}{}{\lfbcom{single context subsitution}}}

\newcommand{\lfbSSub}{
\lfbrulehead{\bar{\sigma}}{::=}{\lfbcom{series subsitution}}\lfbprodnewline
\lfbfirstprodline{|}{ \bullet }{}{}{}{\lfbcom{empty series subsitution}}\lfbprodnewline
\lfbprodline{|}{\bar{\sigma}  \lfbsym{,}  \mathbb{x}  \coloneqq  \overrightarrow{y}}{}{}{}{\lfbcom{append series variable subsitution}}\lfbprodnewline
\lfbprodline{|}{\bar{\sigma}  \lfbsym{,} \, \text{\faLock}}{}{}{}{\lfbcom{next world}}\lfbprodnewline
\lfbprodline{|}{ \mathbb{x} \coloneqq \overrightarrow{x} }{}{}{}{}\lfbprodnewline
\lfbprodline{|}{ \bar{\sigma}_{{\mathrm{1}}} \lfbsym{,} \bar{\sigma}_{{\mathrm{2}}} } {\textsf{M}}{}{}{\lfbcom{append variable subsitution}}\lfbprodnewline
\lfbprodline{|}{\mathsf{head} \, \lfbsym{(}  \bar{\sigma}  \lfbsym{)}} {\textsf{M}}{}{}{}\lfbprodnewline
\lfbprodline{|}{\bar{\sigma}  \uparrow  \lfbnt{k}} {\textsf{M}}{}{}{\lfbcom{pop transitions by k}}\lfbprodnewline
\lfbprodline{|}{ \mathsf{destruct} \lfbsym{(} \Gamma \lfbsym{;} \Sigma \lfbsym{)} } {\textsf{M}}{}{}{\lfbcom{destruct series variables}}\lfbprodnewline
\lfbprodline{|}{ \mathsf{destruct} _{ G } \lfbsym{(} \Gamma \lfbsym{;} \Sigma \lfbsym{)} } {\textsf{M}}{}{}{\lfbcom{destruct series variables with explicit variable generator}}\lfbprodnewline
\lfbprodline{|}{\lfbsym{(}  \bar{\sigma}  \lfbsym{)}} {\textsf{S}}{}{}{}}

\newcommand{\lfbVGen}{
\lfbrulehead{G}{::=}{\lfbcom{variable generator}}\lfbprodnewline
\lfbfirstprodline{|}{\lfbsym{(}  G_v  \lfbsym{,}  G_s  \lfbsym{)}}{}{}{}{}}

\newcommand{\lfbVGenV}{
\lfbrulehead{G_v}{::=}{\lfbcom{stream of variables}}\lfbprodnewline
\lfbfirstprodline{|}{\lfbmv{x}  \lfbsym{,}  G_v}{}{}{}{}}

\newcommand{\lfbVGenS}{
\lfbrulehead{G_s}{::=}{\lfbcom{stream of series variables}}\lfbprodnewline
\lfbfirstprodline{|}{\mathbb{x}  \lfbsym{,}  G_s}{}{}{}{}}

\newcommand{\lfbVars}{
\lfbrulehead{V}{::=}{\lfbcom{set of variables}}\lfbprodnewline
\lfbfirstprodline{|}{\lfbsym{\{}  \lfbmv{x}  \lfbsym{\}}}{}{}{}{\lfbcom{set of variables with a single element}}\lfbprodnewline
\lfbprodline{|}{\lfbsym{\{}  \mathbb{x}  \lfbsym{\}}}{}{}{}{\lfbcom{set of variables with a single series variable}}\lfbprodnewline
\lfbprodline{|}{\lfbsym{\{}  \overrightarrow{x}  \lfbsym{\}}}{}{}{}{\lfbcom{set of variables from a variable series}}\lfbprodnewline
\lfbprodline{|}{ \mathsf{FV} _{ \lfbnt{k} }( \lfbnt{M} ) }{}{}{}{\lfbcom{Free variabels of a term}}\lfbprodnewline
\lfbprodline{|}{ \mathsf{FV} _{ \lfbnt{k} }( \theta ) }{}{}{}{\lfbcom{Free variabels of a explicit subsitution}}\lfbprodnewline
\lfbprodline{|}{ \mathsf{FV} _{ \lfbnt{k} }( \sigma ) }{}{}{}{\lfbcom{Free variables of a subsitution}}\lfbprodnewline
\lfbprodline{|}{V_{{\mathrm{1}}} \, \cup \, V_{{\mathrm{2}}}}{}{}{}{\lfbcom{union of context varibales}}\lfbprodnewline
\lfbprodline{|}{V_{{\mathrm{1}}}  \lfbsym{-}  V_{{\mathrm{2}}}}{}{}{}{\lfbcom{Residual set of context variables}}\lfbprodnewline
\lfbprodline{|}{\mathsf{dom} \, \lfbsym{(}  \Gamma  \lfbsym{)}}{}{}{}{\lfbcom{domain of an environment}}\lfbprodnewline
\lfbprodline{|}{\mathsf{dom} \, \lfbsym{(}  \sigma  \lfbsym{)}}{}{}{}{\lfbcom{domain of a subsitution}}\lfbprodnewline
\lfbprodline{|}{\mathsf{rg} \, \lfbsym{(}  \bar{\sigma}  \lfbsym{)}}{}{}{}{\lfbcom{region of a series subsitution}}}

\newcommand{\lfbCVs}{
\lfbrulehead{\lfbnt{CVs}}{::=}{\lfbcom{set of context vairables}}\lfbprodnewline
\lfbfirstprodline{|}{\lfbsym{\{}  \gamma  \lfbsym{\}}}{}{}{}{\lfbcom{set of context variables with single element}}\lfbprodnewline
\lfbprodline{|}{\mathsf{FCV} \, \lfbsym{(}  \lfbnt{T}  \lfbsym{)}}{}{}{}{\lfbcom{Free context variables of a type}}\lfbprodnewline
\lfbprodline{|}{\mathsf{FCV} \, \lfbsym{(}  \lfbnt{C}  \lfbsym{)}}{}{}{}{\lfbcom{Free context variables of a context}}\lfbprodnewline
\lfbprodline{|}{\mathsf{FCV} \, \lfbsym{(}  \lfbnt{M}  \lfbsym{)}}{}{}{}{\lfbcom{Free context variabels of a term}}\lfbprodnewline
\lfbprodline{|}{\mathsf{FCV} \, \lfbsym{(}  \theta  \lfbsym{)}}{}{}{}{\lfbcom{Free context variabels of a explicit subsitution}}\lfbprodnewline
\lfbprodline{|}{\mathsf{FCV} \, \lfbsym{(}  \Gamma  \lfbsym{)}}{}{}{}{\lfbcom{Free context variables of a environment}}\lfbprodnewline
\lfbprodline{|}{\mathsf{FCV} \, \lfbsym{(}  \sigma  \lfbsym{)}}{}{}{}{\lfbcom{Free context variables of a subsitution}}\lfbprodnewline
\lfbprodline{|}{\mathsf{FCV} \, \lfbsym{(}  \Sigma  \lfbsym{)}}{}{}{}{\lfbcom{Free context variables of a context substitution}}\lfbprodnewline
\lfbprodline{|}{\lfbnt{CVs_{{\mathrm{1}}}} \, \cup \, \lfbnt{CVs_{{\mathrm{2}}}}}{}{}{}{\lfbcom{union of context varibales}}\lfbprodnewline
\lfbprodline{|}{\lfbnt{CVs_{{\mathrm{1}}}}  \lfbsym{-}  \lfbnt{CVs_{{\mathrm{2}}}}}{}{}{}{\lfbcom{Residual set of context variables}}\lfbprodnewline
\lfbprodline{|}{\mathsf{dom} \, \lfbsym{(}  \Sigma  \lfbsym{)}}{}{}{}{\lfbcom{Domain of context substitution}}\lfbprodnewline
\lfbprodline{|}{\mathsf{dom} \, \lfbsym{(}  \tilde{\Sigma}  \lfbsym{)}}{}{}{}{\lfbcom{Domain of RC assignment}}}

\newcommand{\lfbBRed}{
\lfbrulehead{\lfbnt{BRed}}{::=}{\lfbcom{beta reduction}}\lfbprodnewline
\lfbfirstprodline{|}{\lfbnt{M}  \rightarrow_{\beta}  \lfbnt{N}}{}{}{}{}\lfbprodnewline
\lfbprodline{|}{\theta  \rightarrow_{\beta}  \eta}{}{}{}{}\lfbprodnewline
\lfbprodline{|}{\lfbnt{M}  \rightarrow_{\beta}^{*}  \lfbnt{N}}{}{}{}{}\lfbprodnewline
\lfbprodline{|}{\theta  \rightarrow_{\beta}^{*}  \eta}{}{}{}{}}

\newcommand{\lfbRC}{
\lfbrulehead{\mathcal{R}  ,\ \mathcal{S}}{::=}{\lfbcom{reducibility candidate}}\lfbprodnewline
\lfbfirstprodline{|}{ \lfbkw{Red} _{ \lfbnt{T} } \lfbsym{[} \tilde{\Sigma} \lfbsym{]} }{}{}{}{}\lfbprodnewline
\lfbprodline{|}{ \lfbkw{Red} _{ \lfbnt{C} } \lfbsym{[} \tilde{\Sigma} \lfbsym{]} }{}{}{}{}\lfbprodnewline
\lfbprodline{|}{ \lfbkw{Red} _{ \Gamma } \lfbsym{[} \tilde{\Sigma} \lfbsym{,} \bar{\sigma} \lfbsym{]} }{}{}{}{\lfbcom{Parametrized reducibility for subsitutiton}}}

\newcommand{\lfbRSub}{
\lfbrulehead{\tilde{\Sigma}}{::=}{\lfbcom{redicibility candidate assignment}}\lfbprodnewline
\lfbfirstprodline{|}{ \bullet }{}{}{}{\lfbcom{empty}}\lfbprodnewline
\lfbprodline{|}{\tilde{\Sigma}  \lfbsym{,}  \gamma  \colon  \lfbnt{C}  \coloneqq  \mathcal{R}}{}{}{}{\lfbcom{append}}\lfbprodnewline
\lfbprodline{|}{ \gamma \coloneqq \lfbnt{C} } {\textsf{M}}{}{}{\lfbcom{single}}\lfbprodnewline
\lfbprodline{|}{ \tilde{\Sigma}_{{\mathrm{1}}} , \tilde{\Sigma}_{{\mathrm{2}}} } {\textsf{M}}{}{}{}\lfbprodnewline
\lfbprodline{|}{\lfbsym{(}  \tilde{\Sigma}  \lfbsym{)}} {\textsf{S}}{}{}{}}

\newcommand{\lfbPseudoRedSub}{
\lfbrulehead{\lfbnt{PseudoRedSub}}{::=}{\lfbcom{pseudo redicibility on substitution for explanation}}\lfbprodnewline
\lfbfirstprodline{|}{ \lfbkw{Red} _{ \Gamma } \lfbsym{[} \tilde{\Sigma} \lfbsym{]} }{}{}{}{\lfbcom{Parametrized reducibility for subsitutiton}}}

\newcommand{\lfbTLO}{
\lfbrulehead{T^{1}  ,\ S^{1}}{::=}{\lfbcom{level-1 types of lttt}}\lfbprodnewline
\lfbfirstprodline{|}{\iota}{}{}{}{}\lfbprodnewline
\lfbprodline{|}{S^{1}  \rightarrow  T^{1}}{}{}{}{}\lfbprodnewline
\lfbprodline{|}{\lfbsym{(}  T^{1}  \lfbsym{)}} {\textsf{S}}{}{}{}\lfbprodnewline
\lfbprodline{|}{ T^{1} } {\textsf{S}}{}{}{}}

\newcommand{\lfbTLM}{
\lfbrulehead{T^{0}  ,\ S^{0}}{::=}{\lfbcom{level-0 types of lttt}}\lfbprodnewline
\lfbfirstprodline{|}{\iota}{}{}{}{}\lfbprodnewline
\lfbprodline{|}{S^{0}  \rightarrow  T^{0}}{}{}{}{}\lfbprodnewline
\lfbprodline{|}{ \bigcirc  T^{1} }{}{}{}{}\lfbprodnewline
\lfbprodline{|}{\lfbsym{(}  T^{0}  \lfbsym{)}} {\textsf{S}}{}{}{}\lfbprodnewline
\lfbprodline{|}{ T^{0} } {\textsf{S}}{}{}{}}

\newcommand{\lfbMLO}{
\lfbrulehead{M^{1}  ,\ N^{1}}{::=}{\lfbcom{level-1 terms of lttt}}\lfbprodnewline
\lfbfirstprodline{|}{\lfbmv{x}}{}{}{}{}\lfbprodnewline
\lfbprodline{|}{ \lambda \lfbmv{x} ^{ T^{1} }. M^{1} }{}{\textsf{bind}\; \lfbmv{x}\; \textsf{in}\; M^{1}}{}{}\lfbprodnewline
\lfbprodline{|}{M^{1} \, N^{1}}{}{}{}{}\lfbprodnewline
\lfbprodline{|}{ \lfbkw{unq}   M^{0} }{}{}{}{}\lfbprodnewline
\lfbprodline{|}{\lfbsym{(}  M^{1}  \lfbsym{)}} {\textsf{S}}{}{}{}}

\newcommand{\lfbMLM}{
\lfbrulehead{M^{0}  ,\ N^{0}}{::=}{\lfbcom{level-0 terms of lttt}}\lfbprodnewline
\lfbfirstprodline{|}{\lfbmv{x}}{}{}{}{}\lfbprodnewline
\lfbprodline{|}{ \lambda \lfbmv{x} ^{ T^{0} }. M^{0} }{}{\textsf{bind}\; \lfbmv{x}\; \textsf{in}\; M^{0}}{}{}\lfbprodnewline
\lfbprodline{|}{M^{0} \, N^{0}}{}{}{}{}\lfbprodnewline
\lfbprodline{|}{ \lfbkw{quo}   M^{1} }{}{}{}{}\lfbprodnewline
\lfbprodline{|}{\lfbsym{(}  M^{0}  \lfbsym{)}} {\textsf{S}}{}{}{}}

\newcommand{\lfbTLi}{
\lfbrulehead{T^{i}  ,\ S^{i}}{::=}{\lfbcom{level-i types of lttt}}\lfbprodnewline
\lfbfirstprodline{|}{\iota}{}{}{}{}\lfbprodnewline
\lfbprodline{|}{S^{i}  \rightarrow  T^{i}}{}{}{}{}\lfbprodnewline
\lfbprodline{|}{\lfbsym{(}  T^{i}  \lfbsym{)}} {\textsf{S}}{}{}{}}

\newcommand{\lfbMLi}{
\lfbrulehead{M^{i}  ,\ N^{i}}{::=}{\lfbcom{level-i terms of lttt}}\lfbprodnewline
\lfbfirstprodline{|}{\lfbmv{x}}{}{}{}{}\lfbprodnewline
\lfbprodline{|}{ \lambda \lfbmv{x} ^{ T^{i} }. M^{i} }{}{\textsf{bind}\; \lfbmv{x}\; \textsf{in}\; M^{i}}{}{}\lfbprodnewline
\lfbprodline{|}{M^{i} \, N^{i}}{}{}{}{}\lfbprodnewline
\lfbprodline{|}{\lfbsym{(}  M^{i}  \lfbsym{)}} {\textsf{S}}{}{}{}}

\newcommand{\lfbGLElm}{
\lfbrulehead{\lfbnt{GLElm}}{::=}{\lfbcom{element of named context of lttt}}\lfbprodnewline
\lfbfirstprodline{|}{ \lfbmv{x}  :^0  T^{0} }{}{}{}{}\lfbprodnewline
\lfbprodline{|}{ \lfbmv{x}  :^1  T^{1} }{}{}{}{}\lfbprodnewline
\lfbprodline{|}{ \lfbmv{x}  :^ \lfbmv{i}   T^{i} }{}{}{}{}}

\newcommand{\lfbGL}{
\lfbrulehead{\Gamma^{\circ}  ,\ \Delta^{\circ}}{::=}{\lfbcom{named context of lttt}}\lfbprodnewline
\lfbfirstprodline{|}{ \cdot }{}{}{}{}\lfbprodnewline
\lfbprodline{|}{\Gamma^{\circ}  \lfbsym{,}  \lfbnt{GLElm}}{}{}{}{}\lfbprodnewline
\lfbprodline{|}{ \Gamma^{\circ}_{{\mathrm{1}}} \lfbsym{,} \Gamma^{\circ}_{{\mathrm{2}}} }{}{}{}{}\lfbprodnewline
\lfbprodline{|}{ \lfbnt{GLElm} }{}{}{}{}}

\newcommand{\lfbGEElm}{
\lfbrulehead{\lfbnt{GEElm}}{::=}{\lfbcom{element of intermediate named context}}\lfbprodnewline
\lfbfirstprodline{|}{ \lfbmv{x}  :^0 \lfbnt{T} }{}{}{}{}\lfbprodnewline
\lfbprodline{|}{ \lfbmv{x}  :^1 \lfbnt{T} }{}{}{}{}\lfbprodnewline
\lfbprodline{|}{ \mathbb{x}  :^1 \gamma }{}{}{}{}}

\newcommand{\lfbGE}{
\lfbrulehead{\tilde{\Gamma}  ,\ \tilde{\Delta}}{::=}{\lfbcom{intermediate named context for embedding}}\lfbprodnewline
\lfbfirstprodline{|}{ \cdot }{}{}{}{}\lfbprodnewline
\lfbprodline{|}{\tilde{\Gamma}  \lfbsym{,}  \lfbnt{GEElm}}{}{}{}{}\lfbprodnewline
\lfbprodline{|}{ \lfbnt{GEElm} }{}{}{}{}\lfbprodnewline
\lfbprodline{|}{ \tilde{\Gamma} , \tilde{\Delta} } {\textsf{M}}{}{}{}\lfbprodnewline
\lfbprodline{|}{\lfbsym{(}  \tilde{\Gamma}  \lfbsym{)}} {\textsf{S}}{}{}{}}

\newcommand{\lfbCMTTTerm}{
\lfbrulehead{\lfbnt{CMTTTerm}}{::=}{\lfbcom{CMTT term}}\lfbprodnewline
\lfbfirstprodline{|}{\lfbmv{x}}{}{}{}{}\lfbprodnewline
\lfbprodline{|}{\lfbkw{box} \, \lfbsym{(} \, \lfbkw{CMTTNCtx} \, \lfbsym{.}  \lfbnt{CMTTTerm}  \lfbsym{)}}{}{}{}{}}

\newcommand{\lfbCMTTType}{
\lfbrulehead{T}{::=}{\lfbcom{CMTT Type}}\lfbprodnewline
\lfbfirstprodline{|}{\lfbsym{[}  \lfbnt{CMTTNctx}  \lfbsym{]}  T}{}{}{}{}}

\newcommand{\lfbCMTTNctx}{
\lfbrulehead{\lfbnt{CMTTNctx}}{::=}{}\lfbprodnewline
\lfbfirstprodline{|}{ \cdot }{}{}{}{}\lfbprodnewline
\lfbprodline{|}{\lfbnt{CMTTNctx}  \lfbsym{,}  \lfbmv{x}  \colon  T}{}{}{}{}\lfbprodnewline
\lfbprodline{|}{ \lfbmv{x} \colon T } {\textsf{M}}{}{}{}}

\newcommand{\lfbterminals}{
\lfbrulehead{\lfbnt{terminals}}{::=}{}\lfbprodnewline
\lfbfirstprodline{|}{ \colon }{}{}{}{}\lfbprodnewline
\lfbprodline{|}{\lfbsym{.}}{}{}{}{}\lfbprodnewline
\lfbprodline{|}{ \lambda }{}{}{}{}\lfbprodnewline
\lfbprodline{|}{ \rightarrow_{\beta} }{}{}{}{}\lfbprodnewline
\lfbprodline{|}{ \rightarrow_{\beta}^{*} }{}{}{}{}\lfbprodnewline
\lfbprodline{|}{ \rightarrow }{}{}{}{}\lfbprodnewline
\lfbprodline{|}{ \vdash }{}{}{}{}\lfbprodnewline
\lfbprodline{|}{ \in }{}{}{}{}\lfbprodnewline
\lfbprodline{|}{ \not\in }{}{}{}{}\lfbprodnewline
\lfbprodline{|}{ \mathsf{FV} }{}{}{}{}\lfbprodnewline
\lfbprodline{|}{ \mathsf{FCV} }{}{}{}{}\lfbprodnewline
\lfbprodline{|}{ \coloneqq }{}{}{}{}\lfbprodnewline
\lfbprodline{|}{ \langle }{}{}{}{}\lfbprodnewline
\lfbprodline{|}{ \rangle }{}{}{}{}\lfbprodnewline
\lfbprodline{|}{ \mathsf{rg} }{}{}{}{}\lfbprodnewline
\lfbprodline{|}{ \lhd }{}{}{}{}\lfbprodnewline
\lfbprodline{|}{ \forall }{}{}{}{}\lfbprodnewline
\lfbprodline{|}{ \Lambda }{}{}{}{}\lfbprodnewline
\lfbprodline{|}{ \leq }{}{}{}{}\lfbprodnewline
\lfbprodline{|}{ @ }{}{}{}{}\lfbprodnewline
\lfbprodline{|}{ id }{}{}{}{}\lfbprodnewline
\lfbprodline{|}{ \uparrow }{}{}{}{}\lfbprodnewline
\lfbprodline{|}{ ++ }{}{}{}{}\lfbprodnewline
\lfbprodline{|}{ \text{\faLock} }{}{}{}{}\lfbprodnewline
\lfbprodline{|}{ \cup }{}{}{}{}\lfbprodnewline
\lfbprodline{|}{ \mathsf{dom} }{}{}{}{}\lfbprodnewline
\lfbprodline{|}{ \mathsf{head} }{}{}{}{}\lfbprodnewline
\lfbprodline{|}{ \mathsf{count} }{}{}{}{}\lfbprodnewline
\lfbprodline{|}{ \mathsf{destruct} }{}{}{}{}\lfbprodnewline
\lfbprodline{|}{ \mathsf{destruct\%} }{}{}{}{}\lfbprodnewline
\lfbprodline{|}{ \mathsf{gensyms} }{}{}{}{}\lfbprodnewline
\lfbprodline{|}{\lfbkw{quo}}{}{}{}{}\lfbprodnewline
\lfbprodline{|}{\lfbkw{unq}}{}{}{}{}\lfbprodnewline
\lfbprodline{|}{ \mathsf{L0} }{}{}{}{}\lfbprodnewline
\lfbprodline{|}{ \mathsf{L1} }{}{}{}{}\lfbprodnewline
\lfbprodline{|}{ \mathsf{vars} }{}{}{}{}\lfbprodnewline
\lfbprodline{|}{ \mathsf{diff} }{}{}{}{}\lfbprodnewline
\lfbprodline{|}{ \leadsto }{}{}{}{}\lfbprodnewline
\lfbprodline{|}{ \vdash_0 }{}{}{}{}\lfbprodnewline
\lfbprodline{|}{ \vdash_1 }{}{}{}{}\lfbprodnewline
\lfbprodline{|}{ \vdash_i }{}{}{}{}\lfbprodnewline
\lfbprodline{|}{ \textrm{int} }{}{}{}{}\lfbprodnewline
\lfbprodline{|}{ \textrm{str} }{}{}{}{}}

\newcommand{\lfbformula}{
\lfbrulehead{\lfbnt{formula}}{::=}{}\lfbprodnewline
\lfbfirstprodline{|}{\lfbnt{judgement}}{}{}{}{}\lfbprodnewline
\lfbprodline{|}{\lfbmv{x}  \colon  \lfbnt{T} \, \in \, \Gamma}{}{}{}{}\lfbprodnewline
\lfbprodline{|}{\mathbb{x}  \colon  \gamma \, \in \, \Gamma}{}{}{}{}\lfbprodnewline
\lfbprodline{|}{\gamma \, \not\in \, \lfbnt{CVs}}{}{}{}{}\lfbprodnewline
\lfbprodline{|}{\lfbmv{x} \, \not\in \, V}{}{}{}{}\lfbprodnewline
\lfbprodline{|}{\mathbb{x} \, \not\in \, V}{}{}{}{}\lfbprodnewline
\lfbprodline{|}{\Gamma  \leq  \Delta}{}{}{}{}\lfbprodnewline
\lfbprodline{|}{\lfbmv{x}  \coloneqq  \lfbnt{M} \, \in \, \sigma}{}{}{}{}\lfbprodnewline
\lfbprodline{|}{\mathbb{x}  \coloneqq  \mathbb{y} \, \in \, \sigma}{}{}{}{}\lfbprodnewline
\lfbprodline{|}{\gamma  \coloneqq  \lfbnt{C} \, \in \, \Sigma}{}{}{}{}\lfbprodnewline
\lfbprodline{|}{\mathbb{x}  \coloneqq  \overrightarrow{y} \, \in \, \bar{\sigma}}{}{}{}{}\lfbprodnewline
\lfbprodline{|}{\mathsf{dom} \, \lfbsym{(}  \sigma  \lfbsym{)}}{}{}{}{}\lfbprodnewline
\lfbprodline{|}{\lfbmv{x} \, \not\in \, \mathsf{dom} \, \lfbsym{(}  \Delta  \lfbsym{)}}{}{}{}{}\lfbprodnewline
\lfbprodline{|}{\text{\faLock} \, \not\in \, \Delta}{}{}{}{}\lfbprodnewline
\lfbprodline{|}{ \text{\faLock} _{ \lfbnt{k} }  \not\in \sigma }{}{}{}{}\lfbprodnewline
\lfbprodline{|}{\gamma  \colon  \lfbnt{C}  \coloneqq  \mathcal{R} \, \in \, \tilde{\Sigma}}{}{}{}{}\lfbprodnewline
\lfbprodline{|}{\mathcal{R}  \lfbsym{(}  \Gamma  \lfbsym{,}  \lfbnt{M}  \lfbsym{)}}{}{}{}{}\lfbprodnewline
\lfbprodline{|}{\mathcal{R}  \lfbsym{(}  \Gamma  \lfbsym{,}  \theta  \lfbsym{)}}{}{}{}{}\lfbprodnewline
\lfbprodline{|}{\mathcal{R}  \lfbsym{(}  \Delta  \lfbsym{,}  \sigma  \lfbsym{)}}{}{}{}{}\lfbprodnewline
\lfbprodline{|}{ \lfbmv{x}  :^0  T^{0} \in \Gamma^{\circ} }{}{}{}{}\lfbprodnewline
\lfbprodline{|}{ \lfbmv{x}  :^1  T^{1} \in \Gamma^{\circ} }{}{}{}{}\lfbprodnewline
\lfbprodline{|}{ \lfbmv{x}  :^ \lfbmv{i} T^{i} \in \Gamma^{\circ} }{}{}{}{}}

\newcommand{\lfbJKripkeTrans}{
\lfbrulehead{\lfbnt{JKripkeTrans}}{::=}{}\lfbprodnewline
\lfbfirstprodline{|}{\lfbnt{k}  \colon  \Gamma  \lhd  \Delta}{}{}{}{\lfbcom{Kripke Transition}}}

\newcommand{\lfbJTyping}{
\lfbrulehead{\lfbnt{JTyping}}{::=}{}\lfbprodnewline
\lfbfirstprodline{|}{\Gamma  \vdash  \lfbnt{M}  \colon  \lfbnt{T}}{}{}{}{\lfbcom{Type inhavitation}}\lfbprodnewline
\lfbprodline{|}{\Gamma  \vdash  \theta  \colon  \lfbnt{C}}{}{}{}{\lfbcom{Context inhavitation}}\lfbprodnewline
\lfbprodline{|}{\Gamma  \vdash  \sigma  \colon  \Delta}{}{}{}{\lfbcom{Typing on Global Substitution}}\lfbprodnewline
\lfbprodline{|}{ \Gamma^{\circ} \vdash_1 M^{1} \colon T^{1} }{}{}{}{\lfbcom{Type inhavitation for level-1 lttt term}}\lfbprodnewline
\lfbprodline{|}{ \Gamma^{\circ} \vdash_0 M^{0} \colon T^{0} }{}{}{}{\lfbcom{Type inhavitation for level-0 lttt term}}\lfbprodnewline
\lfbprodline{|}{ \Gamma^{\circ} \vdash_i M^{i} \colon T^{i} }{}{}{}{\lfbcom{Type inhavitation for level-i lttt term}}\lfbprodnewline
\lfbprodline{|}{\Gamma^{\circ}  \leadsto  \tilde{\Gamma}}{}{}{}{\lfbcom{embedding relation from lamfb to lttt named context}}}

\newcommand{\lfbjudgement}{
\lfbrulehead{\lfbnt{judgement}}{::=}{}\lfbprodnewline
\lfbfirstprodline{|}{\lfbnt{JKripkeTrans}}{}{}{}{}\lfbprodnewline
\lfbprodline{|}{\lfbnt{JTyping}}{}{}{}{}}

\newcommand{\lfbuserXXsyntax}{
\lfbrulehead{\lfbnt{user\_syntax}}{::=}{}\lfbprodnewline
\lfbfirstprodline{|}{\lfbmv{x}}{}{}{}{}\lfbprodnewline
\lfbprodline{|}{\gamma}{}{}{}{}\lfbprodnewline
\lfbprodline{|}{\mathbb{x}}{}{}{}{}\lfbprodnewline
\lfbprodline{|}{\lfbmv{i}}{}{}{}{}\lfbprodnewline
\lfbprodline{|}{\lfbnt{k}}{}{}{}{}\lfbprodnewline
\lfbprodline{|}{\iota}{}{}{}{}\lfbprodnewline
\lfbprodline{|}{\lfbnt{T}}{}{}{}{}\lfbprodnewline
\lfbprodline{|}{\lfbnt{C}}{}{}{}{}\lfbprodnewline
\lfbprodline{|}{\lfbnt{M}}{}{}{}{}\lfbprodnewline
\lfbprodline{|}{\theta}{}{}{}{}\lfbprodnewline
\lfbprodline{|}{\overrightarrow{x}}{}{}{}{}\lfbprodnewline
\lfbprodline{|}{\lfbnt{GElm}}{}{}{}{}\lfbprodnewline
\lfbprodline{|}{\Gamma}{}{}{}{}\lfbprodnewline
\lfbprodline{|}{\lfbnt{SubElm}}{}{}{}{}\lfbprodnewline
\lfbprodline{|}{\sigma}{}{}{}{}\lfbprodnewline
\lfbprodline{|}{\Sigma}{}{}{}{}\lfbprodnewline
\lfbprodline{|}{\bar{\sigma}}{}{}{}{}\lfbprodnewline
\lfbprodline{|}{G}{}{}{}{}\lfbprodnewline
\lfbprodline{|}{G_v}{}{}{}{}\lfbprodnewline
\lfbprodline{|}{G_s}{}{}{}{}\lfbprodnewline
\lfbprodline{|}{V}{}{}{}{}\lfbprodnewline
\lfbprodline{|}{\lfbnt{CVs}}{}{}{}{}\lfbprodnewline
\lfbprodline{|}{\lfbnt{BRed}}{}{}{}{}\lfbprodnewline
\lfbprodline{|}{\mathcal{R}}{}{}{}{}\lfbprodnewline
\lfbprodline{|}{\tilde{\Sigma}}{}{}{}{}\lfbprodnewline
\lfbprodline{|}{\lfbnt{PseudoRedSub}}{}{}{}{}\lfbprodnewline
\lfbprodline{|}{T^{1}}{}{}{}{}\lfbprodnewline
\lfbprodline{|}{T^{0}}{}{}{}{}\lfbprodnewline
\lfbprodline{|}{M^{1}}{}{}{}{}\lfbprodnewline
\lfbprodline{|}{M^{0}}{}{}{}{}\lfbprodnewline
\lfbprodline{|}{T^{i}}{}{}{}{}\lfbprodnewline
\lfbprodline{|}{M^{i}}{}{}{}{}\lfbprodnewline
\lfbprodline{|}{\lfbnt{GLElm}}{}{}{}{}\lfbprodnewline
\lfbprodline{|}{\Gamma^{\circ}}{}{}{}{}\lfbprodnewline
\lfbprodline{|}{\lfbnt{GEElm}}{}{}{}{}\lfbprodnewline
\lfbprodline{|}{\tilde{\Gamma}}{}{}{}{}\lfbprodnewline
\lfbprodline{|}{\lfbnt{CMTTTerm}}{}{}{}{}\lfbprodnewline
\lfbprodline{|}{T}{}{}{}{}\lfbprodnewline
\lfbprodline{|}{\lfbnt{CMTTNctx}}{}{}{}{}\lfbprodnewline
\lfbprodline{|}{\lfbnt{terminals}}{}{}{}{}\lfbprodnewline
\lfbprodline{|}{\lfbnt{formula}}{}{}{}{}}

\newcommand{\lfbgrammar}{\lfbgrammartabular{
\lfbk\lfbinterrule
\lfbbase\lfbinterrule
\lfbT\lfbinterrule
\lfbC\lfbinterrule
\lfbM\lfbinterrule
\lfbMM\lfbinterrule
\lfbxs\lfbinterrule
\lfbGElm\lfbinterrule
\lfbG\lfbinterrule
\lfbSubElm\lfbinterrule
\lfbSub\lfbinterrule
\lfbCSub\lfbinterrule
\lfbSSub\lfbinterrule
\lfbVGen\lfbinterrule
\lfbVGenV\lfbinterrule
\lfbVGenS\lfbinterrule
\lfbVars\lfbinterrule
\lfbCVs\lfbinterrule
\lfbBRed\lfbinterrule
\lfbRC\lfbinterrule
\lfbRSub\lfbinterrule
\lfbPseudoRedSub\lfbinterrule
\lfbTLO\lfbinterrule
\lfbTLM\lfbinterrule
\lfbMLO\lfbinterrule
\lfbMLM\lfbinterrule
\lfbTLi\lfbinterrule
\lfbMLi\lfbinterrule
\lfbGLElm\lfbinterrule
\lfbGL\lfbinterrule
\lfbGEElm\lfbinterrule
\lfbGE\lfbinterrule
\lfbCMTTTerm\lfbinterrule
\lfbCMTTType\lfbinterrule
\lfbCMTTNctx\lfbinterrule
\lfbterminals\lfbinterrule
\lfbformula\lfbinterrule
\lfbJKripkeTrans\lfbinterrule
\lfbJTyping\lfbinterrule
\lfbjudgement\lfbinterrule
\lfbuserXXsyntax\lfbafterlastrule
}}

% defnss
% defns JKripkeTrans
%% defn kripke_transision
\newcommand{\lfbdrulektransXXempty}[1]{\lfbdrule[#1]{%
}{
\lfbsym{0}  \colon  \Gamma  \lhd  \Gamma}{%
{\lfbdrulename{ktrans\_empty}}{}%
}}

\newcommand{\lfbdrulektransXXweak}[1]{\lfbdrule[#1]{%
\lfbpremise{\lfbnt{k}  \colon  \Gamma  \lhd  \Delta}%
\lfbpremise{\lfbmv{x} \, \not\in \, \mathsf{dom} \, \lfbsym{(}  \Delta  \lfbsym{)}}%
}{
\lfbnt{k}  \colon  \Gamma  \lhd  \Delta  \lfbsym{,}  \lfbmv{x}  \colon  \lfbnt{T}}{%
{\lfbdrulename{ktrans\_weak}}{}%
}}

\newcommand{\lfbdrulektransXXlock}[1]{\lfbdrule[#1]{%
\lfbpremise{\lfbnt{k}  \colon  \Gamma  \lhd  \Delta}%
}{
\lfbnt{k}  \lfbsym{+}  \lfbsym{1}  \colon  \Gamma  \lhd  \Delta  \lfbsym{,}  \text{\faLock}}{%
{\lfbdrulename{ktrans\_lock}}{}%
}}

\newcommand{\lfbdefnkripkeXXtransision}[1]{\begin{lfbdefnblock}[#1]{$\lfbnt{k}  \colon  \Gamma  \lhd  \Delta$}{\lfbcom{Kripke Transition}}
\lfbusedrule{\lfbdrulektransXXempty{}}
\lfbusedrule{\lfbdrulektransXXweak{}}
\lfbusedrule{\lfbdrulektransXXlock{}}
\end{lfbdefnblock}}

\newcommand{\lfbdefnsJKripkeTrans}{
\lfbdefnkripkeXXtransision{}}

% defns JTyping
%% defn typing
\newcommand{\lfbdruletypeXXvar}[1]{\lfbdrule[#1]{%
\lfbpremise{\lfbmv{x}  \colon  \lfbnt{T} \, \in \, \mathsf{head} \, \lfbsym{(}  \Gamma  \lfbsym{)}}%
}{
\Gamma  \vdash  \lfbmv{x}  \colon  \lfbnt{T}}{%
{\lfbdrulename{type\_var}}{}%
}}

\newcommand{\lfbdruletypeXXabs}[1]{\lfbdrule[#1]{%
\lfbpremise{\Gamma  \lfbsym{,}  \lfbmv{x}  \colon  \lfbnt{T_{{\mathrm{1}}}}  \vdash  \lfbnt{M}  \colon  \lfbnt{T_{{\mathrm{2}}}}}%
}{
\Gamma  \vdash   \lambda \lfbmv{x} ^{ \lfbnt{T_{{\mathrm{1}}}} }. \lfbnt{M}   \colon  \lfbnt{T_{{\mathrm{1}}}}  \rightarrow  \lfbnt{T_{{\mathrm{2}}}}}{%
{\lfbdrulename{type\_abs}}{}%
}}

\newcommand{\lfbdruletypeXXapp}[1]{\lfbdrule[#1]{%
\lfbpremise{\Gamma  \vdash  \lfbnt{M_{{\mathrm{1}}}}  \colon  \lfbnt{T_{{\mathrm{1}}}}  \rightarrow  \lfbnt{T_{{\mathrm{2}}}}}%
\lfbpremise{\Gamma  \vdash  \lfbnt{M_{{\mathrm{2}}}}  \colon  \lfbnt{T_{{\mathrm{1}}}}}%
}{
\Gamma  \vdash  \lfbnt{M_{{\mathrm{1}}}} \, \lfbnt{M_{{\mathrm{2}}}}  \colon  \lfbnt{T_{{\mathrm{2}}}}}{%
{\lfbdrulename{type\_app}}{}%
}}

\newcommand{\lfbdruletypeXXquote}[1]{\lfbdrule[#1]{%
\lfbpremise{ \Gamma  \lfbsym{,}  \text{\faLock} \lfbsym{,} \Delta   \vdash  \lfbnt{M}  \colon  \lfbnt{T}}%
\lfbpremise{\text{\faLock} \, \not\in \, \Delta}%
}{
\Gamma  \vdash   \lfbkw{quo} \langle \Delta \rangle \lfbnt{M}   \colon  \lfbsym{[}   \mathsf{rg} ( \Gamma )   \vdash  \lfbnt{T}  \lfbsym{]}}{%
{\lfbdrulename{type\_quote}}{}%
}}

\newcommand{\lfbdruletypeXXunq}[1]{\lfbdrule[#1]{%
\lfbpremise{\Gamma  \vdash  \lfbnt{M}  \colon  \lfbsym{[}  \lfbnt{C}  \vdash  \lfbnt{T}  \lfbsym{]}}%
\lfbpremise{\Delta  \vdash  \theta  \colon  \lfbnt{C}}%
\lfbpremise{\lfbnt{k}  \colon  \Gamma  \lhd  \Delta}%
}{
\Delta  \vdash   \lfbkw{unq} _{ \lfbnt{k} } \lfbnt{M} [  \theta  ]   \colon  \lfbnt{T}}{%
{\lfbdrulename{type\_unq}}{}%
}}

\newcommand{\lfbdruletypeXXctxabs}[1]{\lfbdrule[#1]{%
\lfbpremise{\gamma \, \not\in \, \mathsf{FCV} \, \lfbsym{(}  \Gamma  \lfbsym{)}}%
\lfbpremise{\Gamma  \vdash  \lfbnt{M}  \colon  \lfbnt{T}}%
}{
\Gamma  \vdash   \Lambda \gamma . \lfbnt{M}   \colon   \forall \gamma . \lfbnt{T} }{%
{\lfbdrulename{type\_ctxabs}}{}%
}}

\newcommand{\lfbdruletypeXXctxapp}[1]{\lfbdrule[#1]{%
\lfbpremise{\Gamma  \vdash  \lfbnt{M}  \colon   \forall \gamma . \lfbnt{T} }%
}{
\Gamma  \vdash   \lfbnt{M} @ \lfbnt{C}   \colon  \lfbnt{T}  \lfbsym{[}   \gamma \coloneqq \lfbnt{C}   \lfbsym{]}}{%
{\lfbdrulename{type\_ctxapp}}{}%
}}

\newcommand{\lfbdefntyping}[1]{\begin{lfbdefnblock}[#1]{$\Gamma  \vdash  \lfbnt{M}  \colon  \lfbnt{T}$}{\lfbcom{Type inhavitation}}
\lfbusedrule{\lfbdruletypeXXvar{}}
\lfbusedrule{\lfbdruletypeXXabs{}}
\lfbusedrule{\lfbdruletypeXXapp{}}
\lfbusedrule{\lfbdruletypeXXquote{}}
\lfbusedrule{\lfbdruletypeXXunq{}}
\lfbusedrule{\lfbdruletypeXXctxabs{}}
\lfbusedrule{\lfbdruletypeXXctxapp{}}
\end{lfbdefnblock}}

%% defn contexting
\newcommand{\lfbdrulecontextXXempty}[1]{\lfbdrule[#1]{%
}{
\Gamma  \vdash   \bullet   \colon   \bullet }{%
{\lfbdrulename{context\_empty}}{}%
}}

\newcommand{\lfbdrulecontextXXcons}[1]{\lfbdrule[#1]{%
\lfbpremise{\Gamma  \vdash  \theta  \colon  \lfbnt{C}}%
\lfbpremise{\Gamma  \vdash  \lfbnt{M}  \colon  \lfbnt{T}}%
}{
\Gamma  \vdash  \theta  \lfbsym{,}  \lfbnt{M}  \colon  \lfbnt{C}  \lfbsym{,}  \lfbnt{T}}{%
{\lfbdrulename{context\_cons}}{}%
}}

\newcommand{\lfbdrulecontextXXappendXXpctx}[1]{\lfbdrule[#1]{%
\lfbpremise{\Gamma  \vdash  \theta  \colon  \lfbnt{C}}%
\lfbpremise{\mathbb{x}  \colon  \gamma \, \in \, \Gamma}%
}{
\Gamma  \vdash  \theta  \lfbsym{,}  \mathbb{x}  \colon  \lfbnt{C}  \lfbsym{,}  \gamma}{%
{\lfbdrulename{context\_append\_pctx}}{}%
}}

\newcommand{\lfbdefncontexting}[1]{\begin{lfbdefnblock}[#1]{$\Gamma  \vdash  \theta  \colon  \lfbnt{C}$}{\lfbcom{Context inhavitation}}
\lfbusedrule{\lfbdrulecontextXXempty{}}
\lfbusedrule{\lfbdrulecontextXXcons{}}
\lfbusedrule{\lfbdrulecontextXXappendXXpctx{}}
\end{lfbdefnblock}}

%% defn substitutiontyping
\newcommand{\lfbdrulesubtypXXempty}[1]{\lfbdrule[#1]{%
}{
\Gamma  \vdash   \bullet   \colon   \bullet }{%
{\lfbdrulename{subtyp\_empty}}{}%
}}

\newcommand{\lfbdrulesubtypXXcons}[1]{\lfbdrule[#1]{%
\lfbpremise{\Gamma  \vdash  \sigma  \colon  \Delta}%
\lfbpremise{\Gamma  \vdash  \lfbnt{M}  \colon  \lfbnt{T}}%
\lfbpremise{\lfbmv{x} \, \not\in \, \mathsf{dom} \, \lfbsym{(}  \Delta  \lfbsym{)}}%
}{
\Gamma  \vdash  \sigma  \lfbsym{,}  \lfbmv{x}  \coloneqq  \lfbnt{M}  \colon  \Delta  \lfbsym{,}  \lfbmv{x}  \colon  \lfbnt{T}}{%
{\lfbdrulename{subtyp\_cons}}{}%
}}

\newcommand{\lfbdrulesubtypXXtrans}[1]{\lfbdrule[#1]{%
\lfbpremise{\Gamma_{{\mathrm{1}}}  \vdash  \sigma  \colon  \Delta}%
\lfbpremise{\lfbnt{k}  \colon  \Gamma_{{\mathrm{1}}}  \lhd  \Gamma_{{\mathrm{2}}}}%
}{
\Gamma_{{\mathrm{2}}}  \vdash  \sigma  \lfbsym{,}   \text{\faLock} _{ \lfbnt{k} }   \colon  \Delta  \lfbsym{,}  \text{\faLock}}{%
{\lfbdrulename{subtyp\_trans}}{}%
}}

\newcommand{\lfbdefnsubstitutiontyping}[1]{\begin{lfbdefnblock}[#1]{$\Gamma  \vdash  \sigma  \colon  \Delta$}{\lfbcom{Typing on Global Substitution}}
\lfbusedrule{\lfbdrulesubtypXXempty{}}
\lfbusedrule{\lfbdrulesubtypXXcons{}}
\lfbusedrule{\lfbdrulesubtypXXtrans{}}
\end{lfbdefnblock}}

%% defn lotyping

\newcommand{\lfbdefnlotyping}[1]{\begin{lfbdefnblock}[#1]{$ \Gamma^{\circ} \vdash_1 M^{1} \colon T^{1} $}{\lfbcom{Type inhavitation for level-1 lttt term}}
\end{lfbdefnblock}}

%% defn lmtyping

\newcommand{\lfbdefnlmtyping}[1]{\begin{lfbdefnblock}[#1]{$ \Gamma^{\circ} \vdash_0 M^{0} \colon T^{0} $}{\lfbcom{Type inhavitation for level-0 lttt term}}
\end{lfbdefnblock}}

%% defn lityping

\newcommand{\lfbdefnlityping}[1]{\begin{lfbdefnblock}[#1]{$ \Gamma^{\circ} \vdash_i M^{i} \colon T^{i} $}{\lfbcom{Type inhavitation for level-i lttt term}}
\end{lfbdefnblock}}

%% defn translatesto

\newcommand{\lfbdefntranslatesto}[1]{\begin{lfbdefnblock}[#1]{$\Gamma^{\circ}  \leadsto  \tilde{\Gamma}$}{\lfbcom{embedding relation from lamfb to lttt named context}}
\end{lfbdefnblock}}

\newcommand{\lfbdefnsJTyping}{
\lfbdefntyping{}\lfbdefncontexting{}\lfbdefnsubstitutiontyping{}\lfbdefnlotyping{}\lfbdefnlmtyping{}\lfbdefnlityping{}\lfbdefntranslatesto{}}

\newcommand{\lfbdefnss}{
\lfbdefnsJKripkeTrans
\lfbdefnsJTyping
}

\newcommand{\lfball}{\lfbmetavars\\[0pt]
\lfbgrammar\\[5.0mm]
\lfbdefnss}

%% file: 1-introduction.tex
\section{Introduction}
It is a common technique in metaprogramming to use code as a first-class value to generate, combine, and evaluate code at compile- and run-time.  Type systems for first-class code are known to correspond to proof systems of modal logic under the Curry--Howard isomorphism~\cite{Davies17,Pfenning01,Davies01,TsukadaI10LMCS,Nanevski08}: Modality corresponds to a type constructor for code types, controlling free variables and their types in code values.  Such modal type systems have been proposed for various areas of metaprogramming, including multi-stage computation~\cite{Taha03,Calcango03,Kiselyov14} and syntactic metaprogramming~\cite{Ganz01,Stucki18}, and, more recently, applied to proof assistants~\cite{Cave13,Pientka19,Stampoulis10}.

Modal types come in two flavors: implicit and explicit contexts. On the one hand, modal types with implicit contexts do not show typing contexts---free variables and their types---of code values.  A classical example of a modal type system with implicit contexts is \lamcirc~\cite{Davies17}, in which a code type is expressed by $\bigcirc T$ (``code of $T$''), no matter what variables are referenced in the code. It has been applied to real programming languages for multi-stage programming, such as MetaOCaml~\cite{Calcango03,Kiselyov14}.  Since the type operator $\bigcirc$ is derived from the modality ``next'' in linear-time temporal logic, we call these code types linear-time temporal types.  On the other hand, modal types with explicit contexts show typing contexts in code types.  For example, the type of code \verb|x+2| is expressed by $[x:int]int$, which denotes code of an integer expression that includes free occurrences of an integer variable $x$.  Such types are often called contextual modal types~\cite{Nanevski08}.  Prior work on modal type systems points out that contextual modal types have an advantage over linear-time temporal types in dealing with mutable reference cells and run-time code evaluation~\cite{Kim06,Rhiger12,Kiselyov16}. Contextual modal types have recently been applied to proof assistants~\cite{Pientka10,Cave13,Pientka19,Stampoulis10}, where users can operate on code representation of proof terms with explicit contexts.

Some previous work \cite{Kim06,Nanevski05,Cave13,Pientka19,Puech16} on contextual modal types has suggested \emph{polymorphic contexts}---polymorphism over typing contexts in contextual modal types---to abstract part of typing contexts by context variables $\gamma$: For example, the type $\forall \gamma. [\gamma]T_1 \rightarrow [\gamma]T_2$ denotes functions that take code of type $T_1$ under an arbitrary typing context $\gamma$ and returns code of type $T_2$ under the same typing context $\gamma$.  Although we can see that polymorphic contexts will play an important role in metaprogramming with contextual modal types, its type-theoretic foundations are not fully investigated yet.

\paragraph{Our contributions.}
This paper proposes a novel contextual modal type theory \lamfb that provides a type-theoretic foundation for polymorphic contexts. Our technical contributions are summarized below:
\begin{itemize}
\item We develop contextual modal type theory \lamfb with polymorphic contexts formally: we give its syntax, type system, and operational semantics given by $\beta$-reduction.  A notable feature of \lamfb is that it allows abstraction of multiple parts of a single context with context variables, e.g., $\forall \gamma_1. \forall \gamma_2. [\gamma_1, x:T_1, \gamma_2]T_2$.
\item We prove basic properties of \lamfb: subject reduction, strong normalization, and confluence.  Our strong normalization proof is based on Girard's parametric reducibility method, which is adapted to polymorphic contexts.
\item To demonstrate the expressive power of polymorphic contexts, we give translation from a two-level fragment of \lamcirc~\cite{Davies17} to \lamfb and prove that the translation preserves typing.  To our knowledge, this is the first result that formally describes the relation between linear-time temporal types and contextual modal types. We will see that \lamfb's major advantage that allows multiple abstractions over a single context plays a vital role.
\end{itemize}

\paragraph{Organization of the paper.}
Section~\ref{section:motivation} provides motivating examples from metaprogramming.  Our formal development starts with simple Fitch-style modal type theory \lambra in Section~\ref{section:fcmtt}.  We extend \lambra to \lamfb with polymorphic contexts and prove subject reduction in Section~\ref{section:polyctx}; we prove strong normalization of \lamfb in Section~\ref{section:basicprops}. Section~\ref{section:embedlttt} develops a sound embedding from linear-time temporal types to contextual modal types. Finally, we discuss related work in Section~\ref{section:relatedwork} and give a conclusion in Section~\ref{section:conclusion}.  We often omit straightforward definitions and proofs for brevity; interested readers are referred to Appendix.

%% file: 2-motivation.tex
\section{Motivation} \label{section:motivation}
We provide some metaprogramming examples to illustrate the informal idea behind our type theory and motivate polymorphic contexts.

\subsection{Simple Contextual Modal Types: Specializing Power Function}
First, we show a typical example from staged computation
%% \AI{``staged computation''?}
, the power function, to demonstrate how we can use contextual modal types for staged computation.  The examples in this section are written in a hypothetical OCaml-like language with contextual modal types.

\begin{verbatim}
  (* val pow : int -> [int |- int] *)
  let rec pow n = match n with
    | 0 -> `<x: int> 1
    | n -> let u = pow (n-1) in `<x: int>(x * ,1(u)[x])

  (* val power4 : int -> int *)
  let power4 = ,0(`<>( fun x:int -> ,1(pow 4)[x] ))[]
\end{verbatim}

The function \verb|pow| generates a piece of code \verb|x * ... * x * 1| that multiplies variable \verb|x| \verb|n| times; the function \verb|power4| puts the code generated by \verb|pow| under function abstraction and evaluates the code at run-time to obtain a function value to compute \(x^4\) without recursion.

This example uses two constructs for code manipulation: \emph{quote} of the form \verb|`<|$\Gamma$\verb|>|$M$ and \emph{unquote} of the form \verb|,|$n$\verb|(|$M$\verb|)[|$M_1,\ldots,M_k$\verb|]|.  The former, which is similar to quasi-quotation in Lisp, generates code of an expression $M$ paired with a variable environment $\Gamma$ under which the code is evaluated. In the example, the quote \verb|`<x: int> 1| is code of constant \verb|1| with the environment with single integer variable \verb|x|. The quote has a contextual modal type \verb=[int |- int]=, where the premise (\verb|int| on the left of \verb=|-=) corresponds to the environment \verb|x:int| and the succedent (\verb|int| on the right) to the code body.

The type for the environment, which we call a \textit{context}, is a sequence of types and does not involve variables. Similarly to de Bruijn indices, we identify variables in a context by their position rather than by their names. For instance, two quotes, \verb|`<x:int, y:int>x| and \verb|`<z:int, w:int>z|, are considered $\alpha$-equivalent because both use the first variable in the environment even though the variable names in the two environments are different. Both terms have the same contextual modal type \verb=[int, int |- int]=.

An unquote \verb|,|$n$\verb|(|$M$\verb|)[|$M_1,\ldots,M_k$\verb|]| is used to expand a code value $M$. For example, \verb|,1(u)[x]| expands \verb|u| of type \verb=[int |- int]=.  In addition to the code to be expanded, an unquote involves two annotations, an \textit{explicit substitution} (\verb|[|$M_1, \ldots,M_k$\verb|]|) and a \textit{stage transition} $n$.  An explicit substitution provides the definitions of the variables in the environment of a quote value.  In the example code, \verb|,1(u)[x]| supplies an explicit substitution \verb|[x]| as the definition for a single-variable context \verb|int|. If \verb|u| is \verb|`<y:int>y * 1|, then the unquote will expand to \verb|x * 1|, replacing \verb|y| with its definition \verb|x|.  Roughly speaking, a stage transition represents the number of nested quotes surrounding $M$.  The expression \verb|,1(u)[x]| applies the explicit substitution to \verb|u|, and splice the obtained code into the surrounding quote.  Thus, \verb|`<x:int>(...)| adds ``\verb|x *|'' to the code denoted by \verb|u|.  On the contrary, the unquote \verb|,0(|$M$\verb|)[]| computes $M$ (to obtain the code value \verb|fun x:int -> x * x * x * x * 1| with the empty environment) and expands it; since there is no surrounding quote, the expansion amounts to running the code.  In this sense, the unquote in this language can be considered as unquote in Lisp-like languages if the stage transition is 1 and as \verb|eval| function if it is 0.

\subsection{Polymorphic Contexts: Macro \texttt{repeat}}
Secondly, consider a macro called \verb|repeat|, which repeats a given piece of code $n$ times.  For example, we expect Lisp code \verb|(repeat 2 | \verb|(print "hello"))| to show \verb|hello| two times. We can imitate such a macro as follows:
\begin{verbatim}
  (* val repeat : int -> [string -> unit |- unit]
                      -> [string -> unit |- unit]  *)
  let rec repeat n body = match n with
    | 0 -> `<pr: string -> unit>(())
    | n -> let u = repeat (n-1) body in
           `<pr: string -> unit>(,1(u)[pr]; ,1(body)[pr])
\end{verbatim}
This function \verb|repeat| takes an integer \verb|n| for the number of repetitions and a code value to be repeated. For example, a macro call in Lisp \verb|(repeat 2| \verb|(print "hello"))| can be represented below.
\begin{center}
    \verb|,1(repeat 2 `<pr:string -> unit>(pr "hello"))[print]|
\end{center}
Note that the environment \verb|pr:string -> unit| is expected to be function \verb|print|. After applying the function \verb|repeat|, we obtain the following code.
\begin{center}
  \verb|,1(`<pr:string -> unit>(pr "hello"; pr "hello"; ()))[print]| \\
\end{center}
Finally, by using unquote, the code is fully expanded (with substituting library function \verb|print| for \verb|pr|) to
\begin{center}
    \verb|print "hello"; print "hello"; ()| .
\end{center}

A problem with the function \verb|repeat| is that it accepts code values with an environment that consists of a single variable of type \verb|string -> unit|.  We rather expect the function to accept code values with various patterns of contexts and to have multiple types that differ only in contexts: e.g.,
\begin{itemize}
  \item % original one with \verb|pr|: \\
        \verb=int -> [string -> unit |- unit] -> [string -> unit |- unit]=,
  \item % a context with \verb|pr| and two integer variables: \\
        \verb=int -> [string -> unit, int, int |- unit]= \\
        \hspace{3em}\verb=-> [string -> unit, int, int |- unit]=, and
  \item % a context with another function of type \verb|unit -> unit|: \\
        \verb=int -> [unit -> unit |- unit] -> [unit -> unit |- unit]= .
\end{itemize}

We will resolve this issue by abstracting the context part of the function with a \textit{context variable} \verb|G|. As a result, we obtain the type for generic \verb|repeat|: \verb=forall G. int -> [G |- int] -> [G |- int]=. We call the type starting with \verb|forall G.| a \textit{polymorphic context type}, which means that we can instantiate the context variable \verb|G| with any context. We can implement this generic function \verb|poly_rep| by using a context variable as follows.

\begin{verbatim}
  (* val poly_rep : forall G. int -> [G |- unit] -> [G |- unit] *)
  let rec poly_rep G n body =
    match n with
    | 0 -> `<xs: G>(())
    | n -> let u = poly_rep G (n-1) body in
           `<xs: G>(,1(u)[xs]; ,1(body)[xs])
\end{verbatim}

This function takes an additional context argument \verb|G|, which is used in quotes.  The \verb|xs| is called a \textit{series variable}, which is a novelty in this paper. A series variable stands for a sequence of program variables---corresponding to the fact that a context variable stands for a sequence of types---and forms an environment by pairing with a context variable. For example, \verb|xs:G| will represent environment \verb|x:int, y:string| if we substitute \verb|x, y| for \verb|xs|, and \verb|int, string| for \verb|G|. We can also use series variables for explicit substitution. If we use a series variable in an explicit substitution, as in \verb|,1(u)[xs]|, it will be replaced with a sequence of variables. For instance, if \verb|xs:G| expands to \verb|x:int, y:string|, then \verb|,1(u)[xs]| also expands to \verb|,1(u)[x,y]|.  Series variables in explicit substitutions work like identity substitutions~\cite{Stampoulis10,Cave13,Pientka19,Puech16}, which pass variables from an environment to explicit substitutions as-is.

Using \verb|poly_rep|, we can repeat code with two variables as follows:
\begin{verbatim}
  poly_rep (unit->int, int->unit) 3
    (`<rand:unit->int, printInt:int->unit>(printInt(rand())))
\end{verbatim}
We apply to the context \verb|unit->int, int->unit| in order to instantiate the context variable \verb|G|.  It is worth noting that the series variables accompanied by \verb|G| will also be replaced automatically with fresh variables.  In this case, the quote
  \verb|`<xs: G>(,u[xs]; ,body[xs])|
will turn into 
\begin{center}
  \verb|`<x: unit->int, y:int->unit>(,u[x,y]; ,body[x,y])|
\end{center}
where the series variable \verb|xs| is replaced with fresh variables \verb|x,y|.
This way, a mapping between variables and types in typing contexts is maintained.

\subsection{More Polymorphic Contexts: Combining Different Environments}
Sometimes, we might want to use pieces of code with different environments. Consider a function \verb|generic_plus|, which takes two pieces of code as arguments and returns a piece of code that sums the values of the two arguments. We can implement such a function with ease.
\begin{verbatim}
 (* val generic_plus:
         forall G H. [G |- int] -> [H |- int] -> [G, H |- int] *)
 let generic_plus G H x y = `<xs: G, ys: H>(,1(x)[xs] + ,1(y)[ys])
\end{verbatim}
The implementation is quite simple. It takes two context variables \verb|G| and \verb|H| and puts them together in the same context. As a result, we can use variables from both contexts.  Our contextual modal type theory is novel in that it permits multiple context variables in the same context, as in \verb=[G, H |- int]=.  As far as we understand, previous work that supports context polymorphism cannot express it.  We discuss differences from previous work in Section~\ref{section:relatedwork}.

%% file: 3-fcmtt.tex
\section{Simple Fitch-Style Contextual Modal Type Theory} \label{section:fcmtt}
As an introduction to contextual modal types, this section formulates simple contextual type theory \lambra without polymorphic contexts.
Nanevski et al.~\cite{Nanevski08} formulated their original contextual modal type theory in dual-context style~\cite{Pfenning01,Davies01,Kavvos17}, which has judgments with two-level contexts. In contrast, we formulate \lambra in so-called Fitch- or Kripke-style~\cite{Clouston18,Borghuis94,Martini96,Davies01,Valliappan21}. We choose this design because the Fitch-style formulation provides Lisp-like quote/unquote syntax, which is akin to that in linear-temporal type theories~\cite{Davies17,TsukadaI10LMCS}, and hence it is easier to compare these two type theories. We demonstrate a formal comparison in Section~\ref{section:embedlttt}.

We obtain \lambra by extending S4 Fitch-style modal calculus with contextual modal type theory. This combination is somewhat novel although it is fairly straightforward. For the base S4 Fitch-style modal calculus, we follow Valliappan et al.~\cite{Valliappan21}. We follow Nanevski et al.~\cite{Nanevski08} for the contextual part, while we tweak definitions for an extension to polymorphic contexts in Section~\ref{section:polyctx}.

\begin{figure}[bpt]
    \begin{tabular}{lcl}
        \textbf{Types }                 & $\lfbnt{S},\lfbnt{T}$  & $\Coloneqq \iota \mid \lfbnt{S}  \rightarrow  \lfbnt{T} \mid \lfbsym{[}  \lfbnt{C}  \vdash  \lfbnt{T}  \lfbsym{]}$                      \\
        \textbf{Contexts }              & $\lfbnt{C},\lfbnt{D}$  & $\Coloneqq  \bullet  \mid \lfbnt{C}  \lfbsym{,}  \lfbnt{T}$                                      \\
        \textbf{Stage transitions }     & $\lfbnt{k}$        & $\in \mathbb{N}$                                                                \\
        \textbf{Terms }                 & $\lfbnt{M}, \lfbnt{N}$ & $\Coloneqq \lfbmv{x} \mid  \lambda \lfbmv{x} ^{ \lfbnt{T} }. \lfbnt{M}  \mid MN \mid  \lfbkw{quo} \langle \Gamma \rangle \lfbnt{M}  \ (\text{where }\text{\faLock} \, \not\in \, \Gamma) \mid  \lfbkw{unq} _{ \lfbnt{k} } \lfbnt{M} [  \theta  ] $ \\
        \textbf{Explicit Subst.} & $\theta$       & $\Coloneqq  \bullet  \mid \theta  \lfbsym{,}  \lfbnt{M}$                                        \\
        \textbf{Named Contexts }          & $\Gamma, \Delta$ & $\Coloneqq  \bullet   \mid \Gamma  \lfbsym{,}  \lfbmv{x}  \colon  \lfbnt{T} \mid \Gamma  \lfbsym{,}  \text{\faLock}$
    \end{tabular}
    \caption{Syntax of \lambra} \label{fig:lamfbobjs}
\end{figure}

\subsection{Syntax and Type System}

Types and terms in \lambra are shown in Fig.~\ref{fig:lamfbobjs}. Types consist of base types, ranged over by $\iota$, function types $\lfbnt{S}  \rightarrow  \lfbnt{T}$, and contextual modal types $\lfbsym{[}  \lfbnt{C}  \vdash  \lfbnt{T}  \lfbsym{]}$. A contextual modal type $\lfbsym{[}  \lfbnt{C}  \vdash  \lfbnt{T}  \lfbsym{]}$ generalizes an S4 modal type $\Box\lfbnt{T}$ by adding a \textit{context} $\lfbnt{C}$, which is a finite sequence of types. It describes code of type $\lfbnt{T}$ with free variables whose types are $\lfbnt{C}$. Note that a contextual modal type with the empty context $\lfbsym{[}   \bullet   \vdash  \lfbnt{T}  \lfbsym{]}$ has the same meaning as $\Box\lfbnt{T}$, which denotes closed code of type $\lfbnt{T}$. In addition to standard terms of simply typed lambda calculus, \lambra has two forms, quote $ \lfbkw{quo} \langle \Gamma \rangle \lfbnt{M} $ and unquote $ \lfbkw{unq} _{ \lfbnt{k} } \lfbnt{M} [  \theta  ] $.
% We gave an informal idea of those terms in Section~\ref{section:motivation} while we used different notations.
We define stage transitions as natural numbers, and explicit substitutions as sequences of terms.

We often use the word \emph{named contexts} for typing contexts with variables and use ``contexts'' for type-only ones. Similarly to other Fitch-style formulations, \lambra extends named contexts with a lock operator \faLock{} that delimits variables declared inside quotes. A named context is well formed iff the variables in it do not have duplication; we assume that all named contexts are well formed.  We also assume that the named context in a quote to be single-level, i.e., not to contain $ \text{\faLock} $.  We denote $ \mathsf{rg} ( \Gamma ) $ for a context that we get by forgetting variables in $\Gamma$ where $\text{\faLock} \, \not\in \, \Gamma$, and $\mathsf{dom} \, \lfbsym{(}  \Gamma  \lfbsym{)}$ for the set of variables in $\Gamma$ (a lock can appear in $\Gamma$). We also define the weakening relation $\Gamma_{{\mathrm{1}}}  \leq  \Gamma_{{\mathrm{2}}}$ as follows.

\begin{center}
  \bottomAlignProof
  \AxiomC{}
    \UnaryInfC{$ \bullet   \leq   \bullet $}
    \DisplayProof
    \quad
  \bottomAlignProof
    \AxiomC{$\Gamma_{{\mathrm{1}}}  \leq  \Gamma_{{\mathrm{2}}}$}
    \UnaryInfC{$\Gamma_{{\mathrm{1}}}  \lfbsym{,}  \lfbmv{x}  \colon  \lfbnt{T}  \leq  \Gamma_{{\mathrm{2}}}  \lfbsym{,}  \lfbmv{x}  \colon  \lfbnt{T}$}
    \DisplayProof
    \quad
  \bottomAlignProof
    \AxiomC{$\Gamma_{{\mathrm{1}}}  \leq  \Gamma_{{\mathrm{2}}}$}
    \UnaryInfC{$\Gamma_{{\mathrm{1}}}  \leq  \Gamma_{{\mathrm{2}}}  \lfbsym{,}  \lfbmv{x}  \colon  \lfbnt{T}$}
    \DisplayProof
    \quad
  \bottomAlignProof
    \AxiomC{$\Gamma_{{\mathrm{1}}}  \leq  \Gamma_{{\mathrm{2}}}$}
    \UnaryInfC{$\Gamma_{{\mathrm{1}}}  \lfbsym{,}  \text{\faLock}  \leq  \Gamma_{{\mathrm{2}}}  \lfbsym{,}  \text{\faLock}$}
    \DisplayProof
\end{center}

\begin{figure}[bpt]
    \framebox{\mbox{$ \mathsf{FV} _{ \lfbnt{k} }( \lfbnt{M} ) $}} \framebox{\mbox{$ \mathsf{FV} _{ \lfbnt{k} }( \theta ) $}}
    \begin{align*}
         \mathsf{FV} _{ \lfbnt{k} }( \lfbmv{x} )            & = \begin{cases}
                                      {\lfbmv{x}} & \text{if $\lfbnt{k} = 0$} \\
                                      \emptyset   & \text{otherwise}
                                  \end{cases}      \\
         \mathsf{FV} _{ \lfbnt{k} }(  \lambda \lfbmv{x} ^{ \lfbnt{T} }. \lfbnt{M}  )       & = \begin{cases}
                                       \mathsf{FV} _{ \lfbnt{k} }( \lfbnt{M} )   \lfbsym{-}  \lfbsym{\{}  \lfbmv{x}  \lfbsym{\}} & \text{if $\lfbnt{k} = 0$} \\
                                       \mathsf{FV} _{ \lfbnt{k} }( \lfbnt{M} )         & \text{otherwise}
                                  \end{cases}                             \\
         \mathsf{FV} _{ \lfbnt{k} }( \lfbnt{M} \, \lfbnt{N} )          & =  \mathsf{FV} _{ \lfbnt{k} }( \lfbnt{M} )  \, \cup \,  \mathsf{FV} _{ \lfbnt{k} }( \lfbnt{N} )                                           \\
         \mathsf{FV} _{ \lfbnt{k} }(  \lfbkw{quo} \langle \Gamma \rangle \lfbnt{M}  )        & =  \mathsf{FV} _{ \lfbnt{k}  \lfbsym{+}  \lfbsym{1} }( \lfbnt{M} )                                                       \\
         \mathsf{FV} _{ \lfbnt{k_{{\mathrm{2}}}} }(  \lfbkw{unq} _{ \lfbnt{k_{{\mathrm{1}}}} } \lfbnt{M} [  \theta  ]  )  & =  \mathsf{FV} _{ \lfbnt{k_{{\mathrm{2}}}}  \lfbsym{-}  \lfbnt{k_{{\mathrm{1}}}} }( \lfbnt{M} )  \, \cup \,  \mathsf{FV} _{ \lfbnt{k_{{\mathrm{2}}}} }( \theta )  \\[1ex]
         \mathsf{FV} _{ \lfbnt{k} }(  \bullet  )  & = \emptyset                  \qquad\qquad
         \mathsf{FV} _{ \lfbnt{k} }( \theta  \lfbsym{,}  \lfbnt{M} )      =  \mathsf{FV} _{ \lfbnt{k} }( \theta )  \, \cup \,  \mathsf{FV} _{ \lfbnt{k} }( \lfbnt{M} ) 
    \end{align*}
    \caption{Free variables} \label{fig:deffv}
\end{figure}

Similarly to other Fitch-style formulations, \lamfb has a somewhat complex binding structure. We show the definition of free variables in Fig.~\ref{fig:deffv}. The notion of free variables is extended to have levels. For a term $\lfbnt{M}$ and integer $\lfbnt{k}$, $ \mathsf{FV} _{ \lfbnt{k} }( \lfbnt{M} ) $ is a set of free variables in $\lfbnt{M}$ at level $\lfbnt{k}$, which roughly stands for the number of quotes surrounding $\lfbnt{M}$.  Since an unquote $ \lfbkw{unq} _{ \lfbnt{k_{{\mathrm{1}}}} } \lfbnt{M} [  \theta  ] $ cancels $k$ surrounding quotes, the level is lowered by $\lfbnt{k_{{\mathrm{1}}}}$.  \lambra has two binding forms: A lambda abstraction $ \lambda \lfbmv{x} ^{ \lfbnt{T} }. \lfbnt{M} $ binds all level-0 free occurrences of $\lfbmv{x}$ in $\lfbnt{M}$ and a quote $ \lfbkw{quo} \langle \Gamma \rangle \lfbnt{M} $ (where $\Gamma$ is \faLock-free) binds all level-0 free variables from $\Gamma$ in $\lfbnt{M}$.  According to these binding forms, we  define $\alpha$-equivalence (but omit its definition).  For example, $ \lambda \lfbmv{x} ^{ \lfbsym{[}   \lfbnt{T_{{\mathrm{1}}}}   \vdash  \lfbnt{T_{{\mathrm{2}}}}  \lfbsym{]} }.  \lfbkw{quo} \langle  \lfbmv{x}  \colon  \lfbnt{T_{{\mathrm{3}}}}  \rangle \lfbsym{(}  \lfbmv{x} \, \lfbsym{(}   \lfbkw{unq} _{ \lfbsym{1} } \lfbsym{(}  \lfbmv{x}  \lfbsym{)} [   \lfbmv{y}   ]   \lfbsym{)}  \lfbsym{)}  $ is $\alpha$-equivalent to
$ \lambda z ^{ \lfbsym{[}   \lfbnt{T_{{\mathrm{1}}}}   \vdash  \lfbnt{T_{{\mathrm{2}}}}  \lfbsym{]} }.  \lfbkw{quo} \langle  \lfbmv{x}  \colon  \lfbnt{T_{{\mathrm{3}}}}  \rangle \lfbsym{(}  \lfbmv{x} \, \lfbsym{(}   \lfbkw{unq} _{ \lfbsym{1} } \lfbsym{(}  z  \lfbsym{)} [   \lfbmv{y}   ]   \lfbsym{)}  \lfbsym{)}  $. As we shall see later, the typing rules of \lambra enforces well typed terms to be closed with regard to negative-level free variables. Thus, we only care about positive-level free variables in this paper and assume that the meta variable $k$ ranges over natural numbers.
% Note that typing rules enforce all level-0 free variables in $\lfbnt{M}$ to be bound by $\Gamma$ and capture-avoiding substitutions are applied only to well typed terms. Thus, we do not have to consider free variables of negative levels.

Typing rules are given in Fig.~\ref{fig:typingrulesoffcmtt}. The judgment $\lfbnt{k}  \colon  \Gamma  \lhd  \Delta$ states that there is a stage transition $\lfbnt{k}$ between two named contexts $\Gamma$ and $\Delta$. The rules mean that $\lfbnt{k}$ is the number of locks between $\Gamma$ and $\Delta$, e.g., $\lfbsym{0}  \colon   \lfbmv{x}  \colon  \lfbnt{T}   \lhd   \lfbmv{x}  \colon  \lfbnt{T} $ and $\lfbsym{2}  \colon   \lfbmv{y}  \colon  \lfbnt{T_{{\mathrm{1}}}}   \lhd   \lfbmv{y}  \colon  \lfbnt{T_{{\mathrm{1}}}}   \lfbsym{,}  \text{\faLock}  \lfbsym{,}  \text{\faLock}  \lfbsym{,}  z  \colon  \lfbnt{T_{{\mathrm{2}}}}$. The judgments $\Gamma  \vdash  \lfbnt{M}  \colon  \lfbnt{T}$ and $\Gamma  \vdash  \theta  \colon  \lfbnt{C}$ states that term $\lfbnt{M}$ has type $\lfbnt{T}$, or explicit substitution $\theta$ has context $\lfbnt{C}$ under named context $\Gamma$, respectively. The rules for variable $\lfbmv{x}$, lambda abstraction $ \lambda \lfbmv{x} ^{ \lfbnt{T} }. \lfbnt{M} $, and application $\lfbnt{M_{{\mathrm{1}}}} \, \lfbnt{M_{{\mathrm{2}}}}$ are almost the same as those in simply typed lambda calculus, except that we only care about variables from $\mathsf{head} \, \lfbsym{(}  \Gamma  \lfbsym{)}$, the level-0 part of $\Gamma$.  The type of a quote $ \lfbkw{quo} \langle \Gamma \rangle \lfbnt{M} $ is derived by popping all level-0 variables in the named context.  (Recall $\text{\faLock} \, \not\in \, \Gamma$.) Thus, $\Gamma$ binds all level-0 free variables in $\lfbnt{M}$. An unquote $ \lfbkw{unq} _{ \lfbnt{k} } \lfbnt{M} [  \theta  ] $ uses $\theta$ as a substitution for the context $\lfbnt{C}$, and $\lfbnt{k}$ as the stage transitions between $\lfbnt{M}$ and $\theta$. We call a judgment is \textit{evident} when it is derived from these typing rules. We assume that judgments in this paper are evident if not stated explicitly.

\begin{figure}[bpt]
    \framebox{\mbox{$\lfbnt{k}  \colon  \Gamma  \lhd  \Delta$}}
    \[
  \bottomAlignProof
        \AxiomC{}
        \UnaryInfC{$\lfbsym{0}  \colon  \Gamma  \lhd  \Gamma$}
        \DisplayProof
        \quad
  \bottomAlignProof
        \AxiomC{$\lfbnt{k}  \colon  \Gamma  \lhd  \Delta$}
        \UnaryInfC{$\lfbnt{k}  \colon  \Gamma  \lhd  \Delta  \lfbsym{,}  \lfbmv{x}  \colon  \lfbnt{T}$}
        \DisplayProof
  \bottomAlignProof
        \quad
        \AxiomC{$\lfbnt{k}  \colon  \Gamma  \lhd  \Delta$}
        \UnaryInfC{$\lfbnt{k}  \lfbsym{+}  \lfbsym{1}  \colon  \Gamma  \lhd  \Delta  \lfbsym{,}  \text{\faLock}$}
        \DisplayProof
    \]

    \framebox{\mbox{$\Gamma  \vdash  \lfbnt{M}  \colon  \lfbnt{T}$}}
    \framebox{\mbox{$\Gamma  \vdash  \theta  \colon  \lfbnt{C}$}}
    \[
        \AxiomC{$\lfbmv{x}  \colon  \lfbnt{T} \, \in \, \mathsf{head} \, \lfbsym{(}  \Gamma  \lfbsym{)}$}
        \UnaryInfC{$\Gamma  \vdash  \lfbmv{x}  \colon  \lfbnt{T}$}
        \DisplayProof
        \quad
        \AxiomC{$\Gamma  \lfbsym{,}  \lfbmv{x}  \colon  \lfbnt{T_{{\mathrm{1}}}}  \vdash  \lfbnt{M}  \colon  \lfbnt{T_{{\mathrm{2}}}}$}
        \UnaryInfC{$\Gamma  \vdash   \lambda \lfbmv{x} ^{ \lfbnt{T_{{\mathrm{1}}}} }. \lfbnt{M}   \colon  \lfbnt{T_{{\mathrm{1}}}}  \rightarrow  \lfbnt{T_{{\mathrm{2}}}}$}
        \DisplayProof
        \quad
        \AxiomC{$\Gamma  \vdash  \lfbnt{M_{{\mathrm{1}}}}  \colon  \lfbnt{T_{{\mathrm{1}}}}  \rightarrow  \lfbnt{T_{{\mathrm{2}}}}$}
        \AxiomC{$\Gamma  \vdash  \lfbnt{M_{{\mathrm{2}}}}  \colon  \lfbnt{T_{{\mathrm{1}}}}$}
        \BinaryInfC{$\Gamma  \vdash  \lfbnt{M_{{\mathrm{1}}}} \, \lfbnt{M_{{\mathrm{2}}}}  \colon  \lfbnt{T_{{\mathrm{2}}}}$}
        \DisplayProof
        \]
        \[
  \bottomAlignProof
        \AxiomC{$ \Gamma  \lfbsym{,}  \text{\faLock} \lfbsym{,} \Delta   \vdash  \lfbnt{M}  \colon  \lfbnt{T}$}
%        \AxiomC{$\text{\faLock} \, \not\in \, \Delta$}
        \UnaryInfC{$\Gamma  \vdash   \lfbkw{quo} \langle \Delta \rangle \lfbnt{M}   \colon  \lfbsym{[}   \mathsf{rg} ( \Delta )   \vdash  \lfbnt{T}  \lfbsym{]}$}
        \DisplayProof
        \quad
  \bottomAlignProof
        \AxiomC{$\Gamma  \vdash  \lfbnt{M}  \colon  \lfbsym{[}  \lfbnt{C}  \vdash  \lfbnt{T}  \lfbsym{]}$}
        \AxiomC{$\Delta  \vdash  \theta  \colon  \lfbnt{C}$}
        \AxiomC{$\lfbnt{k}  \colon  \Gamma  \lhd  \Delta$}
        \TrinaryInfC{$\Delta  \vdash   \lfbkw{unq} _{ \lfbnt{k} } \lfbnt{M} [  \theta  ]   \colon  \lfbnt{T}$}
        \DisplayProof
    \]
    \[
  \bottomAlignProof
        \AxiomC{}
        \UnaryInfC{$\Gamma  \vdash   \bullet   \colon   \bullet $}
        \DisplayProof
        \quad
  \bottomAlignProof
        \AxiomC{$\Gamma  \vdash  \theta  \colon  \lfbnt{C}$}
        \AxiomC{$\Gamma  \vdash  \lfbnt{M}  \colon  \lfbnt{T}$}
        \BinaryInfC{$\Gamma  \vdash  \theta  \lfbsym{,}  \lfbnt{M}  \colon  \lfbnt{C}  \lfbsym{,}  \lfbnt{T}$}
        \DisplayProof
    \]

    \textbf{Auxiliary function}
    \begin{align*}
        \mathsf{head} \, \lfbsym{(}   \bullet   \lfbsym{)} &=  \bullet   &
        \mathsf{head} \, \lfbsym{(}  \Gamma  \lfbsym{,}  \text{\faLock}  \lfbsym{)} &=  \bullet    &
        \mathsf{head} \, \lfbsym{(}  \Gamma  \lfbsym{,}  \lfbmv{x}  \colon  \lfbnt{T}  \lfbsym{)} &= \mathsf{head} \, \lfbsym{(}  \Gamma  \lfbsym{)}  \lfbsym{,}  \lfbmv{x}  \colon  \lfbnt{T}
    \end{align*}

    \caption{Typing rules of \lambra}
    \label{fig:typingrulesoffcmtt}
\end{figure}

\subsection{Substitution}
\begin{figure}[tb]
    \begin{tabular}{lll}
        \textbf{Substitution} & $\sigma$ & $\Coloneqq  \bullet  \mid \sigma  \lfbsym{,}  \lfbmv{x}  \coloneqq  \lfbnt{M} \mid \sigma  \lfbsym{,}   \text{\faLock} _{ \lfbnt{k} } $
    \end{tabular}
    \vspace{0em}

    \framebox{\mbox{$\Gamma  \vdash  \sigma  \colon  \Delta$}}
    \[
  \bottomAlignProof
        \AxiomC{}
        \UnaryInfC{$\Gamma  \vdash   \bullet   \colon   \bullet $}
        \DisplayProof
        \quad
  \bottomAlignProof
        \AxiomC{$\Gamma_{{\mathrm{1}}}  \vdash  \sigma  \colon  \Delta$}
        \AxiomC{$\lfbnt{k}  \colon  \Gamma_{{\mathrm{1}}}  \lhd  \Gamma_{{\mathrm{2}}}$}
        \BinaryInfC{$\Gamma_{{\mathrm{2}}}  \vdash  \sigma  \lfbsym{,}   \text{\faLock} _{ \lfbnt{k} }   \colon  \Delta  \lfbsym{,}  \text{\faLock}$}
        \DisplayProof
        \quad
  \bottomAlignProof
        \AxiomC{$\Gamma  \vdash  \sigma  \colon  \Delta$}
        \AxiomC{$\Gamma  \vdash  \lfbnt{M}  \colon  \lfbnt{T}$}
        \BinaryInfC{$\Gamma  \vdash  \sigma  \lfbsym{,}  \lfbmv{x}  \coloneqq  \lfbnt{M}  \colon  \Delta  \lfbsym{,}  \lfbmv{x}  \colon  \lfbnt{T}$}
        \DisplayProof
    \]

    \mbox{\fbox{$\lfbnt{M}  \lfbsym{[}  \sigma  \lfbsym{]}$} \fbox{$\theta  \lfbsym{[}  \sigma  \lfbsym{]}$}}
    \begin{align*}
        \lfbmv{x}  \lfbsym{[}  \sigma  \lfbsym{]}            & = \begin{cases}
                                        M & \text{if $\lfbmv{x}  \coloneqq  \lfbnt{M} \, \in \, \mathsf{head} \, \lfbsym{(}  \sigma  \lfbsym{)}$} \\
                                        x & \text{otherwise}                      \\
                                    \end{cases}                                              \\
        \lfbsym{(}   \lambda \lfbmv{x} ^{ \lfbnt{T} }. \lfbnt{M}   \lfbsym{)}  \lfbsym{[}  \sigma  \lfbsym{]}    & =  \lambda \lfbmv{x} ^{ \lfbnt{T} }. \lfbsym{(}  \lfbnt{M}  \lfbsym{[}  \sigma  \lfbsym{]}  \lfbsym{)}  \text{ \textbf{where} $\lfbmv{x} \, \not\in \, \mathsf{dom} \, \lfbsym{(}  \mathsf{head} \, \lfbsym{(}  \sigma  \lfbsym{)}  \lfbsym{)}$ and $ \lfbmv{x} \, \not\in \,  \mathsf{FV} _{ \lfbsym{0} }( \sigma )  $} \\
        \lfbsym{(}  \lfbnt{M} \, \lfbnt{N}  \lfbsym{)}  \lfbsym{[}  \sigma  \lfbsym{]}        & = \lfbsym{(}  \lfbnt{M}  \lfbsym{[}  \sigma  \lfbsym{]}  \lfbsym{)} \, \lfbsym{(}  \lfbnt{N}  \lfbsym{[}  \sigma  \lfbsym{]}  \lfbsym{)}                                                                                \\
        \lfbsym{(}   \lfbkw{quo} \langle \Gamma \rangle \lfbnt{M}   \lfbsym{)}  \lfbsym{[}  \sigma  \lfbsym{]}      & =  \lfbkw{quo} \langle \Gamma \rangle \lfbsym{(}  \lfbnt{M}  \lfbsym{[}   \sigma  \lfbsym{,}   \text{\faLock} _{ \lfbsym{1} }  \lfbsym{,}  id _{ \Gamma }    \lfbsym{]}  \lfbsym{)}                                                                 \\
        \lfbsym{(}   \lfbkw{unq} _{ \lfbnt{k} } \lfbnt{M} [  \theta  ]   \lfbsym{)}  \lfbsym{[}  \sigma  \lfbsym{]} & =  \lfbkw{unq} _{ \lfbsym{(}   \mathsf{count} ( \lfbnt{k} , \sigma )   \lfbsym{)} } \lfbsym{(}  \lfbnt{M}  \lfbsym{[}  \sigma  \uparrow  \lfbnt{k}  \lfbsym{]}  \lfbsym{)} [  \theta  \lfbsym{[}  \sigma  \lfbsym{]}  ]  \\[1ex]
         \bullet   \lfbsym{[}  \sigma  \lfbsym{]} & =  \bullet     \qquad\qquad
        \lfbsym{(}  \theta  \lfbsym{,}  \lfbnt{M}  \lfbsym{)}  \lfbsym{[}  \sigma  \lfbsym{]}    = \theta  \lfbsym{[}  \sigma  \lfbsym{]}  \lfbsym{,}  \lfbnt{M}  \lfbsym{[}  \sigma  \lfbsym{]}
    \end{align*}

    \mbox{\textbf{Auxiliary functions}}

    \centering
    \begin{tabular}{cc}
        $ \mathsf{FV} _{ \lfbnt{k} }( \sigma  \lfbsym{,}  \lfbmv{x}  \coloneqq  \lfbnt{M} )  =  \mathsf{FV} _{ \lfbnt{k} }( \sigma )  \, \cup \,  \mathsf{FV} _{ \lfbnt{k} }( \lfbnt{M} ) $ & {
            $ \mathsf{FV} _{ \lfbnt{k_{{\mathrm{2}}}} }( \sigma  \lfbsym{,}   \text{\faLock} _{ \lfbnt{k_{{\mathrm{1}}}} }  )  = \begin{cases}
                 \mathsf{FV} _{ \lfbnt{k_{{\mathrm{2}}}}  \lfbsym{-}  \lfbnt{k_{{\mathrm{1}}}} }( \sigma )  & \text{if $\lfbnt{k_{{\mathrm{2}}}} \ge \lfbnt{k_{{\mathrm{1}}}}$} \\
                \emptyset             & \text{otherwise}
            \end{cases}$
        }
    \end{tabular}

    \begin{minipage}[t]{0.45\columnwidth}\small
    \centering
        \begin{align*}
            \mathsf{head} \, \lfbsym{(}  \sigma  \lfbsym{,}  \lfbmv{x}  \coloneqq  \lfbnt{M}  \lfbsym{)}            & = \mathsf{head} \, \lfbsym{(}  \sigma  \lfbsym{)}  \lfbsym{,}  \lfbmv{x}  \coloneqq  \lfbnt{M} \\
            \mathsf{head} \, \lfbsym{(}  \sigma  \lfbsym{,}   \text{\faLock} _{ \lfbnt{k} }   \lfbsym{)}            & =  \bullet  \\
             \mathsf{count} ( \lfbsym{0} , \sigma )                   & = \lfbsym{0}                 \\
             \mathsf{count} ( \lfbsym{(}  \lfbnt{k_{{\mathrm{1}}}}  \lfbsym{+}  \lfbsym{1}  \lfbsym{)} ,  \bullet  )      & = \lfbnt{k_{{\mathrm{1}}}}  \lfbsym{+}  \lfbsym{1}              \\
             \mathsf{count} ( \lfbsym{(}  \lfbnt{k}  \lfbsym{+}  \lfbsym{1}  \lfbsym{)} , \lfbsym{(}  \sigma  \lfbsym{,}  \lfbmv{x}  \coloneqq  \lfbnt{M}  \lfbsym{)} )    & =  \mathsf{count} ( \lfbnt{k} , \sigma )        \\
             \mathsf{count} ( \lfbsym{(}  \lfbnt{k_{{\mathrm{1}}}}  \lfbsym{+}  \lfbsym{1}  \lfbsym{)} , \lfbsym{(}  \sigma  \lfbsym{,}   \text{\faLock} _{ \lfbnt{k_{{\mathrm{2}}}} }   \lfbsym{)} )  & =  \mathsf{count} ( \lfbnt{k_{{\mathrm{1}}}} , \sigma )  + \lfbnt{k_{{\mathrm{2}}}}
        \end{align*}
    \end{minipage}
    \hfill
    \begin{minipage}[t]{0.45\columnwidth}\small
        \centering
        \begin{align*}
             id _{  \bullet  }  & =  \bullet     \\
             id _{ \Gamma  \lfbsym{,}  \lfbmv{x}  \colon  \lfbnt{T} }  & =  id _{ \Gamma }   \lfbsym{,}  \lfbmv{x}  \coloneqq  \lfbmv{x} \\
             id _{ \Gamma  \lfbsym{,}  \text{\faLock} }   & =  id _{ \Gamma }   \lfbsym{,}   \text{\faLock} _{ \lfbsym{1} }  \\
            \sigma  \uparrow  \lfbsym{0}                      & = \sigma               \\
             \bullet   \uparrow  \lfbsym{(}  \lfbnt{k}  \lfbsym{+}  \lfbsym{1}  \lfbsym{)}         & =  \bullet           \\
            \lfbsym{(}  \sigma  \lfbsym{,}  \lfbmv{x}  \coloneqq  \lfbnt{M}  \lfbsym{)}  \uparrow  \lfbsym{(}  \lfbnt{k}  \lfbsym{+}  \lfbsym{1}  \lfbsym{)}      & = \sigma  \uparrow  \lfbsym{(}  \lfbnt{k}  \lfbsym{+}  \lfbsym{1}  \lfbsym{)}     \\
            \lfbsym{(}  \sigma  \lfbsym{,}   \text{\faLock} _{ \lfbnt{k_{{\mathrm{1}}}} }   \lfbsym{)}  \uparrow  \lfbsym{(}  \lfbnt{k_{{\mathrm{2}}}}  \lfbsym{+}  \lfbsym{1}  \lfbsym{)}    & = \sigma  \uparrow  \lfbnt{k_{{\mathrm{2}}}}
        \end{align*}
    \end{minipage}

    \caption{Substitution}
    \label{fig:defsubstitution}
\end{figure}

We define substitution on terms and explicit substitutions. We follow the style of Valliappan et al.~\cite{Valliappan21}, which propose simultaneous substitution on all free variables with any level. We provide definitions related to substitutions in Fig.~\ref{fig:defsubstitution}.

A substitution typing judgment $\Gamma  \vdash  \sigma  \colon  \Delta$ denotes that we can replace a named context $\Delta$ with another $\Gamma$ by applying a substitution $\sigma$. A lock substitution $ \text{\faLock} _{ \lfbnt{k} } $ has two roles. First, they provide information on the level of free variables to be substituted. For example, if $\sigma =  \sigma_{{\mathrm{1}}}  \lfbsym{,}   \text{\faLock} _{ \lfbnt{k} }  \lfbsym{,} \sigma_{{\mathrm{2}}} $ where $\sigma_{{\mathrm{2}}}$ does not have lock substitutions, $\sigma_{{\mathrm{2}}}$ substitutes level-0 free variables, and $\sigma_{{\mathrm{1}}}$ substitutes higher-level free variables. Second, they replace lock themselves. If $\sigma$ has a lock substitution $ \text{\faLock} _{ \lfbnt{k} } $, it means that it replaces a lock in $\Delta$ into $\lfbnt{k}$ locks in $\Gamma$.

Substitution application on terms $\lfbnt{M}  \lfbsym{[}  \sigma  \lfbsym{]}$ on explicit substitutions $\theta  \lfbsym{[}  \sigma  \lfbsym{]}$ performs actual substitution operations. They are defined to satisfy the following lemma, which is expected by the intuition of the substitution typing.

\begin{lemma}[Substitution Lemma]
    If $\Gamma  \vdash  \lfbnt{M}  \colon  \lfbnt{T}$ and $\Delta  \vdash  \sigma  \colon  \Gamma$, then $\Delta  \vdash  \lfbnt{M}  \lfbsym{[}  \sigma  \lfbsym{]}  \colon  \lfbnt{T}$.
\end{lemma}

Let us consider an example of substituting variables in $\Gamma  \vdash  \lfbsym{(}   \lfbkw{unq} _{ \lfbsym{1} } \lfbsym{(}  \lfbmv{x}  \lfbsym{)} [   \lfbmv{y}   ]   \lfbsym{)} \, \lfbmv{y}  \colon  \lfbnt{T}$, where $\Gamma =  \lfbmv{x}  \colon  \lfbsym{[}   \lfbnt{S}   \vdash  \lfbnt{S}  \rightarrow  \lfbnt{T}  \lfbsym{]}   \lfbsym{,}  \text{\faLock}  \lfbsym{,}  \lfbmv{y}  \colon  \lfbnt{S}$. We can construct the following substitution that provides a term for each variable in $\Gamma$.
\[
     \lfbmv{x'}  \colon  \lfbsym{[}   \lfbnt{S}   \vdash  \lfbnt{S}  \rightarrow  \lfbnt{T}  \lfbsym{]}   \lfbsym{,}  z  \colon  \lfbnt{S}  \rightarrow  \lfbnt{S}  \lfbsym{,}  w  \colon  \lfbnt{S}  \vdash  \lfbsym{(}   \lfbmv{x}  \coloneqq  \lfbmv{x'}   \lfbsym{,}   \text{\faLock} _{ \lfbsym{0} }   \lfbsym{,}  \lfbmv{y}  \coloneqq  z \, w  \lfbsym{)}  \colon  \Gamma
\]
This substitution replaces level-0 occurrences of $\lfbmv{y}$ to $z \, w$ and level-1 occurrences of $\lfbmv{x}$ to $\lfbmv{x'}$. $  \text{\faLock} _{ \lfbsym{0} }  $ in the substitution denotes that level-1 free variables of target terms are mapped to level-0 terms; that is why the level-0 term $\lfbmv{x'}$ is supplied for the level-1 variable $\lfbmv{x}$. We can observe that the substitution is applied as follows.
\begin{eqnarray}
   & & \lfbsym{(}  \lfbsym{(}   \lfbkw{unq} _{ \lfbsym{1} } \lfbsym{(}  \lfbmv{x}  \lfbsym{)} [   \lfbmv{y}   ]   \lfbsym{)} \, \lfbmv{y}  \lfbsym{)}  \lfbsym{[}   \lfbmv{x}  \coloneqq  \lfbmv{x'}   \lfbsym{,}   \text{\faLock} _{ \lfbsym{0} }   \lfbsym{,}  \lfbmv{y}  \coloneqq  z \, w  \lfbsym{]} \\
   & = & \lfbsym{(}   \lfbkw{unq} _{ \lfbsym{1} } \lfbsym{(}  \lfbmv{x}  \lfbsym{)} [   \lfbmv{y}   ]   \lfbsym{)}  \lfbsym{[}   \lfbmv{x}  \coloneqq  \lfbmv{x'}   \lfbsym{,}   \text{\faLock} _{ \lfbsym{0} }   \lfbsym{,}  \lfbmv{y}  \coloneqq  z \, w  \lfbsym{]} \, \lfbsym{(}  \lfbmv{y}  \lfbsym{[}   \lfbmv{x}  \coloneqq  \lfbmv{x'}   \lfbsym{,}   \text{\faLock} _{ \lfbsym{0} }   \lfbsym{,}  \lfbmv{y}  \coloneqq  z \, w  \lfbsym{]}  \lfbsym{)} \\
   & = & \lfbsym{(}   \lfbkw{unq} _{ \lfbsym{0} } \lfbsym{(}  \lfbmv{x}  \lfbsym{[}   \lfbmv{x}  \coloneqq  \lfbmv{x'}   \lfbsym{]}  \lfbsym{)} [    \lfbmv{y}  \lfbsym{[}   \lfbmv{x}  \coloneqq  \lfbmv{x'}   \lfbsym{,}   \text{\faLock} _{ \lfbsym{0} }   \lfbsym{,}  \lfbmv{y}  \coloneqq  z \, w  \lfbsym{]}    ]   \lfbsym{)} \,  \lfbsym{(}  \lfbmv{y}  \lfbsym{[}   \lfbmv{x}  \coloneqq  \lfbmv{x'}   \lfbsym{,}   \text{\faLock} _{ \lfbsym{0} }   \lfbsym{,}  \lfbmv{y}  \coloneqq  z \, w  \lfbsym{]}  \lfbsym{)}  \\
   & = & \lfbsym{(}   \lfbkw{unq} _{ \lfbsym{0} } \lfbsym{(}  \lfbmv{x'}  \lfbsym{)} [   z \, w   ]   \lfbsym{)} \, \lfbsym{(}  z \, w  \lfbsym{)}
\end{eqnarray}
The most interesting equation is the one from (2) to (3). The substitution to $\lfbmv{x}$ is shifted by 1 level and the stage transition of the unquote changes from $\lfbsym{1}$ to $\lfbsym{0}$ to align staging levels. The resulting term is given type $\lfbnt{T}$ under the new named context as the substitution lemma states.

We can confirm that an identity substitution $ id _{ \Gamma } $ does not affect the result of substitution. We take advantage of this property to define reduction later.

\begin{lemma}\label{lem:idsub}
    $\lfbnt{M}  \lfbsym{[}  \sigma  \lfbsym{]} = \lfbnt{M}  \lfbsym{[}    id _{ \Gamma }  \lfbsym{,} \sigma   \lfbsym{]}$ for any $\Gamma$.
\end{lemma}

\subsection{Local Soundness/Completeness and Reduction}
According to Pfenning and Davies~\cite{Pfenning01}, the introduction and elimination rules for a type constructor should satisfy local soundness and local completeness. We confirm that contextual modal types satisfy those conditions.

Local soundness states that the elimination rule is not too strong. For the case of contextual modal types, we can witness it by the following local reduction where we obtain the derivation $\mathcal{D}'$ by application of the substitution $[  id _{ \Gamma }   \lfbsym{,}   \text{\faLock} _{ \lfbnt{k} }   \lfbsym{,}  \Delta  \coloneqq  \theta ]$, which we obtain from $\mathcal{E}$ and $\lfbnt{k}  \colon  \Gamma  \lhd  \Gamma'$. Here, $\overrightarrow{x}$ denotes a sequence of variables; $\overrightarrow{x}  \colon  \lfbnt{C}$ and $\overrightarrow{x}  \coloneqq  \theta$ denote a \faLock-free named context that maps each variable in $\overrightarrow{x}$ to each type in $\lfbnt{C}$ and a substitution that maps each variable in $\overrightarrow{x}$ to each term in $\theta$, respectively.

\begin{center}
    {\small
    \def\ScoreOverhang{1pt}
    \bottomAlignProof
    \AxiomC{$\mathcal{D}$}
    \noLine
    \UnaryInfC{$\Gamma  \lfbsym{,}  \text{\faLock}  \lfbsym{,}  \overrightarrow{x}  \colon  \lfbnt{C}  \vdash  \lfbnt{M}  \colon  \lfbnt{T}$}
    \UnaryInfC{$\Gamma  \vdash   \lfbkw{quo} \langle  \overrightarrow{x}  \colon  \lfbnt{C}  \rangle \lfbnt{M}   \colon  \lfbsym{[}  \lfbnt{C}  \vdash  \lfbnt{T}  \lfbsym{]}$}
    \AxiomC{$\mathcal{E}$}
    \noLine
    \UnaryInfC{$\Gamma'  \vdash  \theta  \colon  \lfbnt{C}$}
    \AxiomC{$\lfbnt{k}  \colon  \Gamma  \lhd  \Gamma'$}
    \insertBetweenHyps{\hskip 1em}
    \TrinaryInfC{$\Gamma'  \vdash   \lfbkw{unq} _{ \lfbnt{k} } \lfbsym{(}   \lfbkw{quo} \langle  \overrightarrow{x}  \colon  \lfbnt{C}  \rangle \lfbnt{M}   \lfbsym{)} [  \theta  ]   \colon  \lfbnt{T}$}
    \DisplayProof
    $\Rightarrow$
    \bottomAlignProof
    \AxiomC{$\mathcal{D}'$}
    \noLine
    \UnaryInfC{$\Gamma'  \vdash  \lfbnt{M}  \lfbsym{[}   id _{ \Gamma }   \lfbsym{,}   \text{\faLock} _{ \lfbnt{k} }   \lfbsym{,}  \overrightarrow{x}  \coloneqq  \theta  \lfbsym{]}  \colon  \lfbnt{T}$}
    \DisplayProof
}
\end{center}

Local completeness states that the elimination rule is sufficiently strong. We can confirm this condition by the following local expansion.

\begin{center}
    {\small
    \def\ScoreOverhang{1pt}
    \bottomAlignProof
    \AxiomC{$\mathcal{D}$}
    \noLine
    \UnaryInfC{$\Gamma  \vdash  \lfbnt{M}  \colon  \lfbsym{[}  \lfbnt{C}  \vdash  \lfbnt{T}  \lfbsym{]}$}
    \DisplayProof
    $\Rightarrow$
    \bottomAlignProof
    \AxiomC{$\mathcal{D}$}
    \noLine
    \UnaryInfC{$\Gamma  \vdash  \lfbnt{M}  \colon  \lfbsym{[}  \lfbnt{C}  \vdash  \lfbnt{T}  \lfbsym{]}$}
    \AxiomC{\vdots}
    \noLine
    \UnaryInfC{$\Gamma  \lfbsym{,}  \text{\faLock}  \lfbsym{,}  \overrightarrow{x}  \colon  \lfbnt{C}  \vdash   \overrightarrow{x}   \colon  \lfbnt{C}$}
    \AxiomC{\vdots}
    \noLine
    \UnaryInfC{$\lfbsym{1}  \colon  \Gamma  \lhd  \Gamma  \lfbsym{,}  \text{\faLock}  \lfbsym{,}  \overrightarrow{x}  \colon  \lfbnt{C}$}
    \insertBetweenHyps{\hskip 1em}
    \TrinaryInfC{$\Gamma  \lfbsym{,}  \text{\faLock}  \lfbsym{,}  \overrightarrow{x}  \colon  \lfbnt{C}  \vdash   \lfbkw{unq} _{ \lfbsym{1} } \lfbnt{M} [   \overrightarrow{x}   ]   \colon  \lfbnt{T}$}
    \UnaryInfC{$\Gamma  \vdash   \lfbkw{quo} \langle  \overrightarrow{x}  \colon  \lfbnt{C}  \rangle  \lfbkw{unq} _{ \lfbsym{1} } \lfbnt{M} [   \overrightarrow{x}   ]    \colon  \lfbsym{[}  \lfbnt{C}  \vdash  \lfbnt{T}  \lfbsym{]}$}
    \DisplayProof
}
\end{center}

These local reduction and expansion lead to $\beta$-reduction and $\eta$-expansion, respectively. This paper focuses on $\beta$-reduction, which we define as follows.

\begin{definition}[$\beta$-reduction]
    We inductively define full reduction relations on terms and explicit substitutions, $ \rightarrow_{\beta} $.  We show main rules other than congruence below. We also define $ \rightarrow_{\beta}^{*} $ as the reflexive transitive closure of $ \rightarrow_{\beta} $.
    \begin{center}
        \bottomAlignProof
        \AxiomC{}
        \UnaryInfC{$\lfbsym{(}   \lambda \lfbmv{x} ^{ \lfbnt{S} }. \lfbnt{M}   \lfbsym{)} \, \lfbnt{N}  \rightarrow_{\beta}  \lfbnt{M}  \lfbsym{[}   \lfbmv{x}  \coloneqq  \lfbnt{N}   \lfbsym{]} $}
        \DisplayProof
        \quad
        \bottomAlignProof
        \AxiomC{}
        \UnaryInfC{$ \lfbkw{unq} _{ \lfbnt{k} } \lfbsym{(}   \lfbkw{quo} \langle  \overrightarrow{x}  \colon  \lfbnt{C}  \rangle \lfbnt{M}   \lfbsym{)} [  \theta  ]   \rightarrow_{\beta}  \lfbnt{M}  \lfbsym{[}    \text{\faLock} _{ \lfbnt{k} }    \lfbsym{,}  \overrightarrow{x}  \coloneqq  \theta  \lfbsym{]} $}
        \DisplayProof
    \end{center}
\end{definition}

These two rules correspond to local reductions of function types and contextual modal types. We safely omit identity substitutions found in these rules, thanks to Lemma~\ref{lem:idsub}.
% , because those substitutions depend on named contexts outside of the terms.
We do not dive into the basic properties of \lambra for now because we discuss those of its extension \lamfb in Sections~\ref{section:polyctx} and \ref{section:basicprops}.

%% file: 4-polyctx.tex
\section{Polymorphic Contexts} \label{section:polyctx}
This section proposes a novel type theory \lamfb that extends \lambra with polymorphic contexts. We quickly go through an overview of its syntax and semantics, focusing on the differences from \lambra. As examples in Section~\ref{section:motivation}, the critical idea of \lamfb is the notion of series variables, which can be considered the term representation for context variables.

\subsection{Syntax, Type System, and Substitution}
\begin{figure}[t]
    \begin{tabular}{lcl}
        \textbf{Types}                  & $\lfbnt{S}, \lfbnt{T} $ & $ \Coloneqq \ldots \mid  \forall \gamma . \lfbnt{T} $             \\
        \textbf{Contexts}               & $\lfbnt{C}, \lfbnt{D} $ & $ \Coloneqq \ldots \mid \lfbnt{C}  \lfbsym{,}  \gamma$             \\
        \textbf{Terms}                  & $\lfbnt{M}, \lfbnt{N}$  & $\Coloneqq \dots \mid  \Lambda \gamma . \lfbnt{M}  \mid  \lfbnt{M} @ \lfbnt{C} $ \\
        \textbf{Explicit Subst.}        & $\theta$        & $\Coloneqq \ldots \mid \theta  \lfbsym{,}  \mathbb{x}$              \\
        \textbf{Named Contexts}         & $\Gamma, \Delta$  & $\Coloneqq \ldots \mid \Gamma  \lfbsym{,}  \mathbb{x}  \colon  \gamma$             \\
    \end{tabular}
    \caption{Syntax of \lamfb} \label{def:syntaxlamfb}
\end{figure}

We provide the syntax of \lamfb in Fig.~\ref{def:syntaxlamfb}.  First, \lamfb has two additional sorts of variables: context variables $\gamma, \delta$, standing for contexts, and series variables $\mathbb{x}, \mathbb{y}$, representing sequences of variables.  \lamfb adds polymorphic context types of the form $ \forall \gamma . \lfbnt{T} $, which binds $\gamma$ in $\lfbnt{T}$.  It represents the set of types obtained by substituting any context $\lfbnt{C}$ for the context variable $\gamma$. Two kinds of terms $ \Lambda \gamma . \lfbnt{M} $ and $ \lfbnt{M} @ \lfbnt{C} $ are added as introduction and elimination for polymorphic context types.  We allow $\lfbnt{C}$ to include polymorphic context types; thus, polymorphism in \lamfb is impredicative.  The definition of contexts means that we can abstract any part of a context with context variables, e.g., $ \forall \gamma_{{\mathrm{1}}} .  \forall \gamma_{{\mathrm{2}}} . \lfbsym{[}   \gamma_{{\mathrm{1}}}   \lfbsym{,}  \iota  \lfbsym{,}  \gamma_{{\mathrm{2}}}  \vdash  \iota  \lfbsym{]}  $. Accordingly, series variables can appear in explicit substitutions, which correspond to context variables. We can also abstract any part of a named context with a pair of a series variable and a context variable $ \mathbb{x}  \colon  \gamma $. $ \mathsf{FV} $ is updated to accommodate series variables but we omit the definition here.

It is worth noting that context variables are not subject to staging. This allows us to use the same context variable across levels---for example, the type $ \forall \gamma . \lfbsym{[}   \gamma   \vdash  \lfbsym{[}   \gamma   \vdash  \lfbnt{T}  \lfbsym{]}  \lfbsym{]} $ binds both occurrences of $\gamma$ although they are in different levels. The definition of free context variables, denoted $ \mathsf{FCV} (-)$, is straightforward and we omit it in this paper.

\begin{figure}[t]
    \framebox{\mbox{$\Gamma  \vdash  \lfbnt{M}  \colon  \lfbnt{T}$}}
    \framebox{\mbox{$\Gamma  \vdash  \theta  \colon  \lfbnt{C}$}}
    \[
        \AxiomC{$\Gamma  \vdash  \lfbnt{M}  \colon  \lfbnt{T}$}
        \AxiomC{$\gamma \, \not\in \, \mathsf{FCV} \, \lfbsym{(}  \Gamma  \lfbsym{)}$}
        \BinaryInfC{$\Gamma  \vdash   \Lambda \gamma . \lfbnt{M}   \colon   \forall \gamma . \lfbnt{T} $}
        \DisplayProof
        \quad
        \AxiomC{$\Gamma  \vdash  \lfbnt{M}  \colon   \forall \gamma . \lfbnt{T} $}
        \UnaryInfC{$\Gamma  \vdash   \lfbnt{M} @ \lfbnt{C}   \colon  \lfbnt{T}  \lfbsym{[}   \gamma \coloneqq \lfbnt{C}   \lfbsym{]}$}
        \DisplayProof
    \]
    \begin{prooftree}
        \AxiomC{$\Gamma  \vdash  \theta  \colon  \lfbnt{C}$}
        \AxiomC{$\mathbb{x}  \colon  \gamma \, \in \, \mathsf{head} \, \lfbsym{(}  \Gamma  \lfbsym{)}$}
        \BinaryInfC{$\Gamma  \vdash  \theta  \lfbsym{,}  \mathbb{x}  \colon  \lfbnt{C}  \lfbsym{,}  \gamma$}
    \end{prooftree}

    \textbf{Substitution} $\sigma \Coloneqq \ldots \mid \sigma  \lfbsym{,}  \mathbb{x}  \coloneqq  \mathbb{y}$
    \vspace{1em}

    \mbox{\fbox{$\lfbnt{M}  \lfbsym{[}  \sigma  \lfbsym{]}$}}
    \mbox{\fbox{$\theta  \lfbsym{[}  \sigma  \lfbsym{]}$}}
    \begin{align*}
        \ldots              &                                                     &
        \lfbsym{(}   \Lambda \gamma . \lfbnt{M}   \lfbsym{)}  \lfbsym{[}  \sigma  \lfbsym{]}   & =  \Lambda \gamma . \lfbsym{(}  \lfbnt{M}  \lfbsym{[}  \sigma  \lfbsym{]}  \lfbsym{)}  \text{ if $\gamma \, \not\in \, \mathsf{FCV} \, \lfbsym{(}  \sigma  \lfbsym{)}$} &
        \lfbsym{(}   \lfbnt{M} @ \lfbnt{C}   \lfbsym{)}  \lfbsym{[}  \sigma  \lfbsym{]} & =  \lfbsym{(}  \lfbnt{M}  \lfbsym{[}  \sigma  \lfbsym{]}  \lfbsym{)} @ \lfbnt{C} 
    \end{align*}
    \begin{align*}
        \ldots              & &
        \lfbsym{(}  \theta  \lfbsym{,}  \mathbb{x}  \lfbsym{)}  \lfbsym{[}  \sigma  \lfbsym{]} & = \begin{cases}
                                    \theta  \lfbsym{[}  \sigma  \lfbsym{]}  \lfbsym{,}  \mathbb{y} & \text{if $\mathbb{x}  \coloneqq  \mathbb{y} \, \in \, \mathsf{head} \, \lfbsym{(}  \sigma  \lfbsym{)}$} \\
                                    \theta  \lfbsym{[}  \sigma  \lfbsym{]}  \lfbsym{,}  \mathbb{x} & \text{else}                        \\
                                \end{cases}
    \end{align*}

    \framebox{\mbox{$\Gamma  \vdash  \sigma  \colon  \Delta$}}
    \begin{prooftree}
        \AxiomC{$\Gamma  \vdash  \sigma  \colon  \Delta$}
        \AxiomC{$\mathbb{y}  \colon  \gamma \, \in \, \mathsf{head} \, \lfbsym{(}  \Gamma  \lfbsym{)}$}
        \AxiomC{$\mathbb{x} \, \not\in \, \mathsf{dom} \, \lfbsym{(}  \Delta  \lfbsym{)}$}
        \TrinaryInfC{$\Gamma  \vdash  \sigma  \lfbsym{,}  \mathbb{x}  \coloneqq  \mathbb{y}  \colon  \Delta  \lfbsym{,}  \mathbb{x}  \colon  \gamma$}
    \end{prooftree}

    \caption{Additional typing rules and definitions of substitutions in \lamfb}
    \label{fig:typingrulesoflamfb}
\end{figure}

\begin{sloppypar}
    We give additional typing rules and defining clauses of substitutions in Fig.~\ref{fig:typingrulesoflamfb}.
    We also extend the auxiliary functions such as $ \mathsf{head} $ to accommodate the new syntax but we omit their definitions.
    The introduction and elimination rules for polymorphic context types are similar to those for the polymorphic types in System~F~\cite{Girard89}.
    The definition of context substitution $\lfbnt{T}  \lfbsym{[}   \gamma \coloneqq \lfbnt{C}   \lfbsym{]}$ for types is straightforward and omitted.
    The other rule for explicit substitutions states that we can add $ \mathbb{x}  \colon  \gamma $ to an explicit substitution if it appears in the level-0 part of $\Gamma$.
    The point of the extension of substitution is that a series variable can only be replaced with another series variable, not an explicit substitution.
    Strictly speaking, we can replace a series variable of context variable $\gamma$ with any explicit substitution of the same context.
    However, only series variables can inhabit $\gamma$, so we define substitution in this way.
\end{sloppypar}

With these extensions, we can confirm that the following substitution lemma holds as expected.

\begin{lemma}[Substitution Lemma]
    \begin{enumerate}
        \item If $\Gamma  \vdash  \lfbnt{M}  \colon  \lfbnt{T}$ and $\Delta  \vdash  \sigma  \colon  \Gamma$, then $\Delta  \vdash  \lfbnt{M}  \lfbsym{[}  \sigma  \lfbsym{]}  \colon  \lfbnt{T}$.
        \item If $\Gamma  \vdash  \theta  \colon  \lfbnt{C}$ and $\Delta  \vdash  \sigma  \colon  \Gamma$, then $\Delta  \vdash  \theta  \lfbsym{[}  \sigma  \lfbsym{]}  \colon  \lfbnt{C}$.
    \end{enumerate}
\end{lemma}

\subsection{Context Substitution}
We also define substitution for context variables, which is the most non-trivial part of \lamfb. To describe the core idea of context substitution, let us consider a term $ \lfbkw{quo} \langle  \mathbb{x}  \colon  \gamma  \rangle \lfbsym{(}   \lfbkw{unq} _{ \lfbsym{1} } \lfbnt{M} [   \mathbb{x}   ]   \lfbsym{)} $. If we naively substitute a context $ \lfbnt{T}   \lfbsym{,}  \delta$ for the context variable $\gamma$ in this term, we would obtain $ \lfbkw{quo} \langle   \mathbb{x} \colon \lfbsym{(}   \lfbnt{T}   \lfbsym{,}  \delta  \lfbsym{)}   \rangle \lfbsym{(}   \lfbkw{unq} _{ \lfbsym{1} } \lfbnt{M} [   \mathbb{x}   ]   \lfbsym{)} $, where $  \mathbb{x} \colon \lfbsym{(}   \lfbnt{T}   \lfbsym{,}  \delta  \lfbsym{)}  $ is simply ill formed as a named context.  Instead, we will take the following steps.
\begin{enumerate}\sloppy
    \item We check the occurrences of $\gamma$ in the named context of the quote $ \lfbkw{quo} \langle  \mathbb{x}  \colon  \gamma  \rangle \lfbsym{(}   \lfbkw{unq} _{ \lfbsym{1} } \lfbnt{M} [   \mathbb{x}   ]   \lfbsym{)} $, and collect series variables that are associated to $\gamma$. In this case, we have only $\mathbb{x}$.
    \item We generate a series of fresh variables to be substituted for $\mathbb{x}$. Each variable corresponds to each element of the new context $ \lfbnt{T}   \lfbsym{,}  \delta$.  Suppose we generate new variables $\lfbmv{x}, \mathbb{y}$ for $ \lfbnt{T}   \lfbsym{,}  \delta$. As a result, we get a \textit{variable series substitution} $ \mathbb{x} \coloneqq  \lfbmv{x}   \lfbsym{,}  \mathbb{y} $.
    \item We apply context substitution $ \gamma \coloneqq  \lfbnt{T}   \lfbsym{,}  \delta $ to the named context $ \mathbb{x}  \colon  \gamma $ along with $ \mathbb{x} \coloneqq  \lfbmv{x}   \lfbsym{,}  \mathbb{y} $. As a result, we get a new named context $ \lfbmv{x}  \colon  \lfbnt{T}   \lfbsym{,}  \mathbb{y}  \colon  \delta$.
    \item We also apply the variable series substitution to $ \lfbkw{unq} _{ \lfbsym{1} } \lfbnt{M} [   \mathbb{x}   ] $ and obtain
       $ \lfbkw{unq} _{ \lfbsym{1} } \lfbnt{M} [   \lfbmv{x}   \lfbsym{,}  \mathbb{y}  ] $.
    \item As a result, we obtain a substituted term $ \lfbkw{quo} \langle  \lfbmv{x}  \colon  \lfbnt{T}   \lfbsym{,}  \mathbb{y}  \colon  \delta \rangle \lfbsym{(}   \lfbkw{unq} _{ \lfbsym{1} } \lfbnt{M} [   \lfbmv{x}   \lfbsym{,}  \mathbb{y}  ]   \lfbsym{)} $.
\end{enumerate}

In this way, substitution for context variables essentially requires three operations (1) to replace context variables with contexts, (2) to generate fresh variables to be substituted for series variables, and (3) to replace series variables with sequences of variables.  We start its formal definition with the following new objects. We denote infinite sequences of ordinary variables and series variables without duplication by $G_v$ and $G_s$, respectively.

\begin{center}
    \begin{tabular}[]{lcl}
        \textbf{Context substitution}         & $\Sigma$       & $\Coloneqq  \bullet  \mid \Sigma  \lfbsym{,}  \gamma  \coloneqq  \lfbnt{C}$                       \\
        \textbf{Variable series}             & $\overrightarrow{x}, \overrightarrow{y}$ & $\Coloneqq  \bullet  \mid \overrightarrow{x}  \lfbsym{,}  \lfbmv{y} \mid \overrightarrow{x}  \lfbsym{,}  \mathbb{y}$                  \\
        \textbf{Variable series substitution} & $\bar{\sigma}$       & $\Coloneqq  \bullet  \mid \bar{\sigma}  \lfbsym{,}  \mathbb{x}  \coloneqq  \overrightarrow{y} \mid \bar{\sigma}  \lfbsym{,} \, \text{\faLock}$ \\
        \textbf{Variable generator}     & $G$  & $\Coloneqq \lfbsym{(}  G_v  \lfbsym{,}  G_s  \lfbsym{)}$
    \end{tabular}
\end{center}

A context substitution $\Sigma$ maps context variables to contexts, and a variable series substitution $\bar{\sigma}$ maps series variables to variable series, that is, sequences of ordinary/series variables. Note that series substitution does not affect stage levels; hence, locks in series substitution are not annotated with stage transitions. A variable generator consists of streams of non-duplicating variables and series variables. We use it to generate fresh variables. We denote $\mathsf{rg} \, \lfbsym{(}  \bar{\sigma}  \lfbsym{)}$ for the variable series obtained from the range of $\bar{\sigma}$.

We define application of context substitution in Fig.~\ref{fig:defcsubstandssubst}. Application of a context substitution to types $\lfbnt{T}  \lfbsym{[}  \Sigma  \lfbsym{]}$ and contexts $\lfbnt{C}  \lfbsym{[}  \Sigma  \lfbsym{]}$ is straightforward; we simply replace context variables in a capture-avoiding manner. We omit their definitions from the figure. On the contrary, context substitution on terms $ \lfbnt{M} \lfbsym{[} \Sigma \lfbsym{;} \bar{\sigma} \lfbsym{]} _{ G } $ and explicit substitutions $ \theta \lfbsym{[} \Sigma \lfbsym{;} \bar{\sigma} \lfbsym{]} _{ G } $ comes with not only $\Sigma$ but also a variable series substitution $\bar{\sigma}$ and a variable generator $G$.
$\Sigma$ is used to replace context variables in types in $\lambda$-abstractions and $\Gamma$ in a quote; $\bar{\sigma}$ is used to substitute series variables in explicit substitutions and $\Gamma$ in a quote.  The most interesting is the case for a quote $ \lfbkw{quo} \langle \Gamma \rangle \lfbnt{M} $: first, a variable series substitution $\bar{\sigma}'$ is generated by the auxiliary function $ \mathsf{destruct} $ (Step 2 above); second, $\Sigma$ and the generated $\bar{\sigma}'$ are applied to $\Gamma$ to yield the new named context (Step 3); finally, we apply $\Sigma$ and $ \bar{\sigma}  \lfbsym{,} \, \text{\faLock} \lfbsym{,} \bar{\sigma}' $ to the body of the quote (Step 4), after removing variables in $\mathsf{dom} \, \lfbsym{(}  \Gamma  \lfbsym{)}$ and generated ones from the generator; here, $\lfbsym{(}  G_v  \lfbsym{,}  G_s  \lfbsym{)} - S$ means $(G_v\setminus S, G_s \setminus S)$.
The auxiliary function $ \mathsf{destruct} _{ G } \lfbsym{(} \Gamma \lfbsym{;} \Sigma \lfbsym{)} $ scans $\Gamma$ to find context variables in the domain of $\Sigma$, generates fresh (ordinary/series) variables by using $ \mathsf{gensyms} $, and returns a variable series substitution.  $ \mathsf{gensyms} _{ G } \lfbsym{(} \lfbnt{C} \lfbsym{;} V \lfbsym{)} $ produces a sequence of ordinary/series variables of the same length as $\lfbnt{C}$; fresh variables are chosen from earlier ones in $G$ but not in $V$.

For example, consider applying $\Sigma =  \gamma \coloneqq  \lfbnt{T_{{\mathrm{1}}}}   \lfbsym{,}  \gamma' $ and the empty variable series substitution
to $\lfbnt{M} =  \lfbkw{quo} \langle  \mathbb{x}  \colon  \gamma   \lfbsym{,}  \lfbmv{x}  \colon  \iota  \lfbsym{,}  \mathbb{y}  \colon  \gamma \rangle \lfbnt{M_{{\mathrm{0}}}} $.
$ \mathsf{destruct} _{ G } \lfbsym{(} \lfbsym{(}   \mathbb{x}  \colon  \gamma   \lfbsym{,}  \lfbmv{x}  \colon  \iota  \lfbsym{,}  \mathbb{y}  \colon  \gamma  \lfbsym{)} \lfbsym{;} \lfbsym{(}   \gamma \coloneqq  \lfbnt{T_{{\mathrm{1}}}}   \lfbsym{,}  \gamma'   \lfbsym{)} \lfbsym{)} $ returns $ \mathbb{x} \coloneqq  (   \lfbmv{x'}   \lfbsym{,}  \mathbb{x}'  )    \lfbsym{,}  \mathbb{y}  \coloneqq   (   \lfbmv{y'}   \lfbsym{,}  \mathbb{y}'  ) $ for some fresh $\lfbmv{x'}, \mathbb{x}', \lfbmv{y'}$, and $\mathbb{y}'$ (with respect to $G$) and, thus, 
$ \lfbnt{M} \lfbsym{[} \Sigma \lfbsym{;}  \bullet  \lfbsym{]} _{ G } $ is 
$ \lfbkw{quo} \langle  \lfbmv{x'}  \colon  \lfbnt{T_{{\mathrm{1}}}}   \lfbsym{,}  \mathbb{x}'  \colon  \gamma'  \lfbsym{,}  \lfbmv{x}  \colon  \iota  \lfbsym{,}  \lfbmv{y'}  \colon  \lfbnt{T_{{\mathrm{1}}}}  \lfbsym{,}  \mathbb{y}'  \colon  \gamma' \rangle \lfbnt{M'_{{\mathrm{0}}}} $ where
$\lfbnt{M'_{{\mathrm{0}}}} =  \lfbnt{M_{{\mathrm{0}}}} \lfbsym{[} \Sigma \lfbsym{;} \lfbsym{(}   \bullet   \lfbsym{,} \, \text{\faLock}  \lfbsym{,}  \mathbb{x}  \coloneqq   (   \lfbmv{x'}   \lfbsym{,}  \mathbb{x}'  )   \lfbsym{,}  \mathbb{y}  \coloneqq   (   \lfbmv{y'}   \lfbsym{,}  \mathbb{y}'  )   \lfbsym{)} \lfbsym{]} _{ G' } $ and
$G' = G - \{\mathbb{x}, \lfbmv{x}, \mathbb{y}, \lfbmv{x'}, \mathbb{x}', \lfbmv{y'}, \mathbb{y}'\}$.

% In addition to replacing context variables (as in $\lfbnt{T}  \lfbsym{[}  \Sigma  \lfbsym{]}$ in the rule for abstractions), it replaces series variables in explicit substitutions using the variable series substitution, and extends the variable series substitution while generating with fresh variables, if it finds context variables in the named context in a quote. The auxiliary function $ \mathsf{destruct} $ is responsible for generating a new variable series substitution. The function generates fresh variables from a variable generator to avoid conflict with the given named context and already generated variables.

\begin{figure}
    \framebox{\mbox{$ \lfbnt{M} \lfbsym{[} \Sigma \lfbsym{;} \bar{\sigma} \lfbsym{]} _{ G } $}}
    \begin{align*}
         \lfbmv{x} \lfbsym{[} \Sigma \lfbsym{;} \bar{\sigma} \lfbsym{]} _{ G }           & = \lfbmv{x}                                                              \\
         \lfbsym{(}   \lambda \lfbmv{x} ^{ \lfbnt{T} }. \lfbnt{M}   \lfbsym{)} \lfbsym{[} \Sigma \lfbsym{;} \bar{\sigma} \lfbsym{]} _{ G }     & =  \lambda \lfbmv{x} ^{ \lfbsym{(}  \lfbnt{T}  \lfbsym{[}  \Sigma  \lfbsym{]}  \lfbsym{)} }. \lfbsym{(}   \lfbnt{M} \lfbsym{[} \Sigma \lfbsym{;} \bar{\sigma} \lfbsym{]} _{ G }   \lfbsym{)}                          \\
         \lfbsym{(}  \lfbnt{M} \, \lfbnt{N}  \lfbsym{)} \lfbsym{[} \Sigma \lfbsym{;} \bar{\sigma} \lfbsym{]} _{ G }        & = \lfbsym{(}   \lfbnt{M} \lfbsym{[} \Sigma \lfbsym{;} \bar{\sigma} \lfbsym{]} _{ G }   \lfbsym{)} \, \lfbsym{(}   \lfbnt{N} \lfbsym{[} \Sigma \lfbsym{;} \bar{\sigma} \lfbsym{]} _{ G }   \lfbsym{)}             \\
         \lfbsym{(}   \lfbkw{quo} \langle \Gamma \rangle \lfbnt{M}   \lfbsym{)} \lfbsym{[} \Sigma \lfbsym{;} \bar{\sigma} \lfbsym{]} _{ G }     & =  \lfbkw{quo} \langle  \Gamma   \lfbsym{[}   \Sigma   \lfbsym{;}   \bar{\sigma}'   \lfbsym{]}  \rangle \lfbsym{(}   \lfbnt{M} \lfbsym{[} \Sigma \lfbsym{;} \lfbsym{(}   \bar{\sigma}  \lfbsym{,} \, \text{\faLock} \lfbsym{,} \bar{\sigma}'   \lfbsym{)} \lfbsym{]} _{ G' }   \lfbsym{)}  \\
                                              & \qquad \text{where $\bar{\sigma}' =  \mathsf{destruct} _{ G } \lfbsym{(} \Gamma \lfbsym{;} \Sigma \lfbsym{)} $}    \\
                                              & \qquad \text{and $G' = G - (\mathsf{dom} \, \lfbsym{(}  \Gamma  \lfbsym{)} \, \cup \, \mathsf{rg} \, \lfbsym{(}  \bar{\sigma}'  \lfbsym{)})$} \\
         \lfbsym{(}   \lfbkw{unq} _{ \lfbnt{k} } \lfbnt{M} [  \theta  ]   \lfbsym{)} \lfbsym{[} \Sigma \lfbsym{;} \bar{\sigma} \lfbsym{]} _{ G }  & =  \lfbkw{unq} _{ \lfbnt{k} } \lfbsym{(}   \lfbnt{M} \lfbsym{[} \Sigma \lfbsym{;} \bar{\sigma}  \uparrow  \lfbnt{k} \lfbsym{]} _{ G }   \lfbsym{)} [   \theta \lfbsym{[} \Sigma \lfbsym{;} \bar{\sigma} \lfbsym{]} _{ G }   ]  \\
       \lfbsym{(}   \Lambda \gamma . \lfbnt{M}   \lfbsym{)} \lfbsym{[} \Sigma \lfbsym{;} \bar{\sigma} \lfbsym{]} _{ G }  & =  \Lambda \gamma . \lfbsym{(}   \lfbnt{M} \lfbsym{[} \Sigma \lfbsym{;} \bar{\sigma} \lfbsym{]} _{ G }   \lfbsym{)}   \qquad \text{ if $\gamma \, \not\in \, \mathsf{dom} \, \lfbsym{(}  \Sigma  \lfbsym{)}$ and $ \gamma \, \not\in \, \mathsf{FCV} \, \lfbsym{(}  \Sigma  \lfbsym{)} $} \\
       \lfbsym{(}   \lfbnt{M} @ \lfbnt{C}   \lfbsym{)} \lfbsym{[} \Sigma \lfbsym{;} \bar{\sigma} \lfbsym{]} _{ G }  & =    \lfbsym{(}   \lfbnt{M} \lfbsym{[} \Sigma \lfbsym{;} \bar{\sigma} \lfbsym{]} _{ G }   \lfbsym{)} @ \lfbsym{(}  \lfbnt{C}  \lfbsym{[}  \Sigma  \lfbsym{]}  \lfbsym{)}  
    \end{align*}

%    \vspace{-1.0em}
    \begin{tabularx}{\textwidth}{ X  X }
        {
            \framebox{\mbox{$ \theta \lfbsym{[} \Sigma \lfbsym{;} \bar{\sigma} \lfbsym{]} _{ G } $}}
            \begin{align*}
                  \bullet  \lfbsym{[} \Sigma \lfbsym{;} \bar{\sigma} \lfbsym{]} _{ G }  & =  \bullet                                                                                   \\
                 \lfbsym{(}  \theta  \lfbsym{,}  \lfbnt{M}  \lfbsym{)} \lfbsym{[} \Sigma \lfbsym{;} \bar{\sigma} \lfbsym{]} _{ G }    & = \lfbsym{(}   \theta \lfbsym{[} \Sigma \lfbsym{;} \bar{\sigma} \lfbsym{]} _{ G }   \lfbsym{)}  \lfbsym{,}  \lfbsym{(}   \lfbnt{M} \lfbsym{[} \Sigma \lfbsym{;} \bar{\sigma} \lfbsym{]} _{ G }   \lfbsym{)}                                       \\
                 \lfbsym{(}  \theta  \lfbsym{,}  \mathbb{x}  \lfbsym{)} \lfbsym{[} \Sigma \lfbsym{;} \bar{\sigma} \lfbsym{]} _{ G }   & =\begin{cases}
                                                             \lfbsym{(}   \theta \lfbsym{[} \Sigma \lfbsym{;} \bar{\sigma} \lfbsym{]} _{ G }   \lfbsym{)} \lfbsym{,} \overrightarrow{y}  \\
                                                            \quad \text{if $\mathbb{x}  \coloneqq  \overrightarrow{y} \, \in \, \mathsf{head} \, \lfbsym{(}  \bar{\sigma}  \lfbsym{)}$} \\
                                                            \lfbsym{(}   \theta \lfbsym{[} \Sigma \lfbsym{;} \bar{\sigma} \lfbsym{]} _{ G }   \lfbsym{)}  \lfbsym{,}  \mathbb{x} \quad \text{otherwise}
                                                        \end{cases}
            \end{align*}
        } & {
            \framebox{\mbox{$ \Gamma   \lfbsym{[}   \Sigma   \lfbsym{;}   \bar{\sigma}   \lfbsym{]} $}}
            \begin{align*}
                  \bullet    \lfbsym{[}   \Sigma   \lfbsym{;}   \bar{\sigma}   \lfbsym{]}    & =  \bullet                                                                                                                                    \\
                 \lfbsym{(}  \Gamma  \lfbsym{,}  \lfbmv{x}  \colon  \lfbnt{T}  \lfbsym{)}   \lfbsym{[}   \Sigma   \lfbsym{;}   \bar{\sigma}   \lfbsym{]}   & =  \Gamma   \lfbsym{[}   \Sigma   \lfbsym{;}   \bar{\sigma}   \lfbsym{]}   \lfbsym{,}  \lfbmv{x}  \colon  \lfbnt{T}  \lfbsym{[}  \Sigma  \lfbsym{]}                                                                                                                  \\
                 \lfbsym{(}  \Gamma  \lfbsym{,}  \mathbb{x}  \colon  \gamma  \lfbsym{)}   \lfbsym{[}   \Sigma   \lfbsym{;}   \bar{\sigma}   \lfbsym{]}  & = \begin{cases}
                                                     \Gamma   \lfbsym{[}   \Sigma   \lfbsym{;}   \bar{\sigma}   \lfbsym{]}   \lfbsym{,}  \overrightarrow{y}  \colon  \lfbnt{C} \\
                                                    \quad \text{if $\mathbb{x}  \coloneqq  \overrightarrow{y} \, \in \, \mathsf{head} \, \lfbsym{(}  \bar{\sigma}  \lfbsym{)}$} \\
                                                    \quad  \text{and $\gamma  \coloneqq  \lfbnt{C} \, \in \, \Sigma$} \\
                                                     \Gamma   \lfbsym{[}   \Sigma   \lfbsym{;}   \bar{\sigma}   \lfbsym{]}   \lfbsym{,}  \mathbb{x}  \colon  \gamma \quad \text{else}
                                                \end{cases} \\
                 \lfbsym{(}  \Gamma  \lfbsym{,}  \text{\faLock}  \lfbsym{)}   \lfbsym{[}   \Sigma   \lfbsym{;}   \bar{\sigma}   \lfbsym{]}    & =  \Gamma   \lfbsym{[}   \Sigma   \lfbsym{;}   \bar{\sigma}  \uparrow  \lfbsym{1}   \lfbsym{]}   \lfbsym{,}  \text{\faLock}
            \end{align*}
        }
    \end{tabularx}

%    \vspace{-2.0em}
    \mbox{\textbf{Auxiliary functions}}
    \begin{align*}
         \mathsf{destruct} _{ G } \lfbsym{(} \lfbsym{(}  \Gamma  \lfbsym{,}  \lfbmv{x}  \colon  \lfbnt{T}  \lfbsym{)} \lfbsym{;} \Sigma \lfbsym{)}    & =  \mathsf{destruct} _{ G } \lfbsym{(} \Gamma \lfbsym{;} \Sigma \lfbsym{)}  \\
         \mathsf{destruct} _{ G } \lfbsym{(} \lfbsym{(}  \Gamma  \lfbsym{,}  \mathbb{x}  \colon  \gamma  \lfbsym{)} \lfbsym{;} \Sigma \lfbsym{)}   & =
        \begin{cases}
          \bar{\sigma}  \lfbsym{,}  \mathbb{x}  \coloneqq  \overrightarrow{x}              & \text{if $\gamma  \coloneqq  \lfbnt{C} \, \in \, \Sigma$}     \\
            \multicolumn{2}{l}{\text{ where $\bar{\sigma} =  \mathsf{destruct} _{ G } \lfbsym{(} \Gamma \lfbsym{;} \Sigma \lfbsym{)} $}}         \\
            \multicolumn{2}{l}{\text{ and $\overrightarrow{x} =  \mathsf{gensyms} _{ G } \lfbsym{(} \lfbnt{C} \lfbsym{;} \mathsf{dom} \, \lfbsym{(}  \Gamma  \lfbsym{)} \, \cup \, \mathsf{rg} \, \lfbsym{(}  \bar{\sigma}  \lfbsym{)} \lfbsym{)} $}} \\
             \mathsf{destruct} _{ G } \lfbsym{(} \Gamma \lfbsym{;} \Sigma \lfbsym{)}  & \text{otherwise}                      \\
        \end{cases} \\
         \mathsf{destruct} _{ G } \lfbsym{(} \lfbsym{(}  \Gamma  \lfbsym{,}  \text{\faLock}  \lfbsym{)} \lfbsym{;} \Sigma \lfbsym{)}     & =  \mathsf{destruct} _{ G } \lfbsym{(} \Gamma \lfbsym{;} \Sigma \lfbsym{)}   \lfbsym{,} \, \text{\faLock} \\
         \mathsf{gensyms} _{ \lfbsym{(}  G_v  \lfbsym{,}  G_s  \lfbsym{)} } \lfbsym{(}  \bullet  \lfbsym{;} V \lfbsym{)}  & =  \bullet  \\
         \mathsf{gensyms} _{ \lfbsym{(}  G_v  \lfbsym{,}  G_s  \lfbsym{)} } \lfbsym{(} \lfbsym{(}  \lfbnt{C}  \lfbsym{,}  \lfbnt{T}  \lfbsym{)} \lfbsym{;} V \lfbsym{)}    & =  \mathsf{gensyms} _{ \lfbsym{(}  G_v  \lfbsym{,}  G_s  \lfbsym{)} } \lfbsym{(} \lfbnt{C} \lfbsym{;} V \, \cup \, \lfbsym{\{}  \lfbmv{x}  \lfbsym{\}} \lfbsym{)}   \lfbsym{,}  \lfbmv{x} \\
                                                      & \qquad \text{where $\lfbmv{x}$ is the first element of $G_v$ such that $\lfbmv{x} \, \not\in \, V$} \\
         \mathsf{gensyms} _{ \lfbsym{(}  G_v  \lfbsym{,}  G_s  \lfbsym{)} } \lfbsym{(} \lfbsym{(}  \lfbnt{C}  \lfbsym{,}  \gamma  \lfbsym{)} \lfbsym{;} V \lfbsym{)}    & =  \mathsf{gensyms} _{ \lfbsym{(}  G_v  \lfbsym{,}  G_s  \lfbsym{)} } \lfbsym{(} \lfbnt{C} \lfbsym{;} V \, \cup \, \lfbsym{\{}  \mathbb{x}  \lfbsym{\}} \lfbsym{)}   \lfbsym{,}  \mathbb{x} \\
                                                      & \qquad \text{where $\mathbb{x}$ is the first element of $G_s$ such that $\mathbb{x} \, \not\in \, V$}
    \end{align*}
    \caption{Context substitutions and variable series substitutions}
    \label{fig:defcsubstandssubst}
\end{figure}

We can confirm that context substitution preserves evident judgments, which is stated in the following context substitution lemma.

\begin{lemma}[Context Substitution Lemma]
    \begin{enumerate}
        \item If $\Gamma  \vdash  \lfbnt{M}  \colon  \lfbnt{T}$ then $ \Gamma   \lfbsym{[}   \Sigma   \lfbsym{;}   \bar{\sigma}   \lfbsym{]}   \vdash   \lfbnt{M} \lfbsym{[} \Sigma \lfbsym{;} \bar{\sigma} \lfbsym{]} _{ G' }   \colon  \lfbnt{T}  \lfbsym{[}  \Sigma  \lfbsym{]}$ where $\bar{\sigma} =  \mathsf{destruct} _{ G } \lfbsym{(} \Gamma \lfbsym{;} \Sigma \lfbsym{)} $ and $G' = G - (\mathsf{dom} \, \lfbsym{(}  \Gamma  \lfbsym{)} \, \cup \, \mathsf{rg} \, \lfbsym{(}  \bar{\sigma}  \lfbsym{)})$ for any $\Sigma$ and $G$.
        \item If $\Gamma  \vdash  \theta  \colon  \lfbnt{C}$ then $ \Gamma   \lfbsym{[}   \Sigma   \lfbsym{;}   \bar{\sigma}   \lfbsym{]}   \vdash   \theta \lfbsym{[} \Sigma \lfbsym{;} \bar{\sigma} \lfbsym{]} _{ G' }   \colon  \lfbnt{C}  \lfbsym{[}  \Sigma  \lfbsym{]}$ where $\bar{\sigma} =  \mathsf{destruct} _{ G } \lfbsym{(} \Gamma \lfbsym{;} \Sigma \lfbsym{)} $ and $G' = G - (\mathsf{dom} \, \lfbsym{(}  \Gamma  \lfbsym{)} \, \cup \, \mathsf{rg} \, \lfbsym{(}  \bar{\sigma}  \lfbsym{)})$ for any $\Sigma$ and $G$.
    \end{enumerate}
\end{lemma}

Although we use variable generators to get fresh variables, the result of context substitution should be equivalent under renaming. We can confirm this intuition by the following lemma.

\begin{lemma} \label{lem:ctxsubst}
    If $\Gamma  \vdash  \lfbnt{M}  \colon  \lfbnt{T}$, $\bar{\sigma}_{{\mathrm{1}}} =  \mathsf{destruct} _{ G_{{\mathrm{1}}} } \lfbsym{(} \Gamma \lfbsym{;} \Sigma \lfbsym{)} $ and $\bar{\sigma}_{{\mathrm{2}}} =  \mathsf{destruct} _{ G_{{\mathrm{2}}} } \lfbsym{(} \Gamma \lfbsym{;} \Sigma \lfbsym{)} $, then there is a renaming substitution $\sigma$ such that $ \Gamma   \lfbsym{[}   \Sigma   \lfbsym{;}   \bar{\sigma}_{{\mathrm{1}}}   \lfbsym{]}   \vdash   \lfbnt{M} \lfbsym{[} \Sigma \lfbsym{;} \bar{\sigma}_{{\mathrm{2}}} \lfbsym{]} _{ G'_{{\mathrm{1}}} }   \lfbsym{[}  \sigma  \lfbsym{]}  \colon  \lfbnt{T}  \lfbsym{[}  \Sigma  \lfbsym{]}$ with some $G'_{{\mathrm{1}}}$.
\end{lemma}

\begin{corollary}
    If $\mathsf{dom} \, \lfbsym{(}  \Sigma  \lfbsym{)} \cap \mathsf{FCV} \, \lfbsym{(}  \Gamma  \lfbsym{)} = \emptyset$ and $\Gamma  \vdash  \lfbnt{M}  \colon  \lfbnt{T}$, then $ \lfbnt{M} \lfbsym{[} \Sigma \lfbsym{;}  \bullet  \lfbsym{]} _{ G_{{\mathrm{1}}} }  =_{\alpha}  \lfbnt{M} \lfbsym{[} \Sigma \lfbsym{;}  \bullet  \lfbsym{]} _{ G_{{\mathrm{2}}} } $.
\end{corollary}

Based on this nature of context substitution, we may omit variable generators from context substitution applications.

\subsection{Local Soundness and Completeness}
Local soundness and local completeness are extended to polymorphic context types as follows. We use context substitution to obtain $\mathcal{D'}$ in the local reduction pattern. In this pattern, we observe $ \mathsf{destruct} \lfbsym{(} \Gamma \lfbsym{;}  \gamma \coloneqq \lfbnt{C}  \lfbsym{)}  =  \bullet $ because $\gamma \, \not\in \, \mathsf{FCV} \, \lfbsym{(}  \Gamma  \lfbsym{)}$, and hence we get $\Gamma  \vdash   \lfbnt{M} \lfbsym{[}  \gamma \coloneqq \lfbnt{C}  \lfbsym{;}  \bullet  \lfbsym{]}   \colon  \lfbnt{T}  \lfbsym{[}   \gamma \coloneqq \lfbnt{C}   \lfbsym{]}$. For the local expansion pattern, we have to pick a context variable $\delta$ that is fresh against $\Gamma$.

    \vspace{1em}Local Soundness
    \[
        \bottomAlignProof
        \AxiomC{$\mathcal{D}$}
        \noLine
        \UnaryInfC{$\Gamma  \vdash  \lfbnt{M}  \colon  \lfbnt{T}$}
        \AxiomC{$\gamma \, \not\in \, \mathsf{FCV} \, \lfbsym{(}  \Gamma  \lfbsym{)}$}
        \BinaryInfC{$\Gamma  \vdash   \Lambda \gamma . \lfbnt{M}   \colon   \forall \gamma . \lfbnt{T} $}
        \UnaryInfC{$\Gamma  \vdash   \lfbsym{(}   \Lambda \gamma . \lfbnt{M}   \lfbsym{)} @ \lfbnt{C}   \colon  \lfbnt{T}  \lfbsym{[}   \gamma \coloneqq \lfbnt{C}   \lfbsym{]}$}
        \DisplayProof
        \Longrightarrow
        \bottomAlignProof
        \AxiomC{$\mathcal{D}'$}
        \noLine
        \UnaryInfC{$\Gamma  \vdash   \lfbnt{M} \lfbsym{[}  \gamma \coloneqq \lfbnt{C}  \lfbsym{;}  \bullet  \lfbsym{]}   \colon  \lfbnt{T}  \lfbsym{[}   \gamma \coloneqq \lfbnt{C}   \lfbsym{]}$}
        \DisplayProof
    \]

    \vspace{0.5em}
    Local Completeness
    \vspace{-1em}
    \[
        \bottomAlignProof
        \AxiomC{$\mathcal{D}$}
        \noLine
        \UnaryInfC{$\Gamma  \vdash  \lfbnt{M}  \colon   \forall \gamma . \lfbnt{T} $}
        \DisplayProof
        \Longrightarrow
        \bottomAlignProof
        \AxiomC{$\mathcal{D}'$}
        \noLine
        \UnaryInfC{$\Gamma  \vdash  \lfbnt{M}  \colon   \forall \gamma . \lfbnt{T} $}
        \UnaryInfC{$\Gamma  \vdash   \lfbnt{M} @  \delta    \colon  \lfbnt{T}  \lfbsym{[}   \gamma \coloneqq  \delta    \lfbsym{]}$}
        \AxiomC{$\delta \, \not\in \, \mathsf{FCV} \, \lfbsym{(}  \Gamma  \lfbsym{)}$}
        \BinaryInfC{$\Gamma  \vdash   \Lambda \delta . \lfbsym{(}   \lfbnt{M} @  \delta    \lfbsym{)}   \colon   \forall \delta . \lfbsym{(}  \lfbnt{T}  \lfbsym{[}   \gamma \coloneqq  \delta    \lfbsym{]}  \lfbsym{)} $}
        \DisplayProof
    \]

As a result, we obtain an additional reduction rule for $ \rightarrow_{\beta} $ below. We omit additional congruence rules.
\[
    \AxiomC{}
    \UnaryInfC{$ \lfbsym{(}   \Lambda \gamma . \lfbnt{M}   \lfbsym{)} @ \lfbnt{C}   \rightarrow_{\beta}   \lfbnt{M} \lfbsym{[}  \gamma \coloneqq \lfbnt{C}  \lfbsym{;}  \bullet  \lfbsym{]} $}
    \DisplayProof
\]
By using the substitution and context substitution lemmas, it is not hard to show subject reduction with regard to this $\beta$-reduction.

\begin{theorem}[Subject Reduction]
    \begin{enumerate}
        \item If $\Gamma  \vdash  \lfbnt{M}  \colon  \lfbnt{T}$ and $\lfbnt{M}  \rightarrow_{\beta}  \lfbnt{M'}$, then $\Gamma  \vdash  \lfbnt{M'}  \colon  \lfbnt{T}$.
        \item If $\Gamma  \vdash  \theta  \colon  \lfbnt{C}$ and $\theta  \rightarrow_{\beta}  \theta'$, then $\Gamma  \vdash  \theta'  \colon  \lfbnt{C}$.
    \end{enumerate}
\end{theorem}

Furthermore, $\beta$-reduction satisfies strong normalization and confluence. We only refer to confluence here because we will prove strong normalization in the next section.

\begin{theorem}[Confluence]
    If $\Gamma  \vdash  \lfbnt{M}  \colon  \lfbnt{T}$, $\lfbnt{M}  \rightarrow_{\beta}^{*}  \lfbnt{N_{{\mathrm{1}}}}$ and $\lfbnt{M}  \rightarrow_{\beta}^{*}  \lfbnt{N_{{\mathrm{2}}}}$, then there exists a term $\lfbnt{N_{{\mathrm{3}}}}$ such that $\lfbnt{N_{{\mathrm{1}}}}  \rightarrow_{\beta}^{*}  \lfbnt{N_{{\mathrm{3}}}}$ and $\lfbnt{N_{{\mathrm{2}}}}  \rightarrow_{\beta}^{*}  \lfbnt{N_{{\mathrm{3}}}}$. The same holds also for well-typed explicit substitutions.
\end{theorem}

\begin{proof}
    We use Newmann's lemma~\cite{Sorensen06}. We have strong normalizaiton from Theorem~\ref{thm:sn} (in Section~\ref{section:basicprops}) and weak confluence is easy to show.
\end{proof}

%% file: 5-basicprops.tex
\section{Parametric Reducibility and Strong Normalization} \label{section:basicprops}
This section provides a proof of strong normalization of $\beta$-reduction in \lamfb.  A common approach to proving strong normalization of a modal calculus is to provide a reduction-preserving translation to another strongly normalizing calculus such as simply typed lambda calculi~\cite{Martini96,Borghuis94}. We tried this approach, reducing strong normalization of \lamfb to that of System F~\cite{Girard89}. However, it turned out not to be straightforward.
Instead, we directly prove strong normalization of \lamfb using reducibility in this paper. We follow Girard's parametric reducibility~\cite{Girard89} to define reducibility with polymorphic contexts. We also adopted techniques from logical relation for Fitch-style modal calculi proposed by Valliappan et al.~\cite{Valliappan21} to extend reducibility to our Fitch-style modal type theory. Along with these existing methods, our approach requires several non-trivial extensions of reducibility for contextual modal types, which we detail in this section.

We start with the definition of neutral terms and explicit substitutions.

\begin{definition}[Neutral Terms and Explicit Substitutions]
    \begin{enumerate}
        \item A term $\lfbnt{M}$ is \emph{neutral} iff $\lfbnt{M}$ is either of a variable, application, unquote, or context application.
        \item An explicit substitution $\theta$ is \emph{neutral} iff it can be derived from the rules below.
    \end{enumerate}
    \[
    \bottomAlignProof
        \AxiomC{}
        \UnaryInfC{$ \bullet $ is neutral}
        \DisplayProof
        \quad
    \bottomAlignProof
        \AxiomC{$\theta$ is neutral}
        \AxiomC{$\lfbnt{M}$ is neutral}
        \BinaryInfC{$\theta  \lfbsym{,}  \lfbnt{M}$ is neutral}
        \DisplayProof
        \quad
    \bottomAlignProof
        \AxiomC{$\theta$ is neutral}
        \UnaryInfC{$\theta  \lfbsym{,}  \mathbb{x}$ is neutral}
        \DisplayProof
    \]
\end{definition}

The definition of neutral terms is standard, while the one for neutral explicit substitutions is somewhat specific to \lamfb but straightforward: $\theta$ is neutral iff all terms in $\theta$ are neutral. Then, we define reducibility candidates.

\begin{definition}[Reducibility Candidates]
    Given a type $\lfbnt{T}$, let $\mathcal{R}$ be a set of evident judgments of type $\lfbnt{T}$. We write $\mathcal{R}  \lfbsym{(}  \Gamma  \lfbsym{,}  \lfbnt{M}  \lfbsym{)}$ iff $\Gamma  \vdash  \lfbnt{M}  \colon  \lfbnt{T} \in \mathcal{R}$. $\mathcal{R}$ is a \emph{reducibility candidate of $\lfbnt{T}$} iff it satisfies all of the following properties.
    \begin{description}
        \item[CR0] If $\mathcal{R}  \lfbsym{(}  \Gamma  \lfbsym{,}  \lfbnt{M}  \lfbsym{)}$ and $\Gamma  \leq  \Gamma'$, then $\mathcal{R}  \lfbsym{(}  \Gamma'  \lfbsym{,}  \lfbnt{M}  \lfbsym{)}$.
        \item[CR1] If $\mathcal{R}  \lfbsym{(}  \Gamma  \lfbsym{,}  \lfbnt{M}  \lfbsym{)}$, then $\lfbnt{M}$ is strongly normalizing with regard to $ \rightarrow_{\beta} $.
        \item[CR2] If $\mathcal{R}  \lfbsym{(}  \Gamma  \lfbsym{,}  \lfbnt{M}  \lfbsym{)}$ and $\lfbnt{M}  \rightarrow_{\beta}  \lfbnt{M'}$, then $\mathcal{R}  \lfbsym{(}  \Gamma  \lfbsym{,}  \lfbnt{M'}  \lfbsym{)}$.
        \item[CR3] If $\lfbnt{M}$ is neutral, $\Gamma  \vdash  \lfbnt{M}  \colon  \lfbnt{T}$, and $\mathcal{R}  \lfbsym{(}  \Gamma  \lfbsym{,}  \lfbnt{M'}  \lfbsym{)}$ for all $\lfbnt{M'}$ such that $\lfbnt{M}  \rightarrow_{\beta}  \lfbnt{M'}$, then $\mathcal{R}  \lfbsym{(}  \Gamma  \lfbsym{,}  \lfbnt{M}  \lfbsym{)}$.
    \end{description}
    We also define a reducibility candidate of context $\lfbnt{C}$ similarly.
\end{definition}

We abbreviate reducibility candidate as RC. As a next step, we define \textit{reducibility candidate assignments} to define reducibility with parameters. We only need to care about reducibility candidates of contexts because \lamfb do not have polymorphic types.

\begin{center}
    \textbf{RC assignment} $\tilde{\Sigma} \Coloneqq  \bullet  \mid \tilde{\Sigma}  \lfbsym{,}  \gamma  \colon  \lfbnt{C}  \coloneqq  \mathcal{R}$ (where $\mathcal{R}$ is an RC of $\lfbnt{C}$)
\end{center}

$\tilde{\Sigma}$ is well-formed if it does not have duplicating context variables in it. We assume that all reducibility candidate assignments are well-formed. We write $\mathsf{dom} \, \lfbsym{(}  \tilde{\Sigma}  \lfbsym{)}$ for the set of context variables on the left side of $:=$ in $\tilde{\Sigma}$, and $\Sigma$ for the context substitution that we can obtain by forgetting RCs in $\tilde{\Sigma}$.

On top of that, we define reducibility with parameters.

\begin{definition}[Parametric Reducibility] \label{def:redparam}
    Given an RC assignment $\tilde{\Sigma}$, a type $\lfbnt{T}$, and a context $\lfbnt{C}$ where $\mathsf{FCV} \, \lfbsym{(}  \lfbnt{T}  \lfbsym{)}  \subseteq \mathsf{dom} \, \lfbsym{(}  \tilde{\Sigma}  \lfbsym{)}$ and $\mathsf{FCV} \, \lfbsym{(}  \lfbnt{C}  \lfbsym{)}  \subseteq \mathsf{dom} \, \lfbsym{(}  \tilde{\Sigma}  \lfbsym{)}$, we define $ \lfbkw{Red} _{ \lfbnt{T} } \lfbsym{[} \tilde{\Sigma} \lfbsym{]} $ and $ \lfbkw{Red} _{ \lfbnt{C} } \lfbsym{[} \tilde{\Sigma} \lfbsym{]} $, a set of evident judgments of a type $\lfbnt{T}  \lfbsym{[}  \Sigma  \lfbsym{]}$ and a context $\lfbnt{C}  \lfbsym{[}  \Sigma  \lfbsym{]}$, respectively, as follows. We write $ \lfbkw{Red} _{ \lfbnt{T} } \lfbsym{[} \tilde{\Sigma} \lfbsym{]}   \lfbsym{(}  \Gamma  \lfbsym{,}  \lfbnt{M}  \lfbsym{)}$ iff $\Gamma  \vdash  \lfbnt{M}  \colon  \lfbnt{T} \in  \lfbkw{Red} _{ \lfbnt{T} } \lfbsym{[} \tilde{\Sigma} \lfbsym{]} $; similarly for $ \lfbkw{Red} _{ \lfbnt{C} } \lfbsym{[} \tilde{\Sigma} \lfbsym{]}   \lfbsym{(}  \Gamma  \lfbsym{,}  \theta  \lfbsym{)}$.

    \begin{itemize}
        \item If $\lfbnt{T} = \iota$, $ \lfbkw{Red} _{ \lfbnt{T} } \lfbsym{[} \tilde{\Sigma} \lfbsym{]}   \lfbsym{(}  \Gamma  \lfbsym{,}  \lfbnt{M}  \lfbsym{)}$ iff $\lfbnt{M}$ is strongly normalizing with regard to $ \rightarrow_{\beta} $.
        \item If $\lfbnt{T} = \lfbnt{T_{{\mathrm{1}}}}  \rightarrow  \lfbnt{T_{{\mathrm{2}}}}$, $ \lfbkw{Red} _{ \lfbnt{T} } \lfbsym{[} \tilde{\Sigma} \lfbsym{]}   \lfbsym{(}  \Gamma  \lfbsym{,}  \lfbnt{M}  \lfbsym{)}$ iff $ \lfbkw{Red} _{ \lfbnt{T_{{\mathrm{2}}}} } \lfbsym{[} \tilde{\Sigma} \lfbsym{]}   \lfbsym{(}  \Delta  \lfbsym{,}  \lfbnt{M} \, \lfbnt{N}  \lfbsym{)}$ for any $\Delta$ and $\lfbnt{N}$ such that $\Gamma  \leq  \Delta$ and $ \lfbkw{Red} _{ \lfbnt{T_{{\mathrm{1}}}} } \lfbsym{[} \tilde{\Sigma} \lfbsym{]}   \lfbsym{(}  \Delta  \lfbsym{,}  \lfbnt{N}  \lfbsym{)}$.
        \item If $\lfbnt{T} = \lfbsym{[}  \lfbnt{C}  \vdash  \lfbnt{T'}  \lfbsym{]}$, $ \lfbkw{Red} _{ \lfbnt{T} } \lfbsym{[} \tilde{\Sigma} \lfbsym{]}   \lfbsym{(}  \Gamma  \lfbsym{,}  \lfbnt{M}  \lfbsym{)}$ iff $ \lfbkw{Red} _{ \lfbnt{T'} } \lfbsym{[} \tilde{\Sigma} \lfbsym{]}   \lfbsym{(}  \Delta'  \lfbsym{,}   \lfbkw{unq} _{ \lfbnt{k} } \lfbnt{M} [  \theta  ]   \lfbsym{)}$ for any $\Delta$, $\Delta'$, $\lfbnt{k}$ and $\theta$ such that $\Gamma  \leq  \Delta$, $\lfbnt{k}  \colon  \Delta  \lhd  \Delta'$ and $ \lfbkw{Red} _{ \lfbnt{C} } \lfbsym{[} \tilde{\Sigma} \lfbsym{]}   \lfbsym{(}  \Delta'  \lfbsym{,}  \theta  \lfbsym{)}$.
        \item If $\lfbnt{T} =  \forall \gamma . \lfbnt{T'} $, $ \lfbkw{Red} _{ \lfbnt{T} } \lfbsym{[} \tilde{\Sigma} \lfbsym{]}   \lfbsym{(}  \Gamma  \lfbsym{,}  \lfbnt{M}  \lfbsym{)}$ iff $ \lfbkw{Red} _{ \lfbnt{T'} } \lfbsym{[} \tilde{\Sigma}  \lfbsym{,}  \gamma  \colon  \lfbnt{C}  \coloneqq  \mathcal{R} \lfbsym{]}   \lfbsym{(}  \Gamma  \lfbsym{,}   \lfbnt{M} @ \lfbnt{C}   \lfbsym{)}$ for any $\lfbnt{C}$ and an RC $\mathcal{R}$ of $\lfbnt{C}$.
        \item If $\lfbnt{C} =  \bullet $, $ \lfbkw{Red} _{ \lfbnt{C} } \lfbsym{[} \tilde{\Sigma} \lfbsym{]}   \lfbsym{(}  \Gamma  \lfbsym{,}  \theta  \lfbsym{)}$ always holds (where $\theta$ is always $ \bullet $).
        \item If $\lfbnt{C} = \lfbnt{C'}  \lfbsym{,}  \lfbnt{T}$, $ \lfbkw{Red} _{ \lfbnt{C} } \lfbsym{[} \tilde{\Sigma} \lfbsym{]}   \lfbsym{(}  \Gamma  \lfbsym{,}  \theta  \lfbsym{)}$ iff $ \lfbkw{Red} _{ \lfbnt{C'} } \lfbsym{[} \tilde{\Sigma} \lfbsym{]}   \lfbsym{(}  \Gamma  \lfbsym{,}  \theta'  \lfbsym{)}$ and $ \lfbkw{Red} _{ \lfbnt{T} } \lfbsym{[} \tilde{\Sigma} \lfbsym{]}   \lfbsym{(}  \Gamma  \lfbsym{,}  \lfbnt{M}  \lfbsym{)}$ where $\theta = \theta', \lfbnt{M}$.
        \item If $\lfbnt{C} = \lfbnt{C'}  \lfbsym{,}  \gamma$: $ \lfbkw{Red} _{ \lfbnt{C} } \lfbsym{[} \tilde{\Sigma} \lfbsym{]}   \lfbsym{(}  \Gamma  \lfbsym{,}  \theta  \lfbsym{)}$ iff $ \lfbkw{Red} _{ \lfbnt{C'} } \lfbsym{[} \tilde{\Sigma} \lfbsym{]}   \lfbsym{(}  \Gamma  \lfbsym{,}  \theta_{{\mathrm{1}}}  \lfbsym{)}$ and $\mathcal{R}  \lfbsym{(}  \Gamma  \lfbsym{,}  \theta_{{\mathrm{2}}}  \lfbsym{)}$ for some $\theta_{{\mathrm{1}}}$, $\theta_{{\mathrm{2}}}$, and $\mathcal{R}$ such that $\theta =  \theta_{{\mathrm{1}}} \lfbsym{,} \theta_{{\mathrm{2}}} $ and $\gamma  \colon  \lfbnt{D}  \coloneqq  \mathcal{R} \, \in \, \tilde{\Sigma}$.
    \end{itemize}
\end{definition}

The definition for context variables is somewhat complicated. As $\lfbsym{(}  \lfbnt{C'}  \lfbsym{,}  \gamma  \lfbsym{)}  \lfbsym{[}  \Sigma  \lfbsym{]} =  \lfbnt{C'}  \lfbsym{[}  \Sigma  \lfbsym{]} \lfbsym{,} \lfbnt{D} $, we need two reducible explicit substitutions $\theta_{{\mathrm{1}}}$ and $\theta_{{\mathrm{2}}}$ where $\theta_{{\mathrm{1}}}$ is for $\lfbnt{C'}  \lfbsym{[}  \Sigma  \lfbsym{]}$ and $\theta_{{\mathrm{2}}}$ for $\lfbnt{D}$. Because $\lfbnt{D}$ comes from the context variable $\gamma$, we use the RC $\mathcal{R}$ from $\tilde{\Sigma}$ to confirm that $\theta_{{\mathrm{2}}}$ is reducible.

The parametric reducibility is a reducibility candidate in fact, stated as the following lemma.

\begin{lemma}
    \begin{enumerate}
        \item $ \lfbkw{Red} _{ \lfbnt{T} } \lfbsym{[} \tilde{\Sigma} \lfbsym{]} $ is an RC of $\lfbnt{T}$.
        \item $ \lfbkw{Red} _{ \lfbnt{C} } \lfbsym{[} \tilde{\Sigma} \lfbsym{]} $ is an RC of $\lfbnt{C}$.
    \end{enumerate}
\end{lemma}

\begin{longver}
    \begin{proof}
        \begin{description}
            \item[(CR0)] For types, we can prove by cases for each top-level form of $\lfbnt{T}$. For contexts we can prove by induction on the form of contexts.
            \item[(CR1)] We prove it by induction with regard the structure of $\lfbnt{T}$ and $\lfbnt{C}$.
                \begin{itemize}
                    \item If $\lfbnt{T} = \iota$, $\lfbnt{M}$ is strongly normalizing from the definition.
                    \item If $\lfbnt{T} = \lfbnt{S_{{\mathrm{1}}}}  \rightarrow  \lfbnt{S_{{\mathrm{2}}}}$, $ \lfbkw{Red} _{ \lfbnt{S_{{\mathrm{1}}}} } \lfbsym{[} \tilde{\Sigma} \lfbsym{]}   \lfbsym{(}  \lfbsym{(}  \Gamma  \lfbsym{,}  \lfbmv{x}  \colon  \lfbnt{S_{{\mathrm{1}}}}  \lfbsym{)}  \lfbsym{,}  \lfbmv{x}  \lfbsym{)}$ holds by the induction hypothesis of CR3, and hence $ \lfbkw{Red} _{ \lfbnt{S_{{\mathrm{2}}}} } \lfbsym{[} \tilde{\Sigma} \lfbsym{]}   \lfbsym{(}  \lfbsym{(}  \Gamma  \lfbsym{,}  \lfbmv{x}  \colon  \lfbnt{S_{{\mathrm{1}}}}  \lfbsym{)}  \lfbsym{,}  \lfbnt{M} \, \lfbmv{x}  \lfbsym{)}$ holds. We can then derive $\lfbnt{M} \, \lfbmv{x}$ is strongly normalizing from the induction hypothesis of CR1, and $\lfbnt{M}$ also is.
                    \item If $\lfbnt{T} = \lfbsym{[}  \lfbnt{C}  \vdash  \lfbnt{S}  \lfbsym{]}$, $ \lfbkw{Red} _{ \lfbnt{S} } \lfbsym{[} \tilde{\Sigma} \lfbsym{]}   \lfbsym{(}  \lfbsym{(}  \Gamma  \lfbsym{,}  \text{\faLock}  \lfbsym{,}  \overrightarrow{x}  \colon  \lfbnt{C}  \lfbsym{)}  \lfbsym{,}   \overrightarrow{x}   \lfbsym{)}$ holds by the induction hypothesis of CR3, and hence $ \lfbkw{Red} _{ \lfbnt{S_{{\mathrm{2}}}} } \lfbsym{[} \tilde{\Sigma} \lfbsym{]}   \lfbsym{(}  \lfbsym{(}  \Gamma  \lfbsym{,}  \text{\faLock}  \lfbsym{,}  \overrightarrow{x}  \colon  \lfbnt{C}  \lfbsym{)}  \lfbsym{,}   \lfbkw{unq} _{ \lfbsym{1} } \lfbnt{M} [   \overrightarrow{x}   ]   \lfbsym{)}$ holds. We can then derive $ \lfbkw{unq} _{ \lfbsym{1} } \lfbnt{M} [   \overrightarrow{x}   ] $ is strongly normalizing from the induction hypothesis of CR1, and $\lfbnt{M}$ also is.
                    \item If $\lfbnt{T} =  \forall \gamma . \lfbnt{S} $, $ \lfbkw{Red} _{ \lfbnt{S} } \lfbsym{[} \tilde{\Sigma}  \lfbsym{,}  \gamma  \colon  \lfbnt{C}  \coloneqq  \mathcal{R} \lfbsym{]}   \lfbsym{(}  \Gamma  \lfbsym{,}   \lfbnt{M} @ \lfbnt{C}   \lfbsym{)}$ for any $\lfbnt{C}$ and $\mathcal{R}$. We fix $\lfbnt{C}$ and $\mathcal{R}$ to one of them. Then $ \lfbnt{M} @ \lfbnt{C} $ is strongly normalizing by the induction hypothesis of CR1, and $\lfbnt{M}$ also is.
                    \item If $\lfbnt{C} =  \bullet $, $\theta$ is $ \bullet $ and already normal.
                    \item If $\lfbnt{C} = \lfbnt{C'}  \lfbsym{,}  \lfbnt{T}$, $ \lfbkw{Red} _{ \lfbnt{C'} } \lfbsym{[} \tilde{\Sigma} \lfbsym{]}   \lfbsym{(}  \Gamma  \lfbsym{,}  \theta'  \lfbsym{)}$ and $ \lfbkw{Red} _{ \lfbnt{T} } \lfbsym{[} \tilde{\Sigma} \lfbsym{]}   \lfbsym{(}  \Gamma  \lfbsym{,}  \lfbnt{M}  \lfbsym{)}$ holds where $\theta = \theta'  \lfbsym{,}  \lfbnt{M}$. $\theta'$ and $\lfbnt{M}$ are strongly normalizing by the induction hypothesis of CR1, and hence $\theta$ also is.
                    \item If $\lfbnt{C} = \lfbnt{C'}  \lfbsym{,}  \gamma$, $ \lfbkw{Red} _{ \lfbnt{C'} } \lfbsym{[} \tilde{\Sigma} \lfbsym{]}   \lfbsym{(}  \Gamma  \lfbsym{,}  \theta_{{\mathrm{1}}}  \lfbsym{)}$ and $\mathcal{R}  \lfbsym{(}  \Gamma  \lfbsym{,}  \theta_{{\mathrm{2}}}  \lfbsym{)}$  holds for some $\theta_{{\mathrm{1}}}$, and $\theta_{{\mathrm{2}}}$ and $\mathcal{R}$ such that $\theta =  \theta_{{\mathrm{1}}} \lfbsym{,} \theta_{{\mathrm{2}}} $ and $\mathcal{R}  \lfbsym{(}  \Gamma  \lfbsym{,}  \theta_{{\mathrm{2}}}  \lfbsym{)}$. $\theta_{{\mathrm{1}}}$ is strongly normalizing by the induction hypothesis of CR1. $\theta_{{\mathrm{2}}}$ is strongly normalizing because $\mathcal{R}  \lfbsym{(}  \Gamma  \lfbsym{,}  \theta_{{\mathrm{2}}}  \lfbsym{)}$. Therefore $\theta$ is also strongly normalizing.
                \end{itemize}
            \item[(CR2)] We can prove it by simple induction with regard to Definition~\ref{def:redparam}.
            \item[(CR3)] We prove it by induction with regard the structure of $\lfbnt{T}$ and $\lfbnt{C}$.
                \begin{itemize}
                    \item If $\lfbnt{T} = \iota$, $\lfbnt{M}$ is strongly normalizing from the hypothesis of CR3, and hence $ \lfbkw{Red} _{ \lfbnt{T} } \lfbsym{[} \tilde{\Sigma} \lfbsym{]}   \lfbsym{(}  \Gamma  \lfbsym{,}  \lfbnt{M}  \lfbsym{)}$ holds.
                    \item If $\lfbnt{T} = \lfbnt{S_{{\mathrm{1}}}}  \rightarrow  \lfbnt{S_{{\mathrm{2}}}}$, it suffices to show that $ \lfbkw{Red} _{ \lfbnt{S_{{\mathrm{2}}}} } \lfbsym{[} \tilde{\Sigma} \lfbsym{]}   \lfbsym{(}  \Gamma'  \lfbsym{,}  \lfbnt{M} \, \lfbnt{N}  \lfbsym{)}$ for any $\Gamma$ and $\lfbnt{N}$ such that $\Gamma  \leq  \Gamma'$ and $ \lfbkw{Red} _{ \lfbnt{S_{{\mathrm{1}}}} } \lfbsym{[} \tilde{\Sigma} \lfbsym{]}   \lfbsym{(}  \Gamma'  \lfbsym{,}  \lfbnt{N}  \lfbsym{)}$. We prove it as a result of the following sublemma.
                          \begin{enumerate}
                              \item Let $\Gamma$, $\lfbnt{M_{{\mathrm{1}}}}$, $\lfbnt{M_{{\mathrm{2}}}}$ be arbitrary named context and terms such that $\Gamma  \vdash  \lfbnt{M_{{\mathrm{1}}}}  \colon  \lfbnt{S_{{\mathrm{1}}}}  \rightarrow  \lfbnt{S_{{\mathrm{2}}}}$ and $ \lfbkw{Red} _{ \lfbnt{S_{{\mathrm{2}}}} } \lfbsym{[} \tilde{\Sigma} \lfbsym{]}   \lfbsym{(}  \Gamma  \lfbsym{,}  \lfbnt{M_{{\mathrm{2}}}}  \lfbsym{)}$. If $ \lfbkw{Red} _{ \lfbnt{S_{{\mathrm{1}}}}  \rightarrow  \lfbnt{S_{{\mathrm{2}}}} } \lfbsym{[} \tilde{\Sigma} \lfbsym{]}   \lfbsym{(}  \Gamma  \lfbsym{,}  \lfbnt{M'_{{\mathrm{1}}}}  \lfbsym{)}$ holds for any $\lfbnt{M'_{{\mathrm{1}}}}$ such that $\lfbnt{M_{{\mathrm{1}}}}  \rightarrow_{\beta}  \lfbnt{M'_{{\mathrm{1}}}}$, then $ \lfbkw{Red} _{ \lfbnt{S_{{\mathrm{2}}}} } \lfbsym{[} \tilde{\Sigma} \lfbsym{]}   \lfbsym{(}  \Gamma  \lfbsym{,}  \lfbnt{M_{{\mathrm{1}}}} \, \lfbnt{M_{{\mathrm{2}}}}  \lfbsym{)}$ holds.
                          \end{enumerate}
                          We can say that $\lfbnt{N}$ is strongly normalizing by the induction hypothesis of CR1, and hence we can prove this sublemma by induction on reduction steps of $\lfbnt{M_{{\mathrm{2}}}}$.

                          To prove $ \lfbkw{Red} _{ \lfbnt{S_{{\mathrm{2}}}} } \lfbsym{[} \tilde{\Sigma} \lfbsym{]}   \lfbsym{(}  \Gamma  \lfbsym{,}  \lfbnt{M_{{\mathrm{1}}}} \, \lfbnt{M_{{\mathrm{2}}}}  \lfbsym{)}$, it suffices to show that $ \lfbkw{Red} _{ \lfbnt{S_{{\mathrm{2}}}} } \lfbsym{[} \tilde{\Sigma} \lfbsym{]}   \lfbsym{(}  \Gamma  \lfbsym{,}  \lfbnt{N}  \lfbsym{)}$ for any $\lfbnt{N}$ such that $\lfbnt{M_{{\mathrm{1}}}} \, \lfbnt{M_{{\mathrm{2}}}}  \rightarrow_{\beta}  \lfbnt{N}$ according to the induction hypothesis of CR3. There are two cases for the forms of $\lfbnt{N}$:
                          \begin{itemize}
                              \item If $\lfbnt{N} = \lfbnt{M'_{{\mathrm{1}}}} \, \lfbnt{M_{{\mathrm{2}}}}$ where $\lfbnt{M_{{\mathrm{1}}}}  \rightarrow_{\beta}  \lfbnt{M'_{{\mathrm{1}}}}$, then $ \lfbkw{Red} _{ \lfbnt{S_{{\mathrm{1}}}}  \rightarrow  \lfbnt{S_{{\mathrm{2}}}} } \lfbsym{[} \tilde{\Sigma} \lfbsym{]}   \lfbsym{(}  \Gamma  \lfbsym{,}  \lfbnt{M'_{{\mathrm{1}}}}  \lfbsym{)}$ holds by the hypothesis of the sublemma. Therefore $ \lfbkw{Red} _{ \lfbnt{S_{{\mathrm{2}}}} } \lfbsym{[} \tilde{\Sigma} \lfbsym{]}   \lfbsym{(}  \Gamma  \lfbsym{,}  \lfbnt{M'_{{\mathrm{1}}}} \, \lfbnt{M_{{\mathrm{2}}}}  \lfbsym{)}$ holds because $ \lfbkw{Red} _{ \lfbnt{S_{{\mathrm{1}}}} } \lfbsym{[} \tilde{\Sigma} \lfbsym{]}   \lfbsym{(}  \Gamma  \lfbsym{,}  \lfbnt{M_{{\mathrm{2}}}}  \lfbsym{)}$ holds from the hypothesis of the sublemma.
                              \item If $\lfbnt{N} = \lfbnt{M_{{\mathrm{1}}}} \, \lfbnt{M'_{{\mathrm{2}}}}$ where $\lfbnt{M_{{\mathrm{2}}}}  \rightarrow_{\beta}  \lfbnt{M'_{{\mathrm{2}}}}$, then $ \lfbkw{Red} _{ \lfbnt{S_{{\mathrm{2}}}} } \lfbsym{[} \tilde{\Sigma} \lfbsym{]}   \lfbsym{(}  \Gamma  \lfbsym{,}  \lfbnt{M_{{\mathrm{1}}}} \, \lfbnt{M'_{{\mathrm{2}}}}  \lfbsym{)}$ holds from the induction hypothesis of the sublemma.
                          \end{itemize}
                          Note that $\lfbnt{M_{{\mathrm{1}}}} \, \lfbnt{M_{{\mathrm{2}}}}$ will not be a redex because $\lfbnt{M_{{\mathrm{1}}}}$ is neutral. Also, we only need to care the first case for the base case because $\lfbnt{M_{{\mathrm{2}}}}$ is normal form.

                    \item If $\lfbnt{T} = \lfbsym{[}  \lfbnt{C}  \vdash  \lfbnt{S}  \lfbsym{]}$, it suffices to show that $ \lfbkw{Red} _{ \lfbnt{C} } \lfbsym{[} \tilde{\Sigma} \lfbsym{]}   \lfbsym{(}  \Gamma''  \lfbsym{,}   \lfbkw{unq} _{ \lfbnt{k} } \lfbnt{M} [  \theta  ]   \lfbsym{)}$ for any $\Gamma'$, $\Gamma''$, $\lfbnt{k}$, $\theta$ such that $\Gamma  \leq  \Gamma'$, $\lfbnt{k}  \colon  \Gamma'  \lhd  \Gamma''$ and $ \lfbkw{Red} _{ \lfbnt{C} } \lfbsym{[} \tilde{\Sigma} \lfbsym{]}   \lfbsym{(}  \Gamma''  \lfbsym{,}  \theta  \lfbsym{)}$.  We prove it as a direct result of the following sublemma.
                          \begin{enumerate}
                              \item Assume that $\Gamma  \vdash  \lfbnt{M}  \colon  \lfbsym{[}  \lfbnt{C}  \vdash  \lfbnt{S}  \lfbsym{]}$ and $ \lfbkw{Red} _{ \lfbnt{C} } \lfbsym{[} \tilde{\Sigma} \lfbsym{]}   \lfbsym{(}  \Gamma_{{\mathrm{2}}}  \lfbsym{,}  \theta  \lfbsym{)}$ where $\Gamma  \leq  \Gamma_{{\mathrm{1}}}$ and $\lfbnt{k}  \colon  \Gamma_{{\mathrm{1}}}  \lhd  \Gamma_{{\mathrm{2}}}$ for some $\Gamma_{{\mathrm{1}}}$ and $\lfbnt{k}$. If $ \lfbkw{Red} _{ \lfbsym{[}  \lfbnt{C}  \vdash  \lfbnt{S}  \lfbsym{]} } \lfbsym{[} \tilde{\Sigma} \lfbsym{]}   \lfbsym{(}  \Gamma  \lfbsym{,}  \lfbnt{M'}  \lfbsym{)}$ holds for any $\lfbnt{M'}$ such that $\lfbnt{M}  \rightarrow_{\beta}  \lfbnt{M'}$, then $ \lfbkw{Red} _{ \lfbnt{S_{{\mathrm{2}}}} } \lfbsym{[} \tilde{\Sigma} \lfbsym{]}   \lfbsym{(}  \Gamma  \lfbsym{,}   \lfbkw{unq} _{ \lfbnt{k} } \lfbnt{M} [  \theta  ]   \lfbsym{)}$ holds.
                          \end{enumerate}
                          We can say that $\theta$ is strongly normalizing by the induction hypothesis of CR1, and hence we prove this sublemma by induction on reduction steps of $\theta$.

                          To prove $ \lfbkw{Red} _{ \lfbnt{S_{{\mathrm{2}}}} } \lfbsym{[} \tilde{\Sigma} \lfbsym{]}   \lfbsym{(}  \Gamma  \lfbsym{,}   \lfbkw{unq} _{ \lfbnt{k} } \lfbnt{M} [  \theta  ]   \lfbsym{)}$, it suffices to show that $ \lfbkw{Red} _{ \lfbnt{S_{{\mathrm{2}}}} } \lfbsym{[} \tilde{\Sigma} \lfbsym{]}   \lfbsym{(}  \Gamma  \lfbsym{,}  \lfbnt{N}  \lfbsym{)}$ for any $\lfbnt{N}$ such that $ \lfbkw{unq} _{ \lfbnt{k} } \lfbnt{M} [  \theta  ]   \rightarrow_{\beta}  \lfbnt{N}$ according to the induction hypothesis of CR3. There are two cases for the forms of $\lfbnt{N}$:
                          \begin{itemize}
                              \item If $\lfbnt{N} =  \lfbkw{unq} _{ \lfbnt{k} } \lfbnt{M'} [  \theta  ] $ where $\lfbnt{M}  \rightarrow_{\beta}  \lfbnt{M'}$, then $ \lfbkw{Red} _{ \lfbsym{[}  \lfbnt{C}  \vdash  \lfbnt{S}  \lfbsym{]} } \lfbsym{[} \tilde{\Sigma} \lfbsym{]}   \lfbsym{(}  \Gamma  \lfbsym{,}  \lfbnt{M'}  \lfbsym{)}$ holds by the hypothesis of the sublemma. Therefore $ \lfbkw{Red} _{ \lfbnt{S_{{\mathrm{2}}}} } \lfbsym{[} \tilde{\Sigma} \lfbsym{]}   \lfbsym{(}  \Gamma  \lfbsym{,}   \lfbkw{unq} _{ \lfbnt{k} } \lfbnt{M'} [  \theta  ]   \lfbsym{)}$ holds because $ \lfbkw{Red} _{ \lfbnt{C} } \lfbsym{[} \tilde{\Sigma} \lfbsym{]}   \lfbsym{(}  \Gamma  \lfbsym{,}  \theta  \lfbsym{)}$ holds from the hypothesis of the sublemma.
                              \item If $\lfbnt{N} =  \lfbkw{unq} _{ \lfbnt{k} } \lfbnt{M} [  \theta'  ] $ where $\theta  \rightarrow_{\beta}  \theta'$, then $ \lfbkw{Red} _{ \lfbnt{S} } \lfbsym{[} \tilde{\Sigma} \lfbsym{]}   \lfbsym{(}  \Gamma  \lfbsym{,}   \lfbkw{unq} _{ \lfbnt{k} } \lfbnt{M} [  \theta'  ]   \lfbsym{)}$ holds from the induction hypothesis of the sublemma.
                          \end{itemize}
                          Note that $ \lfbkw{unq} _{ \lfbnt{k} } \lfbnt{M} [  \theta'  ] $ will not be a redex because $\lfbnt{M}$ is neutral. Also, we only need to care the first case for the base case because $\theta$ is normal form.

                    \item If $\lfbnt{T} =  \forall \gamma . \lfbnt{S} $, it suffices to show that $ \lfbkw{Red} _{ \lfbnt{S} } \lfbsym{[} \tilde{\Sigma}  \lfbsym{,}  \gamma  \colon  \lfbnt{C}  \coloneqq  \mathcal{R} \lfbsym{]}   \lfbsym{(}  \Gamma  \lfbsym{,}   \lfbnt{M} @ \lfbnt{C}   \lfbsym{)}$ for any $\lfbnt{C}$ and $\mathcal{R}$.

                          For any $\lfbnt{M'}$ such that $ \lfbnt{M} @ \lfbnt{C}   \rightarrow_{\beta}  \lfbnt{M'}$, $\lfbnt{M'} =  \lfbnt{M''} @ \lfbnt{C} $ where $\lfbnt{M}  \rightarrow_{\beta}  \lfbnt{M''}$ because $\lfbnt{M}$ is neutral. We have $ \lfbkw{Red} _{ \lfbnt{T} } \lfbsym{[} \tilde{\Sigma} \lfbsym{]}   \lfbsym{(}  \Gamma  \lfbsym{,}  \lfbnt{M''}  \lfbsym{)}$ from the hypothesis, and hence $ \lfbkw{Red} _{ \lfbnt{S} } \lfbsym{[} \tilde{\Sigma}  \lfbsym{,}  \gamma  \colon  \lfbnt{C}  \coloneqq  \mathcal{R} \lfbsym{]}   \lfbsym{(}  \Gamma  \lfbsym{,}   \lfbnt{M''} @ \lfbnt{C}   \lfbsym{)}$. By the induction hypothesis of CR3, we finally derive that $ \lfbkw{Red} _{ \lfbnt{S} } \lfbsym{[} \tilde{\Sigma}  \lfbsym{,}  \gamma  \colon  \lfbnt{C}  \coloneqq  \mathcal{R} \lfbsym{]}   \lfbsym{(}  \Gamma  \lfbsym{,}   \lfbnt{M} @ \lfbnt{C}   \lfbsym{)}$.

                    \item For contexts, we can reduce the hypothesis to hypotheses for each elements of the contexts.

                    \item If $\lfbnt{C} =  \bullet $, CR3 holds because $\theta =  \bullet $, which is normal.

                    \item If $\lfbnt{C} = \lfbnt{C'}  \lfbsym{,}  \lfbnt{T}$, it suffices to show that $ \lfbkw{Red} _{ \lfbnt{C'} } \lfbsym{[} \tilde{\Sigma} \lfbsym{]}   \lfbsym{(}  \Gamma  \lfbsym{,}  \theta'  \lfbsym{)}$ and $ \lfbkw{Red} _{ \lfbnt{T} } \lfbsym{[} \tilde{\Sigma} \lfbsym{]}   \lfbsym{(}  \Gamma  \lfbsym{,}  \lfbnt{M}  \lfbsym{)}$ where $\theta = \theta'  \lfbsym{,}  \lfbnt{M}$.

                          For any $\lfbnt{M'}$ such that $\lfbnt{M}  \rightarrow_{\beta}  \lfbnt{M'}$, $ \lfbkw{Red} _{ \lfbnt{T} } \lfbsym{[} \tilde{\Sigma} \lfbsym{]}   \lfbsym{(}  \Gamma  \lfbsym{,}  \lfbnt{M'}  \lfbsym{)}$ holds by the hypothesis. Therefore $ \lfbkw{Red} _{ \lfbnt{T} } \lfbsym{[} \tilde{\Sigma} \lfbsym{]}   \lfbsym{(}  \Gamma  \lfbsym{,}  \lfbnt{M}  \lfbsym{)}$ holds by the induction hypothesis of CR3. We can also show that $ \lfbkw{Red} _{ \lfbnt{C'} } \lfbsym{[} \tilde{\Sigma} \lfbsym{]}   \lfbsym{(}  \Gamma  \lfbsym{,}  \lfbnt{M'}  \lfbsym{)}$ in the similar way.

                    \item If $\lfbnt{C} = \lfbnt{C'}  \lfbsym{,}  \gamma$, we can show CR3 in the similar way as the case above.
                \end{itemize}
        \end{description}
    \end{proof}

\end{longver}

We prove a few more auxiliary lemmas for the basic lemma. Firstly, we confirm that context substitution on types or context can be lifted to reducibility assignment.

\begin{lemma} \label{lem:redcsubstlift}
    \begin{enumerate}
        \item $ \lfbkw{Red} _{ \lfbnt{T}  \lfbsym{[}   \gamma \coloneqq \lfbnt{C}   \lfbsym{]} } \lfbsym{[} \tilde{\Sigma} \lfbsym{]}  =  \lfbkw{Red} _{ \lfbnt{T} } \lfbsym{[} \tilde{\Sigma}  \lfbsym{,}  \gamma  \colon  \lfbnt{C}  \lfbsym{[}  \Sigma  \lfbsym{]}  \coloneqq   \lfbkw{Red} _{ \lfbnt{C} } \lfbsym{[} \tilde{\Sigma} \lfbsym{]}  \lfbsym{]} $.
        \item $ \lfbkw{Red} _{ \lfbnt{D}  \lfbsym{[}   \gamma \coloneqq \lfbnt{C}   \lfbsym{]} } \lfbsym{[} \tilde{\Sigma} \lfbsym{]}  =  \lfbkw{Red} _{ \lfbnt{D} } \lfbsym{[} \tilde{\Sigma}  \lfbsym{,}  \gamma  \colon  \lfbnt{C}  \lfbsym{[}  \Sigma  \lfbsym{]}  \coloneqq   \lfbkw{Red} _{ \lfbnt{C} } \lfbsym{[} \tilde{\Sigma} \lfbsym{]}  \lfbsym{]} $.
    \end{enumerate}
\end{lemma}

\begin{longver}
    \begin{proof}
        We prove this by induction on the structure of $\lfbnt{T}$ and $\lfbnt{D}$.

        \begin{itemize}
            \item If $\lfbnt{T} = \iota$,
                  \begin{align*}
                       \lfbkw{Red} _{ \lfbnt{T}  \lfbsym{[}   \gamma \coloneqq \lfbnt{C}   \lfbsym{]} } \lfbsym{[} \tilde{\Sigma} \lfbsym{]}   \lfbsym{(}  \Gamma  \lfbsym{,}  \lfbnt{M}  \lfbsym{)} & \Leftrightarrow  \lfbkw{Red} _{ \iota } \lfbsym{[} \tilde{\Sigma} \lfbsym{]}   \lfbsym{(}  \Gamma  \lfbsym{,}  \lfbnt{M}  \lfbsym{)}                               \\                   \\
                                                             & \Leftrightarrow  \lfbkw{Red} _{ \iota } \lfbsym{[} \tilde{\Sigma}  \lfbsym{,}  \gamma  \colon  \lfbnt{C}  \lfbsym{[}  \Sigma  \lfbsym{]}  \coloneqq   \lfbkw{Red} _{ \lfbnt{C} } \lfbsym{[} \tilde{\Sigma} \lfbsym{]}  \lfbsym{]}   \lfbsym{(}  \Gamma  \lfbsym{,}  \lfbnt{M}  \lfbsym{)}
                  \end{align*}
            \item If $\lfbnt{T} = \lfbnt{S_{{\mathrm{1}}}}  \rightarrow  \lfbnt{S_{{\mathrm{2}}}}$,
                  \begin{align*}
                                      &  \lfbkw{Red} _{ \lfbnt{T}  \lfbsym{[}   \gamma \coloneqq \lfbnt{C}   \lfbsym{]} } \lfbsym{[} \tilde{\Sigma} \lfbsym{]}   \lfbsym{(}  \Gamma  \lfbsym{,}  \lfbnt{M}  \lfbsym{)}                                                                                  \\
                      \Leftrightarrow &  \lfbkw{Red} _{ \lfbnt{S_{{\mathrm{1}}}}  \lfbsym{[}   \gamma \coloneqq \lfbnt{C}   \lfbsym{]}  \rightarrow  \lfbnt{S_{{\mathrm{2}}}}  \lfbsym{[}   \gamma \coloneqq \lfbnt{C}   \lfbsym{]} } \lfbsym{[} \tilde{\Sigma} \lfbsym{]}   \lfbsym{(}  \Gamma  \lfbsym{,}  \lfbnt{M}  \lfbsym{)}                                                             \\
                      \Leftrightarrow &  \lfbkw{Red} _{ \lfbnt{S_{{\mathrm{2}}}}  \lfbsym{[}   \gamma \coloneqq \lfbnt{C}   \lfbsym{]} } \lfbsym{[} \tilde{\Sigma} \lfbsym{]}   \lfbsym{(}  \Gamma'  \lfbsym{,}  \lfbnt{M} \, \lfbnt{N}  \lfbsym{)}                                                                              \\
                                      & \text{ for any $\Gamma'$ and $\lfbnt{N}$ s.t. $\Gamma  \leq  \Gamma'$ and  $ \lfbkw{Red} _{ \lfbnt{S_{{\mathrm{1}}}}  \lfbsym{[}   \gamma \coloneqq \lfbnt{C}   \lfbsym{]} } \lfbsym{[} \tilde{\Sigma} \lfbsym{]}   \lfbsym{(}  \Gamma'  \lfbsym{,}  \lfbnt{N}  \lfbsym{)}$}                \\
                      \Leftrightarrow &  \lfbkw{Red} _{ \lfbnt{S_{{\mathrm{2}}}} } \lfbsym{[} \tilde{\Sigma}  \lfbsym{,}  \gamma  \colon  \lfbnt{C}  \lfbsym{[}  \Sigma  \lfbsym{]}  \coloneqq   \lfbkw{Red} _{ \lfbnt{C} } \lfbsym{[} \tilde{\Sigma} \lfbsym{]}  \lfbsym{]}   \lfbsym{(}  \Gamma'  \lfbsym{,}  \lfbnt{M} \, \lfbnt{N}  \lfbsym{)}                                                              \\
                                      & \text{ for any $\Gamma'$ and $\lfbnt{N}$ s.t. $\Gamma  \leq  \Gamma'$ and  $ \lfbkw{Red} _{ \lfbnt{S_{{\mathrm{1}}}} } \lfbsym{[} \tilde{\Sigma}  \lfbsym{,}  \gamma  \colon  \lfbnt{C}  \lfbsym{[}  \Sigma  \lfbsym{]}  \coloneqq   \lfbkw{Red} _{ \lfbnt{C} } \lfbsym{[} \tilde{\Sigma} \lfbsym{]}  \lfbsym{]}   \lfbsym{(}  \Gamma'  \lfbsym{,}  \lfbnt{N}  \lfbsym{)}$} \\
                                      & \text{ (by the induction hypothesis) }                                                                                  \\
                      \Leftrightarrow &  \lfbkw{Red} _{ \lfbnt{T} } \lfbsym{[} \tilde{\Sigma}  \lfbsym{,}  \gamma  \colon  \lfbnt{C}  \lfbsym{[}  \Sigma  \lfbsym{]}  \coloneqq   \lfbkw{Red} _{ \lfbnt{C} } \lfbsym{[} \tilde{\Sigma} \lfbsym{]}  \lfbsym{]}   \lfbsym{(}  \Gamma  \lfbsym{,}  \lfbnt{M}  \lfbsym{)}                                                                  \\
                  \end{align*}
            \item If $\lfbnt{T} = \lfbsym{[}  \lfbnt{D}  \vdash  \lfbnt{S}  \lfbsym{]}$,
                  \begin{align*}
                                      &  \lfbkw{Red} _{ \lfbnt{T}  \lfbsym{[}   \gamma \coloneqq \lfbnt{C}   \lfbsym{]} } \lfbsym{[} \tilde{\Sigma} \lfbsym{]}   \lfbsym{(}  \Gamma  \lfbsym{,}  \lfbnt{M}  \lfbsym{)}                                                                              \\
                      \Leftrightarrow &  \lfbkw{Red} _{ \lfbsym{[}  \lfbnt{D}  \lfbsym{[}   \gamma \coloneqq \lfbnt{C}   \lfbsym{]}  \vdash  \lfbnt{S}  \lfbsym{[}   \gamma \coloneqq \lfbnt{C}   \lfbsym{]}  \lfbsym{]} } \lfbsym{[} \tilde{\Sigma} \lfbsym{]}   \lfbsym{(}  \Gamma  \lfbsym{,}  \lfbnt{M}  \lfbsym{)}                                                        \\
                      \Leftrightarrow &  \lfbkw{Red} _{ \lfbnt{S}  \lfbsym{[}   \gamma \coloneqq \lfbnt{C}   \lfbsym{]} } \lfbsym{[} \tilde{\Sigma} \lfbsym{]}   \lfbsym{(}  \Gamma''  \lfbsym{,}   \lfbkw{unq} _{ \lfbnt{k} } \lfbnt{M} [  \theta  ]   \lfbsym{)}                                                                    \\
                                      & \text{ for any $\Gamma'$, $\Gamma''$, $\lfbnt{k}$ and $\theta$}                                                           \\
                                      & \text{ s.t. $\Gamma  \leq  \Gamma'$, $\lfbnt{k}  \colon  \Gamma'  \lhd  \Gamma''$ and  $ \lfbkw{Red} _{ \lfbnt{D}  \lfbsym{[}   \gamma \coloneqq \lfbnt{C}   \lfbsym{]} } \lfbsym{[} \tilde{\Sigma} \lfbsym{]}   \lfbsym{(}  \Gamma''  \lfbsym{,}  \theta  \lfbsym{)}$}                  \\
                      \Leftrightarrow &  \lfbkw{Red} _{ \lfbnt{S} } \lfbsym{[} \tilde{\Sigma}  \lfbsym{,}  \gamma  \colon  \lfbnt{C}  \lfbsym{[}  \Sigma  \lfbsym{]}  \coloneqq   \lfbkw{Red} _{ \lfbnt{C} } \lfbsym{[} \tilde{\Sigma} \lfbsym{]}  \lfbsym{]}   \lfbsym{(}  \Gamma''  \lfbsym{,}   \lfbkw{unq} _{ \lfbnt{k} } \lfbnt{M} [  \theta  ]   \lfbsym{)}                                                    \\
                                      & \text{ for any $\Gamma'$, $\Gamma''$, $\lfbnt{k}$ and $\theta$}                                                           \\
                                      & \text{ s.t. $\Gamma  \leq  \Gamma'$, $\lfbnt{k}  \colon  \Gamma'  \lhd  \Gamma''$ and  $ \lfbkw{Red} _{ \lfbnt{D} } \lfbsym{[} \tilde{\Sigma}  \lfbsym{,}  \gamma  \colon  \lfbnt{C}  \lfbsym{[}  \Sigma  \lfbsym{]}  \coloneqq   \lfbkw{Red} _{ \lfbnt{C} } \lfbsym{[} \tilde{\Sigma} \lfbsym{]}  \lfbsym{]}   \lfbsym{(}  \Gamma''  \lfbsym{,}  \theta  \lfbsym{)} $} \\
                                      & \text{ (by the induction hypothesis) }                                                                              \\
                      \Leftrightarrow &  \lfbkw{Red} _{ \lfbnt{T} } \lfbsym{[} \tilde{\Sigma}  \lfbsym{,}  \gamma  \colon  \lfbnt{C}  \lfbsym{[}  \Sigma  \lfbsym{]}  \coloneqq   \lfbkw{Red} _{ \lfbnt{C} } \lfbsym{[} \tilde{\Sigma} \lfbsym{]}  \lfbsym{]}   \lfbsym{(}  \Gamma  \lfbsym{,}  \lfbnt{M}  \lfbsym{)}                                                              \\
                  \end{align*}

            \item If $\lfbnt{T} =  \forall \delta . \lfbnt{S} $,
                  \begin{align*}
                                      &  \lfbkw{Red} _{ \lfbnt{T}  \lfbsym{[}   \gamma \coloneqq \lfbnt{C}   \lfbsym{]} } \lfbsym{[} \tilde{\Sigma} \lfbsym{]}   \lfbsym{(}  \Gamma  \lfbsym{,}  \lfbnt{M}  \lfbsym{)}                                          \\
                      \Leftrightarrow &  \lfbkw{Red} _{  \forall \delta . \lfbsym{(}  \lfbnt{S}  \lfbsym{[}   \gamma \coloneqq \lfbnt{C}   \lfbsym{]}  \lfbsym{)}  } \lfbsym{[} \tilde{\Sigma} \lfbsym{]}   \lfbsym{(}  \Gamma  \lfbsym{,}  \lfbnt{M}  \lfbsym{)}                                    \\
                                      & \text{where $\delta \, \not\in \, \mathsf{FCV} \, \lfbsym{(}  \lfbnt{C}  \lfbsym{)}$}                                                 \\
                      \Leftrightarrow &  \lfbkw{Red} _{ \lfbnt{S}  \lfbsym{[}   \gamma \coloneqq \lfbnt{C}   \lfbsym{]} } \lfbsym{[} \tilde{\Sigma}  \lfbsym{,}  \delta  \colon  \lfbnt{D}  \coloneqq  \mathcal{R} \lfbsym{]}   \lfbsym{(}  \Gamma  \lfbsym{,}   \lfbnt{M} @ \lfbnt{D}   \lfbsym{)}                        \\
                                      & \text{ for any $\lfbnt{D}$ and $\mathcal{R}$ }                                           \\
                      \Leftrightarrow &  \lfbkw{Red} _{ \lfbnt{S} } \lfbsym{[} \tilde{\Sigma}  \lfbsym{,}  \delta  \colon  \lfbnt{D}  \coloneqq  \mathcal{R}  \lfbsym{,}  \gamma  \colon  \lfbnt{C}  \coloneqq   \lfbkw{Red} _{ \lfbnt{C} } \lfbsym{[} \tilde{\Sigma}  \lfbsym{,}  \delta  \colon  \lfbnt{D}  \coloneqq  \mathcal{R} \lfbsym{]}  \lfbsym{]}   \lfbsym{(}  \Gamma  \lfbsym{,}   \lfbnt{M} @ \lfbnt{D}   \lfbsym{)} \\
                                      & \text{ for any $\lfbnt{D}$ and $\mathcal{R}$ }                                           \\
                                      & \text{ (by the induction hypothesis) }                                          \\
                      \Leftrightarrow &  \lfbkw{Red} _{ \lfbnt{S} } \lfbsym{[} \tilde{\Sigma}  \lfbsym{,}  \delta  \colon  \lfbnt{D}  \coloneqq  \mathcal{R}  \lfbsym{,}  \gamma  \colon  \lfbnt{C}  \lfbsym{[}  \Sigma  \lfbsym{]}  \coloneqq   \lfbkw{Red} _{ \lfbnt{C} } \lfbsym{[} \tilde{\Sigma} \lfbsym{]}  \lfbsym{]}   \lfbsym{(}  \Gamma  \lfbsym{,}   \lfbnt{M} @ \lfbnt{D}   \lfbsym{)}        \\
                                      & \text{ for any $\lfbnt{D}$ and $\mathcal{R}$ }                                           \\
                                      & \text{ (because $\delta$ does not occur in $\lfbnt{C}$) }                        \\
                      \Leftrightarrow &  \lfbkw{Red} _{ \lfbnt{T} } \lfbsym{[} \tilde{\Sigma}  \lfbsym{,}  \gamma  \colon  \lfbnt{C}  \lfbsym{[}  \Sigma  \lfbsym{]}  \coloneqq   \lfbkw{Red} _{ \lfbnt{C} } \lfbsym{[} \tilde{\Sigma} \lfbsym{]}  \lfbsym{]}   \lfbsym{(}  \Gamma  \lfbsym{,}  \lfbnt{M}  \lfbsym{)}                          \\
                  \end{align*}

            \item If $\lfbnt{D} =  \bullet $, (trivial).

            \item If $\lfbnt{D} = \lfbnt{D'}  \lfbsym{,}  \gamma$,
                  \begin{align*}
                                      &  \lfbkw{Red} _{ \lfbnt{D}  \lfbsym{[}   \gamma \coloneqq \lfbnt{C}   \lfbsym{]} } \lfbsym{[} \tilde{\Sigma} \lfbsym{]}   \lfbsym{(}  \Gamma  \lfbsym{,}  \theta  \lfbsym{)}                                                           \\
                      \Leftrightarrow &  \lfbkw{Red} _{  \lfbnt{D'}  \lfbsym{[}   \gamma \coloneqq \lfbnt{C}   \lfbsym{]} \lfbsym{,} \lfbnt{C}  } \lfbsym{[} \tilde{\Sigma} \lfbsym{]}   \lfbsym{(}  \Gamma  \lfbsym{,}  \theta  \lfbsym{)}                                                     \\
                      \Leftrightarrow &  \lfbkw{Red} _{ \lfbnt{D'}  \lfbsym{[}   \gamma \coloneqq \lfbnt{C}   \lfbsym{]} } \lfbsym{[} \tilde{\Sigma} \lfbsym{]}   \lfbsym{(}  \Gamma  \lfbsym{,}  \theta_{{\mathrm{1}}}  \lfbsym{)} \text{ and }  \lfbkw{Red} _{ \lfbnt{C} } \lfbsym{[} \tilde{\Sigma} \lfbsym{]}   \lfbsym{(}  \Gamma  \lfbsym{,}  \theta_{{\mathrm{2}}}  \lfbsym{)}                 \\
                                      & \text{ where $\theta =  \theta_{{\mathrm{1}}} \lfbsym{,} \theta_{{\mathrm{2}}} $}                                                       \\
                      \Leftrightarrow &  \lfbkw{Red} _{ \lfbnt{D'} } \lfbsym{[} \tilde{\Sigma}  \lfbsym{,}  \gamma  \colon  \lfbnt{C}  \lfbsym{[}  \Sigma  \lfbsym{]}  \coloneqq   \lfbkw{Red} _{ \lfbnt{C} } \lfbsym{[} \tilde{\Sigma} \lfbsym{]}  \lfbsym{]}   \lfbsym{(}  \Gamma  \lfbsym{,}  \theta_{{\mathrm{1}}}  \lfbsym{)} \text{ and }  \lfbkw{Red} _{ \lfbnt{C} } \lfbsym{[} \tilde{\Sigma} \lfbsym{]}   \lfbsym{(}  \Gamma  \lfbsym{,}  \theta_{{\mathrm{2}}}  \lfbsym{)} \\
                                      & \text{ (by the induction hypothesis) }                                                            \\
                      \Leftrightarrow &  \lfbkw{Red} _{ \lfbnt{D} } \lfbsym{[} \tilde{\Sigma}  \lfbsym{,}  \gamma  \colon  \lfbnt{C}  \lfbsym{[}  \Sigma  \lfbsym{]}  \coloneqq   \lfbkw{Red} _{ \lfbnt{C} } \lfbsym{[} \tilde{\Sigma} \lfbsym{]}  \lfbsym{]}   \lfbsym{(}  \Gamma  \lfbsym{,}  \theta  \lfbsym{)}                                           \\
                  \end{align*}

            \item If $\lfbnt{D} = \lfbnt{D'}  \lfbsym{,}  \delta$ where $\gamma \not = \delta$, (omitted).

        \end{itemize}
    \end{proof}
\end{longver}

Besides, we state three lemmas that correspond to introduction of function types, contextual modal types, and polymorphic context types.

\begin{lemma} \label{lem:redlamintro}
    If $\Gamma  \lfbsym{,}  \lfbmv{x}  \colon  \lfbnt{S}  \lfbsym{[}  \Sigma  \lfbsym{]}  \vdash  \lfbnt{M}  \colon  \lfbnt{T}  \lfbsym{[}  \Sigma  \lfbsym{]}$ and $ \lfbkw{Red} _{ \lfbnt{T} } \lfbsym{[} \tilde{\Sigma} \lfbsym{]}   \lfbsym{(}  \Gamma'  \lfbsym{,}  \lfbnt{M}  \lfbsym{[}   id _{ \Gamma }   \lfbsym{,}  \lfbmv{x}  \coloneqq  \lfbnt{N}  \lfbsym{]}  \lfbsym{)}$ for any $\Gamma'$ and $\lfbnt{N}$ such that $\Gamma  \leq  \Gamma'$ and $ \lfbkw{Red} _{ \lfbnt{S} } \lfbsym{[} \tilde{\Sigma} \lfbsym{]}   \lfbsym{(}  \Gamma'  \lfbsym{,}  \lfbnt{N}  \lfbsym{)}$, then $ \lfbkw{Red} _{ \lfbnt{S}  \rightarrow  \lfbnt{T} } \lfbsym{[} \tilde{\Sigma} \lfbsym{]}   \lfbsym{(}  \Gamma  \lfbsym{,}   \lambda \lfbmv{x} ^{ \lfbnt{S} }. \lfbnt{M}   \lfbsym{)}$.
\end{lemma}

\begin{longver}
    \begin{proof}
        omitted. See the proof of Lemma~\ref{lem:redquointro}.
    \end{proof}
\end{longver}

\begin{lemma}  \label{lem:redquointro}
    If $\Gamma  \lfbsym{,}  \text{\faLock}  \lfbsym{,}  \overrightarrow{x}  \colon  \lfbnt{C}  \lfbsym{[}  \Sigma  \lfbsym{]}  \vdash  \lfbnt{M}  \colon  \lfbnt{T}  \lfbsym{[}  \Sigma  \lfbsym{]}$ and $ \lfbkw{Red} _{ \lfbnt{T} } \lfbsym{[} \tilde{\Sigma} \lfbsym{]}   \lfbsym{(}  \Gamma_{{\mathrm{2}}}  \lfbsym{,}  \lfbnt{M}  \lfbsym{[}   id _{ \Gamma_{{\mathrm{1}}} }   \lfbsym{,}   \text{\faLock} _{ \lfbnt{k} }   \lfbsym{,}  \overrightarrow{x}  \coloneqq  \theta  \lfbsym{]}  \lfbsym{)}$ for any $\Gamma_{{\mathrm{1}}}$, $\Gamma_{{\mathrm{2}}}$, $k$ and $\theta$ such that $\Gamma  \leq  \Gamma_{{\mathrm{1}}}$, $\lfbnt{k}  \colon  \Gamma_{{\mathrm{1}}}  \lhd  \Gamma_{{\mathrm{2}}}$ and $ \lfbkw{Red} _{ \lfbnt{C} } \lfbsym{[} \tilde{\Sigma} \lfbsym{]}   \lfbsym{(}  \Gamma_{{\mathrm{2}}}  \lfbsym{,}  \theta  \lfbsym{)}$, then $ \lfbkw{Red} _{ \lfbsym{[}  \lfbnt{C}  \vdash  \lfbnt{T}  \lfbsym{]} } \lfbsym{[} \tilde{\Sigma} \lfbsym{]}   \lfbsym{(}  \Gamma  \lfbsym{,}   \lfbkw{quo} \langle  \overrightarrow{x}  \colon  \lfbnt{C}  \lfbsym{[}  \Sigma  \lfbsym{]}  \rangle \lfbnt{M}   \lfbsym{)}$.
\end{lemma}

\begin{longver}
    \begin{proof}
        In order to prove $ \lfbkw{Red} _{ \lfbsym{[}  \lfbnt{C}  \vdash  \lfbnt{T}  \lfbsym{]} } \lfbsym{[} \tilde{\Sigma} \lfbsym{]}   \lfbsym{(}  \Gamma  \lfbsym{,}   \lfbkw{quo} \langle  \overrightarrow{x}  \colon  \lfbnt{C}  \lfbsym{[}  \Sigma  \lfbsym{]}  \rangle \lfbnt{M}   \lfbsym{)}$, it suffices to show that $ \lfbkw{Red} _{ \lfbnt{T} } \lfbsym{[} \tilde{\Sigma} \lfbsym{]}   \lfbsym{(}  \Gamma_{{\mathrm{2}}}  \lfbsym{,}   \lfbkw{unq} _{ \lfbnt{k} } \lfbsym{(}   \lfbkw{quo} \langle  \overrightarrow{x}  \colon  \lfbnt{C}  \lfbsym{[}  \Sigma  \lfbsym{]}  \rangle \lfbnt{M}   \lfbsym{)} [  \theta  ]   \lfbsym{)}$ for any $\Gamma_{{\mathrm{1}}}$, $\Gamma_{{\mathrm{2}}}$, $\lfbnt{k}$ and $\theta$ such that $\Gamma  \leq  \Gamma_{{\mathrm{1}}}$, $\lfbnt{k}  \colon  \Gamma_{{\mathrm{1}}}  \lhd  \Gamma_{{\mathrm{2}}}$ and $ \lfbkw{Red} _{ \lfbnt{C} } \lfbsym{[} \tilde{\Sigma} \lfbsym{]}   \lfbsym{(}  \Gamma_{{\mathrm{2}}}  \lfbsym{,}  \theta  \lfbsym{)}$. We get this subgoal as the direct result of the following sublemma:

        \begin{enumerate}
            \item Assume $\Gamma  \lfbsym{,}  \text{\faLock}  \lfbsym{,}  \overrightarrow{x}  \colon  \lfbnt{C}  \lfbsym{[}  \Sigma  \lfbsym{]}  \vdash  \lfbnt{M}  \colon  \lfbnt{T}  \lfbsym{[}  \Sigma  \lfbsym{]}$, $\Gamma  \leq  \Gamma_{{\mathrm{1}}}$, $\lfbnt{k}  \colon  \Gamma_{{\mathrm{1}}}  \lhd  \Gamma_{{\mathrm{2}}}$ and $ \lfbkw{Red} _{ \lfbnt{C} } \lfbsym{[} \tilde{\Sigma} \lfbsym{]}   \lfbsym{(}  \Gamma_{{\mathrm{2}}}  \lfbsym{,}  \theta  \lfbsym{)}$. Then $ \lfbkw{Red} _{ \lfbnt{T} } \lfbsym{[} \tilde{\Sigma} \lfbsym{]}   \lfbsym{(}  \Gamma_{{\mathrm{2}}}  \lfbsym{,}   \lfbkw{unq} _{ \lfbnt{k} } \lfbsym{(}   \lfbkw{quo} \langle  \overrightarrow{x}  \colon  \lfbnt{C}  \lfbsym{[}  \Sigma  \lfbsym{]}  \rangle \lfbnt{M}   \lfbsym{)} [  \theta  ]   \lfbsym{)}$ holds if $ \lfbkw{Red} _{ \lfbnt{T} } \lfbsym{[} \tilde{\Sigma} \lfbsym{]}   \lfbsym{(}  \Gamma'_{{\mathrm{2}}}  \lfbsym{,}  \lfbnt{M}  \lfbsym{[}   id _{ \Gamma'_{{\mathrm{1}}} }   \lfbsym{,}   \text{\faLock} _{ \lfbnt{k'} }   \lfbsym{,}  \overrightarrow{x}  \coloneqq  \theta'  \lfbsym{]}  \lfbsym{)}$ holds for all $\Gamma'_{{\mathrm{1}}}$, $\Gamma'_{{\mathrm{2}}}$, $\lfbnt{k'}$, $\theta'$ such that $\Gamma  \leq  \Gamma'_{{\mathrm{1}}}$, $\lfbnt{k}  \colon  \Gamma'_{{\mathrm{1}}}  \lhd  \Gamma'_{{\mathrm{2}}}$ and $ \lfbkw{Red} _{ \lfbnt{C} } \lfbsym{[} \tilde{\Sigma} \lfbsym{]}   \lfbsym{(}  \Gamma'_{{\mathrm{2}}}  \lfbsym{,}  \theta'  \lfbsym{)}$.
        \end{enumerate}

        $\lfbnt{M}  \lfbsym{[}   id _{ \Gamma'_{{\mathrm{1}}} }   \lfbsym{,}   \text{\faLock} _{ \lfbnt{k'} }   \lfbsym{,}  \overrightarrow{x}  \coloneqq  \theta'  \lfbsym{]}$ and $\theta$ are strongly normalizing from CR1, and hence $\lfbnt{M}$ is also strongly normalizing. Therefore, we can prove this sublemma by induction on reduction steps of $\lfbnt{M}$ and $\theta$.

        It suffices to show that $ \lfbkw{Red} _{ \lfbnt{T} } \lfbsym{[} \tilde{\Sigma} \lfbsym{]}   \lfbsym{(}  \Gamma_{{\mathrm{2}}}  \lfbsym{,}  \lfbnt{N}  \lfbsym{)}$ where $ \lfbkw{unq} _{ \lfbnt{k} } \lfbsym{(}   \lfbkw{quo} \langle  \overrightarrow{x}  \colon  \lfbnt{C}  \lfbsym{[}  \Sigma  \lfbsym{]}  \rangle \lfbnt{M}   \lfbsym{)} [  \theta  ]   \rightarrow_{\beta}  \lfbnt{N}$ by CR3. There are 3 cases with regard to the reduction steps:
        \begin{enumerate}
            \item If $\lfbnt{N} = \lfbnt{M}  \lfbsym{[}   id _{ \Gamma_{{\mathrm{1}}} }   \lfbsym{,}   \text{\faLock} _{ \lfbnt{k} }   \lfbsym{,}  \overrightarrow{x}  \coloneqq  \theta  \lfbsym{]}$, $ \lfbkw{Red} _{ \lfbnt{T} } \lfbsym{[} \tilde{\Sigma} \lfbsym{]}   \lfbsym{(}  \Gamma_{{\mathrm{2}}}  \lfbsym{,}  \lfbnt{N}  \lfbsym{)}$ holds from the hypothesis of the sublemma.
            \item If $\lfbnt{N} =  \lfbkw{unq} _{ \lfbnt{k} } \lfbsym{(}   \lfbkw{quo} \langle  \overrightarrow{x}  \colon  \lfbnt{C}  \lfbsym{[}  \Sigma  \lfbsym{]}  \rangle \lfbnt{M'}   \lfbsym{)} [  \theta  ] $ where $\lfbnt{M}  \rightarrow_{\beta}  \lfbnt{M'}$, we can confirm that $\lfbnt{M'}$ satisfies all preconditions of the sublemma from CR2. And hence we can apply the induction hypothesis of the sublemma to show $ \lfbkw{Red} _{ \lfbnt{T} } \lfbsym{[} \tilde{\Sigma} \lfbsym{]}   \lfbsym{(}  \Gamma_{{\mathrm{2}}}  \lfbsym{,}  \lfbnt{N}  \lfbsym{)}$.
            \item If $\lfbnt{N} =  \lfbkw{unq} _{ \lfbnt{k} } \lfbsym{(}   \lfbkw{quo} \langle  \overrightarrow{x}  \colon  \lfbnt{C}  \lfbsym{[}  \Sigma  \lfbsym{]}  \rangle \lfbnt{M}   \lfbsym{)} [  \theta'  ] $ where $\theta  \rightarrow_{\beta}  \theta'$, same as the case above.
        \end{enumerate}
    \end{proof}
\end{longver}

\begin{lemma} \label{lem:redpolyctxintro}\sloppy
    If $\Gamma  \vdash  \lfbnt{M}  \colon  \lfbnt{T}  \lfbsym{[}  \Sigma  \lfbsym{]}$, $\gamma \, \not\in \, \mathsf{FCV} \, \lfbsym{(}  \Gamma  \lfbsym{)} \, \cup \, \mathsf{FCV} \, \lfbsym{(}  \Sigma  \lfbsym{)} \, \cup \, \mathsf{dom} \, \lfbsym{(}  \Sigma  \lfbsym{)}$,  and $ \lfbkw{Red} _{ \lfbnt{T} } \lfbsym{[} \tilde{\Sigma}  \lfbsym{,}  \gamma  \colon  \lfbnt{C}  \coloneqq  \mathcal{R} \lfbsym{]}   \lfbsym{(}  \Gamma  \lfbsym{,}   \lfbnt{M} \lfbsym{[}  \gamma \coloneqq \lfbnt{C}  \lfbsym{;}  \bullet  \lfbsym{]}   \lfbsym{)}$ for any $\lfbnt{C}$, $\mathcal{R}$ such that $\mathcal{R}$ is an RC of $\lfbnt{C}$, then $ \lfbkw{Red} _{  \forall \gamma . \lfbnt{T}  } \lfbsym{[} \tilde{\Sigma} \lfbsym{]}   \lfbsym{(}  \Gamma  \lfbsym{,}   \Lambda \gamma . \lfbnt{M}   \lfbsym{)}$.
\end{lemma}

\begin{longver}
    \begin{proof}
        It suffices to show that $ \lfbkw{Red} _{ \lfbnt{T} } \lfbsym{[} \tilde{\Sigma}  \lfbsym{,}  \gamma  \colon  \lfbnt{C}  \coloneqq  \mathcal{R} \lfbsym{]}   \lfbsym{(}  \Gamma  \lfbsym{,}   \lfbsym{(}   \Lambda \gamma . \lfbnt{M}   \lfbsym{)} @ \lfbnt{C}   \lfbsym{)}$ for any $\lfbnt{C}$ and $\mathcal{R}$. We fix $\lfbnt{C}$ and $\mathcal{R}$ to arbitrary one. By CR3, it suffices to show that $ \lfbkw{Red} _{ \lfbnt{T} } \lfbsym{[} \tilde{\Sigma}  \lfbsym{,}  \gamma  \colon  \lfbnt{C}  \coloneqq  \mathcal{R} \lfbsym{]}   \lfbsym{(}  \Gamma  \lfbsym{,}  \lfbnt{N}  \lfbsym{)}$ where $ \lfbsym{(}   \Lambda \gamma . \lfbnt{M}   \lfbsym{)} @ \lfbnt{C}   \rightarrow_{\beta}  \lfbnt{N}$ by CR3 (subgoal).

        From the hypothesis and CR1, $ \lfbnt{M} \lfbsym{[}  \gamma \coloneqq \lfbnt{C}  \lfbsym{;}  \bullet  \lfbsym{]} $ is strongly normalizing and hence $\lfbnt{M}$ also is. We prove the subgoal by induction on the number of reduction steps of $\lfbnt{M}$.

        We have three cases with regard to the reduction steps:
        \begin{enumerate}
            \item If $\lfbnt{N} =  \lfbnt{M} \lfbsym{[}  \gamma \coloneqq \lfbnt{C}  \lfbsym{;}  \bullet  \lfbsym{]} $, $ \lfbkw{Red} _{ \lfbnt{T} } \lfbsym{[} \tilde{\Sigma}  \lfbsym{,}  \gamma  \colon  \lfbnt{C}  \coloneqq  \mathcal{R} \lfbsym{]}   \lfbsym{(}  \Gamma  \lfbsym{,}  \lfbnt{N}  \lfbsym{)}$ holds from the hypothesis of this lemma.
            \item If $\lfbnt{N} =  \lfbsym{(}   \Lambda \gamma . \lfbnt{M'}   \lfbsym{)} @ \lfbnt{C} $ where $\lfbnt{M}  \rightarrow_{\beta}  \lfbnt{M'}$,  $ \lfbkw{Red} _{ \lfbnt{T} } \lfbsym{[} \tilde{\Sigma}  \lfbsym{,}  \gamma  \colon  \lfbnt{C}  \coloneqq  \mathcal{R} \lfbsym{]}   \lfbsym{(}  \Gamma  \lfbsym{,}  \lfbnt{N}  \lfbsym{)}$ holds from CR1 and the induction hypothesis of the subgoal.
        \end{enumerate}
    \end{proof}
\end{longver}

\begin{shortver}
    We can prove these lemmas by CR3 and induction on the number of reduction steps of strongly normalizing terms/explicit substitutions.
\end{shortver}

Before the basic lemma, we define reducibility for named contexts. Although we would like something like $ \lfbkw{Red} _{ \Gamma } \lfbsym{[} \tilde{\Sigma} \lfbsym{]} $, this definition does not work because it does not have information on how a named context with series variable $ \mathbb{x}  \colon  \gamma $ will be replaced. Therefore we also need to pass series variables substitution, like $ \lfbkw{Red} _{ \Gamma } \lfbsym{[} \tilde{\Sigma} \lfbsym{,} \bar{\sigma} \lfbsym{]} $ in the same way as context substitution for named contexts.

\begin{definition}[Reducibility for Substitution] \label{def:redsub}
    Given an RC assignment $\tilde{\Sigma}$, a named context $\Gamma$, and a series substitution $\bar{\sigma}$ where $\mathsf{FCV} \, \lfbsym{(}  \Gamma  \lfbsym{)} \subseteq \mathsf{dom} \, \lfbsym{(}  \tilde{\Sigma}  \lfbsym{)}$, we define $ \lfbkw{Red} _{ \Gamma } \lfbsym{[} \tilde{\Sigma} \lfbsym{,} \bar{\sigma} \lfbsym{]} $, a set of evident judgments of a named context $ \Gamma   \lfbsym{[}   \Sigma   \lfbsym{;}   \bar{\sigma}   \lfbsym{]} $, as follows:

    \begin{itemize}\sloppy
        \item If $\Gamma =  \bullet $, then $ \lfbkw{Red} _{ \Gamma } \lfbsym{[} \tilde{\Sigma} \lfbsym{,} \bar{\sigma} \lfbsym{]}   \lfbsym{(}  \Delta  \lfbsym{,}  \sigma  \lfbsym{)}$ always holds (where $\sigma =  \bullet $).
        \item If $\Gamma = \Gamma'  \lfbsym{,}  \lfbmv{x}  \colon  \lfbnt{T}$, then $ \lfbkw{Red} _{ \Gamma } \lfbsym{[} \tilde{\Sigma} \lfbsym{,} \bar{\sigma} \lfbsym{]}   \lfbsym{(}  \Delta  \lfbsym{,}  \sigma  \lfbsym{)}$ iff $ \lfbkw{Red} _{ \Gamma' } \lfbsym{[} \tilde{\Sigma} \lfbsym{,} \bar{\sigma} \lfbsym{]}   \lfbsym{(}  \Delta  \lfbsym{,}  \sigma'  \lfbsym{)}$ and $ \lfbkw{Red} _{ \lfbnt{T} } \lfbsym{[} \tilde{\Sigma} \lfbsym{]}   \lfbsym{(}  \Delta  \lfbsym{,}  \lfbnt{M}  \lfbsym{)}$ for some $\sigma'$, $\lfbnt{M}$ such that $\sigma = \lfbsym{(}  \sigma'  \lfbsym{,}  \lfbmv{x}  \coloneqq  \lfbnt{M}  \lfbsym{)}$.
        \item If $\Gamma = \Gamma'  \lfbsym{,}  \mathbb{x}  \colon  \gamma$, then $ \lfbkw{Red} _{ \Gamma } \lfbsym{[} \tilde{\Sigma} \lfbsym{,} \bar{\sigma} \lfbsym{]}   \lfbsym{(}  \Delta  \lfbsym{,}  \sigma  \lfbsym{)}$ iff $ \lfbkw{Red} _{ \Gamma' } \lfbsym{[} \tilde{\Sigma} \lfbsym{,} \bar{\sigma} \lfbsym{]}   \lfbsym{(}  \Delta  \lfbsym{,}  \sigma'  \lfbsym{)}$ and $\mathcal{R}  \lfbsym{(}  \Delta  \lfbsym{,}  \theta  \lfbsym{)}$ for some $\sigma'$, $\theta$ and $\mathcal{R}$ such that $\gamma  \colon  \lfbnt{C}  \coloneqq  \mathcal{R} \, \in \, \tilde{\Sigma}$, $\sigma = \lfbsym{(}  \sigma'  \lfbsym{,}  \overrightarrow{x}  \coloneqq  \theta  \lfbsym{)}$ and $\mathbb{x}  \coloneqq  \overrightarrow{x} \, \in \, \bar{\sigma} $.
        \item If $\Gamma = \Gamma'  \lfbsym{,}  \text{\faLock}$, then $ \lfbkw{Red} _{ \Gamma } \lfbsym{[} \tilde{\Sigma} \lfbsym{,} \bar{\sigma} \lfbsym{]}   \lfbsym{(}  \Delta  \lfbsym{,}  \sigma  \lfbsym{)}$ iff $ \lfbkw{Red} _{ \Gamma' } \lfbsym{[} \tilde{\Sigma} \lfbsym{,} \bar{\sigma} \lfbsym{]}   \lfbsym{(}  \Delta  \uparrow  \lfbnt{k}  \lfbsym{,}  \sigma'  \lfbsym{)}$ for some $\sigma'$ and $\lfbnt{k}$ such that $\sigma = \lfbsym{(}  \sigma'  \lfbsym{,}   \text{\faLock} _{ \lfbnt{k} }   \lfbsym{)}$.
    \end{itemize}
\end{definition}

We use series variables substitution in the third rule to generate a substitution for $ \lfbsym{(}   \mathbb{x}  \colon  \gamma   \lfbsym{)}   \lfbsym{[}   \Sigma   \lfbsym{;}   \bar{\sigma}   \lfbsym{]}  =  \overrightarrow{x}  \colon  \lfbnt{C} $. Finally, we prove the basic lemma.

\begin{lemma}[Basic Lemma] \label{lem:basic}
    \begin{itemize}\sloppy
        \item If $\Gamma  \vdash  \lfbnt{M}  \colon  \lfbnt{T}$ and $ \lfbkw{Red} _{ \Gamma } \lfbsym{[} \tilde{\Sigma} \lfbsym{,} \bar{\sigma} \lfbsym{]}   \lfbsym{(}  \Delta  \lfbsym{,}  \sigma'  \lfbsym{)}$ where $\bar{\sigma} =  \mathsf{destruct} \lfbsym{(} \Gamma \lfbsym{;} \Sigma \lfbsym{)} $, then $ \lfbkw{Red} _{ \lfbnt{T} } \lfbsym{[} \tilde{\Sigma} \lfbsym{]}   \lfbsym{(}  \Delta  \lfbsym{,}   \lfbnt{M} \lfbsym{[} \Sigma \lfbsym{;} \bar{\sigma} \lfbsym{]}   \lfbsym{[}  \sigma'  \lfbsym{]}  \lfbsym{)}$.
        \item If $\Gamma  \vdash  \theta  \colon  \lfbnt{C}$ and $ \lfbkw{Red} _{ \Gamma } \lfbsym{[} \tilde{\Sigma} \lfbsym{,} \bar{\sigma} \lfbsym{]}   \lfbsym{(}  \Delta  \lfbsym{,}  \sigma'  \lfbsym{)}$ where $\bar{\sigma} =  \mathsf{destruct} \lfbsym{(} \Gamma \lfbsym{;} \Sigma \lfbsym{)} $, then $ \lfbkw{Red} _{ \lfbnt{C} } \lfbsym{[} \tilde{\Sigma} \lfbsym{]}   \lfbsym{(}  \Delta  \lfbsym{,}   \theta \lfbsym{[} \Sigma \lfbsym{;} \bar{\sigma} \lfbsym{]}   \lfbsym{[}  \sigma'  \lfbsym{]}  \lfbsym{)}$.
    \end{itemize}
\end{lemma}

\begin{longver}
    \begin{proof}
        By induction on the derivation of $\Gamma  \vdash  \lfbnt{M}  \colon  \lfbnt{T}$. We cover major cases in this proof.

        \begin{itemize}
            \item If the derivation is quo-intro $\Gamma  \vdash   \lfbkw{quo} \langle \Gamma' \rangle \lfbnt{M'}   \colon  \lfbsym{[}   \mathsf{rg} ( \Gamma' )   \vdash  \lfbnt{T'}  \lfbsym{]}$, it is derived from $ \Gamma  \lfbsym{,}  \text{\faLock} \lfbsym{,} \Gamma'   \vdash  \lfbnt{M'}  \colon  \lfbnt{T'}$.
                  \begin{align*}
                                      &  \lfbkw{Red} _{ \lfbsym{[}   \mathsf{rg} ( \Gamma' )   \vdash  \lfbnt{T'}  \lfbsym{]} } \lfbsym{[} \tilde{\Sigma} \lfbsym{]}   \lfbsym{(}  \Delta  \lfbsym{,}   \lfbsym{(}   \lfbkw{quo} \langle \Gamma' \rangle \lfbnt{M'}   \lfbsym{)} \lfbsym{[} \Sigma \lfbsym{;} \bar{\sigma} \lfbsym{]}   \lfbsym{[}  \sigma'  \lfbsym{]}  \lfbsym{)}                                                               \\
                      \Leftrightarrow &  \lfbkw{Red} _{ \lfbsym{[}   \mathsf{rg} ( \Gamma' )   \vdash  \lfbnt{T'}  \lfbsym{]} } \lfbsym{[} \tilde{\Sigma} \lfbsym{]}   \lfbsym{(}  \Delta  \lfbsym{,}   \lfbkw{quo} \langle  \Gamma'   \lfbsym{[}   \Sigma   \lfbsym{;}   \bar{\sigma}'   \lfbsym{]}  \rangle \lfbsym{(}   \lfbnt{M'} \lfbsym{[} \Sigma \lfbsym{;} \lfbsym{(}   \bar{\sigma}  \lfbsym{,} \, \text{\faLock} \lfbsym{,} \bar{\sigma}'   \lfbsym{)} \lfbsym{]}   \lfbsym{[}   \sigma'  \lfbsym{,}   \text{\faLock} _{ \lfbsym{1} }  \lfbsym{,}  id _{  \Gamma'   \lfbsym{[}   \Sigma   \lfbsym{;}   \bar{\sigma}'   \lfbsym{]}  }    \lfbsym{]}  \lfbsym{)}   \lfbsym{)} \\
                                      & \text{ where $\bar{\sigma}' =  \mathsf{destruct} \lfbsym{(} \Gamma' \lfbsym{;} \Sigma \lfbsym{)} $}                                                                        \\
                  \end{align*}

                  By Lemma~\ref{lem:redquointro}, it suffices to show that $ \lfbkw{Red} _{ \lfbnt{T'} } \lfbsym{[} \tilde{\Sigma} \lfbsym{]}   \lfbsym{(}  \Delta_{{\mathrm{2}}}  \lfbsym{,}   \lfbnt{M'} \lfbsym{[} \Sigma \lfbsym{;} \lfbsym{(}   \bar{\sigma}  \lfbsym{,} \, \text{\faLock} \lfbsym{,} \bar{\sigma}'   \lfbsym{)} \lfbsym{]}   \lfbsym{[}   \sigma'  \lfbsym{,}   \text{\faLock} _{ \lfbsym{1} }  \lfbsym{,}  id _{  \Gamma'   \lfbsym{[}   \Sigma   \lfbsym{;}   \bar{\sigma}'   \lfbsym{]}  }    \lfbsym{]}  \lfbsym{[}   id _{ \Delta }   \lfbsym{,}   \text{\faLock} _{ \lfbsym{(}  \lfbnt{k}  \lfbsym{)} }   \lfbsym{,}   \Gamma'   \lfbsym{[}   \Sigma   \lfbsym{;}   \bar{\sigma}'   \lfbsym{]}   \coloneqq  \theta  \lfbsym{]}  \lfbsym{)}$ for any $\Delta_{{\mathrm{1}}}$, $\Delta_{{\mathrm{2}}}$, $\lfbnt{k}$ and $\theta$ such that $\Delta  \leq  \Delta_{{\mathrm{1}}}$, $\lfbnt{k}  \colon  \Delta_{{\mathrm{1}}}  \lhd  \Delta_{{\mathrm{2}}}$ and $ \lfbkw{Red} _{  \mathsf{rg} ( \Gamma' )  } \lfbsym{[} \tilde{\Sigma} \lfbsym{]}   \lfbsym{(}  \Delta_{{\mathrm{2}}}  \lfbsym{,}  \theta  \lfbsym{)}$ (subgoal).

                  We fix $\Delta_{{\mathrm{1}}}$, $\Delta_{{\mathrm{2}}}$, $\lfbnt{k}$ and $\theta$ to arbitrary ones. From $ \lfbkw{Red} _{ \Gamma } \lfbsym{[} \tilde{\Sigma} \lfbsym{,} \bar{\sigma} \lfbsym{]}   \lfbsym{(}  \Delta  \lfbsym{,}  \sigma'  \lfbsym{)}$, we obtain $ \lfbkw{Red} _{  \Gamma  \lfbsym{,}  \text{\faLock} \lfbsym{,} \Gamma'  } \lfbsym{[} \tilde{\Sigma} \lfbsym{,} \lfbsym{(}   \bar{\sigma}  \lfbsym{,} \, \text{\faLock} \lfbsym{,} \bar{\sigma}'   \lfbsym{)} \lfbsym{]}   \lfbsym{(}  \Delta_{{\mathrm{2}}}  \lfbsym{,}  \lfbsym{(}  \sigma'  \lfbsym{,}   \text{\faLock} _{ \lfbsym{(}  \lfbnt{k}  \lfbsym{)} }   \lfbsym{,}   \Gamma'   \lfbsym{[}   \Sigma   \lfbsym{;}   \bar{\sigma}'   \lfbsym{]}   \coloneqq  \theta  \lfbsym{)}  \lfbsym{)}$.
                  Then we can apply the induction hypothesis to $ \Gamma  \lfbsym{,}  \text{\faLock} \lfbsym{,} \Gamma'   \vdash  \lfbnt{M'}  \colon  \lfbnt{T'}$, and we get $ \lfbkw{Red} _{ \lfbnt{T'} } \lfbsym{[} \tilde{\Sigma} \lfbsym{]}   \lfbsym{(}  \Delta_{{\mathrm{2}}}  \lfbsym{,}   \lfbnt{M'} \lfbsym{[} \Sigma \lfbsym{;} \lfbsym{(}   \bar{\sigma}  \lfbsym{,} \, \text{\faLock} \lfbsym{,} \bar{\sigma}'   \lfbsym{)} \lfbsym{]}   \lfbsym{[}  \sigma'  \lfbsym{,}   \text{\faLock} _{ \lfbsym{(}  \lfbnt{k}  \lfbsym{)} }   \lfbsym{,}   \Gamma'   \lfbsym{[}   \Sigma   \lfbsym{;}   \bar{\sigma}'   \lfbsym{]}   \coloneqq  \theta  \lfbsym{]}  \lfbsym{)}$ (Note that $ \bar{\sigma}  \lfbsym{,} \, \text{\faLock} \lfbsym{,} \bar{\sigma}' $ = $ \mathsf{destruct} \lfbsym{(} \lfbsym{(}   \Gamma  \lfbsym{,}  \text{\faLock} \lfbsym{,} \Gamma'   \lfbsym{)} \lfbsym{;} \Sigma \lfbsym{)} $). This is equal to the subgoal because $ \lfbnt{M'} \lfbsym{[} \Sigma \lfbsym{;} \lfbsym{(}   \bar{\sigma}  \lfbsym{,} \, \text{\faLock} \lfbsym{,} \bar{\sigma}'   \lfbsym{)} \lfbsym{]}   \lfbsym{[}  \sigma'  \lfbsym{,}   \text{\faLock} _{ \lfbsym{(}  \lfbnt{k}  \lfbsym{)} }   \lfbsym{,}   \Gamma'   \lfbsym{[}   \Sigma   \lfbsym{;}   \bar{\sigma}'   \lfbsym{]}   \coloneqq  \theta  \lfbsym{]} =  \lfbnt{M'} \lfbsym{[} \Sigma \lfbsym{;} \lfbsym{(}   \bar{\sigma}  \lfbsym{,} \, \text{\faLock} \lfbsym{,} \bar{\sigma}'   \lfbsym{)} \lfbsym{]}   \lfbsym{[}   \sigma'  \lfbsym{,}   \text{\faLock} _{ \lfbsym{1} }  \lfbsym{,}  id _{  \Gamma'   \lfbsym{[}   \Sigma   \lfbsym{;}   \bar{\sigma}'   \lfbsym{]}  }    \lfbsym{]}  \lfbsym{[}   id _{ \Delta }   \lfbsym{,}   \text{\faLock} _{ \lfbsym{(}  \lfbnt{k}  \lfbsym{)} }   \lfbsym{,}   \Gamma'   \lfbsym{[}   \Sigma   \lfbsym{;}   \bar{\sigma}'   \lfbsym{]}   \coloneqq  \theta  \lfbsym{]}$.

            \item If the derivation is $\forall$-intro $\Gamma  \vdash   \Lambda \gamma . \lfbnt{M'}   \colon   \forall \gamma . \lfbnt{T'} $, it is derived from $\Gamma  \vdash  \lfbnt{M'}  \colon  \lfbnt{T'}$. We rename $\gamma$ to sufficiently fresh one that is $\gamma \, \not\in \, \mathsf{FCV} \, \lfbsym{(}  \Gamma  \lfbsym{)} \, \cup \, \mathsf{FCV} \, \lfbsym{(}  \Sigma  \lfbsym{)} \, \cup \, \mathsf{dom} \, \lfbsym{(}  \Sigma  \lfbsym{)}$.
                  Then we want to show $ \lfbkw{Red} _{  \forall \gamma . \lfbnt{T'}  } \lfbsym{[} \tilde{\Sigma} \lfbsym{]}   \lfbsym{(}  \Delta  \lfbsym{,}   \lfbsym{(}   \Lambda \gamma . \lfbnt{M'}   \lfbsym{)} \lfbsym{[} \Sigma \lfbsym{;} \bar{\sigma} \lfbsym{]}   \lfbsym{[}  \sigma'  \lfbsym{]}  \lfbsym{)} \Leftrightarrow  \lfbkw{Red} _{  \forall \gamma . \lfbnt{T'}  } \lfbsym{[} \tilde{\Sigma} \lfbsym{]}   \lfbsym{(}  \Delta  \lfbsym{,}   \Lambda \gamma . \lfbsym{(}   \lfbnt{M'} \lfbsym{[} \Sigma \lfbsym{;} \bar{\sigma} \lfbsym{]}   \lfbsym{[}  \sigma'  \lfbsym{]}  \lfbsym{)}   \lfbsym{)}$. By Lemma~\ref{lem:redpolyctxintro}, it suffices to show that $ \lfbkw{Red} _{ \lfbnt{T'} } \lfbsym{[} \tilde{\Sigma}  \lfbsym{,}  \gamma  \colon  \lfbnt{C}  \coloneqq  \mathcal{R} \lfbsym{]}   \lfbsym{(}  \Delta  \lfbsym{,}    \lfbnt{M'} \lfbsym{[} \Sigma \lfbsym{;} \bar{\sigma} \lfbsym{]}   \lfbsym{[}  \sigma'  \lfbsym{]} \lfbsym{[}  \gamma \coloneqq \lfbnt{C}  \lfbsym{;}  \bullet  \lfbsym{]}   \lfbsym{)}$ for any $\lfbnt{C}$ and $\mathcal{R}$.

                  We fix  $\lfbnt{C}$ and $\mathcal{R}$ to arbitrary one.
                  We have $ \lfbkw{Red} _{ \lfbnt{T'} } \lfbsym{[} \tilde{\Sigma}  \lfbsym{,}  \gamma  \colon  \lfbnt{C}  \coloneqq  \mathcal{R} \lfbsym{]}   \lfbsym{(}  \Delta  \lfbsym{,}    \lfbnt{M'} \lfbsym{[} \Sigma \lfbsym{;} \bar{\sigma} \lfbsym{]}   \lfbsym{[}  \sigma'  \lfbsym{]} \lfbsym{[}  \gamma \coloneqq \lfbnt{C}  \lfbsym{;}  \bullet  \lfbsym{]}   \lfbsym{)} \Leftrightarrow  \lfbkw{Red} _{ \lfbnt{T'} } \lfbsym{[} \tilde{\Sigma}  \lfbsym{,}  \gamma  \colon  \lfbnt{C}  \coloneqq  \mathcal{R} \lfbsym{]}   \lfbsym{(}  \Delta  \lfbsym{,}   \lfbnt{M'} \lfbsym{[} \lfbsym{(}  \Sigma  \lfbsym{,}  \gamma  \coloneqq  \lfbnt{C}  \lfbsym{)} \lfbsym{;} \bar{\sigma} \lfbsym{]}   \lfbsym{[}  \sigma'  \lfbsym{]}  \lfbsym{)}$ by the freshness of $\gamma$. Also  $ \lfbkw{Red} _{ \Gamma } \lfbsym{[} \lfbsym{(}  \tilde{\Sigma}  \lfbsym{,}  \gamma  \colon  \lfbnt{C}  \coloneqq  \mathcal{R}  \lfbsym{)} \lfbsym{,} \bar{\sigma} \lfbsym{]}   \lfbsym{(}  \Delta  \lfbsym{,}  \sigma'  \lfbsym{)}$ and $\bar{\sigma} =  \mathsf{destruct} \lfbsym{(} \Gamma \lfbsym{;} \lfbsym{(}  \Sigma  \lfbsym{,}  \gamma  \coloneqq  \lfbnt{C}  \lfbsym{)} \lfbsym{)} $ holds from the freshness of $\gamma$. Therefore we can apply the induction hypothesis to obtain $ \lfbkw{Red} _{ \lfbnt{T'} } \lfbsym{[} \tilde{\Sigma}  \lfbsym{,}  \gamma  \colon  \lfbnt{C}  \coloneqq  \mathcal{R} \lfbsym{]}   \lfbsym{(}  \Delta  \lfbsym{,}   \lfbnt{M'} \lfbsym{[} \lfbsym{(}  \Sigma  \lfbsym{,}  \gamma  \coloneqq  \lfbnt{C}  \lfbsym{)} \lfbsym{;} \bar{\sigma} \lfbsym{]}   \lfbsym{[}  \sigma'  \lfbsym{]}  \lfbsym{)}$.

            \item If the derivation is $\forall$-elim $\Gamma  \vdash   \lfbnt{M'} @ \lfbnt{C}   \colon  \lfbnt{T'}  \lfbsym{[}   \gamma \coloneqq \lfbnt{C}   \lfbsym{]}$, it is derived from $\Gamma  \vdash  \lfbnt{M'}  \colon   \forall \gamma . \lfbnt{T'} $. We rename $\gamma$ to sufficiently fresh one that is $\gamma \, \not\in \, \mathsf{FCV} \, \lfbsym{(}  \Gamma  \lfbsym{)} \, \cup \, \mathsf{FCV} \, \lfbsym{(}  \Sigma  \lfbsym{)} \, \cup \, \mathsf{dom} \, \lfbsym{(}  \Sigma  \lfbsym{)}$.
                  \begin{align*}
                                      &  \lfbkw{Red} _{ \lfbnt{T'}  \lfbsym{[}   \gamma \coloneqq \lfbnt{C}   \lfbsym{]} } \lfbsym{[} \tilde{\Sigma} \lfbsym{]}   \lfbsym{(}  \Delta  \lfbsym{,}   \lfbsym{(}   \lfbnt{M'} @ \lfbnt{C}   \lfbsym{)} \lfbsym{[} \Sigma \lfbsym{;} \bar{\sigma} \lfbsym{]}   \lfbsym{[}  \sigma'  \lfbsym{]}  \lfbsym{)}                       \\
                      \Leftrightarrow &  \lfbkw{Red} _{ \lfbnt{T'}  \lfbsym{[}   \gamma \coloneqq \lfbnt{C}   \lfbsym{]} } \lfbsym{[} \tilde{\Sigma} \lfbsym{]}   \lfbsym{(}  \Delta  \lfbsym{,}  \lfbsym{(}    \lfbnt{M'} \lfbsym{[} \Sigma \lfbsym{;} \bar{\sigma} \lfbsym{]}   \lfbsym{[}  \sigma'  \lfbsym{]} @ \lfbnt{C}  \lfbsym{[}  \Sigma  \lfbsym{]}   \lfbsym{)}  \lfbsym{)}                 \\
                      \Leftrightarrow &  \lfbkw{Red} _{ \lfbnt{T'} } \lfbsym{[} \tilde{\Sigma}  \lfbsym{,}  \gamma  \colon  \lfbnt{C}  \lfbsym{[}  \Sigma  \lfbsym{]}  \coloneqq   \lfbkw{Red} _{ \lfbnt{C} } \lfbsym{[} \tilde{\Sigma} \lfbsym{]}  \lfbsym{]}   \lfbsym{(}  \Delta  \lfbsym{,}  \lfbsym{(}    \lfbnt{M'} \lfbsym{[} \Sigma \lfbsym{;} \bar{\sigma} \lfbsym{]}   \lfbsym{[}  \sigma'  \lfbsym{]} @ \lfbnt{C}  \lfbsym{[}  \Sigma  \lfbsym{]}   \lfbsym{)}  \lfbsym{)} \\
                                      & \text{ by Lemma~\ref{lem:redcsubstlift}}                                                \\
                  \end{align*}
                  We can derive the last statement from $ \lfbkw{Red} _{  \forall \gamma . \lfbnt{T'}  } \lfbsym{[} \tilde{\Sigma} \lfbsym{]}   \lfbsym{(}  \Delta  \lfbsym{,}  \lfbsym{(}   \lfbnt{M'} \lfbsym{[} \Sigma \lfbsym{;} \bar{\sigma} \lfbsym{]}   \lfbsym{[}  \sigma'  \lfbsym{]}  \lfbsym{)}  \lfbsym{)}$, and it holds by the induction hypothesis.

            \item If the derivation is series variables $\Gamma  \vdash  \theta'  \lfbsym{,}  \mathbb{x}  \colon  \lfbnt{C_{{\mathrm{1}}}}  \lfbsym{,}  \gamma$, it is derived from $\Gamma  \vdash  \theta'  \colon  \lfbnt{C'}$ and $\mathbb{x}  \colon  \gamma \, \in \, \mathsf{head} \, \lfbsym{(}  \Gamma  \lfbsym{)}$.
                  It suffices to show that $ \lfbkw{Red} _{ \lfbnt{C'} } \lfbsym{[} \tilde{\Sigma} \lfbsym{]}   \lfbsym{(}  \Delta  \lfbsym{,}   \theta' \lfbsym{[} \Sigma \lfbsym{;} \bar{\sigma} \lfbsym{]}   \lfbsym{[}  \sigma'  \lfbsym{]}  \lfbsym{)}$ and $\mathcal{R}  \lfbsym{(}  \Delta  \lfbsym{,}    \mathbb{x}  \lfbsym{[} \Sigma \lfbsym{;} \bar{\sigma} \lfbsym{]}   \lfbsym{[}  \sigma'  \lfbsym{]}  \lfbsym{)}$ where $\gamma  \colon  \lfbnt{D}  \coloneqq  \mathcal{R} \, \in \, \tilde{\Sigma}$. We have the former by applying the induction hypothesis to $\Gamma  \vdash  \theta'  \colon  \lfbnt{C'}$. From $\mathbb{x}  \colon  \gamma \, \in \, \mathsf{head} \, \lfbsym{(}  \Gamma  \lfbsym{)}$, we can reconstruct $ \lfbkw{Red} _{  \mathbb{x}  \colon  \gamma  } \lfbsym{[} \tilde{\Sigma} \lfbsym{,} \bar{\sigma} \lfbsym{]}   \lfbsym{(}  \Delta  \lfbsym{,}   \overrightarrow{x}  \coloneqq   \overrightarrow{x}   \lfbsym{[}  \sigma'  \lfbsym{]}   \lfbsym{)}$ from $ \lfbkw{Red} _{ \Gamma } \lfbsym{[} \tilde{\Sigma} \lfbsym{,} \bar{\sigma} \lfbsym{]}   \lfbsym{(}  \Delta  \lfbsym{,}  \sigma'  \lfbsym{)}$ where $\overrightarrow{x} =   \mathbb{x}  \lfbsym{[} \Sigma \lfbsym{;} \bar{\sigma} \lfbsym{]} $. We finally get $\mathcal{R}  \lfbsym{(}  \Delta  \lfbsym{,}   \overrightarrow{x}   \lfbsym{[}  \sigma'  \lfbsym{]}  \lfbsym{)}$ from Definition~\ref{def:redsub}.
        \end{itemize}
    \end{proof}
\end{longver}

Strong normalization is proved as a special case of the basic lemma, where we choose $\Sigma$, $\bar{\sigma}$ and $\sigma'$ as identity substitutions respectively.

\begin{theorem}[Strong Normalization]\label{thm:sn}
    If $\Gamma  \vdash  \lfbnt{M}  \colon  \lfbnt{T}$, then $\lfbnt{M}$ is strongly normalizing with regard to $ \rightarrow_{\beta} $.
\end{theorem}

%% file: 6-embedlttt.tex
\section{Embedding Linear-Time Temporal Type Theory} \label{section:embedlttt}
In this section, we present a type-preserving embedding from a linear-time temporal type theory to \lamfb. Linear-time temporal type theories like \lamcirc by Davies~\cite{Davies17} provide a basis for multi-stage programming languages such as MetaOCaml \cite{Calcango03,Kiselyov14}, and hence is proved to provide benefits for real use cases. Thus, it is important to discuss formal connections between linear-time temporal and contextual modal types. It helps us figure out whether we can use contextual modal types instead of linear temporal types in multi-stage programming languages.

In this section, we first define \lamcirctwo, a two-level fragment of \lamcirc, as a source language to simplify our embedding. Then, we discuss the core insights of our embedding from \lamcirctwo and give a formal definition of our embedding from \lamcirctwo to \lamfb. We also prove its soundness---the embedding preserves typing---while a proof that it also preserves semantics is left for future work.

\begin{figure}[t]
    \begin{tabular}{lcl}
        \textbf{Level-0 Types} & $T^{0}, S^{0}$ & $\coloneqq \iota \mid S^{0}  \rightarrow  T^{0} \mid  \bigcirc  T^{1} $ \\
        \textbf{Level-0 Terms} & $M^{0}, N^{0}$ & $\coloneqq \lfbmv{x} \mid  \lambda \lfbmv{x} ^{ T^{0} }. M^{0}  \mid M^{0} \, N^{0} \mid  \lfbkw{quo}   M^{1} $ \\
        \textbf{Level-1 Types} & $T^{1}, S^{1}$ & $\coloneqq \iota \mid S^{1}  \rightarrow  T^{1}$ \\
        \textbf{Level-1 Terms} & $M^{1}, N^{1}$ & $\coloneqq \lfbmv{x} \mid  \lambda \lfbmv{x} ^{ T^{1} }. M^{1}  \mid M^{1} \, N^{1} \mid  \lfbkw{unq}   M^{0} $ \\
        \textbf{Named Contexts} & $\Gamma^{\circ}, \Delta^{\circ}$ & $\coloneqq  \cdot  \mid \Gamma^{\circ}  \lfbsym{,}   \lfbmv{x}  :^0  T^{0}  \mid \Gamma^{\circ}  \lfbsym{,}   \lfbmv{x}  :^1  T^{1} $
    \end{tabular}
    \vspace{1em}

    \framebox{\mbox{$ \Gamma^{\circ} \vdash_i M^{i} \colon T^{i} $}} $(i \in \{0, 1\})$
    \[
        \AxiomC{$ \lfbmv{x}  :^ \lfbmv{i} T^{i} \in \Gamma^{\circ} $}
        \UnaryInfC{$ \Gamma^{\circ} \vdash_i \lfbmv{x} \colon T^{i} $}
        \DisplayProof
        \quad
        \AxiomC{$ \Gamma^{\circ}  \lfbsym{,}   \lfbmv{x}  :^ \lfbmv{i}   T^{i}_{{\mathrm{1}}}  \vdash_i M^{i} \colon T^{i}_{{\mathrm{2}}} $}
        \UnaryInfC{$ \Gamma^{\circ} \vdash_i  \lambda \lfbmv{x} ^{ T^{i}_{{\mathrm{1}}} }. M^{i}  \colon T^{i}_{{\mathrm{1}}}  \rightarrow  T^{i}_{{\mathrm{2}}} $}
        \DisplayProof
       \quad
        \AxiomC{$ \Gamma^{\circ} \vdash_i M^{i} \colon T^{i}_{{\mathrm{1}}}  \rightarrow  T^{i}_{{\mathrm{2}}} $}
        \AxiomC{$ \Gamma^{\circ} \vdash_i N^{i} \colon T^{i}_{{\mathrm{1}}} $}
        \BinaryInfC{$ \Gamma^{\circ} \vdash_i M^{i} \, N^{i} \colon T^{i}_{{\mathrm{2}}} $}
        \DisplayProof
    \]

    \[
        \AxiomC{$ \Gamma^{\circ} \vdash_1 M^{1} \colon T^{1} $}
        \UnaryInfC{$ \Gamma^{\circ} \vdash_0  \lfbkw{quo}   M^{1}  \colon  \bigcirc  T^{1}  $}
        \DisplayProof
        \quad
        \AxiomC{$ \Gamma^{\circ} \vdash_0 M^{0} \colon  \bigcirc  T^{1}  $}
        \UnaryInfC{$ \Gamma^{\circ} \vdash_1  \lfbkw{unq}   M^{0}  \colon T^{1} $}
        \DisplayProof
    \]

    \caption{Syntax and typing rules of \lamcirctwo} \label{fig:deflamcirctwo}
\end{figure}

\lamcirctwo has two stages: level-1 is the future stage. We define types and terms for each level (and metavariables are indexed by 0 or 1). A temporal type $ \bigcirc  T^{1} $ denotes a code for the future-stage value of $T^{1}$. Unlike contextual modal types, temporal types do not show context explicitly. Instead, typing judgments hold future-stage named contexts that implicitly represent contexts of those code types. A type judgment $ \Gamma^{\circ} \vdash_i M^{i} \colon T^{i} $ (where $i=0,1$) describes typing at the stage $\lfbmv{i}$, where $\Gamma^{\circ}$ includes variables of both levels. Also, it is easy to see that \lamcirctwo also has syntax for quote and unquote as in \lamfb but they are not annotated with
named contexts and explicit substitutions, respectively.  Typing rules do little with named contexts.

These differences lead to the difference in binding structure. For example, consider a \lamcirctwo-term $ \lambda f ^{   \bigcirc  T^{1}_{{\mathrm{1}}}    \rightarrow    \bigcirc  T^{1}_{{\mathrm{2}}}   }.  \lfbkw{quo}   \lfbsym{(}   \lambda \lfbmv{x} ^{ T^{1}_{{\mathrm{1}}} }.  \lfbkw{unq}   \lfbsym{(}  f \,  \lfbkw{quo}   \lfbmv{x}   \lfbsym{)}    \lfbsym{)}  $. In this term, the outer lambda binds the level-0 occurrence of $f$ and the inner lambda binds the level-1 occurrence of $\lfbmv{x}$, although $\lfbkw{quo}$ and $\lfbkw{unq}$ are placed between binders and variable references. To embed \lamcirctwo to \lamfb, we have to emulate this behavior of \lamcirctwo.

We design our embedding from \lamcirctwo to \lamfb based on the following insights. First of all, we naturally embed quote and unquote of \lamcirctwo to those of \lamfb (by recovering missing annotations). Secondly, we can recover a hidden context of code types in \lamcirctwo from the types of level-1 free variables. For example, in the judgment
\[    \lfbmv{x}  :^0   \bigcirc  \textrm{int}     \lfbsym{,}   \lfbmv{y}  :^1  \textrm{int}  \vdash_0  \lfbkw{quo}   \lfbmv{y}  \colon  \bigcirc  \textrm{int}  , \]
the context of the type $ \bigcirc  \textrm{int} $ (of $ \lfbkw{quo}   \lfbmv{y} $) should be $ \textrm{int} $ because the named context has a level-1 binding $  \lfbmv{y}  :^1  \textrm{int}  $. As a result, $ \bigcirc  \textrm{int} $ under
$  \lfbmv{x}  :^0   \bigcirc  \textrm{int}     \lfbsym{,}   \lfbmv{y}  :^1  \textrm{int} $ is embedded into $\lfbsym{[}   \textrm{int}   \vdash  \textrm{int}  \lfbsym{]}$. Thirdly, recovered contexts of code types sometimes need to be extended. Let us consider the following judgment:
\[
     \cdot   \vdash_0   \lambda f ^{   \bigcirc  \textrm{int}    \rightarrow    \bigcirc  \textrm{str}   }.  \lfbkw{quo}   \lfbsym{(}   \lambda \lfbmv{x} ^{ \textrm{int} }.  \lfbkw{unq}   \lfbsym{(}  f \,  \lfbkw{quo}   \lfbmv{x}   \lfbsym{)}    \lfbsym{)}   
    : \lfbsym{(}    \bigcirc  \textrm{int}    \rightarrow    \bigcirc  \textrm{str}    \lfbsym{)}  \rightarrow   \bigcirc  \lfbsym{(}  \textrm{int}  \rightarrow  \textrm{str}  \lfbsym{)} .
\]
The hidden context of the $f$ is empty, and hence the type of $f$ should be $\lfbsym{[}   \bullet   \vdash  \textrm{int}  \lfbsym{]}  \rightarrow  \lfbsym{[}   \bullet   \vdash  \textrm{str}  \lfbsym{]}$. However, $f$ is used inside the level-1 binder $ \lambda  \lfbmv{x}^{int}$, and hence this use of $f$ should be typed as $\lfbsym{[}   \textrm{int}   \vdash  \textrm{str}  \lfbsym{]}  \rightarrow  \lfbsym{[}   \textrm{int}   \vdash  \textrm{str}  \lfbsym{]}$. We need to extend the context of the code type as an abstraction under $\lfbkw{quo}$ extends the level-1 context. Thus, the polymorphic context type $ \forall \gamma .  \lfbsym{[}   \gamma   \vdash  \textrm{int}  \lfbsym{]}  \rightarrow  \lfbsym{[}   \gamma   \vdash  \textrm{str}  \lfbsym{]}  $ is more appropriate for $f$. In this way, polymorphic contexts allow us to extend the context of an argument of code type, according to where the argument is used.

\begin{figure}[t]
    \begin{tabularx}{\textwidth}{ X  X }
        {
            \framebox{\mbox{$ \llbracket  T^{1}  \rrbracket $}}
            \begin{align*}
                 \llbracket  \iota  \rrbracket  &= \iota \\
                 \llbracket  T^{1}_{{\mathrm{1}}}  \rightarrow  T^{1}_{{\mathrm{2}}}  \rrbracket  &=  \llbracket  T^{1}_{{\mathrm{1}}}  \rrbracket   \rightarrow   \llbracket  T^{1}_{{\mathrm{2}}}  \rrbracket 
            \end{align*}
        } & {
            \framebox{\mbox{$ \llbracket  T^{0}  \rrbracket_{ \lfbnt{C} } $}}
            \begin{align*}
                 \llbracket  \iota  \rrbracket_{ \lfbnt{C} }  &= \iota \\
                 \llbracket  T^{0}_{{\mathrm{1}}}  \rightarrow  T^{0}_{{\mathrm{2}}}  \rrbracket_{ \lfbnt{C} }  &= \lfbsym{(}   \forall \gamma .  \llbracket  T^{0}_{{\mathrm{1}}}  \rrbracket_{ \lfbnt{C}  \lfbsym{,}  \gamma }    \lfbsym{)}  \rightarrow   \llbracket  T^{0}_{{\mathrm{2}}}  \rrbracket_{ \lfbnt{C} }  \\
                & \qquad \text{ for fresh $\gamma$} \\
                 \llbracket   \bigcirc  T^{1}   \rrbracket_{ \lfbnt{C} }  &= \lfbsym{[}  \lfbnt{C}  \vdash   \llbracket  T^{1}  \rrbracket   \lfbsym{]}
            \end{align*}
        } \\
        {
            \framebox{\mbox{$ \llbracket  M^{1}  \rrbracket_{ \tilde{\Gamma} } $}}
            \begin{align*}
                 \llbracket  \lfbmv{x}  \rrbracket_{ \tilde{\Gamma} }  &= \lfbmv{x} \\
                 \llbracket   \lambda \lfbmv{x} ^{ T^{1} }. M^{1}   \rrbracket_{ \tilde{\Gamma} }  &=  \lambda \lfbmv{x} ^{  \llbracket  T^{1}  \rrbracket  }.  \llbracket  M^{1}  \rrbracket_{ \tilde{\Gamma}  \lfbsym{,}   \lfbmv{x}  :^1  \llbracket  T^{1}  \rrbracket   }   \\
                 \llbracket  M^{1} \, N^{1}  \rrbracket_{ \tilde{\Gamma} }  &=  \llbracket  M^{1}  \rrbracket_{ \tilde{\Gamma} }  \,  \llbracket  N^{1}  \rrbracket_{ \tilde{\Gamma} }  \\
                 \llbracket   \lfbkw{unq}   M^{0}   \rrbracket_{ \tilde{\Gamma} }  &=  \lfbkw{unq} _{ \lfbsym{1} }  \llbracket  M^{0}  \rrbracket_{ \tilde{\Gamma} }  [   \mathsf{dom} \lfbsym{(}  \lvert  \tilde{\Gamma}  \rvert_1  \lfbsym{)}   ] 
            \end{align*}
        } & {
            \framebox{\mbox{$ \llbracket  M^{0}  \rrbracket_{ \tilde{\Gamma} } $}}
            \begin{align*}
                 \llbracket  \lfbmv{x}  \rrbracket_{ \tilde{\Gamma} }  &=  \lfbmv{x} @ \mathsf{diff} \, \lfbsym{(}  \lfbmv{x}  \lfbsym{,}  \tilde{\Gamma}  \lfbsym{)}  \\
                 \llbracket   \lambda \lfbmv{x} ^{ T^{0} }. M^{0}   \rrbracket_{ \tilde{\Gamma} }  &=  \lambda \lfbmv{x} ^{ \lfbnt{T} }.  \llbracket  M^{0}  \rrbracket_{ \tilde{\Gamma}  \lfbsym{,}   \lfbmv{x}  :^0 \lfbnt{T}  }   \\
                & \qquad \text{where $\lfbnt{T} =  \forall \gamma .  \llbracket  T^{0}  \rrbracket_{  \mathsf{rg} (  \lvert  \tilde{\Gamma}  \rvert_1  )   \lfbsym{,}  \gamma }  $} \\
                 \llbracket  M^{0} \, N^{0}  \rrbracket_{ \tilde{\Gamma} }  &=  \llbracket  M^{1}  \rrbracket_{ \tilde{\Gamma} }  \, \lfbsym{(}   \Lambda \gamma .  \llbracket  N^{1}  \rrbracket_{ \tilde{\Gamma}  \lfbsym{,}   \mathbb{x}  :^1 \gamma  }    \lfbsym{)}\\
                & \qquad \text{ for a fresh $\mathbb{x}$ and $\gamma$} \\
                 \llbracket   \lfbkw{quo}   M^{1}   \rrbracket_{ \tilde{\Gamma} }  &=  \lfbkw{quo} \langle  \lvert  \tilde{\Gamma}  \rvert_1  \rangle  \llbracket  M^{1}  \rrbracket_{ \tilde{\Gamma} }  
            \end{align*}
        }
    \end{tabularx}
    \vspace{-0.5em}

    \textbf{Intermediate Named Context} $\tilde{\Gamma} \coloneqq  \cdot  \mid \tilde{\Gamma}  \lfbsym{,}   \lfbmv{x}  :^0 \lfbnt{T}  \mid \tilde{\Gamma}  \lfbsym{,}   \lfbmv{x}  :^1 \lfbnt{T}  \mid \tilde{\Gamma}  \lfbsym{,}   \mathbb{x}  :^1 \gamma $

    \vspace{0.5em}
    \framebox{\mbox{$\Gamma^{\circ}  \leadsto  \tilde{\Gamma}$}}
    \[
        \AxiomC{}
        \UnaryInfC{$ \cdot   \leadsto   \cdot $}
        \DisplayProof
        \qquad
        \AxiomC{$\Gamma^{\circ}  \leadsto  \tilde{\Gamma}$}
        \UnaryInfC{$\Gamma^{\circ}  \lfbsym{,}   \lfbmv{x}  :^0  T^{0}   \leadsto  \tilde{\Gamma}  \lfbsym{,}   \lfbmv{x}  :^0  \forall \gamma .  \llbracket  T^{0}  \rrbracket_{  \mathsf{rg} (  \lvert  \tilde{\Gamma}  \rvert_1  )   \lfbsym{,}  \gamma }   $}
        \DisplayProof
    \]

    \[
        \AxiomC{$\Gamma^{\circ}  \leadsto  \tilde{\Gamma}$}
        \UnaryInfC{$\Gamma^{\circ}  \lfbsym{,}   \lfbmv{x}  :^1  T^{1}   \leadsto  \tilde{\Gamma}  \lfbsym{,}   \lfbmv{x}  :^1  \llbracket  T^{1}  \rrbracket  $}
        \DisplayProof
        \qquad
        \AxiomC{$\Gamma^{\circ}  \leadsto  \tilde{\Gamma}$}
        \UnaryInfC{$\Gamma^{\circ}  \leadsto  \tilde{\Gamma}  \lfbsym{,}   \mathbb{x}  :^1 \gamma $}
        \DisplayProof
    \]

    \caption{Embedding from \lamcirctwo} \label{fig:embedlamcirctwo}

\end{figure}

The definition of our embedding is shown in Figure~\ref{fig:embedlamcirctwo}. Level-1 types do not include code types and are translated to \lamfb types in a straightforward manner; the translation of level-0 types carries a context that is used to signify the context of code types. If it translates a function type, we introduce polymorphic context type to the argument type so that we can extend the context of the type later.  For example, the \lamcirctwo type $\lfbsym{(}   \bigcirc  \textrm{int}   \rightarrow   \bigcirc  \textrm{str}   \lfbsym{)}  \rightarrow   \bigcirc  \lfbsym{(}  \textrm{int}  \rightarrow  \textrm{str}  \lfbsym{)} $ translates to $\lfbsym{(}   \forall \gamma .  \lfbsym{(}   \forall \delta . \lfbsym{[}   \gamma   \lfbsym{,}  \delta  \vdash  \textrm{int}  \lfbsym{]}   \lfbsym{)}  \rightarrow  \lfbsym{[}   \gamma   \vdash  \textrm{str}  \lfbsym{]}    \lfbsym{)}  \rightarrow  \lfbsym{[}   \bullet   \vdash  \textrm{int}  \rightarrow  \textrm{str}  \lfbsym{]}$ under an empty context.

Before discussing term translation, we introduce \textit{intermediate named contexts} $\tilde{\Gamma}$, an intermediate representation of embedded named contexts.  Their structure is similar to named contexts in \lamcirctwo while its elements are variables and types of \lamfb.  We write $ \lvert  \tilde{\Gamma}  \rvert_0 $ for the level-0 fragment of $\tilde{\Gamma}$ and $ \lvert  \tilde{\Gamma}  \rvert_1 $ for the level-1 fragment of $\tilde{\Gamma}$.  The relation $\Gamma^{\circ}  \leadsto  \tilde{\Gamma}$ means that $\Gamma^{\circ}$ can be translated into $\tilde{\Gamma}$. The point is that $\Gamma^{\circ}$ can be translated into multiple intermediate named contexts. For example, the \lamcirctwo named context $  \lfbmv{x}  :^1  T^{1}    \lfbsym{,}   \lfbmv{y}  :^0   \bigcirc  S^{1}    \lfbsym{,}   z  :^0   \bigcirc  S^{1}  $ can be translated to both $  \lfbmv{x}  :^1  \llbracket  T^{1}  \rrbracket     \lfbsym{,}   \lfbmv{y}  :^0 \lfbsym{[}    \llbracket  T^{1}  \rrbracket    \vdash   \llbracket  S^{1}  \rrbracket   \lfbsym{]}   \lfbsym{,}   z  :^0 \lfbsym{[}    \llbracket  T^{1}  \rrbracket    \vdash   \llbracket  S^{1}  \rrbracket   \lfbsym{]} $ and $  \lfbmv{x}  :^1  \llbracket  T^{1}  \rrbracket     \lfbsym{,}   \lfbmv{y}  :^0 \lfbsym{[}    \llbracket  T^{1}  \rrbracket    \vdash   \llbracket  S^{1}  \rrbracket   \lfbsym{]}   \lfbsym{,}   \mathbb{x}  :^1 \gamma   \lfbsym{,}   z  :^0 \lfbsym{[}    \llbracket  T^{1}  \rrbracket    \lfbsym{,}  \gamma  \vdash   \llbracket  S^{1}  \rrbracket   \lfbsym{]} $ due to the last rule of $ \leadsto $. This is necessary to strengthen the induction hypothesis to prove the soundness theorem (Theorem~\ref{thm:embedsound}) below.

Term embedding carries an intermediate named context for two purposes. First, it is used to infer a named context and an explicit substitution for quote and unquote, respectively, as seen in the embedding rules. Second, it is used to know a missing context that we need to extend when using level-0 variables. As described in the rule for named context translation, the types of level-1 variables always translate to polymorphic context types so that we can extend their context when those variables are used. $\mathsf{diff} \, \lfbsym{(}  \lfbmv{x}  \lfbsym{,}  \tilde{\Gamma}  \lfbsym{)}$ determines the missing context, defined as $\mathsf{diff} \, \lfbsym{(}  \lfbmv{x}  \lfbsym{,}  \lfbsym{(}   \tilde{\Gamma}  \lfbsym{,}   \lfbmv{x}  :^0 \lfbnt{T}  , \tilde{\Delta}   \lfbsym{)}  \lfbsym{)} =  \mathsf{rg} (  \lvert  \tilde{\Delta}  \rvert_1  ) $ (or undefined otherwise).

Finally, we prove the soundness of the translation.

\begin{theorem}[Soundness of Embedding from \lamcirctwo] \label{thm:embedsound}
    \begin{itemize}
        \item If $ \Gamma^{\circ} \vdash_0 M^{0} \colon T^{0} $ and $\Gamma^{\circ}  \leadsto  \tilde{\Gamma}$, then $ \lvert  \tilde{\Gamma}  \rvert_0   \vdash   \llbracket  M^{0}  \rrbracket_{ \tilde{\Gamma} }   \colon   \llbracket  T^{0}  \rrbracket_{  \mathsf{rg} (  \lvert  \tilde{\Gamma}  \rvert_1  )  } $.
        \item If $ \Gamma^{\circ} \vdash_1 M^{1} \colon T^{1} $ and $\Gamma^{\circ}  \leadsto  \tilde{\Gamma}$, then $  \lvert  \tilde{\Gamma}  \rvert_0   \lfbsym{,}  \text{\faLock} \lfbsym{,}  \lvert  \tilde{\Gamma}  \rvert_1    \vdash   \llbracket  M^{1}  \rrbracket_{ \tilde{\Gamma} }   \colon   \llbracket  T^{1}  \rrbracket $.
    \end{itemize}
\end{theorem}

\begin{proof}[Sketch]
    \sloppy
    By mutual induction on derivation of \lamcirctwo.

    We focus on the case of level-0 application. If $M^{0} = M^{0}_{{\mathrm{1}}} \, M^{0}_{{\mathrm{2}}}$, then $ \Gamma^{\circ} \vdash_0 M^{0}_{{\mathrm{1}}} \colon S^{0}  \rightarrow  T^{0} $ and $ \Gamma^{\circ} \vdash_0 M^{0}_{{\mathrm{2}}} \colon S^{0} $ for some $S^{0}$. By the induction hypothesis, we have the two \lamfb judgments below.
    \begin{itemize}
        \item $ \lvert  \tilde{\Gamma}  \rvert_0   \vdash   \llbracket  M^{0}_{{\mathrm{1}}}  \rrbracket_{ \tilde{\Gamma} }   \colon  \lfbsym{(}   \forall \gamma .  \llbracket  S^{0}  \rrbracket_{  \mathsf{rg} (  \lvert  \tilde{\Gamma}  \rvert_1  )   \lfbsym{,}  \gamma }    \lfbsym{)}  \rightarrow   \llbracket  T^{0}  \rrbracket_{  \mathsf{rg} (  \lvert  \tilde{\Gamma}  \rvert_1  )  } $
        \item $ \lvert  \tilde{\Gamma}  \lfbsym{,}   \mathbb{x}  :^1 \gamma   \rvert_0   \vdash   \llbracket  M^{0}_{{\mathrm{2}}}  \rrbracket_{ \tilde{\Gamma}  \lfbsym{,}   \mathbb{x}  :^1 \gamma  }   \colon   \llbracket  S^{0}  \rrbracket_{  \mathsf{rg} (  \lvert  \tilde{\Gamma}  \lfbsym{,}   \mathbb{x}  :^1 \gamma   \rvert_1  )  } $
    \end{itemize}
    The second judgment holds because $\Gamma^{\circ}  \leadsto  \tilde{\Gamma}  \lfbsym{,}   \mathbb{x}  :^1 \gamma $ can be derived from $\Gamma^{\circ}  \leadsto  \tilde{\Gamma}$. We can derive $ \lvert  \tilde{\Gamma}  \rvert_0   \vdash   \Lambda \gamma .  \llbracket  M^{0}_{{\mathrm{2}}}  \rrbracket_{ \tilde{\Gamma}  \lfbsym{,}   \mathbb{x}  :^1 \gamma  }    \colon   \forall \gamma .  \llbracket  S^{0}  \rrbracket_{  \mathsf{rg} (  \lvert  \tilde{\Gamma}  \lfbsym{,}   \mathbb{x}  :^1 \gamma   \rvert_1  )  }  $ from the second judgment considering that $ \lvert  \tilde{\Gamma}  \lfbsym{,}   \mathbb{x}  :^1 \gamma   \rvert_0  =  \lvert  \tilde{\Gamma}  \rvert_0 $. Then we can apply this judgment to the first judgment, and we obtain $ \lvert  \tilde{\Gamma}  \rvert_0   \vdash   \llbracket  M^{0}_{{\mathrm{1}}}  \rrbracket_{ \tilde{\Gamma} }  \, \lfbsym{(}   \Lambda \gamma .  \llbracket  M^{0}_{{\mathrm{2}}}  \rrbracket_{ \tilde{\Gamma}  \lfbsym{,}   \mathbb{x}  :^1 \gamma  }    \lfbsym{)}  \colon   \llbracket  T^{0}  \rrbracket_{  \mathsf{rg} (  \lvert  \tilde{\Gamma}  \rvert_1  )  } $. $\qed$
\end{proof}

This embedding requires multiple abstractions of a context: As we have seen, $\lfbsym{(}   \bigcirc  \textrm{int}   \rightarrow   \bigcirc  \textrm{str}   \lfbsym{)}  \rightarrow   \bigcirc  \lfbsym{(}  \textrm{int}  \rightarrow  \textrm{str}  \lfbsym{)} $ translates to $\lfbsym{(}   \forall \gamma .  \lfbsym{(}   \forall \delta . \lfbsym{[}   \gamma   \lfbsym{,}  \delta  \vdash  \textrm{int}  \lfbsym{]}   \lfbsym{)}  \rightarrow  \lfbsym{[}   \gamma   \vdash  \textrm{str}  \lfbsym{]}    \lfbsym{)}  \rightarrow  \lfbsym{[}   \bullet   \vdash  \textrm{int}  \rightarrow  \textrm{str}  \lfbsym{]}$, where the type $\lfbsym{[}   \gamma   \lfbsym{,}  \delta  \vdash  \textrm{int}  \lfbsym{]}$ uses two context variables. This fact strongly suggests that context variables in \lamfb are essential for embedding linear-time temporal types, and hence also for staged computation.

%% \todo{reductionを素直には保存しないことに言及}

%% file: 7-relatedwork.tex
\section{Related Work} \label{section:relatedwork}
\paragraph*{Contextual Modal Type Theory.}
Early work on calculi for metaprogramming with explicit contexts include \lampolyopen by Kim et al.~\cite{Kim06} and \nubox by Nanevski and Pfenning~\cite{Nanevski05}. On the one hand, \lampolyopen has a Fitch-style-like modal type system with explicit contexts and is type safe in the presence of mutable reference and run-time evaluation. On the other hand, \nubox has a dual-context-like modal type system that is type sound with run-time evaluation. Both calculi use symbolic representation for named contexts of quoted code. As a result, names in quoted code are not subject to $\alpha$-conversion. It is worth noting that both calculi discuss some forms of context polymorphism to achieve flexibility for computation with contexts: \lampolyopen employs let-polymorphism for context variables and \nubox proposes support polymorphism that allows contexts to include arbitrary names.

\begin{sloppypar}
    Nanevski and Pfenning refined \nubox to contextual modal type theory (CMTT)~\cite{Nanevski08}, allowing $\alpha$-conversion for variables in quoted code. It is very close to our \lambra while they employed dual-context style formulation. We believe it is not difficult to apply polymorphic context types to dual-context CMTT, although we do not explore it in this paper. One notable difference between CMTT and \lambra is that CMTT has a \emph{named} context inside a contextual modal type, instead of an (unnamed) context. This approach makes $\alpha$-conversion somewhat complicated: a CMTT term $box(x \colon T . x)$ has a type $[x \colon T]T$ while an $\alpha$-equivalent term $box(y \colon T . y)$ has a bit different type $[y \colon T]T$. Instead, \lambra omits names from contexts in contextual modal types by identifying variables in a context by their positions; hence $\alpha$-equivalent terms always have the same type in \lambra.
\end{sloppypar}

\paragraph*{Prior Work on Polymorphic Contexts.}
Contextual modal type systems have been applied to proof assistants~\cite{Pientka10,Cave13,Pientka19,Stampoulis10}. Those proof assistants are designed to allow users to inspect code representation of proof terms using contextual modal types. In particular, Beluga~\cite{Pientka10,Cave13} allows users to perform pattern match against code with polymorphic contexts whereas \lamfb allows only for generative metaprogramming. The prior proposals used an identity substitution $id_{\phi}$ as a term representation of a context variable $\phi$, whereas we use series variables for that purpose. Type-theoretic formalization with identity substitutions is examined by Puech's unpublished work~\cite{Puech16}. He proposes \lamctx and \lamctxi that are dual-context and Fitch-style contextual modal type theories, respectively, with polymorphic types and identity substitutions, the latter of which is very similar to \lamfb in fact. However, a formalization with identity substitutions introduces a significant restriction that we can only abstract a single part of a context. If we abstract multiple parts in a context by identity substitutions, we have a term like $\mathbf{quo}\langle\gamma,\gamma\rangle id_{\gamma}$ that is ill scoped because we do not know which $\gamma$ is referred to by $id_{\gamma}$. One might consider introducing a restriction that a context variable does not duplicate in a context. However, it is not straightforward to avoid ill-scoped terms like $(\Lambda \delta.\mathbf{quo}\langle\gamma,\delta\rangle id_{\gamma})@\gamma$, which reduces to the previous term. This is why we introduce series variables, which are convenient for multiple abstractions of contexts.

\paragraph*{Context Subtyping.}
Rhiger~\cite{Rhiger12} proposed a Fitch-style contextual modal type system \lambrasub that achieves safe code operation with mutable reference and run-time evaluation. An interesting point of \lambrasub is that it employs linear-time flavored named contexts where a quote does not discard a future-stage context, and achieves flexibility of computation with context by introducing structural subtyping for contexts. Therefore, the relation between context subtyping of \lambrasub and polymorphic context of \lamfb can be considered as that between subtyping and row-polymorphism for record calculi. Kiselyov et al. proposed a type system \verb|<NJ>| with a notion of \textit{refined environment classifiers}~\cite{Kiselyov16}, which can be interpreted as encapsulated representation of contexts. \verb|<NJ>| is similar to \lambrasub in the sense that it employs classifier subtyping while it is closer to nominal subtyping. It is worth mentioning that \verb|<NJ>| permits bounded polymorphism over classifiers, and hence one can write a type like $\forall \gamma. (\forall \delta \succ \gamma. \langle T_1 \rangle^{\delta} \rightarrow \langle T_2 \rangle^{\delta}) \rightarrow \langle T_1 \rightarrow T_2 \rangle^{\gamma}$. Their bounded polymorphism is likely as expressive as polymorphic contexts of \lamfb, and we are interested in developing the relation between \verb|<NJ>| and \lamfb.

\paragraph*{Relation to Polymorphic Types.}
\lamfb does not support ordinary polymorphic types because our primary goal in this paper was to reason context polymorphism. We should mention to M\oe{}bius~\cite{Jang22} as notable prior work, which provides a contextual modal type system for metaprogramming capable of pattern matching against open code with polymorphic types. Their type variables can appear as a binder in contextual modal type, e.g., \verb=['a : *, v : list 'a |- 'a]=, while \lamfb does not allow such occurrences of context variables. Further investigation is necessary to figure out where this difference comes from.

\paragraph*{Modal Types for Algebraic Effects and Handlers.}
ECMTT~\cite{Zyuzin21} is an interesting application of contextual modal types to algebraic effects and handlers~\cite{DBLP:conf/esop/PlotkinP09}. It uses contexts to track effects of computations and use explicit substitutions to supply effect handlers. The authors mention that ECMTT needs some form of context polymorphism to support effect polymorphism. We expect the polymorphic context types in \lamfb will provide a basis for such an extension. In particular, the polymorphic context types in \lamfb can abstract contexts multiple times; hence, we can describe a function that combines computations with different effects, e.g., $\forall \gamma,\delta. [\gamma \vdash T] \rightarrow [\delta \vdash T] \rightarrow [\gamma, \delta \vdash T]$.

\paragraph*{Linear-Time Temporal Types.}
There are several attempts at revealing the relation between explicit contexts of contextual modal type theory and implicit contexts of linear-time temporal type theory. However, we need to point out that not all of them achieve their goal. For example, Davies~\cite{Davies17} pointed out that the translation from \lampolyopen to \lamcirc, proposed by Kim et al.~\cite{Kim06}, was not sound for some cases. Puech~\cite{Puech16} also claimed a sound translation from \lamctxi to \lamalpha~\cite{Taha03}, which is an extension of \lamcirc with environment classifiers, but it did not work for some cases, either. His translation does not use polymorphic contexts to infer hidden contexts. Instead, it infers hidden contexts by introducing logic variables for unknown contexts and collecting constraints on those logic variables through typing derivations. Consequently, the following judgment fails to translate because $f$ is used in two different scopes, and hence contradicting constraints for $f$ is generated.
\begin{align*}
    & f :^0 \bigcirc T \rightarrow \bigcirc T, g :^0 \bigcirc T \rightarrow \bigcirc T \rightarrow \bigcirc T, z :^1 T  \\
    & \hspace{6em} \vdash g\ (\mathbf{quo}((\lambda x : T. \mathbf{unq}(f \mathbf{quo}x)) z) (f \mathbf{quo}z))
\end{align*}

These failing translations conversely indicate that the hypothesis by Davies~\cite{Davies17} is right: a sound translation from \lamcirc requires a full form of context polymorphism as in our \lamfb. Kameyama et al.~\cite{Kameyama08} provided a sound translation from a 2-level fragment of \lamalpha to System F with products and a fixed point operator. Their translation uses polymorphic types to represent unknown contexts, similarly to our approach. However, their translation takes an approach different from ours. For example, a \lamcirc type $\bigcirc T \rightarrow \bigcirc T \rightarrow \bigcirc T$ is encoded to $ \forall \gamma . \lfbsym{(}  \lfbsym{[}   \gamma   \vdash  \lfbnt{T}  \lfbsym{]}  \rightarrow   \forall \delta . \lfbsym{(}  \lfbsym{[}   \gamma   \lfbsym{,}  \delta  \vdash  \lfbnt{T}  \lfbsym{]}  \rightarrow  \lfbsym{[}   \gamma   \lfbsym{,}  \delta  \vdash  \lfbnt{T}  \lfbsym{]}  \lfbsym{)}   \lfbsym{)} $ if we apply their approach to \lamfb, whereas the same type is encoded to $\lfbsym{(}   \forall \gamma . \lfbsym{[}   \gamma   \vdash  \lfbnt{T}  \lfbsym{]}   \lfbsym{)}  \rightarrow  \lfbsym{(}   \forall \gamma . \lfbsym{[}   \gamma   \vdash  \lfbnt{T}  \lfbsym{]}   \lfbsym{)}  \rightarrow  \lfbsym{[}   \bullet   \vdash  \lfbnt{T}  \lfbsym{]}$ by the approach discussed in Section~\ref{section:embedlttt}. There are two major differences between their approach and ours. Firstly, their translation needs to insert \textit{coercion} functions that extend contexts in types in conjunction with polymorphic types. On the contrary, our approach achieves the same goal purely by polymorphic contexts, making the translation much more concise. Secondly, their source language supports richer expressions than \lamcirctwo, including run-time evaluation and fixpoint. It is left for future work to figure out whether our approach can also embed such features of \lamalpha to \lamfb.

%% file: 8-conclusion.tex
\section{Conclusion} \label{section:conclusion}
This paper has proposed a novel contextual modal type theory \lamfb with polymorphic contexts. It is novel in that it supports full-fledged polymorphic contexts and allows us to abstract any parts of a context multiple times. We have given its semantics by $\beta$-reduction and proved subject reduction, strong normalization, and confluence. We have also demonstrated a sound embedding from linear-time temporal type theory. We expect that this result shows that \lamfb endows expressiveness sufficient to describe programs with staged computation.

We regard this work as a first step to establishing a mature modal type theory that reasons hygienic binding operations provided by procedural macros of Scheme, Racket, and several languages. Future work includes formal reasoning of the relation between contextual modal types and refined environment classifiers and developing contextual modal type theory that can express first-class variable names.

%% file: fulldef.tex
\section{Full Definition of \lamfb}

\subsection{Syntax}

    \begin{tabular}{lcl}
        \textbf{Types }                 & $\lfbnt{S},\lfbnt{T}$  & $\Coloneqq \iota \mid \lfbnt{S}  \rightarrow  \lfbnt{T} \mid \lfbsym{[}  \lfbnt{C}  \vdash  \lfbnt{T}  \lfbsym{]} \mid  \forall \gamma . \lfbnt{T} $             \\ 
        \textbf{Contexts }              & $\lfbnt{C},\lfbnt{D}$  & $\Coloneqq  \bullet  \mid \lfbnt{C}  \lfbsym{,}  \lfbnt{T} \mid \lfbnt{C}  \lfbsym{,}  \gamma$             \\ 
        \textbf{Stage transitions }     & $\lfbnt{k}$        & $\in \mathbb{N}$                                                                \\
        \textbf{Terms }                 & $\lfbnt{M}, \lfbnt{N}$ & $\Coloneqq \lfbmv{x} \mid  \lambda \lfbmv{x} ^{ \lfbnt{T} }. \lfbnt{M}  \mid MN \mid  \lfbkw{quo} \langle \Gamma \rangle \lfbnt{M}  \mid  \lfbkw{unq} _{ \lfbnt{k} } \lfbnt{M} [  \theta  ]  \mid  \Lambda \gamma . \lfbnt{M}  \mid  \lfbnt{M} @ \lfbnt{C} $ \\
        \textbf{Explicit Substitutions} & $\theta$       & $\Coloneqq  \bullet  \mid \theta  \lfbsym{,}  \lfbnt{M} \mid \theta  \lfbsym{,}  \mathbb{x}$\\
        \textbf{Named Contexts }          & $\Gamma, \Delta$ & $\Coloneqq  \bullet   \mid \Gamma  \lfbsym{,}  \lfbmv{x}  \colon  \lfbnt{T} \mid \Gamma  \lfbsym{,}  \mathbb{x}  \colon  \gamma \mid \Gamma  \lfbsym{,}  \text{\faLock}$ \\
    \end{tabular}

\noindent
\framebox{\mbox{$ \mathsf{FV} _{ \lfbnt{k} }( \lfbnt{M} ) $}}
\begin{align*}
   \mathsf{FV} _{ \lfbnt{k} }( \lfbmv{x} )            & = \begin{cases}
    \lfbmv{x} & \text{if $\lfbnt{k} = 0$} \\
    \emptyset   & \text{otherwise}
  \end{cases}      \\
   \mathsf{FV} _{ \lfbnt{k} }(  \lambda \lfbmv{x} ^{ \lfbnt{T} }. \lfbnt{M}  )       & = \begin{cases}
     \mathsf{FV} _{ \lfbnt{k} }( \lfbnt{M} )   \lfbsym{-}  \lfbsym{\{}  \lfbmv{x}  \lfbsym{\}} & \text{if $\lfbnt{k} = 0$} \\
     \mathsf{FV} _{ \lfbnt{k} }( \lfbnt{M} )         & \text{otherwise}
  \end{cases}                             \\
   \mathsf{FV} _{ \lfbnt{k} }( \lfbnt{M} \, \lfbnt{N} )          & =  \mathsf{FV} _{ \lfbnt{k} }( \lfbnt{M} )  \, \cup \,  \mathsf{FV} _{ \lfbnt{k} }( \lfbnt{N} )                                           \\
   \mathsf{FV} _{ \lfbnt{k} }(  \lfbkw{quo} \langle \Gamma \rangle \lfbnt{M}  )        & =  \mathsf{FV} _{ \lfbnt{k}  \lfbsym{+}  \lfbsym{1} }( \lfbnt{M} )                                                       \\
   \mathsf{FV} _{ \lfbnt{k_{{\mathrm{2}}}} }(  \lfbkw{unq} _{ \lfbnt{k_{{\mathrm{1}}}} } \lfbnt{M} [  \theta  ]  )  & = \begin{cases}
     \mathsf{FV} _{ \lfbnt{k_{{\mathrm{2}}}}  \lfbsym{-}  \lfbnt{k_{{\mathrm{1}}}} }( \lfbnt{M} )  \, \cup \,  \mathsf{FV} _{ \lfbnt{k_{{\mathrm{2}}}} }( \theta )  & \text{if $\lfbnt{k_{{\mathrm{2}}}} \geq \lfbnt{k_{{\mathrm{1}}}}$} \\
     \mathsf{FV} _{ \lfbnt{k_{{\mathrm{2}}}} }( \theta )                    & \text{otherwise}
  \end{cases} \\
   \mathsf{FV} _{ \lfbnt{k} }(  \Lambda \gamma . \lfbnt{M}  )   & =  \mathsf{FV} _{ \lfbnt{k} }( \lfbnt{M} )  \\
   \mathsf{FV} _{ \lfbnt{k} }(  \lfbnt{M} @ \lfbnt{C}  )  & =  \mathsf{FV} _{ \lfbnt{k} }( \lfbnt{M} ) 
\end{align*}

\noindent
\framebox{\mbox{$ \mathsf{FV} _{ \lfbnt{k} }( \theta ) $}}
\begin{align*}
   \mathsf{FV} _{ \lfbnt{k} }(  \bullet  )  & = \emptyset                  \\
   \mathsf{FV} _{ \lfbnt{k} }( \theta  \lfbsym{,}  \lfbnt{M} )      & =  \mathsf{FV} _{ \lfbnt{k} }( \theta )  \, \cup \,  \mathsf{FV} _{ \lfbnt{k} }( \lfbnt{M} )  \\
   \mathsf{FV} _{ \lfbnt{k} }( \theta  \lfbsym{,}  \mathbb{x} )  & = \begin{cases}
     \mathsf{FV} _{ \lfbnt{k} }( \theta )  \, \cup \, \lfbsym{\{}  \mathbb{x}  \lfbsym{\}} & \text{if $k=0$}  \\
     \mathsf{FV} _{ \lfbnt{k} }( \theta )             & \text{otherwise} \\
  \end{cases} \\
\end{align*}

\noindent
\framebox{\mbox{$\mathsf{FCV} \, \lfbsym{(}  \lfbnt{T}  \lfbsym{)}$}}
\begin{align*}
  \mathsf{FCV} \, \lfbsym{(}  \iota  \lfbsym{)}     & = \emptyset               \\
  \mathsf{FCV} \, \lfbsym{(}  \lfbnt{S}  \rightarrow  \lfbnt{T}  \lfbsym{)}   & = \mathsf{FCV} \, \lfbsym{(}  \lfbnt{S}  \lfbsym{)} \, \cup \, \mathsf{FCV} \, \lfbsym{(}  \lfbnt{T}  \lfbsym{)} \\
  \mathsf{FCV} \, \lfbsym{(}  \lfbsym{[}  \lfbnt{C}  \vdash  \lfbnt{T}  \lfbsym{]}  \lfbsym{)} & = \mathsf{FCV} \, \lfbsym{(}  \lfbnt{C}  \lfbsym{)} \, \cup \, \mathsf{FCV} \, \lfbsym{(}  \lfbnt{T}  \lfbsym{)} \\
  \mathsf{FCV} \, \lfbsym{(}   \forall \gamma . \lfbnt{T}   \lfbsym{)}     & = \mathsf{FCV} \, \lfbsym{(}  \lfbnt{T}  \lfbsym{)}  \lfbsym{-}  \lfbsym{\{}  \gamma  \lfbsym{\}}
\end{align*}

\noindent
\framebox{\mbox{$\mathsf{FCV} \, \lfbsym{(}  \lfbnt{C}  \lfbsym{)}$}}
\begin{align*}
  \mathsf{FCV} \, \lfbsym{(}   \bullet   \lfbsym{)} & = \emptyset               \\
  \mathsf{FCV} \, \lfbsym{(}  \lfbnt{C}  \lfbsym{,}  \lfbnt{T}  \lfbsym{)}     & = \mathsf{FCV} \, \lfbsym{(}  \lfbnt{C}  \lfbsym{)} \, \cup \, \mathsf{FCV} \, \lfbsym{(}  \lfbnt{T}  \lfbsym{)} \\
  \mathsf{FCV} \, \lfbsym{(}  \lfbnt{C}  \lfbsym{,}  \gamma  \lfbsym{)}     & = \mathsf{FCV} \, \lfbsym{(}  \lfbnt{C}  \lfbsym{)} \, \cup \, \lfbsym{\{}  \gamma  \lfbsym{\}}
\end{align*}

\noindent
\framebox{\mbox{$\mathsf{FCV} \, \lfbsym{(}  \Gamma  \lfbsym{)}$}}
\begin{align*}
  \mathsf{FCV} \, \lfbsym{(}   \bullet   \lfbsym{)} & = \emptyset               \\
  \mathsf{FCV} \, \lfbsym{(}  \Gamma  \lfbsym{,}  \lfbmv{x}  \colon  \lfbnt{T}  \lfbsym{)} & = \mathsf{FCV} \, \lfbsym{(}  \Gamma  \lfbsym{)} \, \cup \, \mathsf{FCV} \, \lfbsym{(}  \lfbnt{T}  \lfbsym{)} \\
  \mathsf{FCV} \, \lfbsym{(}  \Gamma  \lfbsym{,}  \mathbb{x}  \colon  \gamma  \lfbsym{)} & = \mathsf{FCV} \, \lfbsym{(}  \Gamma  \lfbsym{)} \, \cup \, \lfbsym{\{}  \gamma  \lfbsym{\}} \\
  \mathsf{FCV} \, \lfbsym{(}  \Gamma  \lfbsym{,}  \text{\faLock}  \lfbsym{)}  & = \mathsf{FCV} \, \lfbsym{(}  \Gamma  \lfbsym{)}
\end{align*}

\noindent
\framebox{\mbox{$\mathsf{FCV} \, \lfbsym{(}  \lfbnt{M}  \lfbsym{)}$}}
\begin{align*}
  \mathsf{FCV} \, \lfbsym{(}  \lfbmv{x}  \lfbsym{)}         & = \emptyset                \\
  \mathsf{FCV} \, \lfbsym{(}   \lambda \lfbmv{x} ^{ \lfbnt{T} }. \lfbnt{M}   \lfbsym{)}   & = \mathsf{FCV} \, \lfbsym{(}  \lfbnt{T}  \lfbsym{)} \, \cup \, \mathsf{FCV} \, \lfbsym{(}  \lfbnt{M}  \lfbsym{)}  \\
  \mathsf{FCV} \, \lfbsym{(}  \lfbnt{M} \, \lfbnt{N}  \lfbsym{)}       & = \mathsf{FCV} \, \lfbsym{(}  \lfbnt{M}  \lfbsym{)} \, \cup \, \mathsf{FCV} \, \lfbsym{(}  \lfbnt{N}  \lfbsym{)}  \\
  \mathsf{FCV} \, \lfbsym{(}   \lfbkw{quo} \langle \Gamma \rangle \lfbnt{M}   \lfbsym{)}     & = \mathsf{FCV} \, \lfbsym{(}  \Gamma  \lfbsym{)} \, \cup \, \mathsf{FCV} \, \lfbsym{(}  \lfbnt{M}  \lfbsym{)}  \\
  \mathsf{FCV} \, \lfbsym{(}   \lfbkw{unq} _{ \lfbnt{k} } \lfbnt{M} [  \theta  ]   \lfbsym{)} & = \mathsf{FCV} \, \lfbsym{(}  \lfbnt{M}  \lfbsym{)} \, \cup \, \mathsf{FCV} \, \lfbsym{(}  \theta  \lfbsym{)} \\
  \mathsf{FCV} \, \lfbsym{(}   \Lambda \gamma . \lfbnt{M}   \lfbsym{)}      & = \mathsf{FCV} \, \lfbsym{(}  \lfbnt{M}  \lfbsym{)}  \lfbsym{-}  \lfbsym{\{}  \gamma  \lfbsym{\}} \\
  \mathsf{FCV} \, \lfbsym{(}   \lfbnt{M} @ \lfbnt{C}   \lfbsym{)}      & = \mathsf{FCV} \, \lfbsym{(}  \lfbnt{M}  \lfbsym{)} \, \cup \, \mathsf{FCV} \, \lfbsym{(}  \lfbnt{C}  \lfbsym{)}
\end{align*}

\noindent
\framebox{\mbox{$\mathsf{FCV} \, \lfbsym{(}  \theta  \lfbsym{)}$}}
\begin{align*}
  \mathsf{FCV} \, \lfbsym{(}   \bullet   \lfbsym{)} & = \emptyset                \\
  \mathsf{FCV} \, \lfbsym{(}  \theta  \lfbsym{,}  \lfbnt{M}  \lfbsym{)}     & = \mathsf{FCV} \, \lfbsym{(}  \theta  \lfbsym{)} \, \cup \, \mathsf{FCV} \, \lfbsym{(}  \lfbnt{M}  \lfbsym{)} \\
  \mathsf{FCV} \, \lfbsym{(}  \theta  \lfbsym{,}  \mathbb{x}  \lfbsym{)}    & = \mathsf{FCV} \, \lfbsym{(}  \theta  \lfbsym{)}
\end{align*}

\subsection{Type System}

\begin{align*}
  \mathsf{head} \, \lfbsym{(}   \bullet   \lfbsym{)} &=  \bullet   &
  \mathsf{head} \, \lfbsym{(}  \Gamma  \lfbsym{,}  \text{\faLock}  \lfbsym{)} &=  \bullet    \\
  \mathsf{head} \, \lfbsym{(}  \Gamma  \lfbsym{,}  \lfbmv{x}  \colon  \lfbnt{T}  \lfbsym{)} &= \mathsf{head} \, \lfbsym{(}  \Gamma  \lfbsym{)}  \lfbsym{,}  \lfbmv{x}  \colon  \lfbnt{T} &
  \mathsf{head} \, \lfbsym{(}  \Gamma  \lfbsym{,}  \mathbb{x}  \colon  \gamma  \lfbsym{)} &= \mathsf{head} \, \lfbsym{(}  \Gamma  \lfbsym{)}  \lfbsym{,}  \mathbb{x}  \colon  \gamma
\end{align*}

\noindent
\framebox{\mbox{$\Gamma_{{\mathrm{1}}}  \leq  \Gamma_{{\mathrm{2}}}$}}
\[
    \AxiomC{}
    \UnaryInfC{$ \bullet   \leq   \bullet $}
    \DisplayProof
    \quad
    \AxiomC{$\Gamma_{{\mathrm{1}}}  \leq  \Gamma_{{\mathrm{2}}}$}
    \UnaryInfC{$\Gamma_{{\mathrm{1}}}  \lfbsym{,}  \lfbmv{x}  \colon  \lfbnt{T}  \leq  \Gamma_{{\mathrm{2}}}  \lfbsym{,}  \lfbmv{x}  \colon  \lfbnt{T}$}
    \DisplayProof
    \quad
    \AxiomC{$\Gamma_{{\mathrm{1}}}  \leq  \Gamma_{{\mathrm{2}}}$}
    \UnaryInfC{$\Gamma_{{\mathrm{1}}}  \leq  \Gamma_{{\mathrm{2}}}  \lfbsym{,}  \lfbmv{x}  \colon  \lfbnt{T}$}
    \DisplayProof
  \]
  \[
    \AxiomC{$\Gamma_{{\mathrm{1}}}  \leq  \Gamma_{{\mathrm{2}}}$}
    \UnaryInfC{$\Gamma_{{\mathrm{1}}}  \lfbsym{,}  \mathbb{x}  \colon  \gamma  \leq  \Gamma_{{\mathrm{2}}}  \lfbsym{,}  \mathbb{x}  \colon  \gamma$}
    \DisplayProof
    \quad
    \AxiomC{$\Gamma_{{\mathrm{1}}}  \leq  \Gamma_{{\mathrm{2}}}$}
    \UnaryInfC{$\Gamma_{{\mathrm{1}}}  \leq  \Gamma_{{\mathrm{2}}}  \lfbsym{,}  \mathbb{x}  \colon  \gamma$}
    \DisplayProof
    \quad
    \AxiomC{$\Gamma_{{\mathrm{1}}}  \leq  \Gamma_{{\mathrm{2}}}$}
    \UnaryInfC{$\Gamma_{{\mathrm{1}}}  \lfbsym{,}  \text{\faLock}  \leq  \Gamma_{{\mathrm{2}}}  \lfbsym{,}  \text{\faLock}$}
    \DisplayProof
\]

\noindent
\framebox{\mbox{$\lfbnt{k}  \colon  \Gamma  \lhd  \Delta$}}
\[
  \AxiomC{}
  \UnaryInfC{$\lfbsym{0}  \colon  \Gamma  \lhd  \Gamma$}
  \DisplayProof
  \quad
  \AxiomC{$\lfbnt{k}  \colon  \Gamma  \lhd  \Delta$}
  \UnaryInfC{$\lfbnt{k}  \colon  \Gamma  \lhd  \Delta  \lfbsym{,}  \lfbmv{x}  \colon  \lfbnt{T}$}
  \DisplayProof
  \quad
  \AxiomC{$\lfbnt{k}  \colon  \Gamma  \lhd  \Delta$}
  \UnaryInfC{$\lfbnt{k}  \colon  \Gamma  \lhd  \Delta  \lfbsym{,}  \mathbb{x}  \colon  \gamma$}
  \DisplayProof
  \quad
  \AxiomC{$\lfbnt{k}  \colon  \Gamma  \lhd  \Delta$}
  \UnaryInfC{$\lfbnt{k}  \lfbsym{+}  \lfbsym{1}  \colon  \Gamma  \lhd  \Delta  \lfbsym{,}  \text{\faLock}$}
  \DisplayProof
\]

\noindent
\framebox{\mbox{$\Gamma  \vdash  \lfbnt{M}  \colon  \lfbnt{T}$}}
\[
  \AxiomC{$\lfbmv{x}  \colon  \lfbnt{T} \, \in \, \mathsf{head} \, \lfbsym{(}  \Gamma  \lfbsym{)}$}
  \UnaryInfC{$\Gamma  \vdash  \lfbmv{x}  \colon  \lfbnt{T}$}
  \DisplayProof
  \quad
  \AxiomC{$\Gamma  \lfbsym{,}  \lfbmv{x}  \colon  \lfbnt{T_{{\mathrm{1}}}}  \vdash  \lfbnt{M}  \colon  \lfbnt{T_{{\mathrm{2}}}}$}
  \UnaryInfC{$\Gamma  \vdash   \lambda \lfbmv{x} ^{ \lfbnt{T_{{\mathrm{1}}}} }. \lfbnt{M}   \colon  \lfbnt{T_{{\mathrm{1}}}}  \rightarrow  \lfbnt{T_{{\mathrm{2}}}}$}
  \DisplayProof
\]

\[
  \AxiomC{$\Gamma  \vdash  \lfbnt{M_{{\mathrm{1}}}}  \colon  \lfbnt{T_{{\mathrm{1}}}}  \rightarrow  \lfbnt{T_{{\mathrm{2}}}}$}
  \AxiomC{$\Gamma  \vdash  \lfbnt{M_{{\mathrm{2}}}}  \colon  \lfbnt{T_{{\mathrm{1}}}}$}
  \BinaryInfC{$\Gamma  \vdash  \lfbnt{M_{{\mathrm{1}}}} \, \lfbnt{M_{{\mathrm{2}}}}  \colon  \lfbnt{T_{{\mathrm{2}}}}$}
  \DisplayProof
  \quad
  \AxiomC{$ \Gamma  \lfbsym{,}  \text{\faLock} \lfbsym{,} \Delta   \vdash  \lfbnt{M}  \colon  \lfbnt{T}$}
  \AxiomC{$\text{\faLock} \, \not\in \, \Delta$}
  \BinaryInfC{$\Gamma  \vdash   \lfbkw{quo} \langle \Delta \rangle \lfbnt{M}   \colon  \lfbsym{[}   \mathsf{rg} ( \Delta )   \vdash  \lfbnt{T}  \lfbsym{]}$}
  \DisplayProof
\]

\[
  \AxiomC{$\Gamma  \vdash  \lfbnt{M}  \colon  \lfbsym{[}  \lfbnt{C}  \vdash  \lfbnt{T}  \lfbsym{]}$}
  \AxiomC{$\Delta  \vdash  \theta  \colon  \lfbnt{C}$}
  \AxiomC{$\lfbnt{k}  \colon  \Gamma  \lhd  \Delta$}
  \TrinaryInfC{$\Delta  \vdash   \lfbkw{unq} _{ \lfbnt{k} } \lfbnt{M} [  \theta  ]   \colon  \lfbnt{T}$}
  \DisplayProof
\]

\[
  \AxiomC{$\Gamma  \vdash  \lfbnt{M}  \colon  \lfbnt{T}$}
  \AxiomC{$\gamma \, \not\in \, \mathsf{FCV} \, \lfbsym{(}  \Gamma  \lfbsym{)}$}
  \BinaryInfC{$\Gamma  \vdash   \Lambda \gamma . \lfbnt{M}   \colon   \forall \gamma . \lfbnt{T} $}
  \DisplayProof
  \quad
  \AxiomC{$\Gamma  \vdash  \lfbnt{M}  \colon   \forall \gamma . \lfbnt{T} $}
  \UnaryInfC{$\Gamma  \vdash   \lfbnt{M} @ \lfbnt{C}   \colon  \lfbnt{T}  \lfbsym{[}   \gamma \coloneqq \lfbnt{C}   \lfbsym{]}$}
  \DisplayProof
\]

\noindent
\framebox{\mbox{$\Gamma  \vdash  \theta  \colon  \lfbnt{C}$}}
\[
  \AxiomC{}
  \UnaryInfC{$\Gamma  \vdash   \bullet   \colon   \bullet $}
  \DisplayProof
  \quad
  \AxiomC{$\Gamma  \vdash  \theta  \colon  \lfbnt{C}$}
  \AxiomC{$\Gamma  \vdash  \lfbnt{M}  \colon  \lfbnt{T}$}
  \BinaryInfC{$\Gamma  \vdash  \theta  \lfbsym{,}  \lfbnt{M}  \colon  \lfbnt{C}  \lfbsym{,}  \lfbnt{T}$}
  \DisplayProof
  \quad
  \AxiomC{$\Gamma  \vdash  \theta  \colon  \lfbnt{C}$}
  \AxiomC{$\mathbb{x}  \colon  \gamma \, \in \, \mathsf{head} \, \lfbsym{(}  \Gamma  \lfbsym{)}$}
  \BinaryInfC{$\Gamma  \vdash  \lfbsym{(}  \theta  \lfbsym{,}  \mathbb{x}  \lfbsym{)}  \colon  \lfbsym{(}  \lfbnt{C}  \lfbsym{,}  \gamma  \lfbsym{)}$}
  \DisplayProof
\]

\subsection{Substitution}
\[
  \sigma \Coloneqq  \bullet  \mid \sigma  \lfbsym{,}  \lfbmv{x}  \coloneqq  \lfbnt{M} \mid \sigma  \lfbsym{,}  \mathbb{x}  \coloneqq  \mathbb{y} \mid \sigma  \lfbsym{,}   \text{\faLock} _{ \lfbnt{k} } 
\]

% \noindent
% \mbox{\textbf{Auxiliary functions}}
\begin{align*}
   \mathsf{FV} _{ \lfbnt{k} }( \sigma  \lfbsym{,}  \lfbmv{x}  \coloneqq  \lfbnt{M} )    & =  \mathsf{FV} _{ \lfbnt{k} }( \sigma )  \, \cup \,  \mathsf{FV} _{ \lfbnt{k} }( \lfbnt{M} )                                                                                         \\
   \mathsf{FV} _{ \lfbnt{k_{{\mathrm{2}}}} }( \sigma  \lfbsym{,}   \text{\faLock} _{ \lfbnt{k_{{\mathrm{1}}}} }  )  & = \begin{cases}
     \mathsf{FV} _{ \lfbnt{k_{{\mathrm{2}}}}  \lfbsym{-}  \lfbnt{k_{{\mathrm{1}}}} }( \sigma )  & \text{if $\lfbnt{k_{{\mathrm{2}}}} \ge \lfbnt{k_{{\mathrm{1}}}}$} \\
    \emptyset             & \text{otherwise}
  \end{cases}
\end{align*}
\begin{align*}
   id _{  \bullet  }  & =  \bullet     &
   id _{ \Gamma  \lfbsym{,}  \lfbmv{x}  \colon  \lfbnt{T} }  & =  id _{ \Gamma }   \lfbsym{,}  \lfbmv{x}  \coloneqq  \lfbmv{x} &
   id _{ \Gamma  \lfbsym{,}  \mathbb{x}  \colon  \gamma }  & =  id _{ \Gamma }   \lfbsym{,}  \mathbb{x}  \coloneqq  \mathbb{x} &
   id _{ \Gamma  \lfbsym{,}  \text{\faLock} }   & =  id _{ \Gamma }   \lfbsym{,}   \text{\faLock} _{ \lfbsym{1} } 
\end{align*}
\begin{align*}
  \mathsf{head} \, \lfbsym{(}   \bullet   \lfbsym{)}             & =  \bullet  \\
  \mathsf{head} \, \lfbsym{(}  \sigma  \lfbsym{,}  \lfbmv{x}  \coloneqq  \lfbnt{M}  \lfbsym{)}            & = \mathsf{head} \, \lfbsym{(}  \sigma  \lfbsym{)}  \lfbsym{,}  \lfbmv{x}  \coloneqq  \lfbnt{M} \\
  \mathsf{head} \, \lfbsym{(}  \sigma  \lfbsym{,}  \mathbb{x}  \coloneqq  \mathbb{y}  \lfbsym{)}          & = \mathsf{head} \, \lfbsym{(}  \sigma  \lfbsym{)}  \lfbsym{,}  \mathbb{x}  \coloneqq  \mathbb{y} \\
  \mathsf{head} \, \lfbsym{(}  \sigma  \lfbsym{,}   \text{\faLock} _{ \lfbnt{k} }   \lfbsym{)}            & =  \bullet           \\[1ex]
   \mathsf{count} ( \lfbsym{0} , \sigma )                   & = \lfbsym{0}                 \\
   \mathsf{count} ( \lfbsym{(}  \lfbnt{k_{{\mathrm{1}}}}  \lfbsym{+}  \lfbsym{1}  \lfbsym{)} ,  \bullet  )      & = \lfbnt{k_{{\mathrm{1}}}}  \lfbsym{+}  \lfbsym{1}              \\
   \mathsf{count} ( \lfbsym{(}  \lfbnt{k}  \lfbsym{+}  \lfbsym{1}  \lfbsym{)} , \lfbsym{(}  \sigma  \lfbsym{,}  \lfbmv{x}  \coloneqq  \lfbnt{M}  \lfbsym{)} )    & =  \mathsf{count} ( \lfbnt{k} , \sigma )        \\
   \mathsf{count} ( \lfbsym{(}  \lfbnt{k}  \lfbsym{+}  \lfbsym{1}  \lfbsym{)} , \lfbsym{(}  \sigma  \lfbsym{,}  \mathbb{x}  \coloneqq  \mathbb{y}  \lfbsym{)} )  & =  \mathsf{count} ( \lfbnt{k} , \sigma )        \\
   \mathsf{count} ( \lfbsym{(}  \lfbnt{k_{{\mathrm{1}}}}  \lfbsym{+}  \lfbsym{1}  \lfbsym{)} , \lfbsym{(}  \sigma  \lfbsym{,}   \text{\faLock} _{ \lfbnt{k_{{\mathrm{2}}}} }   \lfbsym{)} )  & =  \mathsf{count} ( \lfbnt{k_{{\mathrm{1}}}} , \sigma )  + \lfbnt{k_{{\mathrm{2}}}}    \\[1ex]
  \sigma  \uparrow  \lfbsym{0}                      & = \sigma               \\
   \bullet   \uparrow  \lfbsym{(}  \lfbnt{k}  \lfbsym{+}  \lfbsym{1}  \lfbsym{)}         & =  \bullet           \\
  \lfbsym{(}  \sigma  \lfbsym{,}  \lfbmv{x}  \coloneqq  \lfbnt{M}  \lfbsym{)}  \uparrow  \lfbsym{(}  \lfbnt{k}  \lfbsym{+}  \lfbsym{1}  \lfbsym{)}      & = \sigma  \uparrow  \lfbsym{(}  \lfbnt{k}  \lfbsym{+}  \lfbsym{1}  \lfbsym{)}     \\
  \lfbsym{(}  \sigma  \lfbsym{,}  \mathbb{x}  \coloneqq  \mathbb{y}  \lfbsym{)}  \uparrow  \lfbsym{(}  \lfbnt{k}  \lfbsym{+}  \lfbsym{1}  \lfbsym{)}      & = \sigma  \uparrow  \lfbsym{(}  \lfbnt{k}  \lfbsym{+}  \lfbsym{1}  \lfbsym{)}     \\
  \lfbsym{(}  \sigma  \lfbsym{,}   \text{\faLock} _{ \lfbnt{k_{{\mathrm{1}}}} }   \lfbsym{)}  \uparrow  \lfbsym{(}  \lfbnt{k_{{\mathrm{2}}}}  \lfbsym{+}  \lfbsym{1}  \lfbsym{)}    & = \sigma  \uparrow  \lfbnt{k_{{\mathrm{2}}}}
\end{align*}

\noindent
\framebox{\mbox{$\Gamma  \vdash  \sigma  \colon  \Delta$}}
\[
  \AxiomC{}
  \UnaryInfC{$\Gamma  \vdash   \bullet   \colon   \bullet $}
  \DisplayProof
  \quad
  \AxiomC{$\Gamma  \vdash  \sigma  \colon  \Delta$}
  \AxiomC{$\Gamma  \vdash  \lfbnt{M}  \colon  \lfbnt{T}$}
  \BinaryInfC{$\Gamma  \vdash  \sigma  \lfbsym{,}  \lfbmv{x}  \coloneqq  \lfbnt{M}  \colon  \Delta  \lfbsym{,}  \lfbmv{x}  \colon  \lfbnt{T}$}
  \DisplayProof
  \quad
  \AxiomC{$\Gamma_{{\mathrm{1}}}  \vdash  \sigma  \colon  \Delta$}
  \AxiomC{$\lfbnt{k}  \colon  \Gamma_{{\mathrm{1}}}  \lhd  \Gamma_{{\mathrm{2}}}$}
  \BinaryInfC{$\Gamma_{{\mathrm{2}}}  \vdash  \sigma  \lfbsym{,}   \text{\faLock} _{ \lfbnt{k} }   \colon  \Delta  \lfbsym{,}  \text{\faLock}$}
  \DisplayProof
\]

\begin{prooftree}
  \AxiomC{$\Gamma  \vdash  \sigma  \colon  \Delta$}
  \AxiomC{$\mathbb{y}  \colon  \gamma \, \in \, \mathsf{head} \, \lfbsym{(}  \Gamma  \lfbsym{)}$}
  \AxiomC{$\mathbb{x} \, \not\in \, \mathsf{dom} \, \lfbsym{(}  \Delta  \lfbsym{)}$}
  \TrinaryInfC{$\Gamma  \vdash  \sigma  \lfbsym{,}  \mathbb{x}  \coloneqq  \mathbb{y}  \colon  \Delta  \lfbsym{,}  \mathbb{x}  \colon  \gamma$}
\end{prooftree}

\noindent
\mbox{\fbox{$\lfbnt{M}  \lfbsym{[}  \sigma  \lfbsym{]}$}}
\begin{align*}
  \lfbmv{x}  \lfbsym{[}  \sigma  \lfbsym{]}            & = \begin{cases}
    M & \text{if $\lfbmv{x}  \coloneqq  \lfbnt{M} \, \in \, \mathsf{head} \, \lfbsym{(}  \sigma  \lfbsym{)}$} \\
    x & \text{otherwise}                      \\
  \end{cases}                                              \\
  \lfbsym{(}   \lambda \lfbmv{x} ^{ \lfbnt{T} }. \lfbnt{M}   \lfbsym{)}  \lfbsym{[}  \sigma  \lfbsym{]}    & =  \lambda \lfbmv{x} ^{ \lfbnt{T} }. \lfbsym{(}  \lfbnt{M}  \lfbsym{[}  \sigma  \lfbsym{]}  \lfbsym{)}  \text{ \textbf{where} $\lfbmv{x} \, \not\in \, \mathsf{dom} \, \lfbsym{(}  \mathsf{head} \, \lfbsym{(}  \sigma  \lfbsym{)}  \lfbsym{)}$ and $ \lfbmv{x} \, \not\in \,  \mathsf{FV} _{ \lfbsym{0} }( \sigma )  $} \\
  \lfbsym{(}  \lfbnt{M} \, \lfbnt{N}  \lfbsym{)}  \lfbsym{[}  \sigma  \lfbsym{]}        & = \lfbsym{(}  \lfbnt{M}  \lfbsym{[}  \sigma  \lfbsym{]}  \lfbsym{)} \, \lfbsym{(}  \lfbnt{N}  \lfbsym{[}  \sigma  \lfbsym{]}  \lfbsym{)}                                                                                \\
  \lfbsym{(}   \lfbkw{quo} \langle \Gamma \rangle \lfbnt{M}   \lfbsym{)}  \lfbsym{[}  \sigma  \lfbsym{]}      & =  \lfbkw{quo} \langle \Gamma \rangle \lfbsym{(}  \lfbnt{M}  \lfbsym{[}   \sigma  \lfbsym{,}   \text{\faLock} _{ \lfbsym{1} }  \lfbsym{,}  id _{ \Gamma }    \lfbsym{]}  \lfbsym{)}                                                                 \\
  \lfbsym{(}   \lfbkw{unq} _{ \lfbnt{k} } \lfbnt{M} [  \theta  ]   \lfbsym{)}  \lfbsym{[}  \sigma  \lfbsym{]} & =  \lfbkw{unq} _{ \lfbsym{(}   \mathsf{count} ( \lfbnt{k} , \sigma )   \lfbsym{)} } \lfbsym{(}  \lfbnt{M}  \lfbsym{[}  \sigma  \uparrow  \lfbnt{k}  \lfbsym{]}  \lfbsym{)} [  \theta  \lfbsym{[}  \sigma  \lfbsym{]}  ]  \\
  \lfbsym{(}   \Lambda \gamma . \lfbnt{M}   \lfbsym{)}  \lfbsym{[}  \sigma  \lfbsym{]}   & =  \Lambda \gamma . \lfbsym{(}  \lfbnt{M}  \lfbsym{[}  \sigma  \lfbsym{]}  \lfbsym{)}  \text{ if $\gamma \, \not\in \, \mathsf{FCV} \, \lfbsym{(}  \sigma  \lfbsym{)}$} \\
  \lfbsym{(}   \lfbnt{M} @ \lfbnt{C}   \lfbsym{)}  \lfbsym{[}  \sigma  \lfbsym{]} & =  \lfbsym{(}  \lfbnt{M}  \lfbsym{[}  \sigma  \lfbsym{]}  \lfbsym{)} @ \lfbnt{C} 
\end{align*}

\noindent
\mbox{\fbox{$\theta  \lfbsym{[}  \sigma  \lfbsym{]}$}}
\begin{align*}
   \bullet   \lfbsym{[}  \sigma  \lfbsym{]} & =  \bullet        \\
  \lfbsym{(}  \theta  \lfbsym{,}  \lfbnt{M}  \lfbsym{)}  \lfbsym{[}  \sigma  \lfbsym{]}   & = \theta  \lfbsym{[}  \sigma  \lfbsym{]}  \lfbsym{,}  \lfbnt{M}  \lfbsym{[}  \sigma  \lfbsym{]} \\
  \lfbsym{(}  \theta  \lfbsym{,}  \mathbb{x}  \lfbsym{)}  \lfbsym{[}  \sigma  \lfbsym{]} & = \begin{cases}
                            \theta  \lfbsym{[}  \sigma  \lfbsym{]}  \lfbsym{,}  \mathbb{y} & \text{if $\mathbb{x}  \coloneqq  \mathbb{y} \, \in \, \mathsf{head} \, \lfbsym{(}  \sigma  \lfbsym{)}$} \\
                            \theta  \lfbsym{[}  \sigma  \lfbsym{]}  \lfbsym{,}  \mathbb{x} & \text{otherwise}                        \\
                          \end{cases}
\end{align*}

\subsection{Context Substitution}

\begin{tabular}[]{lcl}
  \textbf{Context substitution}         & $\Sigma$       & $\Coloneqq  \bullet  \mid \Sigma  \lfbsym{,}  \gamma  \coloneqq  \lfbnt{C}$                       \\
  \textbf{Variable series}             & $\overrightarrow{x}, \overrightarrow{y}$ & $\Coloneqq  \bullet  \mid \overrightarrow{x}  \lfbsym{,}  \lfbmv{y} \mid \overrightarrow{x}  \lfbsym{,}  \mathbb{y}$                  \\
  \textbf{Variable series substitution} & $\bar{\sigma}$       & $\Coloneqq  \bullet  \mid \bar{\sigma}  \lfbsym{,}  \mathbb{x}  \coloneqq  \overrightarrow{y} \mid \bar{\sigma}  \lfbsym{,} \, \text{\faLock}$ \\
  \textbf{Variable generator}     & $G$  & $\Coloneqq \lfbsym{(}  G_v  \lfbsym{,}  G_s  \lfbsym{)}$
\end{tabular}

\begin{align*}
   \mathsf{destruct} _{ G } \lfbsym{(} \lfbsym{(}  \Gamma  \lfbsym{,}  \lfbmv{x}  \colon  \lfbnt{T}  \lfbsym{)} \lfbsym{;} \Sigma \lfbsym{)}    & =  \mathsf{destruct} _{ G } \lfbsym{(} \Gamma \lfbsym{;} \Sigma \lfbsym{)}  \\
   \mathsf{destruct} _{ G } \lfbsym{(} \lfbsym{(}  \Gamma  \lfbsym{,}  \mathbb{x}  \colon  \gamma  \lfbsym{)} \lfbsym{;} \Sigma \lfbsym{)}   & =
                                                  \begin{cases}
                                                    \bar{\sigma}  \lfbsym{,}  \mathbb{x}  \coloneqq  \overrightarrow{x} & \text{if $\gamma  \coloneqq  \lfbnt{C} \, \in \, \Sigma$} \\
                                                      \multicolumn{2}{l}{\quad \text{where $\bar{\sigma} =  \mathsf{destruct} _{ G } \lfbsym{(} \Gamma \lfbsym{;} \Sigma \lfbsym{)} $}}         \\
                                                      \multicolumn{2}{l}{\quad \text{and $\overrightarrow{x} =  \mathsf{gensyms} _{ G } \lfbsym{(} \lfbnt{C} \lfbsym{;} \mathsf{dom} \, \lfbsym{(}  \Gamma  \lfbsym{)} \, \cup \, \mathsf{rg} \, \lfbsym{(}  \bar{\sigma}  \lfbsym{)} \lfbsym{)} $}} \\
                                                     \mathsf{destruct} _{ G } \lfbsym{(} \Gamma \lfbsym{;} \Sigma \lfbsym{)}  & \text{otherwise}                                                      \\
                                                  \end{cases} \\
   \mathsf{destruct} _{ G } \lfbsym{(} \lfbsym{(}  \Gamma  \lfbsym{,}  \text{\faLock}  \lfbsym{)} \lfbsym{;} \Sigma \lfbsym{)}     & =  \mathsf{destruct} _{ G } \lfbsym{(} \Gamma \lfbsym{;} \Sigma \lfbsym{)}   \lfbsym{,} \, \text{\faLock} \\[1ex]
   \mathsf{gensyms} _{ \lfbsym{(}  G_v  \lfbsym{,}  G_s  \lfbsym{)} } \lfbsym{(}  \bullet  \lfbsym{;} V \lfbsym{)}  & =  \bullet  \\
   \mathsf{gensyms} _{ \lfbsym{(}  G_v  \lfbsym{,}  G_s  \lfbsym{)} } \lfbsym{(} \lfbsym{(}  \lfbnt{C}  \lfbsym{,}  \lfbnt{T}  \lfbsym{)} \lfbsym{;} V \lfbsym{)}    & =  \mathsf{gensyms} _{ \lfbsym{(}  G_v  \lfbsym{,}  G_s  \lfbsym{)} } \lfbsym{(} \lfbnt{C} \lfbsym{;} V \, \cup \, \lfbsym{\{}  \lfbmv{x}  \lfbsym{\}} \lfbsym{)}   \lfbsym{,}  \lfbmv{x} \\
                                                & \qquad \text{where $\lfbmv{x}$ is the first element of $G_v$ such that $\lfbmv{x} \, \not\in \, V$} \\
   \mathsf{gensyms} _{ \lfbsym{(}  G_v  \lfbsym{,}  G_s  \lfbsym{)} } \lfbsym{(} \lfbsym{(}  \lfbnt{C}  \lfbsym{,}  \gamma  \lfbsym{)} \lfbsym{;} V \lfbsym{)}    & =  \mathsf{gensyms} _{ \lfbsym{(}  G_v  \lfbsym{,}  G_s  \lfbsym{)} } \lfbsym{(} \lfbnt{C} \lfbsym{;} V \, \cup \, \lfbsym{\{}  \mathbb{x}  \lfbsym{\}} \lfbsym{)}   \lfbsym{,}  \mathbb{x} \\
                                                & \qquad \text{where $\mathbb{x}$ is the first element of $G_s$ such that $\mathbb{x} \, \not\in \, V$}
\end{align*}

\noindent
\framebox{\mbox{$\lfbnt{T}  \lfbsym{[}  \Sigma  \lfbsym{]}$}}
\begin{align*}
  \iota  \lfbsym{[}  \Sigma  \lfbsym{]}       & = \iota                   \\
  \lfbsym{(}  \lfbnt{S}  \rightarrow  \lfbnt{T}  \lfbsym{)}  \lfbsym{[}  \Sigma  \lfbsym{]}   & = \lfbnt{S}  \lfbsym{[}  \Sigma  \lfbsym{]}  \rightarrow  \lfbnt{T}  \lfbsym{[}  \Sigma  \lfbsym{]}     \\
  \lfbsym{[}  \lfbnt{C}  \vdash  \lfbnt{T}  \lfbsym{]}  \lfbsym{[}  \Sigma  \lfbsym{]} & = \lfbsym{[}  \lfbnt{C}  \lfbsym{[}  \Sigma  \lfbsym{]}  \vdash  \lfbnt{T}  \lfbsym{[}  \Sigma  \lfbsym{]}  \lfbsym{]} \\
  \lfbsym{(}   \forall \gamma . \lfbnt{T}   \lfbsym{)}  \lfbsym{[}  \Sigma  \lfbsym{]}     & =  \forall \gamma . \lfbsym{(}  \lfbnt{T}  \lfbsym{[}  \Sigma  \lfbsym{]}  \lfbsym{)}   \text{ if $\gamma \, \not\in \, \mathsf{dom} \, \lfbsym{(}  \Sigma  \lfbsym{)}$ and $\gamma \, \not\in \, \mathsf{FCV} \, \lfbsym{(}  \Sigma  \lfbsym{)}$ }
\end{align*}

\noindent
\framebox{\mbox{$\lfbnt{C}  \lfbsym{[}  \Sigma  \lfbsym{]}$}}
\begin{align*}
   \bullet   \lfbsym{[}  \Sigma  \lfbsym{]} & =  \bullet  \\
  \lfbsym{(}  \lfbnt{C}  \lfbsym{,}  \lfbnt{T}  \lfbsym{)}  \lfbsym{[}  \Sigma  \lfbsym{]}   & = \lfbnt{C}  \lfbsym{[}  \Sigma  \lfbsym{]}  \lfbsym{,}  \lfbnt{T}  \lfbsym{[}  \Sigma  \lfbsym{]} \\
  \lfbsym{(}  \lfbnt{C}  \lfbsym{,}  \gamma  \lfbsym{)}  \lfbsym{[}  \Sigma  \lfbsym{]}   & = \begin{cases}
                              \lfbnt{C}  \lfbsym{[}  \Sigma  \lfbsym{]} \lfbsym{,} \lfbnt{D}  & \text{if $\gamma  \coloneqq  \lfbnt{D} \, \in \, \Sigma$} \\
                             \lfbnt{C}  \lfbsym{[}  \Sigma  \lfbsym{]}  \lfbsym{,}  \gamma & \text{otherwise}
  \end{cases}
\end{align*}

\noindent
\framebox{\mbox{$ \lfbnt{M} \lfbsym{[} \Sigma \lfbsym{;} \bar{\sigma} \lfbsym{]} _{ G } $}}
\begin{align*}
   \lfbmv{x} \lfbsym{[} \Sigma \lfbsym{;} \bar{\sigma} \lfbsym{]} _{ G }           & = \lfbmv{x}                                                              \\
   \lfbsym{(}   \lambda \lfbmv{x} ^{ \lfbnt{T} }. \lfbnt{M}   \lfbsym{)} \lfbsym{[} \Sigma \lfbsym{;} \bar{\sigma} \lfbsym{]} _{ G }     & =  \lambda \lfbmv{x} ^{ \lfbsym{(}  \lfbnt{T}  \lfbsym{[}  \Sigma  \lfbsym{]}  \lfbsym{)} }. \lfbsym{(}   \lfbnt{M} \lfbsym{[} \Sigma \lfbsym{;} \bar{\sigma} \lfbsym{]} _{ G }   \lfbsym{)}                          \\
   \lfbsym{(}  \lfbnt{M} \, \lfbnt{N}  \lfbsym{)} \lfbsym{[} \Sigma \lfbsym{;} \bar{\sigma} \lfbsym{]} _{ G }        & = \lfbsym{(}   \lfbnt{M} \lfbsym{[} \Sigma \lfbsym{;} \bar{\sigma} \lfbsym{]} _{ G }   \lfbsym{)} \, \lfbsym{(}   \lfbnt{N} \lfbsym{[} \Sigma \lfbsym{;} \bar{\sigma} \lfbsym{]} _{ G }   \lfbsym{)}             \\
   \lfbsym{(}   \lfbkw{quo} \langle \Gamma \rangle \lfbnt{M}   \lfbsym{)} \lfbsym{[} \Sigma \lfbsym{;} \bar{\sigma} \lfbsym{]} _{ G }     & =  \lfbkw{quo} \langle  \Gamma   \lfbsym{[}   \Sigma   \lfbsym{;}   \bar{\sigma}'   \lfbsym{]}  \rangle \lfbsym{(}   \lfbnt{M} \lfbsym{[} \Sigma \lfbsym{;} \lfbsym{(}   \bar{\sigma}  \lfbsym{,} \, \text{\faLock} \lfbsym{,} \bar{\sigma}'   \lfbsym{)} \lfbsym{]} _{ G' }   \lfbsym{)}  \\
                                        & \qquad \text{where $\bar{\sigma}' =  \mathsf{destruct} _{ G } \lfbsym{(} \Gamma \lfbsym{;} \Sigma \lfbsym{)} $} \\
                                        & \qquad \text{and $G' = G - (\mathsf{dom} \, \lfbsym{(}  \Gamma  \lfbsym{)} \, \cup \, \mathsf{rg} \, \lfbsym{(}  \bar{\sigma}'  \lfbsym{)})$} \\
   \lfbsym{(}   \lfbkw{unq} _{ \lfbnt{k} } \lfbnt{M} [  \theta  ]   \lfbsym{)} \lfbsym{[} \Sigma \lfbsym{;} \bar{\sigma} \lfbsym{]} _{ G }  & =  \lfbkw{unq} _{ \lfbnt{k} } \lfbsym{(}   \lfbnt{M} \lfbsym{[} \Sigma \lfbsym{;} \bar{\sigma}  \uparrow  \lfbnt{k} \lfbsym{]} _{ G }   \lfbsym{)} [   \theta \lfbsym{[} \Sigma \lfbsym{;} \bar{\sigma} \lfbsym{]} _{ G }   ]  \\
   \lfbsym{(}   \Lambda \gamma . \lfbnt{M}   \lfbsym{)} \lfbsym{[} \Sigma \lfbsym{;} \bar{\sigma} \lfbsym{]} _{ G }  & =  \Lambda \gamma . \lfbsym{(}   \lfbnt{M} \lfbsym{[} \Sigma \lfbsym{;} \bar{\sigma} \lfbsym{]} _{ G }   \lfbsym{)}   \qquad \text{ if $\gamma \, \not\in \, \mathsf{dom} \, \lfbsym{(}  \Sigma  \lfbsym{)}$ and $ \gamma \, \not\in \, \mathsf{FCV} \, \lfbsym{(}  \Sigma  \lfbsym{)} $} \\
   \lfbsym{(}   \lfbnt{M} @ \lfbnt{C}   \lfbsym{)} \lfbsym{[} \Sigma \lfbsym{;} \bar{\sigma} \lfbsym{]} _{ G }  & =    \lfbsym{(}   \lfbnt{M} \lfbsym{[} \Sigma \lfbsym{;} \bar{\sigma} \lfbsym{]} _{ G }   \lfbsym{)} @ \lfbsym{(}  \lfbnt{C}  \lfbsym{[}  \Sigma  \lfbsym{]}  \lfbsym{)}  
\end{align*}

\noindent
\framebox{\mbox{$ \theta \lfbsym{[} \Sigma \lfbsym{;} \bar{\sigma} \lfbsym{]} _{ G } $}}
\begin{align*}
    \bullet  \lfbsym{[} \Sigma \lfbsym{;} \bar{\sigma} \lfbsym{]} _{ G }  & =  \bullet                                                                                   \\
   \lfbsym{(}  \theta  \lfbsym{,}  \lfbnt{M}  \lfbsym{)} \lfbsym{[} \Sigma \lfbsym{;} \bar{\sigma} \lfbsym{]} _{ G }    & = \lfbsym{(}   \theta \lfbsym{[} \Sigma \lfbsym{;} \bar{\sigma} \lfbsym{]} _{ G }   \lfbsym{)}  \lfbsym{,}  \lfbsym{(}   \lfbnt{M} \lfbsym{[} \Sigma \lfbsym{;} \bar{\sigma} \lfbsym{]} _{ G }   \lfbsym{)}                                       \\
   \lfbsym{(}  \theta  \lfbsym{,}  \mathbb{x}  \lfbsym{)} \lfbsym{[} \Sigma \lfbsym{;} \bar{\sigma} \lfbsym{]} _{ G }   & =\begin{cases}
     \lfbsym{(}   \theta \lfbsym{[} \Sigma \lfbsym{;} \bar{\sigma} \lfbsym{]} _{ G }   \lfbsym{)} \lfbsym{,} \overrightarrow{y}  & \text{if $\mathbb{x}  \coloneqq  \overrightarrow{y} \, \in \, \mathsf{head} \, \lfbsym{(}  \bar{\sigma}  \lfbsym{)}$} \\
    \lfbsym{(}   \theta \lfbsym{[} \Sigma \lfbsym{;} \bar{\sigma} \lfbsym{]} _{ G }   \lfbsym{)}  \lfbsym{,}  \mathbb{x} & \text{otherwise}
  \end{cases}
\end{align*}

\noindent
\framebox{\mbox{$ \Gamma   \lfbsym{[}   \Sigma   \lfbsym{;}   \bar{\sigma}   \lfbsym{]} $}}
\begin{align*}
    \bullet    \lfbsym{[}   \Sigma   \lfbsym{;}   \bar{\sigma}   \lfbsym{]}    & =  \bullet                                                                                                                                    \\
   \lfbsym{(}  \Gamma  \lfbsym{,}  \lfbmv{x}  \colon  \lfbnt{T}  \lfbsym{)}   \lfbsym{[}   \Sigma   \lfbsym{;}   \bar{\sigma}   \lfbsym{]}   & =  \Gamma   \lfbsym{[}   \Sigma   \lfbsym{;}   \bar{\sigma}   \lfbsym{]}   \lfbsym{,}  \lfbmv{x}  \colon  \lfbnt{T}  \lfbsym{[}  \Sigma  \lfbsym{]}                                                                                                                  \\
   \lfbsym{(}  \Gamma  \lfbsym{,}  \mathbb{x}  \colon  \gamma  \lfbsym{)}   \lfbsym{[}   \Sigma   \lfbsym{;}   \bar{\sigma}   \lfbsym{]}  & = \begin{cases}
     \Gamma   \lfbsym{[}   \Sigma   \lfbsym{;}   \bar{\sigma}   \lfbsym{]}   \lfbsym{,}  \overrightarrow{y}  \colon  \lfbnt{C} 
    & \text{if $\mathbb{x}  \coloneqq  \overrightarrow{y} \, \in \, \mathsf{head} \, \lfbsym{(}  \bar{\sigma}  \lfbsym{)}$ and $\gamma  \coloneqq  \lfbnt{C} \, \in \, \Sigma$} \\
     \Gamma   \lfbsym{[}   \Sigma   \lfbsym{;}   \bar{\sigma}   \lfbsym{]}   \lfbsym{,}  \mathbb{x}  \colon  \gamma & \text{otherwise}
  \end{cases} \\
   \lfbsym{(}  \Gamma  \lfbsym{,}  \text{\faLock}  \lfbsym{)}   \lfbsym{[}   \Sigma   \lfbsym{;}   \bar{\sigma}   \lfbsym{]}    & =  \Gamma   \lfbsym{[}   \Sigma   \lfbsym{;}   \bar{\sigma}  \uparrow  \lfbsym{1}   \lfbsym{]}   \lfbsym{,}  \text{\faLock}
\end{align*}

\subsection{Reduction}
\framebox{\mbox{$\lfbnt{M_{{\mathrm{1}}}}  \rightarrow_{\beta}  \lfbnt{M_{{\mathrm{1}}}}$}}
\[
  \AxiomC{}
  \UnaryInfC{$\lfbsym{(}   \lambda \lfbmv{x} ^{ \lfbnt{S} }. \lfbnt{M}   \lfbsym{)} \, \lfbnt{N}  \rightarrow_{\beta}  \lfbnt{M}  \lfbsym{[}   \lfbmv{x}  \coloneqq  \lfbnt{N}   \lfbsym{]} $}
  \DisplayProof
  \quad
  \AxiomC{}
  \UnaryInfC{$ \lfbkw{unq} _{ \lfbnt{k} } \lfbsym{(}   \lfbkw{quo} \langle  \overrightarrow{x}  \colon  \lfbnt{C}  \rangle \lfbnt{M}   \lfbsym{)} [  \theta  ]   \rightarrow_{\beta}  \lfbnt{M}  \lfbsym{[}    \text{\faLock} _{ \lfbnt{k} }    \lfbsym{,}  \overrightarrow{x}  \coloneqq  \theta  \lfbsym{]} $}
  \DisplayProof
\]

\[
  \AxiomC{}
  \UnaryInfC{$ \lfbsym{(}   \Lambda \gamma . \lfbnt{M}   \lfbsym{)} @ \lfbnt{C}   \rightarrow_{\beta}   \lfbnt{M} \lfbsym{[}  \gamma \coloneqq \lfbnt{C}  \lfbsym{;}  \bullet  \lfbsym{]} $}
  \DisplayProof
  \quad
\]
\[
  \AxiomC{$\lfbnt{M_{{\mathrm{1}}}}  \rightarrow_{\beta}  \lfbnt{M_{{\mathrm{2}}}}$}
  \UnaryInfC{$ \lambda \lfbmv{x} ^{ \lfbnt{T} }. \lfbnt{M_{{\mathrm{1}}}}   \rightarrow_{\beta}   \lambda \lfbmv{x} ^{ \lfbnt{T} }. \lfbnt{M_{{\mathrm{2}}}} $}
  \DisplayProof
  \quad
  \AxiomC{$\lfbnt{M_{{\mathrm{1}}}}  \rightarrow_{\beta}  \lfbnt{M_{{\mathrm{2}}}}$}
  \UnaryInfC{$\lfbnt{M_{{\mathrm{1}}}} \, \lfbnt{N}  \rightarrow_{\beta}  \lfbnt{M_{{\mathrm{2}}}} \, \lfbnt{N}$}
  \DisplayProof
  \quad
  \AxiomC{$\lfbnt{M_{{\mathrm{1}}}}  \rightarrow_{\beta}  \lfbnt{M_{{\mathrm{2}}}}$}
  \UnaryInfC{$\lfbnt{N} \, \lfbnt{M_{{\mathrm{1}}}}  \rightarrow_{\beta}  \lfbnt{N} \, \lfbnt{M_{{\mathrm{2}}}}$}
  \DisplayProof
\]

\[
  \AxiomC{$\lfbnt{M_{{\mathrm{1}}}}  \rightarrow_{\beta}  \lfbnt{M_{{\mathrm{2}}}}$}
  \UnaryInfC{$ \lfbkw{quo} \langle \Gamma \rangle \lfbnt{M_{{\mathrm{1}}}}   \rightarrow_{\beta}   \lfbkw{quo} \langle \Gamma \rangle \lfbnt{M_{{\mathrm{2}}}} $}
  \DisplayProof
  \quad
  \AxiomC{$\lfbnt{M_{{\mathrm{1}}}}  \rightarrow_{\beta}  \lfbnt{M_{{\mathrm{2}}}}$}
  \UnaryInfC{$ \lfbkw{unq} _{ \lfbnt{k} } \lfbnt{M_{{\mathrm{1}}}} [  \theta  ]   \rightarrow_{\beta}   \lfbkw{unq} _{ \lfbnt{k} } \lfbnt{M_{{\mathrm{2}}}} [  \theta  ] $}
  \DisplayProof
  \quad
  \AxiomC{$\theta_{{\mathrm{1}}}  \rightarrow_{\beta}  \theta_{{\mathrm{2}}}$}
  \UnaryInfC{$ \lfbkw{unq} _{ \lfbnt{k} } \lfbnt{M} [  \theta_{{\mathrm{1}}}  ]   \rightarrow_{\beta}   \lfbkw{unq} _{ \lfbnt{k} } \lfbnt{M} [  \theta_{{\mathrm{2}}}  ] $}
  \DisplayProof
\]

\[
  \AxiomC{$\lfbnt{M_{{\mathrm{1}}}}  \rightarrow_{\beta}  \lfbnt{M_{{\mathrm{2}}}}$}
  \UnaryInfC{$ \Lambda \gamma . \lfbnt{M_{{\mathrm{1}}}}   \rightarrow_{\beta}   \Lambda \gamma . \lfbnt{M_{{\mathrm{2}}}} $}
  \DisplayProof
  \quad
  \AxiomC{$\lfbnt{M_{{\mathrm{1}}}}  \rightarrow_{\beta}  \lfbnt{M_{{\mathrm{2}}}}$}
  \UnaryInfC{$ \lfbnt{M_{{\mathrm{1}}}} @ \lfbnt{C}   \rightarrow_{\beta}   \lfbnt{M_{{\mathrm{2}}}} @ \lfbnt{C} $}
  \DisplayProof
\]

\noindent
\framebox{\mbox{$\theta_{{\mathrm{1}}}  \rightarrow_{\beta}  \theta_{{\mathrm{2}}}$}}
\[
  \AxiomC{$\lfbnt{M_{{\mathrm{1}}}}  \rightarrow_{\beta}  \lfbnt{M_{{\mathrm{2}}}}$}
  \UnaryInfC{$\theta  \lfbsym{,}  \lfbnt{M_{{\mathrm{1}}}}  \rightarrow_{\beta}  \theta  \lfbsym{,}  \lfbnt{M_{{\mathrm{2}}}}$}
  \DisplayProof
  \quad
  \AxiomC{$\theta_{{\mathrm{1}}}  \rightarrow_{\beta}  \theta_{{\mathrm{2}}}$}
  \UnaryInfC{$\theta_{{\mathrm{1}}}  \lfbsym{,}  \lfbnt{M}  \rightarrow_{\beta}  \theta_{{\mathrm{2}}}  \lfbsym{,}  \lfbnt{M}$}
  \DisplayProof
  \quad
  \AxiomC{$\theta_{{\mathrm{1}}}  \rightarrow_{\beta}  \theta_{{\mathrm{2}}}$}
  \UnaryInfC{$\theta_{{\mathrm{1}}}  \lfbsym{,}  \mathbb{x}  \rightarrow_{\beta}  \theta_{{\mathrm{2}}}  \lfbsym{,}  \mathbb{x}$}
  \DisplayProof
\]

%% file: artifact.tex
\section{Proof of Lemmas and Theorems}
\subsection{Proof of Lemma 3}
\begin{proof}
    By mutual induction on derivation of $\lfbnt{M}$ and $\theta$. We cover only major cases. Other cases are straightforward.
    \begin{itemize}
        \item Case $\lfbnt{M} = \lfbmv{x}$ and $\lfbmv{x}  \colon  \lfbnt{T} \, \in \, \mathsf{head} \, \lfbsym{(}  \Gamma  \lfbsym{)}$. From the typing rules for substitution, there is a mapping $\lfbmv{x}  \coloneqq  \lfbnt{N} \, \in \, \mathsf{head} \, \lfbsym{(}  \sigma  \lfbsym{)}$ such that $\Delta  \vdash  \lfbnt{N}  \colon  \lfbnt{T}$. This $\Delta  \vdash  \lfbnt{N}  \colon  \lfbnt{T}$ is what we want.
        \item Case $\lfbnt{M} =  \lfbkw{quo} \langle \Gamma' \rangle \lfbnt{M'} $.  We have $ \Gamma  \lfbsym{,}  \text{\faLock} \lfbsym{,} \Gamma'   \vdash  \lfbnt{M'}  \colon  \lfbnt{T'}$ for some $\lfbnt{T'}$ where $\lfbnt{T} = \lfbsym{[}   \mathsf{rg} ( \Gamma' )   \vdash  \lfbnt{T'}  \lfbsym{]}$ and $\text{\faLock} \, \not\in \, \Gamma'$. As we can derive $ \Delta  \lfbsym{,}  \text{\faLock} \lfbsym{,} \Gamma'   \vdash   \sigma  \lfbsym{,}   \text{\faLock} _{ \lfbsym{1} }  \lfbsym{,}  id _{ \Gamma' }    \colon   \Gamma  \lfbsym{,}  \text{\faLock} \lfbsym{,} \Gamma' $, we can apply the induction hypothesis and get $ \Delta  \lfbsym{,}  \text{\faLock} \lfbsym{,} \Gamma'   \vdash  \lfbnt{M'}  \lfbsym{[}   \sigma  \lfbsym{,}   \text{\faLock} _{ \lfbsym{1} }  \lfbsym{,}  id _{ \Gamma' }    \lfbsym{]}  \colon  \lfbnt{T'}$. By introducing quote, we confirm that $\Delta  \vdash   \lfbkw{quo} \langle \Gamma' \rangle \lfbsym{(}  \lfbnt{M'}  \lfbsym{[}   \sigma  \lfbsym{,}   \text{\faLock} _{ \lfbsym{1} }  \lfbsym{,}  id _{ \Gamma' }    \lfbsym{]}  \lfbsym{)}   \colon  \lfbnt{T}$.

        \item Case $\lfbnt{M} =  \lfbkw{unq} _{ \lfbnt{k} } \lfbnt{M'} [  \theta  ] $. We have $\Gamma  \uparrow  \lfbnt{k}  \vdash  \lfbnt{M}  \colon  \lfbsym{[}  \lfbnt{C}  \vdash  \lfbnt{T}  \lfbsym{]}$, $\Gamma  \vdash  \theta  \colon  \lfbnt{C}$ and $\lfbnt{k}  \colon  \Gamma  \uparrow  \lfbnt{k}  \lhd  \Gamma$ for some $\lfbnt{C}$. It is easy to confirm that $\lfbnt{k}  \colon  \Gamma  \uparrow  \lfbnt{k}  \lhd  \Gamma$ holds without assumption from the definition of $\Gamma  \uparrow  \lfbnt{k}$. Then we can confirm that $\Delta  \uparrow   \mathsf{count} ( \lfbnt{k} , \sigma )   \vdash  \sigma  \uparrow  \lfbnt{k}  \colon  \Gamma  \uparrow  \lfbnt{k}$ holds from the typing rules of substitution. Consequently, we can apply the induction hypothesis and get $\Delta  \uparrow   \mathsf{count} ( \lfbnt{k} , \sigma )   \vdash  \lfbnt{M'}  \lfbsym{[}  \sigma  \uparrow  \lfbnt{k}  \lfbsym{]}  \colon  \lfbsym{[}  \lfbnt{C}  \vdash  \lfbnt{T}  \lfbsym{]}$ and $\Delta  \vdash  \theta  \lfbsym{[}  \sigma  \lfbsym{]}  \colon  \lfbnt{C}$. Because $ \mathsf{count} ( \lfbnt{k} , \sigma )   \colon  \Delta  \uparrow   \mathsf{count} ( \lfbnt{k} , \sigma )   \lhd  \Delta$, we can derive $\Delta  \vdash   \lfbkw{unq} _{  \mathsf{count} ( \lfbnt{k} , \sigma )  } \lfbnt{M'} [  \theta  ]   \colon  \lfbnt{T}$.
    \end{itemize}
\end{proof}

\subsection{Proof of Lemma 4}
\begin{proof}\sloppy
    By mutual induction on derivation of $\lfbnt{M}$ and $\theta$. We cover only major cases. Other cases are straightforward.
    \begin{itemize}
        \item Case $\lfbnt{M} =  \lfbkw{quo} \langle \Delta \rangle \lfbnt{M'} $.  We have $ \Gamma  \lfbsym{,}  \text{\faLock} \lfbsym{,} \Delta   \vdash  \lfbnt{M'}  \colon  \lfbnt{T'}$ where $\Delta$ is \faLock-free and $\lfbnt{T} = \lfbsym{[}   \mathsf{rg} ( \Delta )   \vdash  \lfbnt{T'}  \lfbsym{]}$. By induction hypothesis, we have $ \lfbsym{(}   \Gamma  \lfbsym{,}  \text{\faLock} \lfbsym{,} \Delta   \lfbsym{)}   \lfbsym{[}   \Sigma   \lfbsym{;}   \bar{\sigma}   \lfbsym{]}   \vdash   \lfbnt{M'} \lfbsym{[} \Sigma \lfbsym{;} \bar{\sigma} \lfbsym{]} _{ G_{{\mathrm{1}}} }   \colon  \lfbnt{T'}  \lfbsym{[}  \Sigma  \lfbsym{]}$ where $\bar{\sigma} =  \mathsf{destruct} _{ G } \lfbsym{(} \lfbsym{(}   \Gamma  \lfbsym{,}  \text{\faLock} \lfbsym{,} \Delta   \lfbsym{)} \lfbsym{;} \Sigma \lfbsym{)} $ and $G_{{\mathrm{1}}} = G - (\mathsf{dom} \, \lfbsym{(}   \Gamma  \lfbsym{,}  \text{\faLock} \lfbsym{,} \Delta   \lfbsym{)} \, \cup \, \mathsf{rg} \, \lfbsym{(}  \bar{\sigma}  \lfbsym{)})$. Here, $\bar{\sigma}$ is equal to $ \bar{\sigma}_{{\mathrm{1}}}  \lfbsym{,} \, \text{\faLock} \lfbsym{,} \bar{\sigma}_{{\mathrm{2}}} $ for some $\bar{\sigma}_{{\mathrm{1}}}$ and $\bar{\sigma}_{{\mathrm{2}}}$ where $\bar{\sigma}_{{\mathrm{1}}} =  \mathsf{destruct} _{ G } \lfbsym{(} \Gamma \lfbsym{;} \Sigma \lfbsym{)} $, $\bar{\sigma}_{{\mathrm{2}}} =  \mathsf{destruct} _{ G_{{\mathrm{2}}} } \lfbsym{(} \Gamma \lfbsym{;} \Sigma \lfbsym{)} $, and $G_{{\mathrm{2}}}$ = $G - (\mathsf{dom} \, \lfbsym{(}  \Gamma  \lfbsym{)} \, \cup \, \mathsf{rg} \, \lfbsym{(}  \bar{\sigma}_{{\mathrm{1}}}  \lfbsym{)})$. Then we can rewrite the judgment to $  \Gamma   \lfbsym{[}   \Sigma   \lfbsym{;}   \bar{\sigma}_{{\mathrm{1}}}   \lfbsym{]}   \lfbsym{,}  \text{\faLock} \lfbsym{,}   \Delta   \lfbsym{[}   \Sigma   \lfbsym{;}   \bar{\sigma}_{{\mathrm{2}}}   \lfbsym{]}     \vdash   \lfbnt{M'} \lfbsym{[} \Sigma \lfbsym{;} \bar{\sigma} \lfbsym{]} _{ G_{{\mathrm{1}}} }   \colon  \lfbnt{T'}  \lfbsym{[}  \Sigma  \lfbsym{]}$ and introduce quotation as $ \Gamma   \lfbsym{[}   \Sigma   \lfbsym{;}   \bar{\sigma}_{{\mathrm{1}}}   \lfbsym{]}   \vdash   \lfbkw{quo} \langle  \Delta   \lfbsym{[}   \Sigma   \lfbsym{;}   \bar{\sigma}_{{\mathrm{2}}}   \lfbsym{]}  \rangle \lfbsym{(}   \lfbnt{M'} \lfbsym{[} \Sigma \lfbsym{;} \lfbsym{(}   \bar{\sigma}_{{\mathrm{1}}}  \lfbsym{,} \, \text{\faLock} \lfbsym{,} \bar{\sigma}_{{\mathrm{2}}}   \lfbsym{)} \lfbsym{]} _{ G_{{\mathrm{1}}} }   \lfbsym{)}   \colon  \lfbnt{T'}  \lfbsym{[}  \Sigma  \lfbsym{]}$. Finally, we confirm that $G_{{\mathrm{1}}} = G_{{\mathrm{2}}} - (\mathsf{dom} \, \lfbsym{(}  \Delta  \lfbsym{)} \, \cup \, \mathsf{rg} \, \lfbsym{(}  \bar{\sigma}_{{\mathrm{2}}}  \lfbsym{)})$.
        
        \item Case $\lfbnt{M} =  \Lambda \gamma . \lfbnt{M'} $. We have $\Gamma  \vdash  \lfbnt{M'}  \colon  \lfbnt{T'}$ where $\lfbnt{T} =  \forall \gamma . \lfbnt{T'} $ and $\gamma \, \not\in \, \mathsf{FCV} \, \lfbsym{(}  \Gamma  \lfbsym{)}$. By the induction hypothesis, $ \Gamma   \lfbsym{[}   \Sigma   \lfbsym{;}   \bar{\sigma}   \lfbsym{]}   \vdash   \lfbnt{M'} \lfbsym{[} \Sigma \lfbsym{;} \bar{\sigma} \lfbsym{]} _{ G' }   \colon  \lfbnt{T'}  \lfbsym{[}  \Sigma  \lfbsym{]}$. Assuming that $\gamma \, \not\in \, \mathsf{dom} \, \lfbsym{(}  \Sigma  \lfbsym{)}$ and $\gamma \, \not\in \, \mathsf{FCV} \, \lfbsym{(}  \Sigma  \lfbsym{)}$, $\gamma \, \not\in \, \mathsf{FCV} \, \lfbsym{(}   \Gamma   \lfbsym{[}   \Sigma   \lfbsym{;}   \bar{\sigma}   \lfbsym{]}   \lfbsym{)}$ holds and hence $ \Gamma   \lfbsym{[}   \Sigma   \lfbsym{;}   \bar{\sigma}   \lfbsym{]}   \vdash   \Lambda \gamma . \lfbsym{(}   \lfbnt{M'} \lfbsym{[} \Sigma \lfbsym{;} \bar{\sigma} \lfbsym{]} _{ G' }   \lfbsym{)}   \colon   \forall \gamma . \lfbsym{(}  \lfbnt{T'}  \lfbsym{[}  \Sigma  \lfbsym{]}  \lfbsym{)} $ holds.
        
        \item Case $\lfbnt{M} =  \lfbnt{M'} @ \lfbnt{C} $. We have $\Gamma  \vdash  \lfbnt{M'}  \colon   \forall \gamma . \lfbnt{T'} $ where $\lfbnt{T'}  \lfbsym{[}   \gamma \coloneqq \lfbnt{C}   \lfbsym{]} = \lfbnt{T}$, $\gamma \, \not\in \, \mathsf{dom} \, \lfbsym{(}  \Sigma  \lfbsym{)}$ and $\gamma \, \not\in \, \mathsf{FCV} \, \lfbsym{(}  \Sigma  \lfbsym{)}$. By the induction hypothesis, $ \Gamma   \lfbsym{[}   \Sigma   \lfbsym{;}   \bar{\sigma}   \lfbsym{]}   \vdash   \lfbnt{M'} \lfbsym{[} \Sigma \lfbsym{;} \bar{\sigma} \lfbsym{]} _{ G' }   \colon  \lfbsym{(}   \forall \gamma . \lfbnt{T'}   \lfbsym{)}  \lfbsym{[}  \Sigma  \lfbsym{]}$. Because $\lfbsym{(}   \forall \gamma . \lfbnt{T'}   \lfbsym{)}  \lfbsym{[}  \Sigma  \lfbsym{]} =  \forall \gamma . \lfbsym{(}  \lfbnt{T'}  \lfbsym{[}  \Sigma  \lfbsym{]}  \lfbsym{)} $ and $\lfbnt{T'}  \lfbsym{[}  \Sigma  \lfbsym{]}  \lfbsym{[}   \gamma \coloneqq \lfbnt{C}  \lfbsym{[}  \Sigma  \lfbsym{]}   \lfbsym{]} = \lfbnt{T'}  \lfbsym{[}   \gamma \coloneqq \lfbnt{C}   \lfbsym{]}  \lfbsym{[}  \Sigma  \lfbsym{]} = \lfbnt{T}  \lfbsym{[}  \Sigma  \lfbsym{]}$, we confirm that $ \Gamma   \lfbsym{[}   \Sigma   \lfbsym{;}   \bar{\sigma}   \lfbsym{]}   \vdash    \lfbnt{M'} \lfbsym{[} \Sigma \lfbsym{;} \bar{\sigma} \lfbsym{]}  @ \lfbsym{(}  \lfbnt{C}  \lfbsym{[}  \Sigma  \lfbsym{]}  \lfbsym{)}   \colon  \lfbnt{T}  \lfbsym{[}  \Sigma  \lfbsym{]}$.
        
        \item Case $\theta = \theta'  \lfbsym{,}  \mathbb{x}$, $\Gamma  \vdash  \theta  \colon  \lfbnt{C'}$ and $\mathbb{x}  \colon  \gamma \, \in \, \mathsf{head} \, \lfbsym{(}  \Gamma  \lfbsym{)}$. From the induction hypothesis, we have $ \Gamma   \lfbsym{[}   \Sigma   \lfbsym{;}   \bar{\sigma}   \lfbsym{]}   \vdash   \theta \lfbsym{[} \Sigma \lfbsym{;} \bar{\sigma} \lfbsym{]} _{ G' }   \colon  \lfbnt{C'}  \lfbsym{[}  \Sigma  \lfbsym{]}$. We can consider two cases:
        \begin{itemize}
            \item If $\gamma  \coloneqq  \lfbnt{D} \, \in \, \Sigma$, we have $\mathbb{x}  \coloneqq  \overrightarrow{x} \, \in \, \bar{\sigma}$ from the definition of $ \mathsf{destruct} $. Therefore we can derive $ \Gamma   \lfbsym{[}   \Sigma   \lfbsym{;}   \bar{\sigma}   \lfbsym{]}   \vdash   \overrightarrow{x}   \colon  \lfbnt{D}$. As a result, we can derive $ \Gamma   \lfbsym{[}   \Sigma   \lfbsym{;}   \bar{\sigma}   \lfbsym{]}   \vdash  \lfbsym{(}    \theta \lfbsym{[} \Sigma \lfbsym{;} \bar{\sigma} \lfbsym{]}  \lfbsym{,} \overrightarrow{x}   \lfbsym{)}  \colon  \lfbsym{(}   \lfbnt{C'}  \lfbsym{[}  \Sigma  \lfbsym{]} \lfbsym{,} \lfbnt{D}   \lfbsym{)}$.
            \item Otherwise, we simply get $ \Gamma   \lfbsym{[}   \Sigma   \lfbsym{;}   \bar{\sigma}   \lfbsym{]}   \vdash  \lfbsym{(}   \theta \lfbsym{[} \Sigma \lfbsym{;} \bar{\sigma} \lfbsym{]}   \lfbsym{,}  \mathbb{x}  \lfbsym{)}  \colon  \lfbsym{(}  \lfbnt{C'}  \lfbsym{[}  \Sigma  \lfbsym{]}  \lfbsym{,}  \gamma  \lfbsym{)}$
        \end{itemize}
    \end{itemize}
\end{proof}

\subsection{Proof of Lemma 5}
\begin{proof}
    By comparing $\bar{\sigma}_{{\mathrm{1}}}$ and $\bar{\sigma}_{{\mathrm{2}}}$, we can point-wise mapping from variables to variables and series variables to series variables, which is the renaming substitution we want.
\end{proof}

\subsection{Proof of Theorem 1}
\begin{proof}
By induction on the definition of $\beta$-reduction. For base cases, we can apply local reduction patterns for each types, using Substitution.
\end{proof}

\subsection{Proof of Lemma 6}
\begin{proof}
    \begin{description}
        \item[(CR0)] For types, we can prove by cases for each top-level form of $\lfbnt{T}$. For contexts we can prove by induction on the form of contexts.
        \item[(CR1)] We prove it by induction with regard the structure of $\lfbnt{T}$ and $\lfbnt{C}$.
            \begin{itemize}
                \item Case $\lfbnt{T} = \iota$.  $\lfbnt{M}$ is strongly normalizing from the definition.
                \item Case $\lfbnt{T} = \lfbnt{S_{{\mathrm{1}}}}  \rightarrow  \lfbnt{S_{{\mathrm{2}}}}$. $ \lfbkw{Red} _{ \lfbnt{S_{{\mathrm{1}}}} } \lfbsym{[} \tilde{\Sigma} \lfbsym{]}   \lfbsym{(}  \lfbsym{(}  \Gamma  \lfbsym{,}  \lfbmv{x}  \colon  \lfbnt{S_{{\mathrm{1}}}}  \lfbsym{)}  \lfbsym{,}  \lfbmv{x}  \lfbsym{)}$ holds by the induction hypothesis of CR3, and hence $ \lfbkw{Red} _{ \lfbnt{S_{{\mathrm{2}}}} } \lfbsym{[} \tilde{\Sigma} \lfbsym{]}   \lfbsym{(}  \lfbsym{(}  \Gamma  \lfbsym{,}  \lfbmv{x}  \colon  \lfbnt{S_{{\mathrm{1}}}}  \lfbsym{)}  \lfbsym{,}  \lfbnt{M} \, \lfbmv{x}  \lfbsym{)}$ holds. We can then derive $\lfbnt{M} \, \lfbmv{x}$ is strongly normalizing from the induction hypothesis of CR1, and $\lfbnt{M}$ also is.
                \item Case $\lfbnt{T} = \lfbsym{[}  \lfbnt{C}  \vdash  \lfbnt{S}  \lfbsym{]}$. $ \lfbkw{Red} _{ \lfbnt{S} } \lfbsym{[} \tilde{\Sigma} \lfbsym{]}   \lfbsym{(}  \lfbsym{(}  \Gamma  \lfbsym{,}  \text{\faLock}  \lfbsym{,}  \overrightarrow{x}  \colon  \lfbnt{C}  \lfbsym{)}  \lfbsym{,}   \overrightarrow{x}   \lfbsym{)}$ holds by the induction hypothesis of CR3, and hence $ \lfbkw{Red} _{ \lfbnt{S_{{\mathrm{2}}}} } \lfbsym{[} \tilde{\Sigma} \lfbsym{]}   \lfbsym{(}  \lfbsym{(}  \Gamma  \lfbsym{,}  \text{\faLock}  \lfbsym{,}  \overrightarrow{x}  \colon  \lfbnt{C}  \lfbsym{)}  \lfbsym{,}   \lfbkw{unq} _{ \lfbsym{1} } \lfbnt{M} [   \overrightarrow{x}   ]   \lfbsym{)}$ holds. We can then derive $ \lfbkw{unq} _{ \lfbsym{1} } \lfbnt{M} [   \overrightarrow{x}   ] $ is strongly normalizing from the induction hypothesis of CR1, and $\lfbnt{M}$ also is.
                \item Case $\lfbnt{T} =  \forall \gamma . \lfbnt{S} $.  $ \lfbkw{Red} _{ \lfbnt{S} } \lfbsym{[} \tilde{\Sigma}  \lfbsym{,}  \gamma  \colon  \lfbnt{C}  \coloneqq  \mathcal{R} \lfbsym{]}   \lfbsym{(}  \Gamma  \lfbsym{,}   \lfbnt{M} @ \lfbnt{C}   \lfbsym{)}$ for any $\lfbnt{C}$ and $\mathcal{R}$. We fix $\lfbnt{C}$ and $\mathcal{R}$ to one of them. Then $ \lfbnt{M} @ \lfbnt{C} $ is strongly normalizing by the induction hypothesis of CR1, and $\lfbnt{M}$ also is.
                \item Case $\lfbnt{C} =  \bullet $. $\theta$ is $ \bullet $ and already normal.
                \item Case $\lfbnt{C} = \lfbnt{C'}  \lfbsym{,}  \lfbnt{T}$. $ \lfbkw{Red} _{ \lfbnt{C'} } \lfbsym{[} \tilde{\Sigma} \lfbsym{]}   \lfbsym{(}  \Gamma  \lfbsym{,}  \theta'  \lfbsym{)}$ and $ \lfbkw{Red} _{ \lfbnt{T} } \lfbsym{[} \tilde{\Sigma} \lfbsym{]}   \lfbsym{(}  \Gamma  \lfbsym{,}  \lfbnt{M}  \lfbsym{)}$ holds where $\theta = \theta'  \lfbsym{,}  \lfbnt{M}$. $\theta'$ and $\lfbnt{M}$ are strongly normalizing by the induction hypothesis of CR1, and hence $\theta$ also is.
                \item Case $\lfbnt{C} = \lfbnt{C'}  \lfbsym{,}  \gamma$. $ \lfbkw{Red} _{ \lfbnt{C'} } \lfbsym{[} \tilde{\Sigma} \lfbsym{]}   \lfbsym{(}  \Gamma  \lfbsym{,}  \theta_{{\mathrm{1}}}  \lfbsym{)}$ and $\mathcal{R}  \lfbsym{(}  \Gamma  \lfbsym{,}  \theta_{{\mathrm{2}}}  \lfbsym{)}$  holds for some $\theta_{{\mathrm{1}}}$, and $\theta_{{\mathrm{2}}}$ and $\mathcal{R}$ such that $\theta =  \theta_{{\mathrm{1}}} \lfbsym{,} \theta_{{\mathrm{2}}} $ and $\mathcal{R}  \lfbsym{(}  \Gamma  \lfbsym{,}  \theta_{{\mathrm{2}}}  \lfbsym{)}$. $\theta_{{\mathrm{1}}}$ is strongly normalizing by the induction hypothesis of CR1. $\theta_{{\mathrm{2}}}$ is strongly normalizing because $\mathcal{R}  \lfbsym{(}  \Gamma  \lfbsym{,}  \theta_{{\mathrm{2}}}  \lfbsym{)}$. Therefore $\theta$ is also strongly normalizing.
            \end{itemize}
        \item[(CR2)] We can prove it by simple induction with regard to the definition of parametric reducibility.
        \item[(CR3)] We prove it by induction with regard the structure of $\lfbnt{T}$ and $\lfbnt{C}$.
            \begin{itemize}
                \item Case $\lfbnt{T} = \iota$. $\lfbnt{M}$ is strongly normalizing from the hypothesis of CR3, and hence $ \lfbkw{Red} _{ \lfbnt{T} } \lfbsym{[} \tilde{\Sigma} \lfbsym{]}   \lfbsym{(}  \Gamma  \lfbsym{,}  \lfbnt{M}  \lfbsym{)}$ holds.
                \item Case $\lfbnt{T} = \lfbnt{S_{{\mathrm{1}}}}  \rightarrow  \lfbnt{S_{{\mathrm{2}}}}$. It suffices to show that $ \lfbkw{Red} _{ \lfbnt{S_{{\mathrm{2}}}} } \lfbsym{[} \tilde{\Sigma} \lfbsym{]}   \lfbsym{(}  \Gamma'  \lfbsym{,}  \lfbnt{M} \, \lfbnt{N}  \lfbsym{)}$ for any $\Gamma$ and $\lfbnt{N}$ such that $\Gamma  \leq  \Gamma'$ and $ \lfbkw{Red} _{ \lfbnt{S_{{\mathrm{1}}}} } \lfbsym{[} \tilde{\Sigma} \lfbsym{]}   \lfbsym{(}  \Gamma'  \lfbsym{,}  \lfbnt{N}  \lfbsym{)}$. We prove it as a result of the following sublemma.
                      \begin{enumerate}
                          \item Let $\Gamma$, $\lfbnt{M_{{\mathrm{1}}}}$, $\lfbnt{M_{{\mathrm{2}}}}$ be arbitrary named context and terms such that $\Gamma  \vdash  \lfbnt{M_{{\mathrm{1}}}}  \colon  \lfbnt{S_{{\mathrm{1}}}}  \rightarrow  \lfbnt{S_{{\mathrm{2}}}}$ and $ \lfbkw{Red} _{ \lfbnt{S_{{\mathrm{2}}}} } \lfbsym{[} \tilde{\Sigma} \lfbsym{]}   \lfbsym{(}  \Gamma  \lfbsym{,}  \lfbnt{M_{{\mathrm{2}}}}  \lfbsym{)}$. If $ \lfbkw{Red} _{ \lfbnt{S_{{\mathrm{1}}}}  \rightarrow  \lfbnt{S_{{\mathrm{2}}}} } \lfbsym{[} \tilde{\Sigma} \lfbsym{]}   \lfbsym{(}  \Gamma  \lfbsym{,}  \lfbnt{M'_{{\mathrm{1}}}}  \lfbsym{)}$ holds for any $\lfbnt{M'_{{\mathrm{1}}}}$ such that $\lfbnt{M_{{\mathrm{1}}}}  \rightarrow_{\beta}  \lfbnt{M'_{{\mathrm{1}}}}$, then $ \lfbkw{Red} _{ \lfbnt{S_{{\mathrm{2}}}} } \lfbsym{[} \tilde{\Sigma} \lfbsym{]}   \lfbsym{(}  \Gamma  \lfbsym{,}  \lfbnt{M_{{\mathrm{1}}}} \, \lfbnt{M_{{\mathrm{2}}}}  \lfbsym{)}$ holds.
                      \end{enumerate}
                      We can say that $\lfbnt{N}$ is strongly normalizing by the induction hypothesis of CR1, and hence we can prove this sublemma by induction on reduction steps of $\lfbnt{M_{{\mathrm{2}}}}$.

                      To prove $ \lfbkw{Red} _{ \lfbnt{S_{{\mathrm{2}}}} } \lfbsym{[} \tilde{\Sigma} \lfbsym{]}   \lfbsym{(}  \Gamma  \lfbsym{,}  \lfbnt{M_{{\mathrm{1}}}} \, \lfbnt{M_{{\mathrm{2}}}}  \lfbsym{)}$, it suffices to show that $ \lfbkw{Red} _{ \lfbnt{S_{{\mathrm{2}}}} } \lfbsym{[} \tilde{\Sigma} \lfbsym{]}   \lfbsym{(}  \Gamma  \lfbsym{,}  \lfbnt{N}  \lfbsym{)}$ for any $\lfbnt{N}$ such that $\lfbnt{M_{{\mathrm{1}}}} \, \lfbnt{M_{{\mathrm{2}}}}  \rightarrow_{\beta}  \lfbnt{N}$ according to the induction hypothesis of CR3. There are two subcases for the forms of $\lfbnt{N}$:
                      \begin{itemize}
                          \item Subcase $\lfbnt{N} = \lfbnt{M'_{{\mathrm{1}}}} \, \lfbnt{M_{{\mathrm{2}}}}$ where $\lfbnt{M_{{\mathrm{1}}}}  \rightarrow_{\beta}  \lfbnt{M'_{{\mathrm{1}}}}$.   $ \lfbkw{Red} _{ \lfbnt{S_{{\mathrm{1}}}}  \rightarrow  \lfbnt{S_{{\mathrm{2}}}} } \lfbsym{[} \tilde{\Sigma} \lfbsym{]}   \lfbsym{(}  \Gamma  \lfbsym{,}  \lfbnt{M'_{{\mathrm{1}}}}  \lfbsym{)}$ holds by the hypothesis of the sublemma. Therefore $ \lfbkw{Red} _{ \lfbnt{S_{{\mathrm{2}}}} } \lfbsym{[} \tilde{\Sigma} \lfbsym{]}   \lfbsym{(}  \Gamma  \lfbsym{,}  \lfbnt{M'_{{\mathrm{1}}}} \, \lfbnt{M_{{\mathrm{2}}}}  \lfbsym{)}$ holds because $ \lfbkw{Red} _{ \lfbnt{S_{{\mathrm{1}}}} } \lfbsym{[} \tilde{\Sigma} \lfbsym{]}   \lfbsym{(}  \Gamma  \lfbsym{,}  \lfbnt{M_{{\mathrm{2}}}}  \lfbsym{)}$ holds from the hypothesis of the sublemma.
                          \item Subcase $\lfbnt{N} = \lfbnt{M_{{\mathrm{1}}}} \, \lfbnt{M'_{{\mathrm{2}}}}$ where $\lfbnt{M_{{\mathrm{2}}}}  \rightarrow_{\beta}  \lfbnt{M'_{{\mathrm{2}}}}$.  $ \lfbkw{Red} _{ \lfbnt{S_{{\mathrm{2}}}} } \lfbsym{[} \tilde{\Sigma} \lfbsym{]}   \lfbsym{(}  \Gamma  \lfbsym{,}  \lfbnt{M_{{\mathrm{1}}}} \, \lfbnt{M'_{{\mathrm{2}}}}  \lfbsym{)}$ holds from the induction hypothesis of the sublemma.
                      \end{itemize}
                      Note that $\lfbnt{M_{{\mathrm{1}}}} \, \lfbnt{M_{{\mathrm{2}}}}$ will not be a redex because $\lfbnt{M_{{\mathrm{1}}}}$ is neutral. Also, we only need to care the first case for the base case because $\lfbnt{M_{{\mathrm{2}}}}$ is normal form.

                \item Case $\lfbnt{T} = \lfbsym{[}  \lfbnt{C}  \vdash  \lfbnt{S}  \lfbsym{]}$.  It suffices to show that $ \lfbkw{Red} _{ \lfbnt{C} } \lfbsym{[} \tilde{\Sigma} \lfbsym{]}   \lfbsym{(}  \Gamma''  \lfbsym{,}   \lfbkw{unq} _{ \lfbnt{k} } \lfbnt{M} [  \theta  ]   \lfbsym{)}$ for any $\Gamma'$, $\Gamma''$, $\lfbnt{k}$, $\theta$ such that $\Gamma  \leq  \Gamma'$, $\lfbnt{k}  \colon  \Gamma'  \lhd  \Gamma''$ and $ \lfbkw{Red} _{ \lfbnt{C} } \lfbsym{[} \tilde{\Sigma} \lfbsym{]}   \lfbsym{(}  \Gamma''  \lfbsym{,}  \theta  \lfbsym{)}$.  We prove it as a direct result of the following sublemma.
                      \begin{enumerate}
                          \item Assume that $\Gamma  \vdash  \lfbnt{M}  \colon  \lfbsym{[}  \lfbnt{C}  \vdash  \lfbnt{S}  \lfbsym{]}$ and $ \lfbkw{Red} _{ \lfbnt{C} } \lfbsym{[} \tilde{\Sigma} \lfbsym{]}   \lfbsym{(}  \Gamma_{{\mathrm{2}}}  \lfbsym{,}  \theta  \lfbsym{)}$ where $\Gamma  \leq  \Gamma_{{\mathrm{1}}}$ and $\lfbnt{k}  \colon  \Gamma_{{\mathrm{1}}}  \lhd  \Gamma_{{\mathrm{2}}}$ for some $\Gamma_{{\mathrm{1}}}$ and $\lfbnt{k}$. If $ \lfbkw{Red} _{ \lfbsym{[}  \lfbnt{C}  \vdash  \lfbnt{S}  \lfbsym{]} } \lfbsym{[} \tilde{\Sigma} \lfbsym{]}   \lfbsym{(}  \Gamma  \lfbsym{,}  \lfbnt{M'}  \lfbsym{)}$ holds for any $\lfbnt{M'}$ such that $\lfbnt{M}  \rightarrow_{\beta}  \lfbnt{M'}$, then $ \lfbkw{Red} _{ \lfbnt{S_{{\mathrm{2}}}} } \lfbsym{[} \tilde{\Sigma} \lfbsym{]}   \lfbsym{(}  \Gamma  \lfbsym{,}   \lfbkw{unq} _{ \lfbnt{k} } \lfbnt{M} [  \theta  ]   \lfbsym{)}$ holds.
                      \end{enumerate}
                      We can say that $\theta$ is strongly normalizing by the induction hypothesis of CR1, and hence we prove this sublemma by induction on reduction steps of $\theta$.

                      To prove $ \lfbkw{Red} _{ \lfbnt{S_{{\mathrm{2}}}} } \lfbsym{[} \tilde{\Sigma} \lfbsym{]}   \lfbsym{(}  \Gamma  \lfbsym{,}   \lfbkw{unq} _{ \lfbnt{k} } \lfbnt{M} [  \theta  ]   \lfbsym{)}$, it suffices to show that $ \lfbkw{Red} _{ \lfbnt{S_{{\mathrm{2}}}} } \lfbsym{[} \tilde{\Sigma} \lfbsym{]}   \lfbsym{(}  \Gamma  \lfbsym{,}  \lfbnt{N}  \lfbsym{)}$ for any $\lfbnt{N}$ such that $ \lfbkw{unq} _{ \lfbnt{k} } \lfbnt{M} [  \theta  ]   \rightarrow_{\beta}  \lfbnt{N}$ according to the induction hypothesis of CR3. There are two subcases for the forms of $\lfbnt{N}$:
                      \begin{itemize}
                          \item Subcase $\lfbnt{N} =  \lfbkw{unq} _{ \lfbnt{k} } \lfbnt{M'} [  \theta  ] $ where $\lfbnt{M}  \rightarrow_{\beta}  \lfbnt{M'}$.  $ \lfbkw{Red} _{ \lfbsym{[}  \lfbnt{C}  \vdash  \lfbnt{S}  \lfbsym{]} } \lfbsym{[} \tilde{\Sigma} \lfbsym{]}   \lfbsym{(}  \Gamma  \lfbsym{,}  \lfbnt{M'}  \lfbsym{)}$ holds by the hypothesis of the sublemma. Therefore $ \lfbkw{Red} _{ \lfbnt{S_{{\mathrm{2}}}} } \lfbsym{[} \tilde{\Sigma} \lfbsym{]}   \lfbsym{(}  \Gamma  \lfbsym{,}   \lfbkw{unq} _{ \lfbnt{k} } \lfbnt{M'} [  \theta  ]   \lfbsym{)}$ holds because $ \lfbkw{Red} _{ \lfbnt{C} } \lfbsym{[} \tilde{\Sigma} \lfbsym{]}   \lfbsym{(}  \Gamma  \lfbsym{,}  \theta  \lfbsym{)}$ holds from the hypothesis of the sublemma.
                          \item Subcase $\lfbnt{N} =  \lfbkw{unq} _{ \lfbnt{k} } \lfbnt{M} [  \theta'  ] $ where $\theta  \rightarrow_{\beta}  \theta'$. $ \lfbkw{Red} _{ \lfbnt{S} } \lfbsym{[} \tilde{\Sigma} \lfbsym{]}   \lfbsym{(}  \Gamma  \lfbsym{,}   \lfbkw{unq} _{ \lfbnt{k} } \lfbnt{M} [  \theta'  ]   \lfbsym{)}$ holds from the induction hypothesis of the sublemma.
                      \end{itemize}
                      Note that $ \lfbkw{unq} _{ \lfbnt{k} } \lfbnt{M} [  \theta'  ] $ will not be a redex because $\lfbnt{M}$ is neutral. Also, we only need to care the first case for the base case because $\theta$ is normal form.

                \item Case $\lfbnt{T} =  \forall \gamma . \lfbnt{S} $.  It suffices to show that $ \lfbkw{Red} _{ \lfbnt{S} } \lfbsym{[} \tilde{\Sigma}  \lfbsym{,}  \gamma  \colon  \lfbnt{C}  \coloneqq  \mathcal{R} \lfbsym{]}   \lfbsym{(}  \Gamma  \lfbsym{,}   \lfbnt{M} @ \lfbnt{C}   \lfbsym{)}$ for any $\lfbnt{C}$ and $\mathcal{R}$.

                      For any $\lfbnt{M'}$ such that $ \lfbnt{M} @ \lfbnt{C}   \rightarrow_{\beta}  \lfbnt{M'}$, $\lfbnt{M'} =  \lfbnt{M''} @ \lfbnt{C} $ where $\lfbnt{M}  \rightarrow_{\beta}  \lfbnt{M''}$ because $\lfbnt{M}$ is neutral. We have $ \lfbkw{Red} _{ \lfbnt{T} } \lfbsym{[} \tilde{\Sigma} \lfbsym{]}   \lfbsym{(}  \Gamma  \lfbsym{,}  \lfbnt{M''}  \lfbsym{)}$ from the hypothesis, and hence $ \lfbkw{Red} _{ \lfbnt{S} } \lfbsym{[} \tilde{\Sigma}  \lfbsym{,}  \gamma  \colon  \lfbnt{C}  \coloneqq  \mathcal{R} \lfbsym{]}   \lfbsym{(}  \Gamma  \lfbsym{,}   \lfbnt{M''} @ \lfbnt{C}   \lfbsym{)}$. By the induction hypothesis of CR3, we finally derive that $ \lfbkw{Red} _{ \lfbnt{S} } \lfbsym{[} \tilde{\Sigma}  \lfbsym{,}  \gamma  \colon  \lfbnt{C}  \coloneqq  \mathcal{R} \lfbsym{]}   \lfbsym{(}  \Gamma  \lfbsym{,}   \lfbnt{M} @ \lfbnt{C}   \lfbsym{)}$.

                \item For contexts, we can reduce the hypothesis to hypotheses for each elements of the contexts.

                \item Case $\lfbnt{C} =  \bullet $.  CR3 holds because $\theta =  \bullet $, which is normal.

                \item Case $\lfbnt{C} = \lfbnt{C'}  \lfbsym{,}  \lfbnt{T}$.  It suffices to show that $ \lfbkw{Red} _{ \lfbnt{C'} } \lfbsym{[} \tilde{\Sigma} \lfbsym{]}   \lfbsym{(}  \Gamma  \lfbsym{,}  \theta'  \lfbsym{)}$ and $ \lfbkw{Red} _{ \lfbnt{T} } \lfbsym{[} \tilde{\Sigma} \lfbsym{]}   \lfbsym{(}  \Gamma  \lfbsym{,}  \lfbnt{M}  \lfbsym{)}$ where $\theta = \theta'  \lfbsym{,}  \lfbnt{M}$.

                      For any $\lfbnt{M'}$ such that $\lfbnt{M}  \rightarrow_{\beta}  \lfbnt{M'}$, $ \lfbkw{Red} _{ \lfbnt{T} } \lfbsym{[} \tilde{\Sigma} \lfbsym{]}   \lfbsym{(}  \Gamma  \lfbsym{,}  \lfbnt{M'}  \lfbsym{)}$ holds by the hypothesis. Therefore $ \lfbkw{Red} _{ \lfbnt{T} } \lfbsym{[} \tilde{\Sigma} \lfbsym{]}   \lfbsym{(}  \Gamma  \lfbsym{,}  \lfbnt{M}  \lfbsym{)}$ holds by the induction hypothesis of CR3. We can also show that $ \lfbkw{Red} _{ \lfbnt{C'} } \lfbsym{[} \tilde{\Sigma} \lfbsym{]}   \lfbsym{(}  \Gamma  \lfbsym{,}  \lfbnt{M'}  \lfbsym{)}$ in the similar way.

                \item Case $\lfbnt{C} = \lfbnt{C'}  \lfbsym{,}  \gamma$. We can show CR3 in the similar way as the case above.
            \end{itemize}
    \end{description}
\end{proof}

\subsection{Proof of Lemma 7}
\begin{proof}
    We prove this by induction on the structure of $\lfbnt{T}$ and $\lfbnt{D}$.

    \begin{itemize}
        \item Case $\lfbnt{T} = \iota$.
              \begin{align*}
                   \lfbkw{Red} _{ \lfbnt{T}  \lfbsym{[}   \gamma \coloneqq \lfbnt{C}   \lfbsym{]} } \lfbsym{[} \tilde{\Sigma} \lfbsym{]}   \lfbsym{(}  \Gamma  \lfbsym{,}  \lfbnt{M}  \lfbsym{)} & \Leftrightarrow  \lfbkw{Red} _{ \iota } \lfbsym{[} \tilde{\Sigma} \lfbsym{]}   \lfbsym{(}  \Gamma  \lfbsym{,}  \lfbnt{M}  \lfbsym{)}                               \\                   \\
                                                         & \Leftrightarrow  \lfbkw{Red} _{ \iota } \lfbsym{[} \tilde{\Sigma}  \lfbsym{,}  \gamma  \colon  \lfbnt{C}  \lfbsym{[}  \Sigma  \lfbsym{]}  \coloneqq   \lfbkw{Red} _{ \lfbnt{C} } \lfbsym{[} \tilde{\Sigma} \lfbsym{]}  \lfbsym{]}   \lfbsym{(}  \Gamma  \lfbsym{,}  \lfbnt{M}  \lfbsym{)}
              \end{align*}
            \item Case $\lfbnt{T} = \lfbnt{S_{{\mathrm{1}}}}  \rightarrow  \lfbnt{S_{{\mathrm{2}}}}$.
              \begin{align*}
                                  &  \lfbkw{Red} _{ \lfbnt{T}  \lfbsym{[}   \gamma \coloneqq \lfbnt{C}   \lfbsym{]} } \lfbsym{[} \tilde{\Sigma} \lfbsym{]}   \lfbsym{(}  \Gamma  \lfbsym{,}  \lfbnt{M}  \lfbsym{)}                                                                                  \\
                  \Leftrightarrow &  \lfbkw{Red} _{ \lfbnt{S_{{\mathrm{1}}}}  \lfbsym{[}   \gamma \coloneqq \lfbnt{C}   \lfbsym{]}  \rightarrow  \lfbnt{S_{{\mathrm{2}}}}  \lfbsym{[}   \gamma \coloneqq \lfbnt{C}   \lfbsym{]} } \lfbsym{[} \tilde{\Sigma} \lfbsym{]}   \lfbsym{(}  \Gamma  \lfbsym{,}  \lfbnt{M}  \lfbsym{)}                                                             \\
                  \Leftrightarrow &  \lfbkw{Red} _{ \lfbnt{S_{{\mathrm{2}}}}  \lfbsym{[}   \gamma \coloneqq \lfbnt{C}   \lfbsym{]} } \lfbsym{[} \tilde{\Sigma} \lfbsym{]}   \lfbsym{(}  \Gamma'  \lfbsym{,}  \lfbnt{M} \, \lfbnt{N}  \lfbsym{)}                                                                              \\
                                  & \text{ for any $\Gamma'$ and $\lfbnt{N}$ s.t. $\Gamma  \leq  \Gamma'$ and  $ \lfbkw{Red} _{ \lfbnt{S_{{\mathrm{1}}}}  \lfbsym{[}   \gamma \coloneqq \lfbnt{C}   \lfbsym{]} } \lfbsym{[} \tilde{\Sigma} \lfbsym{]}   \lfbsym{(}  \Gamma'  \lfbsym{,}  \lfbnt{N}  \lfbsym{)}$}                \\
                  \Leftrightarrow &  \lfbkw{Red} _{ \lfbnt{S_{{\mathrm{2}}}} } \lfbsym{[} \tilde{\Sigma}  \lfbsym{,}  \gamma  \colon  \lfbnt{C}  \lfbsym{[}  \Sigma  \lfbsym{]}  \coloneqq   \lfbkw{Red} _{ \lfbnt{C} } \lfbsym{[} \tilde{\Sigma} \lfbsym{]}  \lfbsym{]}   \lfbsym{(}  \Gamma'  \lfbsym{,}  \lfbnt{M} \, \lfbnt{N}  \lfbsym{)}                                                              \\
                                  & \text{ for any $\Gamma'$ and $\lfbnt{N}$ s.t. $\Gamma  \leq  \Gamma'$ and  $ \lfbkw{Red} _{ \lfbnt{S_{{\mathrm{1}}}} } \lfbsym{[} \tilde{\Sigma}  \lfbsym{,}  \gamma  \colon  \lfbnt{C}  \lfbsym{[}  \Sigma  \lfbsym{]}  \coloneqq   \lfbkw{Red} _{ \lfbnt{C} } \lfbsym{[} \tilde{\Sigma} \lfbsym{]}  \lfbsym{]}   \lfbsym{(}  \Gamma'  \lfbsym{,}  \lfbnt{N}  \lfbsym{)}$} \\
                                  & \text{ (by the induction hypothesis) }                                                                                  \\
                  \Leftrightarrow &  \lfbkw{Red} _{ \lfbnt{T} } \lfbsym{[} \tilde{\Sigma}  \lfbsym{,}  \gamma  \colon  \lfbnt{C}  \lfbsym{[}  \Sigma  \lfbsym{]}  \coloneqq   \lfbkw{Red} _{ \lfbnt{C} } \lfbsym{[} \tilde{\Sigma} \lfbsym{]}  \lfbsym{]}   \lfbsym{(}  \Gamma  \lfbsym{,}  \lfbnt{M}  \lfbsym{)}                                                                  \\
              \end{align*}
        \item Case $\lfbnt{T} = \lfbsym{[}  \lfbnt{D}  \vdash  \lfbnt{S}  \lfbsym{]}$.
              \begin{align*}
                                  &  \lfbkw{Red} _{ \lfbnt{T}  \lfbsym{[}   \gamma \coloneqq \lfbnt{C}   \lfbsym{]} } \lfbsym{[} \tilde{\Sigma} \lfbsym{]}   \lfbsym{(}  \Gamma  \lfbsym{,}  \lfbnt{M}  \lfbsym{)}                                                                              \\
                  \Leftrightarrow &  \lfbkw{Red} _{ \lfbsym{[}  \lfbnt{D}  \lfbsym{[}   \gamma \coloneqq \lfbnt{C}   \lfbsym{]}  \vdash  \lfbnt{S}  \lfbsym{[}   \gamma \coloneqq \lfbnt{C}   \lfbsym{]}  \lfbsym{]} } \lfbsym{[} \tilde{\Sigma} \lfbsym{]}   \lfbsym{(}  \Gamma  \lfbsym{,}  \lfbnt{M}  \lfbsym{)}                                                        \\
                  \Leftrightarrow &  \lfbkw{Red} _{ \lfbnt{S}  \lfbsym{[}   \gamma \coloneqq \lfbnt{C}   \lfbsym{]} } \lfbsym{[} \tilde{\Sigma} \lfbsym{]}   \lfbsym{(}  \Gamma''  \lfbsym{,}   \lfbkw{unq} _{ \lfbnt{k} } \lfbnt{M} [  \theta  ]   \lfbsym{)}                                                                    \\
                                  & \text{ for any $\Gamma'$, $\Gamma''$, $\lfbnt{k}$ and $\theta$}                                                           \\
                                  & \text{ s.t. $\Gamma  \leq  \Gamma'$, $\lfbnt{k}  \colon  \Gamma'  \lhd  \Gamma''$ and  $ \lfbkw{Red} _{ \lfbnt{D}  \lfbsym{[}   \gamma \coloneqq \lfbnt{C}   \lfbsym{]} } \lfbsym{[} \tilde{\Sigma} \lfbsym{]}   \lfbsym{(}  \Gamma''  \lfbsym{,}  \theta  \lfbsym{)}$}                  \\
                  \Leftrightarrow &  \lfbkw{Red} _{ \lfbnt{S} } \lfbsym{[} \tilde{\Sigma}  \lfbsym{,}  \gamma  \colon  \lfbnt{C}  \lfbsym{[}  \Sigma  \lfbsym{]}  \coloneqq   \lfbkw{Red} _{ \lfbnt{C} } \lfbsym{[} \tilde{\Sigma} \lfbsym{]}  \lfbsym{]}   \lfbsym{(}  \Gamma''  \lfbsym{,}   \lfbkw{unq} _{ \lfbnt{k} } \lfbnt{M} [  \theta  ]   \lfbsym{)}                                                    \\
                                  & \text{ for any $\Gamma'$, $\Gamma''$, $\lfbnt{k}$ and $\theta$}                                                           \\
                                  & \text{ s.t. $\Gamma  \leq  \Gamma'$, $\lfbnt{k}  \colon  \Gamma'  \lhd  \Gamma''$ and  $ \lfbkw{Red} _{ \lfbnt{D} } \lfbsym{[} \tilde{\Sigma}  \lfbsym{,}  \gamma  \colon  \lfbnt{C}  \lfbsym{[}  \Sigma  \lfbsym{]}  \coloneqq   \lfbkw{Red} _{ \lfbnt{C} } \lfbsym{[} \tilde{\Sigma} \lfbsym{]}  \lfbsym{]}   \lfbsym{(}  \Gamma''  \lfbsym{,}  \theta  \lfbsym{)} $} \\
                                  & \text{ (by the induction hypothesis) }                                                                              \\
                  \Leftrightarrow &  \lfbkw{Red} _{ \lfbnt{T} } \lfbsym{[} \tilde{\Sigma}  \lfbsym{,}  \gamma  \colon  \lfbnt{C}  \lfbsym{[}  \Sigma  \lfbsym{]}  \coloneqq   \lfbkw{Red} _{ \lfbnt{C} } \lfbsym{[} \tilde{\Sigma} \lfbsym{]}  \lfbsym{]}   \lfbsym{(}  \Gamma  \lfbsym{,}  \lfbnt{M}  \lfbsym{)}                                                              \\
              \end{align*}

        \item Case $\lfbnt{T} =  \forall \delta . \lfbnt{S} $.
              \begin{align*}
                                  &  \lfbkw{Red} _{ \lfbnt{T}  \lfbsym{[}   \gamma \coloneqq \lfbnt{C}   \lfbsym{]} } \lfbsym{[} \tilde{\Sigma} \lfbsym{]}   \lfbsym{(}  \Gamma  \lfbsym{,}  \lfbnt{M}  \lfbsym{)}                                          \\
                  \Leftrightarrow &  \lfbkw{Red} _{  \forall \delta . \lfbsym{(}  \lfbnt{S}  \lfbsym{[}   \gamma \coloneqq \lfbnt{C}   \lfbsym{]}  \lfbsym{)}  } \lfbsym{[} \tilde{\Sigma} \lfbsym{]}   \lfbsym{(}  \Gamma  \lfbsym{,}  \lfbnt{M}  \lfbsym{)}                                    \\
                                  & \text{where $\delta \, \not\in \, \mathsf{FCV} \, \lfbsym{(}  \lfbnt{C}  \lfbsym{)}$}                                                 \\
                  \Leftrightarrow &  \lfbkw{Red} _{ \lfbnt{S}  \lfbsym{[}   \gamma \coloneqq \lfbnt{C}   \lfbsym{]} } \lfbsym{[} \tilde{\Sigma}  \lfbsym{,}  \delta  \colon  \lfbnt{D}  \coloneqq  \mathcal{R} \lfbsym{]}   \lfbsym{(}  \Gamma  \lfbsym{,}   \lfbnt{M} @ \lfbnt{D}   \lfbsym{)}                        \\
                                  & \text{ for any $\lfbnt{D}$ and $\mathcal{R}$ }                                           \\
                  \Leftrightarrow &  \lfbkw{Red} _{ \lfbnt{S} } \lfbsym{[} \tilde{\Sigma}  \lfbsym{,}  \delta  \colon  \lfbnt{D}  \coloneqq  \mathcal{R}  \lfbsym{,}  \gamma  \colon  \lfbnt{C}  \coloneqq   \lfbkw{Red} _{ \lfbnt{C} } \lfbsym{[} \tilde{\Sigma}  \lfbsym{,}  \delta  \colon  \lfbnt{D}  \coloneqq  \mathcal{R} \lfbsym{]}  \lfbsym{]}   \lfbsym{(}  \Gamma  \lfbsym{,}   \lfbnt{M} @ \lfbnt{D}   \lfbsym{)} \\
                                  & \text{ for any $\lfbnt{D}$ and $\mathcal{R}$ }                                           \\
                                  & \text{ (by the induction hypothesis) }                                          \\
                  \Leftrightarrow &  \lfbkw{Red} _{ \lfbnt{S} } \lfbsym{[} \tilde{\Sigma}  \lfbsym{,}  \delta  \colon  \lfbnt{D}  \coloneqq  \mathcal{R}  \lfbsym{,}  \gamma  \colon  \lfbnt{C}  \lfbsym{[}  \Sigma  \lfbsym{]}  \coloneqq   \lfbkw{Red} _{ \lfbnt{C} } \lfbsym{[} \tilde{\Sigma} \lfbsym{]}  \lfbsym{]}   \lfbsym{(}  \Gamma  \lfbsym{,}   \lfbnt{M} @ \lfbnt{D}   \lfbsym{)}        \\
                                  & \text{ for any $\lfbnt{D}$ and $\mathcal{R}$ }                                           \\
                                  & \text{ (because $\delta$ does not occur in $\lfbnt{C}$) }                        \\
                  \Leftrightarrow &  \lfbkw{Red} _{ \lfbnt{T} } \lfbsym{[} \tilde{\Sigma}  \lfbsym{,}  \gamma  \colon  \lfbnt{C}  \lfbsym{[}  \Sigma  \lfbsym{]}  \coloneqq   \lfbkw{Red} _{ \lfbnt{C} } \lfbsym{[} \tilde{\Sigma} \lfbsym{]}  \lfbsym{]}   \lfbsym{(}  \Gamma  \lfbsym{,}  \lfbnt{M}  \lfbsym{)}                          \\
              \end{align*}

        \item Case $\lfbnt{D} =  \bullet $. Trivial.

        \item Case $\lfbnt{D} = \lfbnt{D'}  \lfbsym{,}  \gamma$.
              \begin{align*}
                                  &  \lfbkw{Red} _{ \lfbnt{D}  \lfbsym{[}   \gamma \coloneqq \lfbnt{C}   \lfbsym{]} } \lfbsym{[} \tilde{\Sigma} \lfbsym{]}   \lfbsym{(}  \Gamma  \lfbsym{,}  \theta  \lfbsym{)}                                                           \\
                  \Leftrightarrow &  \lfbkw{Red} _{  \lfbnt{D'}  \lfbsym{[}   \gamma \coloneqq \lfbnt{C}   \lfbsym{]} \lfbsym{,} \lfbnt{C}  } \lfbsym{[} \tilde{\Sigma} \lfbsym{]}   \lfbsym{(}  \Gamma  \lfbsym{,}  \theta  \lfbsym{)}                                                     \\
                  \Leftrightarrow &  \lfbkw{Red} _{ \lfbnt{D'}  \lfbsym{[}   \gamma \coloneqq \lfbnt{C}   \lfbsym{]} } \lfbsym{[} \tilde{\Sigma} \lfbsym{]}   \lfbsym{(}  \Gamma  \lfbsym{,}  \theta_{{\mathrm{1}}}  \lfbsym{)} \text{ and }  \lfbkw{Red} _{ \lfbnt{C} } \lfbsym{[} \tilde{\Sigma} \lfbsym{]}   \lfbsym{(}  \Gamma  \lfbsym{,}  \theta_{{\mathrm{2}}}  \lfbsym{)}                 \\
                                  & \text{ where $\theta =  \theta_{{\mathrm{1}}} \lfbsym{,} \theta_{{\mathrm{2}}} $}                                                       \\
                  \Leftrightarrow &  \lfbkw{Red} _{ \lfbnt{D'} } \lfbsym{[} \tilde{\Sigma}  \lfbsym{,}  \gamma  \colon  \lfbnt{C}  \lfbsym{[}  \Sigma  \lfbsym{]}  \coloneqq   \lfbkw{Red} _{ \lfbnt{C} } \lfbsym{[} \tilde{\Sigma} \lfbsym{]}  \lfbsym{]}   \lfbsym{(}  \Gamma  \lfbsym{,}  \theta_{{\mathrm{1}}}  \lfbsym{)} \text{ and }  \lfbkw{Red} _{ \lfbnt{C} } \lfbsym{[} \tilde{\Sigma} \lfbsym{]}   \lfbsym{(}  \Gamma  \lfbsym{,}  \theta_{{\mathrm{2}}}  \lfbsym{)} \\
                                  & \text{ (by the induction hypothesis) }                                                            \\
                  \Leftrightarrow &  \lfbkw{Red} _{ \lfbnt{D} } \lfbsym{[} \tilde{\Sigma}  \lfbsym{,}  \gamma  \colon  \lfbnt{C}  \lfbsym{[}  \Sigma  \lfbsym{]}  \coloneqq   \lfbkw{Red} _{ \lfbnt{C} } \lfbsym{[} \tilde{\Sigma} \lfbsym{]}  \lfbsym{]}   \lfbsym{(}  \Gamma  \lfbsym{,}  \theta  \lfbsym{)}                                           \\
              \end{align*}

        \item Case $\lfbnt{D} = \lfbnt{D'}  \lfbsym{,}  \delta$ where $\gamma \not = \delta$. Easy.

    \end{itemize}
\end{proof}

\subsection{Proof of Lemma 8}
Omitted. It takes the almost same approach as the proof of Lemma 9.

\subsection{Proof of Lemma 9}
\begin{proof}\sloppy
    In order to prove $ \lfbkw{Red} _{ \lfbsym{[}  \lfbnt{C}  \vdash  \lfbnt{T}  \lfbsym{]} } \lfbsym{[} \tilde{\Sigma} \lfbsym{]}   \lfbsym{(}  \Gamma  \lfbsym{,}   \lfbkw{quo} \langle  \overrightarrow{x}  \colon  \lfbnt{C}  \lfbsym{[}  \Sigma  \lfbsym{]}  \rangle \lfbnt{M}   \lfbsym{)}$, it suffices to show that $ \lfbkw{Red} _{ \lfbnt{T} } \lfbsym{[} \tilde{\Sigma} \lfbsym{]}   \lfbsym{(}  \Gamma_{{\mathrm{2}}}  \lfbsym{,}   \lfbkw{unq} _{ \lfbnt{k} } \lfbsym{(}   \lfbkw{quo} \langle  \overrightarrow{x}  \colon  \lfbnt{C}  \lfbsym{[}  \Sigma  \lfbsym{]}  \rangle \lfbnt{M}   \lfbsym{)} [  \theta  ]   \lfbsym{)}$ for any $\Gamma_{{\mathrm{1}}}$, $\Gamma_{{\mathrm{2}}}$, $\lfbnt{k}$ and $\theta$ such that $\Gamma  \leq  \Gamma_{{\mathrm{1}}}$, $\lfbnt{k}  \colon  \Gamma_{{\mathrm{1}}}  \lhd  \Gamma_{{\mathrm{2}}}$ and $ \lfbkw{Red} _{ \lfbnt{C} } \lfbsym{[} \tilde{\Sigma} \lfbsym{]}   \lfbsym{(}  \Gamma_{{\mathrm{2}}}  \lfbsym{,}  \theta  \lfbsym{)}$. We get this subgoal as the direct result of the following sublemma:

    \begin{enumerate}
        \item Assume $\Gamma  \lfbsym{,}  \text{\faLock}  \lfbsym{,}  \overrightarrow{x}  \colon  \lfbnt{C}  \lfbsym{[}  \Sigma  \lfbsym{]}  \vdash  \lfbnt{M}  \colon  \lfbnt{T}  \lfbsym{[}  \Sigma  \lfbsym{]}$, $\Gamma  \leq  \Gamma_{{\mathrm{1}}}$, $\lfbnt{k}  \colon  \Gamma_{{\mathrm{1}}}  \lhd  \Gamma_{{\mathrm{2}}}$ and $ \lfbkw{Red} _{ \lfbnt{C} } \lfbsym{[} \tilde{\Sigma} \lfbsym{]}   \lfbsym{(}  \Gamma_{{\mathrm{2}}}  \lfbsym{,}  \theta  \lfbsym{)}$. Then $ \lfbkw{Red} _{ \lfbnt{T} } \lfbsym{[} \tilde{\Sigma} \lfbsym{]}   \lfbsym{(}  \Gamma_{{\mathrm{2}}}  \lfbsym{,}   \lfbkw{unq} _{ \lfbnt{k} } \lfbsym{(}   \lfbkw{quo} \langle  \overrightarrow{x}  \colon  \lfbnt{C}  \lfbsym{[}  \Sigma  \lfbsym{]}  \rangle \lfbnt{M}   \lfbsym{)} [  \theta  ]   \lfbsym{)}$ holds if $ \lfbkw{Red} _{ \lfbnt{T} } \lfbsym{[} \tilde{\Sigma} \lfbsym{]}   \lfbsym{(}  \Gamma'_{{\mathrm{2}}}  \lfbsym{,}  \lfbnt{M}  \lfbsym{[}   id _{ \Gamma'_{{\mathrm{1}}} }   \lfbsym{,}   \text{\faLock} _{ \lfbnt{k'} }   \lfbsym{,}  \overrightarrow{x}  \coloneqq  \theta'  \lfbsym{]}  \lfbsym{)}$ holds for all $\Gamma'_{{\mathrm{1}}}$, $\Gamma'_{{\mathrm{2}}}$, $\lfbnt{k'}$, $\theta'$ such that $\Gamma  \leq  \Gamma'_{{\mathrm{1}}}$, $\lfbnt{k}  \colon  \Gamma'_{{\mathrm{1}}}  \lhd  \Gamma'_{{\mathrm{2}}}$ and $ \lfbkw{Red} _{ \lfbnt{C} } \lfbsym{[} \tilde{\Sigma} \lfbsym{]}   \lfbsym{(}  \Gamma'_{{\mathrm{2}}}  \lfbsym{,}  \theta'  \lfbsym{)}$.
    \end{enumerate}

    $\lfbnt{M}  \lfbsym{[}   id _{ \Gamma'_{{\mathrm{1}}} }   \lfbsym{,}   \text{\faLock} _{ \lfbnt{k'} }   \lfbsym{,}  \overrightarrow{x}  \coloneqq  \theta'  \lfbsym{]}$ and $\theta$ are strongly normalizing from CR1, and hence $\lfbnt{M}$ is also strongly normalizing. Therefore we can prove this sublemma by induction on reduction steps of $\lfbnt{M}$ and $\theta$.

    It suffices to show that $ \lfbkw{Red} _{ \lfbnt{T} } \lfbsym{[} \tilde{\Sigma} \lfbsym{]}   \lfbsym{(}  \Gamma_{{\mathrm{2}}}  \lfbsym{,}  \lfbnt{N}  \lfbsym{)}$ where $ \lfbkw{unq} _{ \lfbnt{k} } \lfbsym{(}   \lfbkw{quo} \langle  \overrightarrow{x}  \colon  \lfbnt{C}  \lfbsym{[}  \Sigma  \lfbsym{]}  \rangle \lfbnt{M}   \lfbsym{)} [  \theta  ]   \rightarrow_{\beta}  \lfbnt{N}$ by CR3. There are three cases with regard to the reduction steps:
    \begin{enumerate}
        \item Case $\lfbnt{N} = \lfbnt{M}  \lfbsym{[}   id _{ \Gamma_{{\mathrm{1}}} }   \lfbsym{,}   \text{\faLock} _{ \lfbnt{k} }   \lfbsym{,}  \overrightarrow{x}  \coloneqq  \theta  \lfbsym{]}$. $ \lfbkw{Red} _{ \lfbnt{T} } \lfbsym{[} \tilde{\Sigma} \lfbsym{]}   \lfbsym{(}  \Gamma_{{\mathrm{2}}}  \lfbsym{,}  \lfbnt{N}  \lfbsym{)}$ holds from the hypothesis of the sublemma.
        \item Case $\lfbnt{N} =  \lfbkw{unq} _{ \lfbnt{k} } \lfbsym{(}   \lfbkw{quo} \langle  \overrightarrow{x}  \colon  \lfbnt{C}  \lfbsym{[}  \Sigma  \lfbsym{]}  \rangle \lfbnt{M'}   \lfbsym{)} [  \theta  ] $ where $\lfbnt{M}  \rightarrow_{\beta}  \lfbnt{M'}$.  We can confirm that $\lfbnt{M'}$ satisfies all preconditions of the sublemma from CR2. And hence we can apply the induction hypothesis of the sublemma to show $ \lfbkw{Red} _{ \lfbnt{T} } \lfbsym{[} \tilde{\Sigma} \lfbsym{]}   \lfbsym{(}  \Gamma_{{\mathrm{2}}}  \lfbsym{,}  \lfbnt{N}  \lfbsym{)}$.
        \item Case $\lfbnt{N} =  \lfbkw{unq} _{ \lfbnt{k} } \lfbsym{(}   \lfbkw{quo} \langle  \overrightarrow{x}  \colon  \lfbnt{C}  \lfbsym{[}  \Sigma  \lfbsym{]}  \rangle \lfbnt{M}   \lfbsym{)} [  \theta'  ] $ where $\theta  \rightarrow_{\beta}  \theta'$.  Similar to the case above.
    \end{enumerate}
\end{proof}

\subsection{Proof of Lemma 10}
\begin{proof}
    It suffices to show that $ \lfbkw{Red} _{ \lfbnt{T} } \lfbsym{[} \tilde{\Sigma}  \lfbsym{,}  \gamma  \colon  \lfbnt{C}  \coloneqq  \mathcal{R} \lfbsym{]}   \lfbsym{(}  \Gamma  \lfbsym{,}   \lfbsym{(}   \Lambda \gamma . \lfbnt{M}   \lfbsym{)} @ \lfbnt{C}   \lfbsym{)}$ for any $\lfbnt{C}$ and $\mathcal{R}$. We fix $\lfbnt{C}$ and $\mathcal{R}$ to arbitrary one. By CR3, it suffices to show that $ \lfbkw{Red} _{ \lfbnt{T} } \lfbsym{[} \tilde{\Sigma}  \lfbsym{,}  \gamma  \colon  \lfbnt{C}  \coloneqq  \mathcal{R} \lfbsym{]}   \lfbsym{(}  \Gamma  \lfbsym{,}  \lfbnt{N}  \lfbsym{)}$ where $ \lfbsym{(}   \Lambda \gamma . \lfbnt{M}   \lfbsym{)} @ \lfbnt{C}   \rightarrow_{\beta}  \lfbnt{N}$ by CR3 (subgoal).

    From the hypothesis and CR1, $ \lfbnt{M} \lfbsym{[}  \gamma \coloneqq \lfbnt{C}  \lfbsym{;}  \bullet  \lfbsym{]} $ is strongly normalizing and hence $\lfbnt{M}$ also is. We prove the subgoal by induction on the number of reduction steps of $\lfbnt{M}$.

    We have two cases with regard to the reduction steps:
    \begin{enumerate}
        \item Case $\lfbnt{N} =  \lfbnt{M} \lfbsym{[}  \gamma \coloneqq \lfbnt{C}  \lfbsym{;}  \bullet  \lfbsym{]} $.  $ \lfbkw{Red} _{ \lfbnt{T} } \lfbsym{[} \tilde{\Sigma}  \lfbsym{,}  \gamma  \colon  \lfbnt{C}  \coloneqq  \mathcal{R} \lfbsym{]}   \lfbsym{(}  \Gamma  \lfbsym{,}  \lfbnt{N}  \lfbsym{)}$ holds from the hypothesis of this lemma.
        \item Case $\lfbnt{N} =  \lfbsym{(}   \Lambda \gamma . \lfbnt{M'}   \lfbsym{)} @ \lfbnt{C} $ where $\lfbnt{M}  \rightarrow_{\beta}  \lfbnt{M'}$.  $ \lfbkw{Red} _{ \lfbnt{T} } \lfbsym{[} \tilde{\Sigma}  \lfbsym{,}  \gamma  \colon  \lfbnt{C}  \coloneqq  \mathcal{R} \lfbsym{]}   \lfbsym{(}  \Gamma  \lfbsym{,}  \lfbnt{N}  \lfbsym{)}$ holds from CR1 and the induction hypothesis of the subgoal.
    \end{enumerate}
\end{proof}

\subsection{Proof of Lemma~\ref{lem:basic}}
\begin{proof}\sloppy
    By induction on the derivation of $\Gamma  \vdash  \lfbnt{M}  \colon  \lfbnt{T}$. We cover major cases in this proof.

    \begin{itemize}
        \item Case where the derivation ends with quo-intro $\Gamma  \vdash   \lfbkw{quo} \langle \Gamma' \rangle \lfbnt{M'}   \colon  \lfbsym{[}   \mathsf{rg} ( \Gamma' )   \vdash  \lfbnt{T'}  \lfbsym{]}$.  It is derived from $ \Gamma  \lfbsym{,}  \text{\faLock} \lfbsym{,} \Gamma'   \vdash  \lfbnt{M'}  \colon  \lfbnt{T'}$.
              \begin{align*}
                                  &  \lfbkw{Red} _{ \lfbsym{[}   \mathsf{rg} ( \Gamma' )   \vdash  \lfbnt{T'}  \lfbsym{]} } \lfbsym{[} \tilde{\Sigma} \lfbsym{]}   \lfbsym{(}  \Delta  \lfbsym{,}   \lfbsym{(}   \lfbkw{quo} \langle \Gamma' \rangle \lfbnt{M'}   \lfbsym{)} \lfbsym{[} \Sigma \lfbsym{;} \bar{\sigma} \lfbsym{]}   \lfbsym{[}  \sigma'  \lfbsym{]}  \lfbsym{)}                                                               \\
                  \Leftrightarrow &  \lfbkw{Red} _{ \lfbsym{[}   \mathsf{rg} ( \Gamma' )   \vdash  \lfbnt{T'}  \lfbsym{]} } \lfbsym{[} \tilde{\Sigma} \lfbsym{]}   \lfbsym{(}  \Delta  \lfbsym{,}   \lfbkw{quo} \langle  \Gamma'   \lfbsym{[}   \Sigma   \lfbsym{;}   \bar{\sigma}'   \lfbsym{]}  \rangle \lfbsym{(}   \lfbnt{M'} \lfbsym{[} \Sigma \lfbsym{;} \lfbsym{(}   \bar{\sigma}  \lfbsym{,} \, \text{\faLock} \lfbsym{,} \bar{\sigma}'   \lfbsym{)} \lfbsym{]}   \lfbsym{[}   \sigma'  \lfbsym{,}   \text{\faLock} _{ \lfbsym{1} }  \lfbsym{,}  id _{  \Gamma'   \lfbsym{[}   \Sigma   \lfbsym{;}   \bar{\sigma}'   \lfbsym{]}  }    \lfbsym{]}  \lfbsym{)}   \lfbsym{)} \\
                                  & \text{ where $\bar{\sigma}' =  \mathsf{destruct} \lfbsym{(} \Gamma' \lfbsym{;} \Sigma \lfbsym{)} $}                                                                        \\
              \end{align*}

              By Lemma 9, it suffices to show that $ \lfbkw{Red} _{ \lfbnt{T'} } \lfbsym{[} \tilde{\Sigma} \lfbsym{]}   \lfbsym{(}  \Delta_{{\mathrm{2}}}  \lfbsym{,}   \lfbnt{M'} \lfbsym{[} \Sigma \lfbsym{;} \lfbsym{(}   \bar{\sigma}  \lfbsym{,} \, \text{\faLock} \lfbsym{,} \bar{\sigma}'   \lfbsym{)} \lfbsym{]}   \lfbsym{[}   \sigma'  \lfbsym{,}   \text{\faLock} _{ \lfbsym{1} }  \lfbsym{,}  id _{  \Gamma'   \lfbsym{[}   \Sigma   \lfbsym{;}   \bar{\sigma}'   \lfbsym{]}  }    \lfbsym{]}  \lfbsym{[}   id _{ \Delta }   \lfbsym{,}   \text{\faLock} _{ \lfbsym{(}  \lfbnt{k}  \lfbsym{)} }   \lfbsym{,}   \Gamma'   \lfbsym{[}   \Sigma   \lfbsym{;}   \bar{\sigma}'   \lfbsym{]}   \coloneqq  \theta  \lfbsym{]}  \lfbsym{)}$ for any $\Delta_{{\mathrm{1}}}$, $\Delta_{{\mathrm{2}}}$, $\lfbnt{k}$ and $\theta$ such that $\Delta  \leq  \Delta_{{\mathrm{1}}}$, $\lfbnt{k}  \colon  \Delta_{{\mathrm{1}}}  \lhd  \Delta_{{\mathrm{2}}}$ and $ \lfbkw{Red} _{  \mathsf{rg} ( \Gamma' )  } \lfbsym{[} \tilde{\Sigma} \lfbsym{]}   \lfbsym{(}  \Delta_{{\mathrm{2}}}  \lfbsym{,}  \theta  \lfbsym{)}$ (subgoal).

              We fix $\Delta_{{\mathrm{1}}}$, $\Delta_{{\mathrm{2}}}$, $\lfbnt{k}$ and $\theta$ to arbitrary ones. From $ \lfbkw{Red} _{ \Gamma } \lfbsym{[} \tilde{\Sigma} \lfbsym{,} \bar{\sigma} \lfbsym{]}   \lfbsym{(}  \Delta  \lfbsym{,}  \sigma'  \lfbsym{)}$, we obtain $ \lfbkw{Red} _{  \Gamma  \lfbsym{,}  \text{\faLock} \lfbsym{,} \Gamma'  } \lfbsym{[} \tilde{\Sigma} \lfbsym{,} \lfbsym{(}   \bar{\sigma}  \lfbsym{,} \, \text{\faLock} \lfbsym{,} \bar{\sigma}'   \lfbsym{)} \lfbsym{]}   \lfbsym{(}  \Delta_{{\mathrm{2}}}  \lfbsym{,}  \lfbsym{(}  \sigma'  \lfbsym{,}   \text{\faLock} _{ \lfbsym{(}  \lfbnt{k}  \lfbsym{)} }   \lfbsym{,}   \Gamma'   \lfbsym{[}   \Sigma   \lfbsym{;}   \bar{\sigma}'   \lfbsym{]}   \coloneqq  \theta  \lfbsym{)}  \lfbsym{)}$.
              Then we can apply the induction hypothesis to $ \Gamma  \lfbsym{,}  \text{\faLock} \lfbsym{,} \Gamma'   \vdash  \lfbnt{M'}  \colon  \lfbnt{T'}$, and we get $ \lfbkw{Red} _{ \lfbnt{T'} } \lfbsym{[} \tilde{\Sigma} \lfbsym{]}   \lfbsym{(}  \Delta_{{\mathrm{2}}}  \lfbsym{,}   \lfbnt{M'} \lfbsym{[} \Sigma \lfbsym{;} \lfbsym{(}   \bar{\sigma}  \lfbsym{,} \, \text{\faLock} \lfbsym{,} \bar{\sigma}'   \lfbsym{)} \lfbsym{]}   \lfbsym{[}  \sigma'  \lfbsym{,}   \text{\faLock} _{ \lfbsym{(}  \lfbnt{k}  \lfbsym{)} }   \lfbsym{,}   \Gamma'   \lfbsym{[}   \Sigma   \lfbsym{;}   \bar{\sigma}'   \lfbsym{]}   \coloneqq  \theta  \lfbsym{]}  \lfbsym{)}$ (Note that $ \bar{\sigma}  \lfbsym{,} \, \text{\faLock} \lfbsym{,} \bar{\sigma}' $ = $ \mathsf{destruct} \lfbsym{(} \lfbsym{(}   \Gamma  \lfbsym{,}  \text{\faLock} \lfbsym{,} \Gamma'   \lfbsym{)} \lfbsym{;} \Sigma \lfbsym{)} $). This is equal to the subgoal because $ \lfbnt{M'} \lfbsym{[} \Sigma \lfbsym{;} \lfbsym{(}   \bar{\sigma}  \lfbsym{,} \, \text{\faLock} \lfbsym{,} \bar{\sigma}'   \lfbsym{)} \lfbsym{]}   \lfbsym{[}  \sigma'  \lfbsym{,}   \text{\faLock} _{ \lfbsym{(}  \lfbnt{k}  \lfbsym{)} }   \lfbsym{,}   \Gamma'   \lfbsym{[}   \Sigma   \lfbsym{;}   \bar{\sigma}'   \lfbsym{]}   \coloneqq  \theta  \lfbsym{]} =  \lfbnt{M'} \lfbsym{[} \Sigma \lfbsym{;} \lfbsym{(}   \bar{\sigma}  \lfbsym{,} \, \text{\faLock} \lfbsym{,} \bar{\sigma}'   \lfbsym{)} \lfbsym{]}   \lfbsym{[}   \sigma'  \lfbsym{,}   \text{\faLock} _{ \lfbsym{1} }  \lfbsym{,}  id _{  \Gamma'   \lfbsym{[}   \Sigma   \lfbsym{;}   \bar{\sigma}'   \lfbsym{]}  }    \lfbsym{]}  \lfbsym{[}   id _{ \Delta }   \lfbsym{,}   \text{\faLock} _{ \lfbsym{(}  \lfbnt{k}  \lfbsym{)} }   \lfbsym{,}   \Gamma'   \lfbsym{[}   \Sigma   \lfbsym{;}   \bar{\sigma}'   \lfbsym{]}   \coloneqq  \theta  \lfbsym{]}$.

        \item Case where the derivation ends with $\forall$-intro $\Gamma  \vdash   \Lambda \gamma . \lfbnt{M'}   \colon   \forall \gamma . \lfbnt{T'} $. It is derived from $\Gamma  \vdash  \lfbnt{M'}  \colon  \lfbnt{T'}$. We rename $\gamma$ to sufficiently fresh one that is $\gamma \, \not\in \, \mathsf{FCV} \, \lfbsym{(}  \Gamma  \lfbsym{)} \, \cup \, \mathsf{FCV} \, \lfbsym{(}  \Sigma  \lfbsym{)} \, \cup \, \mathsf{dom} \, \lfbsym{(}  \Sigma  \lfbsym{)}$.
              Then we want to show $ \lfbkw{Red} _{  \forall \gamma . \lfbnt{T'}  } \lfbsym{[} \tilde{\Sigma} \lfbsym{]}   \lfbsym{(}  \Delta  \lfbsym{,}   \lfbsym{(}   \Lambda \gamma . \lfbnt{M'}   \lfbsym{)} \lfbsym{[} \Sigma \lfbsym{;} \bar{\sigma} \lfbsym{]}   \lfbsym{[}  \sigma'  \lfbsym{]}  \lfbsym{)} \Leftrightarrow  \lfbkw{Red} _{  \forall \gamma . \lfbnt{T'}  } \lfbsym{[} \tilde{\Sigma} \lfbsym{]}   \lfbsym{(}  \Delta  \lfbsym{,}   \Lambda \gamma . \lfbsym{(}   \lfbnt{M'} \lfbsym{[} \Sigma \lfbsym{;} \bar{\sigma} \lfbsym{]}   \lfbsym{[}  \sigma'  \lfbsym{]}  \lfbsym{)}   \lfbsym{)}$. By Lemma 10, it suffices to show that $ \lfbkw{Red} _{ \lfbnt{T'} } \lfbsym{[} \tilde{\Sigma}  \lfbsym{,}  \gamma  \colon  \lfbnt{C}  \coloneqq  \mathcal{R} \lfbsym{]}   \lfbsym{(}  \Delta  \lfbsym{,}    \lfbnt{M'} \lfbsym{[} \Sigma \lfbsym{;} \bar{\sigma} \lfbsym{]}   \lfbsym{[}  \sigma'  \lfbsym{]} \lfbsym{[}  \gamma \coloneqq \lfbnt{C}  \lfbsym{;}  \bullet  \lfbsym{]}   \lfbsym{)}$ for any $\lfbnt{C}$ and $\mathcal{R}$.

              We fix  $\lfbnt{C}$ and $\mathcal{R}$ to arbitrary one.
              We have $ \lfbkw{Red} _{ \lfbnt{T'} } \lfbsym{[} \tilde{\Sigma}  \lfbsym{,}  \gamma  \colon  \lfbnt{C}  \coloneqq  \mathcal{R} \lfbsym{]}   \lfbsym{(}  \Delta  \lfbsym{,}    \lfbnt{M'} \lfbsym{[} \Sigma \lfbsym{;} \bar{\sigma} \lfbsym{]}   \lfbsym{[}  \sigma'  \lfbsym{]} \lfbsym{[}  \gamma \coloneqq \lfbnt{C}  \lfbsym{;}  \bullet  \lfbsym{]}   \lfbsym{)} \Leftrightarrow  \lfbkw{Red} _{ \lfbnt{T'} } \lfbsym{[} \tilde{\Sigma}  \lfbsym{,}  \gamma  \colon  \lfbnt{C}  \coloneqq  \mathcal{R} \lfbsym{]}   \lfbsym{(}  \Delta  \lfbsym{,}   \lfbnt{M'} \lfbsym{[} \lfbsym{(}  \Sigma  \lfbsym{,}  \gamma  \coloneqq  \lfbnt{C}  \lfbsym{)} \lfbsym{;} \bar{\sigma} \lfbsym{]}   \lfbsym{[}  \sigma'  \lfbsym{]}  \lfbsym{)}$ by the freshness of $\gamma$. Also  $ \lfbkw{Red} _{ \Gamma } \lfbsym{[} \lfbsym{(}  \tilde{\Sigma}  \lfbsym{,}  \gamma  \colon  \lfbnt{C}  \coloneqq  \mathcal{R}  \lfbsym{)} \lfbsym{,} \bar{\sigma} \lfbsym{]}   \lfbsym{(}  \Delta  \lfbsym{,}  \sigma'  \lfbsym{)}$ and $\bar{\sigma} =  \mathsf{destruct} \lfbsym{(} \Gamma \lfbsym{;} \lfbsym{(}  \Sigma  \lfbsym{,}  \gamma  \coloneqq  \lfbnt{C}  \lfbsym{)} \lfbsym{)} $ holds from the freshness of $\gamma$. Therefore we can apply the induction hypothesis to obtain $ \lfbkw{Red} _{ \lfbnt{T'} } \lfbsym{[} \tilde{\Sigma}  \lfbsym{,}  \gamma  \colon  \lfbnt{C}  \coloneqq  \mathcal{R} \lfbsym{]}   \lfbsym{(}  \Delta  \lfbsym{,}   \lfbnt{M'} \lfbsym{[} \lfbsym{(}  \Sigma  \lfbsym{,}  \gamma  \coloneqq  \lfbnt{C}  \lfbsym{)} \lfbsym{;} \bar{\sigma} \lfbsym{]}   \lfbsym{[}  \sigma'  \lfbsym{]}  \lfbsym{)}$.

            \item Case where the derivation ends with $\forall$-elim $\Gamma  \vdash   \lfbnt{M'} @ \lfbnt{C}   \colon  \lfbnt{T'}  \lfbsym{[}   \gamma \coloneqq \lfbnt{C}   \lfbsym{]}$. It is derived from $\Gamma  \vdash  \lfbnt{M'}  \colon   \forall \gamma . \lfbnt{T'} $. We rename $\gamma$ to sufficiently fresh one that is $\gamma \, \not\in \, \mathsf{FCV} \, \lfbsym{(}  \Gamma  \lfbsym{)} \, \cup \, \mathsf{FCV} \, \lfbsym{(}  \Sigma  \lfbsym{)} \, \cup \, \mathsf{dom} \, \lfbsym{(}  \Sigma  \lfbsym{)}$.
              \begin{align*}
                                  &  \lfbkw{Red} _{ \lfbnt{T'}  \lfbsym{[}   \gamma \coloneqq \lfbnt{C}   \lfbsym{]} } \lfbsym{[} \tilde{\Sigma} \lfbsym{]}   \lfbsym{(}  \Delta  \lfbsym{,}   \lfbsym{(}   \lfbnt{M'} @ \lfbnt{C}   \lfbsym{)} \lfbsym{[} \Sigma \lfbsym{;} \bar{\sigma} \lfbsym{]}   \lfbsym{[}  \sigma'  \lfbsym{]}  \lfbsym{)}                       \\
                  \Leftrightarrow &  \lfbkw{Red} _{ \lfbnt{T'}  \lfbsym{[}   \gamma \coloneqq \lfbnt{C}   \lfbsym{]} } \lfbsym{[} \tilde{\Sigma} \lfbsym{]}   \lfbsym{(}  \Delta  \lfbsym{,}  \lfbsym{(}    \lfbnt{M'} \lfbsym{[} \Sigma \lfbsym{;} \bar{\sigma} \lfbsym{]}   \lfbsym{[}  \sigma'  \lfbsym{]} @ \lfbnt{C}  \lfbsym{[}  \Sigma  \lfbsym{]}   \lfbsym{)}  \lfbsym{)}                 \\
                  \Leftrightarrow &  \lfbkw{Red} _{ \lfbnt{T'} } \lfbsym{[} \tilde{\Sigma}  \lfbsym{,}  \gamma  \colon  \lfbnt{C}  \lfbsym{[}  \Sigma  \lfbsym{]}  \coloneqq   \lfbkw{Red} _{ \lfbnt{C} } \lfbsym{[} \tilde{\Sigma} \lfbsym{]}  \lfbsym{]}   \lfbsym{(}  \Delta  \lfbsym{,}  \lfbsym{(}    \lfbnt{M'} \lfbsym{[} \Sigma \lfbsym{;} \bar{\sigma} \lfbsym{]}   \lfbsym{[}  \sigma'  \lfbsym{]} @ \lfbnt{C}  \lfbsym{[}  \Sigma  \lfbsym{]}   \lfbsym{)}  \lfbsym{)} \\
                                  & \text{ by Lemma 7}                                                \\
              \end{align*}
              We can derive the last statement from $ \lfbkw{Red} _{  \forall \gamma . \lfbnt{T'}  } \lfbsym{[} \tilde{\Sigma} \lfbsym{]}   \lfbsym{(}  \Delta  \lfbsym{,}  \lfbsym{(}   \lfbnt{M'} \lfbsym{[} \Sigma \lfbsym{;} \bar{\sigma} \lfbsym{]}   \lfbsym{[}  \sigma'  \lfbsym{]}  \lfbsym{)}  \lfbsym{)}$, and it holds by the induction hypothesis.

        \item Case where the derivation ends with $\Gamma  \vdash  \theta'  \lfbsym{,}  \mathbb{x}  \colon  \lfbnt{C_{{\mathrm{1}}}}  \lfbsym{,}  \gamma$. It is derived from $\Gamma  \vdash  \theta'  \colon  \lfbnt{C'}$ and $\mathbb{x}  \colon  \gamma \, \in \, \mathsf{head} \, \lfbsym{(}  \Gamma  \lfbsym{)}$.
              It suffices to show that $ \lfbkw{Red} _{ \lfbnt{C'} } \lfbsym{[} \tilde{\Sigma} \lfbsym{]}   \lfbsym{(}  \Delta  \lfbsym{,}   \theta' \lfbsym{[} \Sigma \lfbsym{;} \bar{\sigma} \lfbsym{]}   \lfbsym{[}  \sigma'  \lfbsym{]}  \lfbsym{)}$ and $\mathcal{R}  \lfbsym{(}  \Delta  \lfbsym{,}    \mathbb{x}  \lfbsym{[} \Sigma \lfbsym{;} \bar{\sigma} \lfbsym{]}   \lfbsym{[}  \sigma'  \lfbsym{]}  \lfbsym{)}$ where $\gamma  \colon  \lfbnt{D}  \coloneqq  \mathcal{R} \, \in \, \tilde{\Sigma}$. We have the former by applying the induction hypothesis to $\Gamma  \vdash  \theta'  \colon  \lfbnt{C'}$. From $\mathbb{x}  \colon  \gamma \, \in \, \mathsf{head} \, \lfbsym{(}  \Gamma  \lfbsym{)}$, we can reconstruct $ \lfbkw{Red} _{  \mathbb{x}  \colon  \gamma  } \lfbsym{[} \tilde{\Sigma} \lfbsym{,} \bar{\sigma} \lfbsym{]}   \lfbsym{(}  \Delta  \lfbsym{,}   \overrightarrow{x}  \coloneqq   \overrightarrow{x}   \lfbsym{[}  \sigma'  \lfbsym{]}   \lfbsym{)}$ from $ \lfbkw{Red} _{ \Gamma } \lfbsym{[} \tilde{\Sigma} \lfbsym{,} \bar{\sigma} \lfbsym{]}   \lfbsym{(}  \Delta  \lfbsym{,}  \sigma'  \lfbsym{)}$ where $\overrightarrow{x} =   \mathbb{x}  \lfbsym{[} \Sigma \lfbsym{;} \bar{\sigma} \lfbsym{]} $. We finally get $\mathcal{R}  \lfbsym{(}  \Delta  \lfbsym{,}   \overrightarrow{x}   \lfbsym{[}  \sigma'  \lfbsym{]}  \lfbsym{)}$ from Definition 5.
    \end{itemize}
\end{proof}

\subsection{Proof of Theorem~\ref{thm:sn}}
\begin{proof}
    As written in the paper, we prove this theorem as a corollary of Lemma 11 by choosing $\sigma$, $\bar{\sigma}$ and $\sigma'$ to be identity subsections.
\end{proof}

\subsection{Proof of Theorem~\ref{thm:embedsound}}
\begin{proof}
    By induction on the derivations of $M^{0}$ and $M^{1}$.

    \begin{itemize}\sloppy
        \item Case $M^{0}$ = $\lfbmv{x}$:  $  \Gamma^{\circ}_{{\mathrm{1}}}  \lfbsym{,}   \lfbmv{x}  :^0  T^{0}  \lfbsym{,} \Gamma^{\circ}_{{\mathrm{2}}}  \vdash_0 \lfbmv{x} \colon T^{0} $ holds for some $\Gamma^{\circ}_{{\mathrm{1}}}$,  $\Gamma^{\circ}_{{\mathrm{2}}}$ such that $\Gamma^{\circ} =  \Gamma^{\circ}_{{\mathrm{1}}}  \lfbsym{,}   \lfbmv{x}  :^0  T^{0}  \lfbsym{,} \Gamma^{\circ}_{{\mathrm{2}}} $. From $\Gamma^{\circ}  \leadsto  \tilde{\Gamma}$, we have $\tilde{\Gamma}_{{\mathrm{1}}}$ and $\tilde{\Gamma}_{{\mathrm{2}}}$ such that $ \tilde{\Gamma} =  \tilde{\Gamma}_{{\mathrm{1}}}  \lfbsym{,}   \lfbmv{x}  :^0  \forall \gamma .  \llbracket  T^{0}  \rrbracket_{  \mathsf{rg} (  \lvert  \tilde{\Gamma}_{{\mathrm{1}}}  \rvert_1  )   \lfbsym{,}  \gamma }    , \tilde{\Gamma}_{{\mathrm{2}}} $; hence we can derive $ \lvert  \tilde{\Gamma}  \rvert_0   \vdash  \lfbmv{x}  \colon   \forall \gamma .  \llbracket  T^{0}  \rrbracket_{  \mathsf{rg} (  \lvert  \tilde{\Gamma}_{{\mathrm{1}}}  \rvert_1  )   \lfbsym{,}  \gamma }  $ and then $ \lvert  \tilde{\Gamma}  \rvert_0   \vdash   \lfbmv{x} @ \mathsf{diff} \, \lfbsym{(}  \lfbmv{x}  \lfbsym{,}  \tilde{\Gamma}  \lfbsym{)}   \colon   \llbracket  T^{0}  \rrbracket_{  \mathsf{rg} (  \lvert  \tilde{\Gamma}_{{\mathrm{1}}}  \rvert_1  )   \lfbsym{,}  \gamma }   \lfbsym{[}   \gamma \coloneqq \mathsf{diff} \, \lfbsym{(}  \lfbmv{x}  \lfbsym{,}  \tilde{\Gamma}  \lfbsym{)}   \lfbsym{]}$.
        Applying $\mathsf{diff} \, \lfbsym{(}  \lfbmv{x}  \lfbsym{,}  \tilde{\Gamma}  \lfbsym{)}=  \mathsf{rg} (  \lvert  \tilde{\Gamma}_{{\mathrm{2}}}  \rvert_1  ) $, we have $ \llbracket  T^{0}  \rrbracket_{  \mathsf{rg} (  \lvert  \tilde{\Gamma}_{{\mathrm{1}}}  \rvert_1  )   \lfbsym{,}  \gamma }   \lfbsym{[}   \gamma \coloneqq  \mathsf{rg} (  \lvert  \tilde{\Gamma}_{{\mathrm{2}}}  \rvert_1  )    \lfbsym{]} =  \llbracket  T^{0}  \rrbracket_{   \mathsf{rg} (  \lvert  \tilde{\Gamma}_{{\mathrm{1}}}  \rvert_1  )  \lfbsym{,}  \mathsf{rg} (  \lvert  \tilde{\Gamma}_{{\mathrm{2}}}  \rvert_1  )   }  =  \llbracket  T^{0}  \rrbracket_{  \mathsf{rg} (  \lvert  \tilde{\Gamma}  \rvert_1  )  } $. Thus, we confirm that $ \lvert  \tilde{\Gamma}  \rvert_0   \vdash   \llbracket  \lfbmv{x}  \rrbracket_{ \tilde{\Gamma} }   \colon   \llbracket  T^{0}  \rrbracket_{  \mathsf{rg} (  \lvert  \tilde{\Gamma}  \rvert_1  )  } $

        \item Case $M^{0}$ = $ \lambda \lfbmv{x} ^{ T^{0}_{{\mathrm{1}}} }. N^{0} $: $ \Gamma^{\circ}  \lfbsym{,}   \lfbmv{x}  :^0  T^{0}_{{\mathrm{1}}}  \vdash_0 N^{0} \colon T^{0}_{{\mathrm{2}}} $ holds for some $T^{0}_{{\mathrm{1}}}, T^{0}_{{\mathrm{2}}}$ such that $T^{0} = T^{0}_{{\mathrm{1}}}  \rightarrow  T^{0}_{{\mathrm{2}}}$. From $\Gamma^{\circ}  \leadsto  \tilde{\Gamma}$, we can derive $\Gamma^{\circ}  \lfbsym{,}   \lfbmv{x}  :^0  T^{0}_{{\mathrm{1}}}   \leadsto  \tilde{\Gamma}'$ where $\tilde{\Gamma}' = \tilde{\Gamma}  \lfbsym{,}   \lfbmv{x}  :^0  \forall \gamma .  \llbracket  T^{0}_{{\mathrm{1}}}  \rrbracket_{  \mathsf{rg} (  \lvert  \tilde{\Gamma}  \rvert_1  )   \lfbsym{,}  \gamma }   $. As a result, we have $ \lvert  \tilde{\Gamma}'  \rvert_0   \vdash   \llbracket  N^{0}  \rrbracket_{ \tilde{\Gamma}' }   \colon   \llbracket  T^{0}_{{\mathrm{2}}}  \rrbracket_{  \mathsf{rg} (  \lvert  \tilde{\Gamma}  \rvert_1  )  } $ from the induction hypothesis. As $ \lvert  \tilde{\Gamma}'  \rvert_0  =  \lvert  \tilde{\Gamma}  \rvert_0   \lfbsym{,}  \lfbmv{x}  \colon   \forall \gamma .  \llbracket  T^{0}_{{\mathrm{1}}}  \rrbracket_{  \mathsf{rg} (  \lvert  \tilde{\Gamma}  \rvert_1  )   \lfbsym{,}  \gamma }  $, we can derive $ \lvert  \tilde{\Gamma}  \rvert_0   \vdash   \lambda \lfbmv{x} ^{  \forall \gamma .  \llbracket  T^{0}_{{\mathrm{1}}}  \rrbracket_{  \mathsf{rg} (  \lvert  \tilde{\Gamma}  \rvert_1  )   \lfbsym{,}  \gamma }   }.  \llbracket  N^{0}  \rrbracket_{ \tilde{\Gamma}' }    \colon   \llbracket  T^{0}_{{\mathrm{2}}}  \rrbracket_{  \mathsf{rg} (  \lvert  \tilde{\Gamma}  \rvert_1  )  } $.

        \item Case $M^{0} = M^{0}_{{\mathrm{1}}} \, M^{0}_{{\mathrm{2}}}$: $ \Gamma^{\circ} \vdash_0 M^{0}_{{\mathrm{1}}} \colon S^{0}  \rightarrow  T^{0} $ and $ \Gamma^{\circ} \vdash_0 M^{0}_{{\mathrm{2}}} \colon S^{0} $ for some $S^{0}$. By the induction hypothesis, we have the two \lamfb judgments below.
        \begin{itemize}
            \item $ \lvert  \tilde{\Gamma}  \rvert_0   \vdash   \llbracket  M^{0}_{{\mathrm{1}}}  \rrbracket_{ \tilde{\Gamma} }   \colon  \lfbsym{(}   \forall \gamma .  \llbracket  S^{0}  \rrbracket_{  \mathsf{rg} (  \lvert  \tilde{\Gamma}  \rvert_1  )   \lfbsym{,}  \gamma }    \lfbsym{)}  \rightarrow   \llbracket  T^{0}  \rrbracket_{  \mathsf{rg} (  \lvert  \tilde{\Gamma}  \rvert_1  )  } $
            \item $ \lvert  \tilde{\Gamma}  \lfbsym{,}   \mathbb{x}  :^1 \gamma   \rvert_0   \vdash   \llbracket  M^{0}_{{\mathrm{2}}}  \rrbracket_{ \tilde{\Gamma}  \lfbsym{,}   \mathbb{x}  :^1 \gamma  }   \colon   \llbracket  S^{0}  \rrbracket_{  \mathsf{rg} (  \lvert  \tilde{\Gamma}  \lfbsym{,}   \mathbb{x}  :^1 \gamma   \rvert_1  )  } $
        \end{itemize}
        The second judgment holds because $\Gamma^{\circ}  \leadsto  \tilde{\Gamma}  \lfbsym{,}   \mathbb{x}  :^1 \gamma $ can be derived from $\Gamma^{\circ}  \leadsto  \tilde{\Gamma}$. We can derive $ \lvert  \tilde{\Gamma}  \rvert_0   \vdash   \Lambda \gamma .  \llbracket  M^{0}_{{\mathrm{2}}}  \rrbracket_{ \tilde{\Gamma}  \lfbsym{,}   \mathbb{x}  :^1 \gamma  }    \colon   \forall \gamma .  \llbracket  S^{0}  \rrbracket_{  \mathsf{rg} (  \lvert  \tilde{\Gamma}  \lfbsym{,}   \mathbb{x}  :^1 \gamma   \rvert_1  )  }  $ from the second judgment considering that $ \lvert  \tilde{\Gamma}  \lfbsym{,}   \mathbb{x}  :^1 \gamma   \rvert_0  =  \lvert  \tilde{\Gamma}  \rvert_0 $. Then we can apply this judgment to the first judgment, and we obtain $ \lvert  \tilde{\Gamma}  \rvert_0   \vdash   \llbracket  M^{0}_{{\mathrm{1}}}  \rrbracket_{ \tilde{\Gamma} }  \, \lfbsym{(}   \Lambda \gamma .  \llbracket  M^{0}_{{\mathrm{2}}}  \rrbracket_{ \tilde{\Gamma}  \lfbsym{,}   \mathbb{x}  :^1 \gamma  }    \lfbsym{)}  \colon   \llbracket  T^{0}  \rrbracket_{  \mathsf{rg} (  \lvert  \tilde{\Gamma}  \rvert_1  )  } $. $\qed$

        \item Case $M^{0} =  \lfbkw{quo}   N^{1} $: We have $ \Gamma^{\circ} \vdash_1 N^{1} \colon S^{1} $ for some $S^{1}$ such that $T^{0} =  \bigcirc  S^{1} $. From the induction hypothesis, there is a embedded judgment $  \lvert  \tilde{\Gamma}  \rvert_0   \lfbsym{,}  \text{\faLock} \lfbsym{,}  \lvert  \tilde{\Gamma}  \rvert_1    \vdash   \llbracket  N^{1}  \rrbracket_{ \tilde{\Gamma} }   \colon   \llbracket  S^{1}  \rrbracket $. By applying a derivation rule for quotation, we obtain $ \lvert  \tilde{\Gamma}  \rvert_0   \vdash   \lfbkw{quo} \langle  \lvert  \tilde{\Gamma}  \rvert_1  \rangle  \llbracket  N^{1}  \rrbracket_{ \tilde{\Gamma} }    \colon  \lfbsym{[}   \mathsf{rg} (  \lvert  \tilde{\Gamma}  \rvert_1  )   \vdash   \llbracket  S^{1}  \rrbracket   \lfbsym{]}$

        \item Case $M^{1} = \lfbmv{x}$: $ \Gamma^{\circ} \vdash_1 \lfbmv{x} \colon T^{1} $ holds where $ \lfbmv{x}  :^1  T^{1} \in \Gamma^{\circ} $. It is easy to confirm that $\lfbmv{x}  \colon   \llbracket  T^{1}  \rrbracket  \, \in \,  \lvert  \tilde{\Gamma}  \rvert_1 $, and hence $  \lvert  \tilde{\Gamma}  \rvert_0   \lfbsym{,}  \text{\faLock} \lfbsym{,}  \lvert  \tilde{\Gamma}  \rvert_1    \vdash  \lfbmv{x}  \colon   \llbracket  T^{1}  \rrbracket $ holds.

        \item Case $M^{1} =  \lambda \lfbmv{x} ^{ T^{1} }. N^{1} $: $ \Gamma^{\circ}  \lfbsym{,}   \lfbmv{x}  :^1  T^{1}_{{\mathrm{1}}}  \vdash_1 N^{1} \colon T^{1}_{{\mathrm{2}}} $ holds for some $T^{1}_{{\mathrm{1}}}, T^{1}_{{\mathrm{2}}}$ such that $T^{1} = T^{1}_{{\mathrm{1}}}  \rightarrow  T^{1}_{{\mathrm{2}}}$. From $\Gamma^{\circ}  \leadsto  \tilde{\Gamma}$, we can derive $\Gamma^{\circ}  \lfbsym{,}   \lfbmv{x}  :^1  T^{1}_{{\mathrm{1}}}   \leadsto  \tilde{\Gamma}'$ where $\tilde{\Gamma}' = \tilde{\Gamma}  \lfbsym{,}   \lfbmv{x}  :^1  \llbracket  T^{1}_{{\mathrm{1}}}  \rrbracket  $. As a result, we have $  \lvert  \tilde{\Gamma}'  \rvert_0   \lfbsym{,}  \text{\faLock} \lfbsym{,}  \lvert  \tilde{\Gamma}'  \rvert_1    \vdash   \llbracket  N^{1}  \rrbracket_{ \tilde{\Gamma}' }   \colon   \llbracket  T^{1}_{{\mathrm{2}}}  \rrbracket $ from the induction hypothesis. As $ \lvert  \tilde{\Gamma}'  \rvert_0  =  \lvert  \tilde{\Gamma}  \rvert_0 $ and $ \lvert  \tilde{\Gamma}'  \rvert_1  =  \lvert  \tilde{\Gamma}  \rvert_1   \lfbsym{,}  \lfbmv{x}  \colon   \llbracket  T^{1}_{{\mathrm{1}}}  \rrbracket $, we can derive $  \lvert  \tilde{\Gamma}  \rvert_0   \lfbsym{,}  \text{\faLock} \lfbsym{,}  \lvert  \tilde{\Gamma}  \rvert_1    \vdash   \lambda \lfbmv{x} ^{  \llbracket  T^{1}_{{\mathrm{1}}}  \rrbracket  }.  \llbracket  N^{1}  \rrbracket_{ \tilde{\Gamma}' }    \colon   \llbracket  T^{1}_{{\mathrm{2}}}  \rrbracket $.

        \item Case $M^{1} = M^{1}_{{\mathrm{1}}} \, M^{1}_{{\mathrm{2}}}$. $ \Gamma^{\circ} \vdash_1 M^{1}_{{\mathrm{1}}} \colon S^{1}  \rightarrow  T^{1} $ and $ \Gamma^{\circ} \vdash_1 M^{1}_{{\mathrm{2}}} \colon S^{1} $ holds for some $S^{1}$. By the induction hypothesis, we have the following \lamfb judgments below.
            \begin{itemize}
                \item $  \lvert  \tilde{\Gamma}  \rvert_0   \lfbsym{,}  \text{\faLock} \lfbsym{,}  \lvert  \tilde{\Gamma}  \rvert_1    \vdash   \llbracket  M^{1}_{{\mathrm{1}}}  \rrbracket_{ \tilde{\Gamma} }   \colon   \llbracket  S^{1}  \rrbracket   \rightarrow   \llbracket  T^{1}  \rrbracket $
                \item $  \lvert  \tilde{\Gamma}  \rvert_0   \lfbsym{,}  \text{\faLock} \lfbsym{,}  \lvert  \tilde{\Gamma}  \rvert_1    \vdash   \llbracket  M^{1}_{{\mathrm{2}}}  \rrbracket_{ \tilde{\Gamma} }   \colon   \llbracket  S^{1}  \rrbracket $
            \end{itemize}
            By applying them, we obtain $  \lvert  \tilde{\Gamma}  \rvert_0   \lfbsym{,}  \text{\faLock} \lfbsym{,}  \lvert  \tilde{\Gamma}  \rvert_1    \vdash   \llbracket  M^{1}_{{\mathrm{1}}}  \rrbracket_{ \tilde{\Gamma} }  \,  \llbracket  M^{1}_{{\mathrm{2}}}  \rrbracket_{ \tilde{\Gamma} }   \colon   \llbracket  T^{1}  \rrbracket $.
        \item Case $M^{1} =  \lfbkw{unq}   N^{0} $: $ \Gamma^{\circ} \vdash_0 N^{0} \colon  \bigcirc  T^{1}  $ holds. From the induction hypothesis, we have $ \lvert  \tilde{\Gamma}  \rvert_0   \vdash   \llbracket  N^{0}  \rrbracket_{ \tilde{\Gamma} }   \colon  \lfbsym{[}   \mathsf{rg} (  \lvert  \tilde{\Gamma}  \rvert_1  )   \vdash   \llbracket  T^{1}  \rrbracket   \lfbsym{]}$. It is easy to derive $  \lvert  \tilde{\Gamma}  \rvert_0   \lfbsym{,}  \text{\faLock} \lfbsym{,}  \lvert  \tilde{\Gamma}  \rvert_1    \vdash   \mathsf{dom} \lfbsym{(}  \lvert  \tilde{\Gamma}  \rvert_1  \lfbsym{)}   \colon   \mathsf{rg} (  \lvert  \tilde{\Gamma}  \rvert_1  ) $, and hence $ \lvert  \tilde{\Gamma}  \rvert_0   \vdash   \lfbkw{unq} _{ \lfbsym{1} } \lfbsym{(}   \llbracket  N^{0}  \rrbracket_{ \tilde{\Gamma} }   \lfbsym{)} [   \mathsf{dom} \lfbsym{(}  \lvert  \tilde{\Gamma}  \rvert_1  \lfbsym{)}   ]   \colon   \llbracket  T^{1}  \rrbracket $ holds.
    \end{itemize}
\end{proof}